\pdfoutput=1
\newif\ifbiblatex
\biblatextrue  
\biblatexfalse 
\newif\ifappendicesonly%
\appendicesonlyfalse%
\newif\ifsubmission%
\submissionfalse%
\newif\iflongversion%
\longversiontrue%
\longversionfalse%
\newif\ifcutvisible%
\cutvisibletrue%
\cutvisiblefalse%
\ifcutvisible%

\else

\fi

\ifbiblatex
\newcommand{\citet}[1]{\citeauthor{#1}~\cite{#1}}
\else
\newcommand{\citet}[1]{\cite{#1}}
\fi

\PassOptionsToPackage{expansion=false}{microtype}

\documentclass{lmcs}
\usepackage{mathrsfs}
\usepackage{amsmath, amssymb}
\theoremstyle{definition}
\declaretheorem[name=Theorem, numberwithin=section]{theorem}

\usepackage{booktabs}
\usepackage{xcolor}
\usepackage{soul}
\usepackage{tcolorbox}

\definecolor{todohigh}{RGB}{255,100,100}
\definecolor{todomedium}{RGB}{255,175,100}
\definecolor{todolow}{RGB}{100,180,255}

\newcommand{\TODO}[2][medium]{%
    \ifstrequal{#1}{high}{%
        \marginpar{\textcolor{todohigh}{\textbf{TODO:} #2}}%
        \sethlcolor{todohigh!30}\hl{TODO: #2}%
    }{%
        \ifstrequal{#1}{low}{%
            \marginpar{\textcolor{todolow}{\textbf{TODO:} #2}}%
            \sethlcolor{todolow!30}\hl{TODO: #2}%
        }{%
            \marginpar{\textcolor{todomedium}{\textbf{TODO:} #2}}%
            \sethlcolor{todomedium!30}\hl{TODO: #2}%
        }%
    }%
}

\newcommand{\TODOcomment}[2][medium]{%
    \ifstrequal{#1}{high}{%
        \marginpar{\textcolor{todohigh}{\textbf{TODO:} #2}}%
    }{%
        \ifstrequal{#1}{low}{%
            \marginpar{\textcolor{todolow}{\textbf{TODO:} #2}}%
        }{%
            \marginpar{\textcolor{todomedium}{\textbf{TODO:} #2}}%
        }%
    }%
}

\newcommand{\todoitem}[2][medium]{%
    \ifstrequal{#1}{high}{%
        \item \textcolor{todohigh}{\textbf{[HIGH]} #2}
    }{%
        \ifstrequal{#1}{low}{%
            \item \textcolor{todolow}{[LOW] #2}
        }{%
            \item \textcolor{todomedium}{[MEDIUM] #2}
        }%
    }%
}

\usepackage{setup}
\usepackage{placeins}
\usepackage{caption}

\newcommand{\cut}[1]{#1}

\newcommand{\diff}{\!\smallsetminus\!}

\newcommand{\BoolSet}{\{\smalltrue,\smallfalse\}}

\usepackage{typing-rules}
\usepackage{guard-rules}

\begin{document}

\title[Guard Analysis and Safe Erasure Gradual Typing: a Type System for Elixir]{Guard Analysis and Safe Erasure Gradual Typing: \texorpdfstring{\\}{} a Type System for Elixir}

\author[G.~Castagna]{Giuseppe Castagna\lmcsorcid{0000-0003-0951-7535}}[a]
\author[G. ~Duboc]{Guillaume Duboc\lmcsorcid{0000-0002-6795-9545}}[a,b]

\address{Institut de Recherche en Informatique Fondamentale, CNRS, Université Paris Cité, France}
\address{Dashbit, Poland}

\ifappendicesonly
\else
\begin{abstract}
		

We formalize a new type system for Elixir, a dynamically typed functional programming language of growing popularity that runs on the Erlang virtual machine. Our system combines gradual typing with semantic subtyping to enable precise, sound, and practical static type analysis, without requiring any changes to Elixir's compilation pipeline or runtime. Type soundness is ensured by leveraging runtime checks---both implicit, from the Erlang VM, and explicit, via developer-written guards.

Central to our approach are two key innovations: the notion of \emph{strong functions}, which can be assigned precise types even when applied to inputs that may fall outside their intended domain; and a fine-grained analysis of guards that enables accurate type refinement for case expressions and guarded function definitions. While type information is erased before execution and not used by the compiler, our \emph{safe erasure} gradual typing strategy maintains soundness and expressiveness without compromising compatibility or performance. This work lays the theoretical foundation for Elixir's new type system, outlines its integration into recent versions of the language, and demonstrates its effectiveness on large-scale industrial codebases.
\end{abstract}

\maketitle





\section{Introduction}\label{sec:intro}
Elixir is an open-source dynamic functional programming language that
runs on the BEAM, the Erlang Virtual Machine~\cite{beam}. It was designed
by José Valim for building scalable and maintainable applications with a high degree
of concurrency and seamless distribution. Its characteristics have
earned it a surging adoption by hundreds of industrial actors like Adobe, Apple, BBC, Discord, PepsiCo, Spotify, and tens
of thousands of developers.
\cut{

Elixir is, in essence, a minimalist language, with most of its constructs
being syntactic sugar for the language's core expressions: functions and pattern matching.
}
Despite being a dynamically-typed language, there exist tools to perform static
analysis on Elixir programs, such as Dialyzer~\cite{Dialyzer} or Gradualizer~\cite{gradualizer}, and attempts have been made at forming a theoretic
basis to type it, but with clear limitations~\cite{cassola2020gradual}.

To answer the developers' demand for a stricter, more expressive, and informative type system for Elixir, five years ago we started a collaboration with José Valim and the Elixir development team to study how set-theoretic types augmented with the dynamic type of gradual typing could be used to introduce static typing into Elixir. The results of this collaboration are presented in two distinct companion articles that address fundamentally different aspects of the type system design.

The first article~\cite{castagna2023design} describes the design principles of the type system for Elixir and the roadmap to progressively integrate it along the various Elixir releases. The approach assumes that the type system will begin by analyzing currently untyped code to catch simple errors, and will then progressively evolve as Elixir programmers start annotating their code with types, gradually adding more and more of them until they reach a fully statically typed program. The second companion article is the present one, which provides the formal theoretical foundation by formalizing a type system that solves the technical difficulties of fitting the system outlined in~\cite{castagna2023design} into a language like Elixir.

These companion articles are designed to work in tandem while serving different purposes and targeting distinct audiences. The first companion article targets Elixir programmers and serves as a sort of preliminary tutorial on the concepts related to types. This second companion article, in contrast, addresses type theorists and language implementers, providing the rigorous formalization of the type system,  various algorithms used for type-checking, some hints to the techniques used to implement types and the typing rules, as well as the characterization and proofs of the type safety results that underpin the practical approach described in its companion. It constitutes an essential resource for language implementers who seek to develop a type system for a dynamic language---particularly, an existing language with substantial codebases---following the design principles outlined in~\cite{castagna2023design}.

This companion article maintains a more focused scope than the first. As stated in the title, we concentrate specifically on the distinguishing characteristics of the system: how to perform precise analysis of guards used in pattern matching and how to maximize the precision of the gradual type system without modifying Elixir compilation (the ``erasure'' in the title) while preserving type soundness (the ``safe'' in the title). Other aspects of the type system which are discussed in the first companion article---such as parametric polymorphism, records and maps, modules and protocols---are deliberately not covered here: they fall outside the scope of this second work, which focuses on guard analysis and gradual typing (see Section~\ref{sec:related} for records, maps, and parametric polymorphism).

\subsection{Distinctive Features of the Type System}
\label{sec:overview}
The main novel idea of our type system may be \emph{strong functions},
which is formalized in Section~\ref{sec:safe-erasure}. In the theory and application of gradual typing,
there is a clear rift between two kinds of gradually typed languages. Firstly, those that
treat the \elix{dynamic()} type of gradual typing (also known as \texttt{any}, \texttt{mixed}, or \texttt{unknown}, according to the language) as a liability that needs to be \emph{checked};
to preserve their soundness, these languages (e.g., Reticulated Python~\cite{retphyton}) use different strategies to insert type-checks into their
runtime to protect statically typed parts of their code from dynamically typed
ones. But for some languages, such as TypeScript~\cite{bierman2014},
this is not an option as the runtime does not check types by default.
Hence, a second way for gradual systems is to resort to \emph{full erasure}:
the type annotations of a TypeScript program leave no trace in the JavaScript emitted by the
compiler.\footnote{The ``erasure'' refers to the removal of \emph{static} type information: types may still be present in the runtime.} In that case the static analysis cannot leverage the runtime to enforce type properties and, therefore, it produces a coarser type approximation.
This often pushes full erasure type systems towards \emph{unsoundness}, since it forces the implementer to make a crucial design choice: either the type system is sound but provides little information, due to a pervasive use of the dynamic type; or the type system is more informative thanks to a more liberal use of the dynamic type, but this flexibility is unsound and may thus lead to runtime crashes.

However, our situation is more nuanced. Although Elixir is dynamically
typed, it is compiled and then executed on the Erlang VM which, itself, is \emph{type-safe}
through explicit runtime type-checks. Furthermore, programmers can introduce such checks
explicitly by writing them into guards. Hence, we
had the opportunity to quantify how much checking the VM actually does, and
integrate that into our plans for a gradual type system. The concept of strong
functions directly comes from that: these are functions whose input and output types
are entirely or partially \emph{checked} by the VM either because of
checks inserted by the programmer or by standard checks performed by
the VM; hence, it is possible even
when applied in uncertain conditions (when dynamically typed code is involved) to give their
result a static type. This approach, that we call \emph{safe-erasure
    gradual typing}, refers to the fact that, although no checks are inserted
into the language related to the types asserted by the typechecker, this
``type erasure'' is safe, because the type-checker knows which parts of the code
are checked by the VM or the programmer, and which are not. Practically, this means that the
typechecker can be more precise in its analysis, and infer a type without any \elix{dynamic()} components, even in places
where a \elix{dynamic()} type was statically used, because it knows that the VM will check the
type of the value at runtime. 
%
%
A requirement for this approach to work is to be able to extract as much
(necessarily, static) type information from Elixir guards as possible, which is the subject
of the technical analysis developed in Section~\ref{sec:guard-analysis}.
We will provide a finer placement of our work with respect to current literature on gradual typing in Section~\ref{sec:related} on related work.
\iflongversion
Using guards is not inherently more expressive than simply matching,
for instance, on a type enriched with some capture variables.
However, it is a popular way to program
and has its boons: it allows programmers to focus on values (combining conditions
that are not necessarily related to the type of the value) and code their
intentions in a very straightforward way. Furthermore, since guards are rendered
as explicit checks in the VM, their analysis can be leveraged to control the
dynamic parts of the program.
\fi

A key \cut{to the success} of our approach is the use of semantic subtyping, which allows us to use set-theoretic type operators (union,
intersection, difference), but also provides us with a decidable subtyping
relation~\cite{FCB08,CX11}, which appears crucial especially in the analysis of guards. Indeed,
this analysis constantly mixes very precise conditions on types (including
singleton types), and uses intersections, differences, and unions to refine these results. It would not
otherwise be possible to guarantee such a level of precision, and the pattern
matching would end up being grossly approximated. 
\iflongversion  
That is why we have thought
important to show how to extend this framework for custom types, which we show
in Section~\ref{sec:arity} for fixed-arity function types.

Inference is interesting for typing Elixir in order to be able to propose type
suggestions to annotate a program. It is also used in order to give precise type
annotations to anonymous functions, in case the programmer just wants to write
them and pass them with no further annotation. This is described in Section~\ref{sec:inference}.
Finally, to wrap-up our work, we present in Section~\ref{sec:fw-elixir} the
way to make the link between Elixir proper and all the features
previously discussed.

\fi 

\subsection{A Walkthrough of the Work}
\label{sec:walkthrough}
The type system of Elixir  described by~\citet{castagna2023design} is
a gradual polymorphic type system based on
the polymorphic type system of CDuce~\cite{polyduce1,polyduce2}. In this work, we describe
the five main novelties that are missing in the CDuce type system in
order to type Elixir programs, presenting them one by one. These
are the techniques of $(i)$ \emph{strong function typing} and $(ii)$ \emph{propagation of
	the \elix{dynamic()} type} necessary for safe-erasure gradual typing,
both described in Section~\ref{sec:safe-erasure}; the $(iii)$ \emph{guard
	analysis} described in Section~\ref{sec:guard-analysis}; the typing
(and subtyping) of $(iv)$ \emph{multi-arity functions} presented in
Section~\ref{sec:arity}; the $(v)$ \emph{type inference for anonymous functions}
described in Section~\ref{sec:inference}.  In this section, we are
going to present them one after the other by giving some small examples that should
help the reader understand the technical developments described in the
following sections.

\subsubsection{Soundness}
The type system we define here satisfies the following
soundness property
\begingroup
\begin{quote}\em
	If a closed expression is of type $t$, then it either
	diverges, or produces a value of type $t$, or fails on a dynamic
	check either of the virtual machine or inserted by the programmer.
\end{quote}
\endgroup\noindent
The formal statement of this property is more articulated and given in Theorem \ref{thm:gradual-soundness} of Section~\ref{sec:safe-erasure}. The system is gradual since the type syntax includes a
\elix{dynamic()} type used to type expressions whose type is unknown at compile
time.\cut{\footnote{In Elixir, type identifiers end with \felix{()}, e.g., \felix{integer()}, \felix{boolean()}, \felix{none()}, \felix{dynamic()}, etc.}} The soundness guarantee above is typical of the
so-called \emph{sound gradual typing} approaches. These approaches
ensure soundness by using typing derivations to insert some suitable dynamic checks
at compile time. Our system, instead, does not modify
Elixir standard compilation: types are not used for compilation and
are erased after type-checking. Our system is, thus, a
\emph{type-sound (i.e., safe) erasure gradual typing system}, the first we are aware
of. In particular, the compiler
does not insert any dynamic check in the code apart from those
explicitly written by the programmer. Therefore, our system
must ensure soundness by considering only the checks written by the
programmer or performed by the BEAM machine.

Writing a sound gradual type system for Elixir is easy: since every Elixir computation
that does not diverge or fail on a dynamic check returns a value
(no stuck terms, thanks to the BEAM), then a system that types
every expression by \elix{dynamic()} is trivially sound \ldots but hardly useful. Therefore, we
need a system that must fulfill two opposite requirements
\begin{enumerate}
	\item it must use \elix{dynamic()}  as little as possible so as to be useful, and
	\item it must use \elix{dynamic()} enough so as not to hinder the versatility of
	      gradual typing
\end{enumerate}
The first requirement is fulfilled by the typing of strong functions,
the second requirement by the propagation of \elix{dynamic()}. We will demonstrate both of these aspects next.

\subsubsection{Strong functions (Section~\ref{sec:safe-erasure})}\label{sub:strong}
Consider the definition in Elixir of a function \elix{second} that selects
the second projection of its argument (\elix{elem(!\(e,n\)!)} selects the
$(n{+}1)$-th projection of the tuple $e$):
\begin{minted}{elixir}
def second(x), do: elem(x,1)
\end{minted}
If the argument of the function is not a tuple with at least two
elements, then the BEAM raises a runtime exception.
The function definition above is untyped. We can declare its type by preceding it by a \texttt{\tp}-prefixed type declaration, as in
\begin{minted}{elixir}
!\tp! dynamic() -> dynamic()
def second(x), do: elem(x,1)
\end{minted}
This is one of the simplest types we can declare
for \elix{second}, since it essentially states that \elix{second} is a
function, and nothing more: it expects an argument of an unknown type and---unless it diverges or fails---returns a
result of an unknown type. We can give \elix{second} a type slightly more
\emph{precise} than the type above (i.e., a type where an occurrence of \elix{dynamic()} is replaced by a more specific type), that is:
\begin{minted}{elixir}
!\tp! {dynamic(), dynamic(), ..} -> dynamic()
\end{minted}
which states that the argument of a function of this type must be a
tuple with \emph{at least} two elements of unknown type (the trailing
``\elix{..}'' indicates that the tuple may have further
elements). With this declaration, the application of \elix{second} to
an argument not of this type will be statically rejected, thus
statically avoiding the runtime raise by the BEAM. We can give to the
function also a type that is non-gradual, that is, a type in
which \elix{dynamic()} does not occur---we call such types \emph{static types}---, such as:
\begin{minted}{elixir}
!\tp! {term(), integer()} -> integer()
\end{minted}
or more generally:
\begin{minted}{elixir}
!\tp! {term(), integer(), ..} -> integer()
\end{minted}
where \elix{term()} is Elixir's top type that types all values. Both
these type declarations state that \elix{second} is a function that takes (in the first type) a pair or (in the second type) more generally a tuple whose second
element is an integer, and returns
an integer. If this is the type declared for \elix{second}, then the
type deduced for the
application \elix{second({true,42})} is, as
expected, \elix{integer()}. If \elix{dyn} is an expression of type \elix{dynamic()}, then the type
deduced for \elix{second(dyn)} will be \elix{dynamic()}: if \elix{dyn}
evaluates into a tuple with at least two elements, then the
application will return a value that can be of any type, thus we
cannot deduce for it a type more precise than \elix{dynamic()}. This
differs from current sound gradual typing approaches, which would deduce \elix{integer()} for this application, but also insert a runtime
check that verifies that the result is indeed an integer. However, this is not the way an Elixir programmer would have written this function.  If the programmer intention is that \elix{second} had type \elix{{term(), integer(), ..} -> integer()}, then
the programmer would rather write it as follows:\footnote{Elixir's type
	system inherits parametric polymorphism from CDuce (see~\cite{castagna2023design}). So a more
	precise type for \felix{second} would use a type variable \felix{a}
	which in Elixir is quantified postfixedly by a \felix{when}
	clause: \felix{{term(), a, ..} -> a when a: term()}. We do not consider
	polymorphism in this work (see Section~\ref{sec:related}). Note that this
	polymorphic type would not be strong: no runtime check exists for a type
	variable, so no guard can check that the second element of the argument
	has type \felix{a}.}
\begin{minted}{elixir}
!\tp! {term(), integer(), ..} -> integer()
def second_strong(x) when is_integer(elem(x,1)), do: elem(x,1) !\elabel{complexguard}!
\end{minted}
This is defensive programming. The programmer inserts a \emph{guard}
(introduced by the keyword \elix{when}) that checks that the argument is a tuple whose
second element is an integer (the analysis of this kind of complex guards is the subject
of Section~\ref{sec:guard-analysis}). Thanks to this check (which makes up for the one
inserted at compile time by other sound gradual typing approaches) we can safely deduce
that \elix{second_strong(dyn)} has type \elix{integer()}.
A function like \elix{second_strong} is called a \emph{strong function}, because the programmer inserted a dynamic
check that ensures that even if the function is applied to an argument not in its declared domain,
it will always return a result in its declared codomain---i.e., \elix{integer()}---or fail. This
allows the system to deduce for  \elix{second_strong(dyn)} the type \elix{integer()}
instead of \elix{dynamic()},
thus fulfilling our first requirement.
\cut{ }
A function can be strong not only because it was defensively programmed, but also thanks
to the checks performed at runtime by the BEAM, as for:
\begin{minted}{elixir}
!\tp!  {term(), integer(), ..} -> integer()
def inc_second(x), do: elem(x,1) + 1
\end{minted}
which is also strong because the BEAM virtual machine dynamically checks that both
arguments of an addition are of type \elix{integer()}. In other terms since \elix{elem(x,1)} is an operand of an addition, the BEAM performs the check \elix{is_integer(elem(x,1))} that in \elix{second_strong} was explicitly added by the programmer. Therefore, also in this case,
we know that if the function returns a value, then this value is an integer. Consequently, we can safely deduce the type \elix{integer()}
for \elix{inc_second(dyn)} and, thus, for instance, that  the
addition \elix{inc_second(dyn) + second_strong(dyn)} is well typed.

To determine whether a function is strong, we define in Section~\ref{sec:safe-erasure}
an auxiliary type system that checks whether the function, when applied to arguments \emph{not} in its
domain, returns results in its codomain or fails.\vspace*{-2.3mm}

\subsubsection{Propagation of \elix{dynamic()} (Section~\ref{sec:safe-erasure})}\label{sub:prop}

In fact, for both the above applications, \elix{inc_second(dyn)} and
\elix{second_strong(dyn)}, our system deduces a type better than (i.e., a subtype of)
\elix{integer()}: it deduces \elix{integer() and dynamic()}.
This is an \emph{intersection type}, meaning that  its expressions have both
type \elix{integer()} and type \elix{dynamic()}. The system propagates
the type \elix{dynamic()} of the argument \elix{dyn} of the applications into the result.
This is meant to preserve the versatility of the gradual typing that originated the
application, thus fulfilling our second requirement: expressions of this type can be used
wherever an integer is expected, but also wherever any strict subtype of \elix{integer()}
(e.g., natural numbers) is. To see the advantages of this propagation, consider
the following example that we also use to introduce more set-theoretic type connectives:
\begin{minted}{elixir}
def negate(x) when is_integer(x), do: -x        !\elabel{neg1}!
def negate(x) when is_boolean(x), do: not x     !\elabel{neg2}!
\end{minted}
The definition of \elix{negate} is given by multiple clauses tested in
the order in which they appear. When \elix{negate} is applied, the
runtime first checks  whether the
argument is an integer, and if so, it executes the body of the first
clause, returning the opposite of \elix{x}; otherwise, it checks
whether it is a Boolean, and if so returns its negation; in any other
case the application fails. Multi-clause definitions, thus, are equivalent to
(type-)case expressions (and indeed, in Elixir they are compiled as
such). The usual static checks of \emph{redundancy}
and \emph{exhaustiveness} that are standard for case expressions apply
here, too. For instance, if we declare \elix{negate} to be of
type \elix{integer() -> integer()}, then the type system warns that the
second clause of \elix{negate} is redundant\footnote{It is just a warning and not an error, since while the declaration states that the function is to be applied only integer expressions, arguments with a dynamic type are possible}; if we declare for it the
type \elix{term() -> term()} instead, then the function is not
well-typed since the clauses are not exhaustive. To type the function
above without any warning, we can use a union type, denoted by \elix{or}:
\begin{minted}{elixir}
!\tp! integer() or boolean() -> integer() or boolean() !\elabel{arrowunion}!
\end{minted}
which states that \elix{negate} can be applied to either an integer or a Boolean
argument and returns either an integer or a Boolean result. Next, let us
consider the following definition\footnote{\label{fn:arity}Notice the difference between the
type \felix{dynamic(), dynamic() -> integer()} below and the type
\felix{{dynamic(), dynamic()} -> integer()}: the former is a type of a function
that takes two arguments, while the latter is a type of a function that takes
one argument of type \felix{{{dynamic(),dynamic()}}} (i.e., a pair).}%
\begin{minted}{elixir}
!\tp! dynamic(), dynamic() -> integer() !\elabel{lnmulti1}!
def subtract(a, b), do: a + negate(b)   !\elabel{lnmulti2}!
\end{minted}
and see whether it type-checks. The type declaration states
that \elix{subtract} is a function that, when applied to two arguments
of unknown type, returns an integer (or diverges, or fails).  Since
the parameter \elix{b} is declared of type \elix{dynamic()}, then the
system deduces that \elix{negate(b)} is of type \elix{(integer() or
	boolean) and dynamic()} (the \elix{dynamic()} in the type of \elix{b} is
propagated into the type of the result). To fulfill local requirements,
the static type system can assume \elix{dynamic()} to become any type at run-time: following the
terminology by~\citet{castagna2019gradual}, we say that \elix{dynamic()}
can \emph{materialize} into any other type. In the case at issue, addition
expects integer arguments. Therefore, the function body is well typed only
if we can deduce \elix{integer()} for \elix{negate(b)}. This is possible
since the type of this expression is \elix{(integer() or boolean) and
	dynamic()} and the system can materialize the \elix{dynamic()} in there to
\elix{integer()} thus deriving (a type equivalent to) \elix{integer()}.

Notice the key role played in this deduction by the propagation of \elix{dynamic()}:
had the system deduced for \elix{negate(b)} just the type \elix{integer() or boolean()},
then the body would have been rejected  by the  type system since additions expect arguments of type
\elix{integer()}, and not of type \elix{integer() or boolean()}.

A similar problem would happen had we declared \elix{subtract} to be of type
\begin{minted}{elixir}
!\tp! integer(), integer() -> integer() !\elabel{lnsubtracttype}!
\end{minted}
In that case, the type \elix{integer() or boolean() -> integer() or boolean()} is not
good enough for \elix{negate}: since we assume \elix{b} to be of
type \elix{integer()}, then the type deduced for \elix{negate(b)} is
again \elix{(integer() or boolean())} which is not accepted for
additions. The solution is to give \elix{negate} a better type by using the
intersection type
\begin{minted}{elixir}
!\tp! (integer() -> integer()) and (boolean() -> boolean()) !\elabel{intarrow}!
\end{minted}
which is a subtype of the previous type in line \ref{arrowunion}, and states
that \elix{negate} is a function that returns an integer when
applied to an integer and a Boolean when applied to a Boolean. This
type allows the type system to deduce the type \elix{integer()}
for \elix{negate(b)} whenever \elix{b} is an integer. This example shows why it
is important to specify (or infer) precise intersection types for functions. The
inference system we present in Section~\ref{sec:inference} will infer
for an untyped definition of \elix{negate} the intersection of arrows
in line~\ref{intarrow} rather than the less precise arrow with unions
of line~\ref{arrowunion}.

Finally, we want to signal that the latest typing of \elix{negate} given in line \ref{intarrow} does not modify
the propagation of \elix{dynamic()}: the type deduced
for \elix{negate(dyn)} with the second type declaration is again \elix{(integer() or boolean) and
	dynamic()}.
\vspace*{-1mm}

\subsubsection{Guard Analysis (Section~\ref{sec:guard-analysis})}\label{sub:ga}

Until now, the guards employed in our examples primarily involve straightforward type checks on function parameters  (e.g., \elix{is_integer(a)}, \elix{is_boolean(x)}).
The system we investigate for
safe-erasure gradual typing in Section~\ref{sec:safe-erasure} exclusively focuses on these kinds of tests. There is a single exception in our examples with a more intricate guard,
specifically \elix{is_integer(elem(x,1))} used in line~\ref{complexguard}. In Elixir, guards can encompass complex conditions, utilizing equality and order relations, selection operations, and Boolean operators. To illustrate the range of possibilities, consider the following (albeit artificial) definition:
\begin{minted}{elixir}
def test(x) when is_integer(elem(x,1)) or elem(x,0) == :int, do: elem(x,1)      !\elabel{test1}!
def test(x) when is_boolean(elem(x,0)) or elem(x,0) == elem(x,1), do: elem(x,0) !\elabel{test2}!
\end{minted}
The first clause of the  definition of \elix{test} executes when the
argument is a tuple where \emph{either} the second element is an integer \emph{or} the first element is the
atom \elix{:int} (in Elixir, atoms are user-defined constants prefixed
by a colon). The second
clause requires its argument to be a tuple in which the first element
is either equal to the second element or is a Boolean.

To type this kind of definitions, the type system needs  to conduct an analysis
characterizing the set of values for which a guard succeeds.
Section~\ref{sec:guard-analysis} presents an analysis that characterizes this
set in terms of types. In some cases, it is possible  to precisely represent
this set with just one type. For example, the set of values that satisfy the guard
\elix{is_integer(elem(x,1))} of the function \elix{second_strong} in line~\ref{complexguard}, corresponds exactly to
the values of type \elix{{term(), integer(), ..}}. Likewise, the arguments that
satisfy the guard of the first clause of \elix{test} in line~\ref{test1} are
precisely those of the union type \elix{{term(), integer(), ..} or {:int,
			term(), ..}}, where \elix{:int} denotes the singleton type for the value
\elix{:int}.\cut{\footnote{We use \felix{{:int, term(), ..}} rather than
	\felix{{:int, ..}}, since the absence of a second element would make the guard
	fail.}} However, such a precision is not always achievable, as
demonstrated by the guard in the second clause of \elix{test}
(line~\ref{test2}). Since it is impossible to characterize by a type all and
only the tuples where the first two elements are equal, we have to approximate
this set. To represent the set of values that satisfy such guards, we use two
types---an underapproximation and an overapproximation---referred to as the
\emph{surely accepted type} (since it contains only values for which the guard
succeeds) and the \emph{possibly accepted type} (since it contains all the
values that have a chance to satisfy the guard).\footnote{Ideally, the
	surely accepted type would be the largest type containing only values that
	satisfy the guard, and the possibly accepted type the smallest type
	containing all of them. Such types do not always exist, since types are
	closed only under finite unions. For the guard in line~\ref{test2}, every
	type contained in the set of accepted values can be enlarged by a type such
	as \felix{{1, 1, ..}}, and every type containing that set can be shrunk by
	removing a type such as \felix{{1, 2, ..}}.
	Section~\ref{sec:guard-analysis} therefore computes an underapproximation
	and an overapproximation.} For the guard in line~\ref{test2}, the surely accepted type
is \elix{{boolean(), ..}} since all tuples whose first element is a Boolean
satisfy the guard; the possibly accepted type, instead, is \elix{{term(),
			term(), ..} or {boolean()}} since the only values that may satisfy the guard are
those with at least two elements, or those with just one element of type
\elix{boolean()}. When the possibly accepted type and the surely accepted type
coincide, they provide a precise characterization of the guard, as demonstrated
in the two previous examples of guards (lines~\ref{complexguard} and
\ref{test1}).

The type system uses the possibly/surely accepted types to type case-expressions and multi-clause
function definitions. In particular, to type
a clause, the system computes all the values that are \emph{possibly} accepted
by its guard, minus all those that are \emph{surely} accepted by a previous clause, and uses
this set of values to type the clause's body. For example, when declaring
\elix{test} to be of type \elix{{term(), term(), ..} -> term()}, the
system deduces that the argument of the first clause has
type \elix{{term(), integer(), ..} or {:int, term(), ..}}. For the
second clause, the system subtracts  the type above from the possibly accepted type of the
second clause's guard (intersected with the input type, i.e., \elix{{term(), term(), ..}}), yielding for \elix{x} the type
\elix{{not(:int), not(integer()), ..}},
that is, all the tuples with at least two
elements where the first is not \elix{:int}
(\elix{not} $t$ denotes a \emph{negation type}, which types all the values that are not
of type $t$) and the second is not an
integer.

If the difference computed for some clause is empty, then the clause
is redundant and a warning is issued. This happens, for instance, for the second clause of \elix{test},
if we declare for the function \elix{test} the type
\elix{{:int, term(), ..} -> term()}: all arguments will be captured
by the first clause.

If the domain of the function (or, for case expressions, the type of
the matched expression) is contained in the union of the \emph{surely}
accepted types of all the clauses, then the definition is exhaustive. For
instance, this
is the case if we declare for \elix{test} the type \elix{{boolean(), term()}
	-> term()}. If, instead, it is contained only in the union of the \emph{possibly}
accepted types, then the definition \emph{may} not be exhaustive, and a
warning is emitted as for declaring \elix{{term(), term(), ..} or {boolean()}
	-> term()}. In all the other cases, the definition is considered
ill-typed, as for a declared type \elix{tuple() -> term()}
(where \elix{tuple()} is the type of all tuples), since a tuple with a single element that is not a Boolean is an argument in the domain that cannot be handled by any clause.

As a matter of fact, the guard analysis we present in Section~\ref{sec:guard-analysis}
produces for each guard a result that is far more refined than just the
possibly and surely accepted types for the guards. For each guard, the analysis partitions
both the possibly accepted and the surely accepted types into smaller types that are then used by the inference
of Section~\ref{sec:inference} to produce a typing for non-annotated
functions. For instance, for the non type-annotated version of \elix{test} given in
lines~\ref{test1}--\ref{test2} the guard analysis will produce four
different input types that the inference of
Section~\ref{sec:inference} will use to deduce  the
following intersection type for \elix{test}:
\begin{minted}{elixir}
!\tp! ({term(), integer(), ..} -> integer()) and                    !\elabel{testty1}!
  ({:int, term(), ..} -> term()) and                                !\elabel{testty2}!
  ({boolean()} or {boolean(), not(integer()), ..} -> boolean()) and !\elabel{testty3}!
  ({not(boolean() or :int), not(integer()), ..} -> not(boolean() or :int)) !\elabel{testty4}!
\end{minted}
Splitting the domain of \elix{test} as  in the code above is not so difficult since its
guards  use the connective \elix{or} and, as we will see, to compute the split,
the system in Section~\ref{sec:guard-analysis} normalizes guards into Boolean
disjunctions. Notice, however, that the analysis must take into account the
order in which the guards are written. If in line~\ref{test2} we use the guard
\elix{elem(x,0) == elem(x,1) or is_boolean(elem(x,0))}, that is, if we  swap the
order of the operands of the guard, then the arrow type in line~\ref{testty3} is
no longer correct, since the application \elix{test({true})} would fail and,
therefore, the type \elix{{boolean()}} must not be included in the domain of the arrow in line~\ref{testty3}.

\subsubsection{Multi-arity Functions (Section~\ref{sec:arity})}\label{sub:multi}
At lines~\ref{lnmulti1}-\ref{lnmulti2} we defined the
function \elix{subtract} which has two parameters. This arity is reflected
in its type, where its domain consists of two comma-separated
types (see also Footnote~\ref{fn:arity}). All the other functions that were hitherto given as examples are unary. While the
distinction between unary and binary functions may seem trivial to a
programmer, it holds significant implications for the type system. The CDuce type system can only
handle unary functions, and simulates $n$-ary functions as unary
functions on $n$-tuples. But this is not sufficient in Elixir. First,
applying a function to two arguments or to a pair involves different syntaxes,  e.g., \elix{subtract(42,42)}
and \elix{test({42,42})}. Second, a programmer can explicitly test
whether a function $f$ has arity $n$
using \elix{is_function(!$f$!,!$n$!)}. In order to precisely characterize the set of values accepted by such a guard, we need a type system
in which it is possible to express the type of
all functions of a given arity. For instance, we may want to give a
type to:
\begin{minted}{elixir}
def !\negspace! curry(f) !\negspace! when is_function(f,2), do: fn a -> fn b -> f.(a,b) end end         !\elabel{curry1}!
def !\negspace! curry(f) !\negspace! when is_function(f,3), do: fn a -> fn b -> fn c -> f.(a,b,c) !\negspace! end !\negspace! end !\negspace! end !\elabel{curry2}!
\end{minted}
but in current systems with semantic subtyping, we can only express
the type of \emph{all} functions, that is, \elix{none() ->
	term()}.\footnote{A value is of type \felix{!$s$! -> !$t$!} iff it is a
	function that when
	applied to an argument of type $s$, it  returns only results of type
	$t$; thus, every function vacuously satisfies the
	constraint \felix{none() -> term()}, \cut{which only requires its values to be
	functions,} as there is no value of
	type \felix{none()}.}   Simulating, say, binary functions with functions
on pairs does not work since \elix{{none(), none()} -> term()} would \emph{not} be
the type of all binary functions: since the product with the empty set
gives the empty set, this type is equivalent to \elix{none() ->
	term()}, the type of \emph{all} functions. This is the reason why we
introduced the syntax \elix{(t!$_1$!,!$\cdots$!,t!$_n$!) -> t} that
outlines the arity of the functions. Now the type of all binary
functions can be written as \elix{(none(), none()) -> term()},\footnote{In the
	interpretation of function types presented in Section~\ref{sec:arity}. In
	its refinement by~\citet{lanvin2021semantic}, which induces the subtyping
	relation of this work, \felix{(none(), none()) -> term()} is the largest
	arrow type of arity 2, and the type of all binary functions is the base
	type $\function_2$ of Section~\ref{sec:arity}, which is strictly larger
	(Remark~\ref{rem:two-models}).} and we
can declare for the function \elix{curry} the following type (though,
type variables or even a gradual type would be more useful than this type).
\begin{minted}{elixir}
!\tp! (((none(), none()) -> term()) -> none() -> none() -> term()) and
  (((none(), none(), none()) -> term()) -> none() -> none() -> none() -> term()) 
\end{minted}
All this requires modifications, both in the interpretation of types
and in the algorithm that decides subtyping, that we describe in Section~\ref{sec:arity}.

\subsubsection{Inference (Section~\ref{sec:inference})}\label{subsec:inference}

In a couple of examples we highlighted our system's ability to deduce the type of a function even in the absence of explicit type declarations.  For instance, we said that our type system can infer
for \elix{negate} (lines \ref{neg1}--\ref{neg2}) the intersection
type in line~\ref{intarrow}, and for \elix{test}
(lines~\ref{test1}--\ref{test2}) the type in lines~\ref{testty1}--\ref{testty4}.
This kind of inference is different from the inference performed for
parametric polymorphism by languages of the ML family. Instead of generating and solving unification constraints to deduce the type of function parameters, it leverages
the guard analysis of Section~\ref{sec:guard-analysis} to derive the
type of guarded functions: it simply considers the guards of the
different clauses of a function definition as implicit type
declarations for the function parameters, and use them for type
inference.

This kind of inference is used when explicit type declarations are
omitted. This is particularly valuable for anonymous functions of which we
saw a couple examples in the definition of \elix{curry}
(lines~\ref{curry1}--\ref{curry2}) where the body of the two clauses
consists of anonymous functions. The goal of this kind of inference is to avoid imposing an obligation on programmers to explicitly annotate anonymous functions as in:
\begin{minted}{elixir}
!\tp! list(integer()) -> list(integer())
def bump(lst), do: List.map(fn x when is_integer(x) -> x + 1 end, lst) !\elabel{bump2}!
\end{minted}
Here, the guard \elix{is_integer(x)} already provides the necessary information, making explicit annotations superfluous. Additionally, we view the use of an untyped or anonymous function as an implicit application of gradual typing. We have seen in \S\ref{sub:prop}, that
whenever gradual typing was explicitly introduced by an
annotation, the system propagated \elix{dynamic()} in all intermediate results so as to
preserve the versatility of the initial gradual typing. We do the same
here and propagate the (implicit use of) \elix{dynamic()} in the results of the anonymous/untyped functions by intersecting their inferred type with an extra arrow of the form \elix{!$t$! -> dynamic()}, where $t$ is the domain inferred for the function. For example, the type inferred for \elix{negate} will be the type in line~\ref{intarrow} intersected with the type \elix{integer() or boolean() -> dynamic()}, while the intersection type in lines \ref{testty1}--\ref{testty4} inferred for \elix{test} will have an extra arrow \elix{{term(), term(), ..} or {boolean()} -> dynamic()}. Likewise, the type \elix{(integer() -> integer()) and (integer() -> dynamic())} will be given to the anonymous function in the body of \elix{bump} (line~\ref{bump2}); this type is equivalent to the simpler type \elix{integer() -> (integer() and dynamic())}. All these concepts are formalized in Section~\ref{sec:inference}.
\vspace*{-1.5mm}

\subsubsection{Featherweight Elixir (Section~\ref{sec:fw-elixir})}

The theoretical developments presented in this work are not directly formalized
for Elixir, but rather for a $\lambda$-calculus enriched with tuples and
case-expressions on patterns and guards. To make explicit the connection between
this $\lambda$-calculus---called \emph{Core Elixir}---and Elixir itself, we
identify in Section \ref{sec:fw-elixir} a subset of Elixir, called
\emph{Featherweight Elixir}, that covers all the examples presented in this
section. While the correspondence between the expressions of Core Elixir and
FW-Elixir is straightforward, the correspondence between patterns and guards is
more subtle. To simplify the type analysis, the type tests of Core Elixir are more
expressive than those in FW-Elixir, while guards are less expressive. In
particular, Core Elixir guards cannot be negated, which is an essential
requirement for the guard analysis of Section~\ref{sec:guard-analysis} to work.
For instance, while in FW-Elixir it is possible to define a function such as
\begin{minted}{elixir}
def first(x) when not(tuple_size(x)==0), do: elem(x,0)   !\elabel{first}!
\end{minted}
in Core Elixir it is not possible to write a negated guard such as \elix{not(tuple_size(x)==0)}. Therefore, in Section~\ref{sec:fw-elixir} we demonstrate that it is possible to compile FW-Elixir guards into equivalent Core Elixir guards and, consequently, that the type analysis defined on Core Elixir easily transposes to FW-Elixir and Elixir itself.

\subsubsection{Implementation in Elixir (Section~\ref{sec:implem})} 
All the features and algorithms presented here have been
progressively included in Elixir, starting with the v1.17 release (June 2024) of the language~\cite{elixir117}:
the front-end of the Elixir compiler types (multi-arity) functions using safe erasure
gradual typing, with strong functions, \elix{dynamic()} propagation,
and guard analysis. The latter is used to perform inference as
described in \S\ref{subsec:inference}. The Elixir v1.19 implementation (released on October 16, 2025~\cite{elixir119}) covers all language constructs and includes
basic, atom, tuple, list, map, and function types, as well as the typing of \emph{protocols} (akin to Haskell type classes). Elixir v1.20, released on June 3, 2026~\cite{elixir120}, completes the implementation of guard analysis and uses it to infer intersections for multi-clause functions.

In the last technical section of this work, Section~\ref{sec:implem}, we present some aspects of this implementation, such as the data structures used to represent types and a more detailed description of the implementation of the typing rules for function application. We also outline the current roadmap for integrating the type system into Elixir. In addition, we present some performance results of the current type-checker on large well-tested codebases, such as the Elixir package manager and the Phoenix web framework. Finally, we discuss the user feedback we have received so far. The implementation is available in the Elixir's official repository~\cite{elixirrepo}.

\subsubsection{Quantitative and Qualitative Comparison (Sections~\ref{sec:eval}, \ref{sec:related})}
We conclude our presentation with an empirical evaluation and a survey of related work. Section~\ref{sec:eval} reports experiments conducted on a corpus of real-world codebases to assess the precision of our type system from multiple angles: internally, by examining specific aspects such as dead-code detection, guard exactness, and return-type inference; and externally, by comparing our system against existing gradually-typed languages on controlled benchmarks. Section~\ref{sec:related} then situates our contributions within the broader research landscape, discussing how the various components of our system relate to and differ from prior work in gradual typing, subtyping and polymorphism, as well as from industrial gradually-typed systems.

\vspace*{-1mm}

\subsection{Contributions and Limitations}\label{sec:contributions}

Our primary contribution is to establish of the theoretical foundations of the Elixir type system, whose  general principles we outlined in~\cite{castagna2023design}. Notably, we define what we believe to be the first gradual type system with safe erasure. More generally, we believe that the formalization and the algorithms presented here are an essential resource for language implementers who wish to integrate set-theoretic types into dynamically typed languages---especially one with large existing codebases---following the design principles of~\cite{castagna2023design}. As discussed in Section \ref{sec:related} on related work, several other dynamic languages, such as Erlang,  Luau, R, and Python are starting to adopt---to different extents---some of the concepts of gradual set-theoretic types; their implementation will benefit from the results presented here.

The technical contributions can be summarized as follows:
\begin{enumerate}[left=0pt]
    \item \textbf{Gradual Typing:}  Formalization of two new typing techniques:  strong functions and the propagation of \elix{dynamic()}.
    \item \textbf{Typing of Guards and Patterns:}  Definition of a new typing technique for guards and patterns based on the concepts of possibly accepted types and surely accepted types, yielding a safe-erasure gradual type system. Noteworthy of this technique is its ability to partition the accepted types, yielding a high degree of precision for the type analysis. 
    \item \textbf{Semantic Subtyping Extension:} An extension of the theory of semantic subtyping to endow multi-arity function spaces.
    \item \textbf{Type Inference Techniques:} Development of type inference
          techniques for anonymous functions, leveraging pattern and guard
          analysis.
    \item \textbf{Properties:}  Definition of three different characterizations of type safety
          and the proofs of these properties for a language equipped with safe-erasure gradual
          typing: ordinary type soundness for the static fragment without uncertainty, warning-aware soundness for
          the static fragment with potentially unsafe projections, and erasure soundness for the
          gradual fragment.
\end{enumerate}
Among the points above, the most challenging aspect proved to be the characterization of the gradual safety properties mentioned in the final point. This required defining a new relation $v\tc t$ between values and types, which is more permissive than the static typing relation $v:t$. Developing this relation necessitated a fundamental reconceptualization and refinement of strong functions beyond what was originally presented in~\cite{castagna2023design}.

This work has several limitations, some intentional and others genuine constraints. The intentional omissions include the absence of the typing of records and maps (defined in a different work by~\citet{Cas23records} and extended with row polymorphism in~\cite{CP25}) and of parametric polymorphism
(defined by~\citet{polyduce1,polyduce2}), whose addition to our system requires no new ideas, only the re-establishment of some of our results (see Section~\ref{sec:related}). 
%
The genuine limitations are more significant, with two standing out:
a constrained application of type narrowing and the absence of type reconstruction à la ML.

Regarding type narrowing, our system incorporates a simplistic form of
it, allowing specialization in the branches of a case-expression of  the
type of variables occurring in the matched expression under specific
conditions (see Section \ref{sec:guard-analysis}, Remark~\ref{rem:narrowing}). However, it lacks the granularity of refinements achieved by~\citet{tobin2008design} and \citet{castagna2022type,castagna2024inf}. 

Concerning
type reconstruction, although recognized as valuable, in particular
for anonymous functions, it was not explored in this work,
with the example of the \elix{curry} function  in \S\ref{sub:multi} highlighting its potential
significance.

Although~\citet{castagna2024inf} provides a theoretical solution addressing both limitations, its computational cost currently prohibits practical integration into Elixir.

\FloatBarrier

\section{Safe Erasure Gradual Typing}\label{sec:safe-erasure}
In this section we start to define the Core Elixir language, that we will extend in the following sections by adding guards and patterns (Section~\ref{sec:guard-analysis}), multi-arity functions (Section~\ref{sec:arity}), and unannotated functions (Section~\ref{sec:inference}). Its syntax is given in Figure~\ref{fig:lang1}. It is a typed $\lambda$-calculus with
constants, ranged over by $c$ (these include tuples of constants, such as $\mathtt{\{0, 1\}}$, etc.); variables, ranged over by $x$;
$\lambda$-abstractions $\lambda^\IFace{x}.{e}$ annotated by interfaces
(ranged over by $\mathbb I$, and which are finite sets of arrows whose
intersection declares the type of the $\lambda$-abstraction); tuples
$\pc{\ov{e}}$ (we use the overbar to denote sequences, that is, $\ov e$ stands for $e_1,...,e_n$); projections
$\proj{e}{e}$ where the projection index subscripting the $\pi$ symbol is an expression that should evaluate to an integer;
type-case expressions $\textsf{case } e \p{\tau_i \rarr e_i}_{i\in I}$; and,
for illustrating the typing of BEAM-checked operators, the arithmetic sum $+$.

\begin{figure}[t]
	\begin{grammarfig}
		\textbf{Expressions} & e &\bnfeq& c \bnfor x\bnfor \mbox{$\lambda^\IFace{x}.{e}$}
		\bnfor\app{e}{e} \bnfor\tuple{\ov{e}} \bnfor \proj{e}{e}
		\bnfor \texttt{case } e\,\, \ov{\tau \rarr e}
		\bnfor e + e \\
		\textbf{Test types} & \tau &\bnfeq& b \bnfor
		\tuple{\ov{\tau}} \bnfor \OpenTuple{\ov{\tau}} \\
		\textbf{Base types} & b
		&\bnfeq& \intTop \bnfor \boolTop \bnfor \atomTop \bnfor \function \bnfor \tupleTop  \\
		\textbf{Types}      & t
		&\bnfeq& b \bnfor c \bnfor \fun{t}{t} \bnfor \tuple{\ov{t}}
		\bnfor \OpenTuple{\ov{t}}
		\bnfor t \lor t \bnfor \neg \,t \bnfor \dyn\\
                \textbf{Interfaces} & \mathbb{I} &\bnfeq& \{t_i\to t_i'\}_{i=1..n}
	\end{grammarfig}
	\caption{Expressions and Types Syntax}
	\label{fig:lang1}
\end{figure}


\begin{figure}[t]
	\center
	\begin{tabular}{lrcll}
		\RuleDef{reduction:app}{\RedRule{App}}
		 & $(\lambda^\IFace{x}.{e})(v)$
		 & $\kern-0.5em\reduces\kern-0.5em$
		 & $e \Subst{x}{v}$
		 &                                                       \\
		\RuleDef{reduction:proj}{\RedRule{Proj}}
		 & $\proj{i}{\tuple{v_0,\ldotsTwo,v_n}}$
		 & $\kern-0.5em\reduces\kern-0.5em$
		 & $v_i$
		 & if $i \in [ 0 \ldotsTwo n ]$                          \\
		\RuleDef{reduction:match}{\RedRule{Match}}
		 & $\Case{v}$
		 & $\reduces$
		 & $e_j$
		 & if $v \in \tau_j$ and $v \not\in \biglor_{i<j}\tau_i$ \\
		\RuleDef{reduction:plus}{\RedRule{Plus}}
		 & $v + v'$
		 & $\reduces$
		 & $v''$
		 & if $v, v'$ are integers and  $v'' = v + v'$        \\
		\RuleDef{reduction:context}{\RedRule{Context}}
		 & $\ContextWith{e}$
		 & $\kern-0.5em\reduces\kern-0.5em$
		 & $\ContextWith{e'}$
		 & if $e \reduces e'$ without rule \RuleRef{reduction:context}{\SmallRedRule{Context}}     \\[4mm]
		\RuleDef{reduction:app-omega}{\RedRule{App$_\omega$}}
		 & $v(v')$
		 & $\kern-0.5em\reduces\kern-0.5em$
		 & $\omegaApp$
		 & if $v \not = \lambda^\IFace{x}.{e}$                   \\
		\RuleDef{reduction:proj-omega-range}{\RedRule{Proj$_{\omega,\textsc{range}}$}}
		\hspace*{-5pt}
		 & $\proj{v}{\tuple{\ov{v}}}$
		 & $\kern-0.5em\reduces\kern-0.5em$
		 & $\omegaOutOfRange$
		 & if $v \neq i$ for $i = 0 \ldotsTwo |\ov{v}|{-}1$       \\
		\RuleDef{reduction:proj-omega-not-tuple}{\RedRule{Proj$_{\omega,\textsc{notTuple}}$}}\hspace*{-4.2em}
		 & $\proj{v'}{v}$
		 & $\kern-0.5em\reduces\kern-0.5em$
		 & $\omegaProjection$
		 & if $v \not = \tuple{\ov{v}}$                          \\
		\RuleDef{reduction:match-omega}{\RedRule{Match$_\omega$}}
		 & $\Case{v}$
		 & $\kern-0.5em\reduces\kern-0.5em$
		 & $\omegaCase$
		 & if $v \not\in \biglor_{i\in I}\tau_i$                 \\
		\RuleDef{reduction:plus-omega}{\RedRule{Plus$_\omega$}}
		 & $v + v'$
		 & $\kern-0.5em\reduces\kern-0.5em$
		 & $\omegaPlus$
		 & if $v$ or $v'$ not integers                           \\
		\RuleDef{reduction:context-omega}{\RedRule{Context$_\omega$}}
		 & $\ContextWith{e}$
		 & $\kern-0.5em\reduces\kern-0.5em$
		 & $\omega_p$
		 & if $e \reduces \omega_p$ without rule \RuleRef{reduction:context-omega}{\SmallRedRule{Context$_\omega$}}
	\end{tabular}
	\caption{Standard and Failure Reductions}
	\label{fig:reductions}
\end{figure}


The language has a strict weak-reduction semantics defined by the
reduction rules in Figure~\ref{fig:reductions}. The
semantics is defined in terms of values $v$ and
evaluation contexts $\Context$:
\begingroup
\setlength{\abovedisplayskip}{5pt}
\setlength{\belowdisplayskip}{5pt}
\[\hspace*{-2mm}
	\begin{array}{l @{\,\,} l}
		\textbf{Values}  & v   \bnfeq c \mid \mbox{$\lambda^\IFace{x}.{e}$} \mid \tuple{\ov{v}} \\
		\textbf{Contexts} & \Context  \bnfeq \ContextHole \mid \app{\Context}{e}
		\mid \app{v}{\Context} \mid \tuple{\ov{v}, \Context, \ov{e}}
		\mid \proj{\Context}{e}
		\mid \proj{v}{\Context}
		\mid \Case{\Context}
		\mid \Context {+} e
		\mid v {+} \Context
	\end{array}
\]
\endgroup
The reduction rules are standard: \RuleRef{reduction:app}{\textsc{[App]}} is the call-by-value beta-reduction where $e [v/x]$ denotes
the capture-free substitution of $x$ with $v$ in $e$, \RuleRef{reduction:proj}{\textsc{[Proj]}} defines tuple projection, and \RuleRef{reduction:match}{\textsc{[Match]}}  implements a first-match reduction strategy for type-case expressions: the reductum $e_j$ is the first branch of the type-case whose type $\tau_j$  ``matches" the value $v$.
Given a value $v$ and a test type $\tau$, we denote by $v \in \tau$  the fact that $v$
belongs to the set represented by the \emph{test type} $\tau$ (e.g., $0 \in \intTop$ and $\tuple{0, 1} \in \tupleTop$)\footnote{The relation is defined by a straightforward induction: see Figure~\ref{app:inductive-def-test-type} in Appendix~\ref{app:dyn}.},
and we write $v \not\in \tau$ if not (e.g., $0 \not\in \boolTop$).
The failure reductions  correspond to explicit runtime errors raised by the Erlang VM,
and they will be used to make the statement of type safety properties more precise, by explicitly identifying
which failure states are prevented in a typed program. Failures are denoted as a labeled
symbol $\omega_p$, where the label $p$ informs of the type of exception raised (e.g., $\omegaPlus$ for
trying to sum non-integer values).

Types are defined in Figure~\ref{fig:lang1}.
Base types include integers, Booleans, the type of all atoms $\atomTop$, the type of all functions $\function,$ the type of all tuples $\tupleTop$, and the dynamic type `$\dyn$'.
We also have open tuple types: $\tuple{\ov{t},\,\ldotsTwo}$ denotes any tuple
starting with a sequence of elements of types $\ov{t}$.
Types include set-theoretic connectives: the connectives union $\vee$ and negation
$\neg$ are represented in the syntax, while intersection is defined as $t_1 \wedge t_2 = \neg\p{\neg \,t_1 \vee \neg\, t_2}$, and difference is defined as $t_1 \setminus t_2 = \,t_1 \wedge \neg\, t_2$.
The top type $\topp$, the type of all values, is defined as
$\topp = \intTop \lor \atomTop \lor \function \lor \tupleTop$,\footnote{In Elixir, $\boolTop$ is contained
	in $\atomTop$, since the truth values are the
	atoms \felix{:true} and \felix{:false}.}
while the bottom type $\bott$ is defined as $\bott = \neg \topp$.
Note that, since constants are included in types, every value that does not contain $\lambda$-abstractions exists as a singleton type. Types are defined coinductively (for type recursion) and, as customary in semantic subtyping, they are contractive (no infinite unions or negations) and regular (a condition necessary for the decidability of the subtyping relation): see, e.g., \cite{FCB08} for details.

Note that type-case expressions do not check all types, but only ``test types", ranged over by $\tau$. These are types that contain no arrow types, no dynamic type `$\dyn$' (i.e., they are \emph{static types}), and no set-theoretic connectives. Arrow types are excluded because Elixir can only test whether a value is a function or not (and its arity, see Section~\ref{sec:arity}), but it cannot test whether it has a given functional type. Gradual types are not tested because it is not clear what the semantics of such tests should be. Finally, tests with set-theoretic connectives are deferred to Section~\ref{sec:guard-analysis}.

\subsection{Static typing}
 The type discipline makes use of three distinct type systems of increasing permissiveness, whose judgments respectively are denoted by $\Gamma \vds e : t$ (for static typing), $\Gamma \vdg e : t$ (for gradual typing), and $\Gamma \vdw e \tc t$ (for weak typing).
 These judgments should be read as carrying different guarantees. The static judgment is the ordinary one: in the absence of `$\dyn$' and warning rules, it is a static type system ensuring that well-typed programs do not go wrong. The gradual judgment extends this discipline to programs that contain `$\dyn$' while preserving the key result property: if a well-typed expression terminates with a value, then that value ``has the shape''  described by the inferred type (e.g., if the inferred type is $\dyn\to\dyn$, then the result is a function). Since the gradual system has an erasure semantics (programs run unchanged, with no casts or wrappers inserted), an unsafe use of dynamic data is not intercepted at the boundary: it can only surface as one of the runtime errors of the source language or of the BEAM virtual machine. The weak judgment is the auxiliary system that models these checks, and it is used to state formally what ``has the shape'' means. It is intentionally permissive and therefore often infers imprecise dynamic types.

Figure~\ref{fig:static-rules} presents the ``static''---i.e., ``non-gradual''---typing rules for Core Elixir. Most rules are standard, except for small variations we comment on here.
Rules marked by a ``$\omega$'' correspond to cases in which the
type-checker emits a warning, since the use of such rules may jeopardize type safety. More precisely, whenever
 \RuleRef{typing:static:proj-omega}{rule (proj$_\omega$)} or \RuleRef{typing:static:proj-top-omega}{(proj$^\topp_\omega$)} are used, the type-checker warns that the expression may generate
an ``index out of range'' exception. There is no corresponding rule (app$_\omega$):
in the static system, a function application either type-checks by \RuleRef{typing:static:app}{rule (app)}, or it
does not type-check at all. The side condition of rule
\RuleRef{typing:static:app}{(app)} requires the domain $t_1$ to be
\emph{static}, that is, `$\dyn$' does not occur in it:
Section~\ref{sec:gradual-typing} explains why, and shows that the rule still
applies functions whose domain is gradual. 
The \RuleRef{typing:static:lambda}{rule
($\lambda$)} for lambda abstractions checks intersections of function types
by checking that a function satisfies every arrow type given
in its annotation. Its side condition concerns only interfaces in which
`$\dyn$' occurs in a domain: Section~\ref{sec:gradual-typing} defines
separated interfaces and explains why the condition is needed.
\RuleRef{typing:static:proj}{Rule $(\text{proj})$} uses the fact that the index can have a finite union of integer types
(since we have finite union types and integer singleton types), thus it types the projection
with the union of all the fields that can be selected. \RuleRef{typing:static:case}{Rule $(\text{case})$} types
only the branches that are attainable. For a branch $\tau_i\to e_i$ being attainable means that the matched expression $e$ must be able to produce a value of type $\tau_i$ that is not captured by a previous branch. In semantic subtyping a type is a set of values, and the type of an expression overapproximates the set of values the expression may produce. Thus, a branch $\tau_i\to e_i$ is attainable if the set of
values that may be produced by $e$ (i.e., those  in the type $t$ of $e$), intersected with the values in the test type $\tau_i$ (i.e., the set of values captured by the branch),
minus all values captured by  previous branches (i.e.,
all values in $\tau_j$ for $j < i$) is non-empty.
The side condition $t \leq \bigvee_{i\in I} \tau_i$ ensures exhaustiveness since
it checks that all values that $e$ may produce (i.e., those in $t$) are
contained in the set of all types checked by the expression. For instance, we
can encode \texttt{if$\,e\,$then$\,e_1\,$else$\,e_2$} as the type-case on singleton types
\texttt{case$\,e\,($true$\to e_1,\,$false$\to e_2)$} and exhaustiveness will check
that $e$ is of type \boolTop{} (i.e., $\tttrue\vee\ttfalse$). The relation
$\leq$ is the subtyping relation defined for gradual types
in~\cite{lanvin2021semantic} of which we give more details later in this
section.
\begin{remark}\label{rem:attainability}\em
The reader may wonder why the presence of a (statically detected)
non-attainable branch does not
yield a type error. The reason is that the actual attainability of a branch cannot be
decided locally. For instance, to deduce
the intersection type $(\intTop\to\intTop)\wedge(\boolTop\to\boolTop)$
for the function \smash{$\lambda^{\{\intTop\to\intTop,\,\boolTop\to\boolTop\}}x.\mathtt{case}\;x\;(\intTop\,{\to}\, x{+}1,\boolTop\,{\to}\, \neg x)$},
the system types
the case-expression twice: once under the assumption $x{:}\intTop$, making the
$\boolTop$ branch unattainable, and once under the assumption $x{:}\boolTop$, making the
$\intTop$ branch unattainable. Thus, each branch is attainable at some point, though not at the same time. The property of being statically attainable
is, thus, a global property, not expressible in a compositional system. The type-checker will check that every
branch of every case is typed at least once, and emit an ``unused
branch'' warning when this condition is not met. Our soundness results do not depend on this property.
\end{remark}

\begingroup
\addtolength{\jot}{.5em}
\begin{figure}
	\begin{equation*}
		\begin{gathered}
			\TRuleCst
			\quad
			\TRuleVar
			\quad
			\TRuleTuple
			\\[0mm]
			\TRuleLambda
			\\[0mm]
			\TRuleApp
			\\[0mm]
			\TRuleCase
			\\[0mm]
			\TRuleProj
			\\[0mm]
			\TRuleProjOmega
			\quad
			\TRuleProjTopOmega
			\\[0mm]
			\TRuleAdd
			\quad
			\TRuleSub
		\end{gathered}
	\end{equation*}
	\caption{Declarative static type system  \label{fig:static-rules}}
\end{figure}
\endgroup


The type system of Figure~\ref{fig:static-rules} is sufficient to type
\emph{non-gradual} Core Elixir programs, that is, programs where the type
`$\dyn$' never appears in the interface of a function; but, as we show next,
it also types some programs where `$\dyn$' does appear. Safe erasure starts from the cases left open by
this system: programs where dynamic information is present, but where the
compiler must still avoid inserting casts, wrappers, or any other runtime
checks.

\subsection{Gradual typing}\label{sec:gradual-typing}
The static system $\vds$ of Figure~\ref{fig:static-rules} can also type
expressions whose interfaces involve the dynamic type `$\dyn$', provided only
subtyping is required. For instance, since $\intTop\leq\dyn\lor\intTop$, every
function of type $(\dyn\lor\intTop)\rarr\intTop$ also has type
$\intTop\rarr\intTop$, whose domain is static, and
\RuleRef{typing:static:app}{rule~\textsc{(app)}} types its application to an
integer. In general, however, typing expressions that involve
`$\dyn$' requires additional rules (Figure~\ref{fig:gradual-rules}) and an
auxiliary system for inferring \emph{strong function types}
(Figure~\ref{fig:strong-rules}).

Together with the rules in Figure~\ref{fig:static-rules}, these two figures
define the overall architecture of Elixir's type system. The static rules
establish the baseline: the type-checker's non-gradual operating mode. The
rules marked with `$?$' in Figure~\ref{fig:gradual-rules} describe how the
checker transitions to a gradual mode of reasoning when static typing is
insufficient. The rules marked with `$\star$' and the auxiliary judgment
$\Gamma \vdw e \tc t$ of Figure~\ref{fig:strong-rules} explain how existing
runtime checks yield more precise types within that gradual mode.

The key issue in gradual mode appears in function application. Consider a function
of type $\intTop\rarr\intTop$ applied to an argument of type `$\dyn$'. At runtime,
that argument may materialize to an integer, but it may also materialize to a value
of another type. Since we use an erasure semantics (i.e., static typing information
is not used for compilation), the type-checker cannot rely on wrappers to enforce
the function's domain. The conservative rule therefore assigns result type `$\dyn$'
to such an application. This is sound, but it discards the information carried by
the function's codomain.

Strong functions are the mechanism we introduce to recover this information when the program itself justifies it.
A function has a strong arrow type when its body contains enough checks to ensure
that, even if it is called with an argument outside its static domain, every normal
return still belongs to the codomain. The function may fail, for instance because
a pattern does not match or because an operation checked by the BEAM VM receives
an invalid argument; but if it returns, then the returned value has the type
recorded in the strong arrow.

Before commenting these rules, we need to define three relations on types:
subtyping~($\leq$), precision~($\precision$), and consistent
subtyping~($\consist$).

The subtyping relation is the standard relation of semantic subtyping, where `$t_1 \leq t_2$' indicates set-theoretic inclusion of values, meaning that every value of type $t_1$ is also a value of type $t_2$. The precision relation captures the fact that a type $t_1$ can be \emph{materialized} into (i.e., at runtime it may turn out to be) a type $t_2$, meaning that $t_2$ can be obtained from $t_1$ by replacing some of its occurrences of `$\dyn$' by other types: in that sense $t_1$ is less \emph{precise} than $t_2$---written $t_1\precision t_2$--- since it offers a larger range of possible types, hence values, at runtime. Finally, consistent subtyping relates two types that materialize into two other types, in which the former is a subtype of the latter, and captures the relation between the domain of a function and the type of its argument, when gradual typing comes into play.

The theory of semantic subtyping for gradual types makes it possible to define all these relations
just in terms of the semantic subtyping relation on \emph{static types} (i.e., types in which `$\dyn$' does not occur). This property was first noticed, formalized, and proved in Lanvin's PhD thesis~\citet{lanvin2021semantic}. Lanvin first defines a set-theoretic interpretation of gradual types which induces a subtyping relation that is a conservative extension of the subtyping on static types. Then,  Theorem 6.10
in~\cite{lanvin2021semantic} shows that every gradual type $t$ is \emph{equivalent} to
(i.e., it is both a subtype and a supertype of ---noted $\simeq$) the type $\dynInf{t}\lor(\dyn \land \dynSup{t})$,
where $\dynSup{t}$ (resp.\ $\dynInf{t}$) is the \emph{static} type obtained from $t$ by replacing all covariant occurrences
of $\dyn$ in it with $\topp$ (resp. $\bott$), and all contravariant
occurrences of $\dyn$  with
$\bott$ (resp.\ $\topp$): they respectively denote the maximal and minimal materialization of $t$.
\begingroup
\setlength{\abovedisplayskip}{4pt}
\setlength{\belowdisplayskip}{3pt}
\[ \textit{Equivalence}:\hspace*{2cm} t \hspace{0.5cm}\simeq\hspace{0.5cm}  \dynInf{t}\lor(\dyn \land \dynSup{t}) \hspace*{3.5cm}\]
\endgroup
For instance, $\tuple{\dyn, \dyn}$ (the type of 2-tuples whose two elements can
be anything at runtime) is equivalent to $\dyn \land \tuple{\topp, \topp}$ (the type of values that can be
of any subtype of 2-tuples). Likewise, $\dyn\rarr\dyn$ (the type of functions whose domain and codomain can be anything) is equivalent to
$\topp \rarr \bott\lor(\dyn \land \bott\rarr\topp)$, that is any arrow type, since $\bott\rarr\topp$ and $\topp \rarr \bott$ are respectively the largest and smallest arrow types.

As an aside, notice that since $\dynInf t\leq\dynSup t$, this equivalence implies that every gradual type $t$ can be represented as an interval of a pair of static types $(\dynInf{t}, \dynSup{t})$: this is precisely the way we implemented gradual types in Elixir, since their introduction in the 1.18 release of the compiler (see Section~\ref{sec:data-structures} for details).

Using this equivalence,~\citet{lanvin2021semantic} proves that the subtyping relation on gradual types can be expressed in terms  of the subtyping relation on the minimal and maximal materializations of the types (both static types), as follows:
\begingroup
\setlength{\abovedisplayskip}{4pt}
\setlength{\belowdisplayskip}{3pt}
\[ \textit{Subtyping}:\qquad  t_1 \, \leq \, t_2 \hspace{0.5cm}\iff\hspace{0.5cm}
	(\infDyn{t_1} \leq \infDyn{t_2}) \quad\textnormal{and}\quad (\supDyn{t_1} \leq \supDyn{t_2}) \]
\endgroup
A type is \emph{less precise} or \emph{materializes} into another if the latter can be obtained by
replacing some of the former's occurrences of $\dyn$ by other types. Again using the equivalence, we can define the precision relation in terms of minimal and maximal materializations, as follows:
\begingroup
\setlength{\abovedisplayskip}{4pt}
\setlength{\belowdisplayskip}{3pt}
\[ \textit{Precision}:\qquad  t_1 \, \preccurlyeq \, t_2 \hspace{0.5cm}\iff\hspace{0.5cm}
	(\infDyn{t_1} \leq \infDyn{t_2}) \quad\textnormal{and}\quad (\supDyn{t_2} \leq \supDyn{t_1}) \]
\endgroup
Finally, consistent subtyping, which relates two types that materialize
into two other types, in which the former is a subtype of the latter, is characterized as follows:
\begingroup
\setlength{\abovedisplayskip}{4pt}
\setlength{\belowdisplayskip}{3pt}
\[ \textit{Consistent subtyping}:\quad  t_1 \, \consist \, t_2
	\hspace{0.5cm}\iff\hspace{0.5cm} \infDyn{t_1} \leq \supDyn{t_2} \hspace*{5.1cm}\]
\endgroup
For example,
$\p{\dyn{\lor}\,\boolTop \rarr \intTop}$  is a consistent subtype of
$\p{\intTop\,{\rarr} \dyn}$ since by materializing in each type $\dyn$
to $\intTop$ we obtain for the former type a type that is a subtype of the type obtained for the latter.

For the sake of this presentation, one can take the three equivalences above as the definition of the three relations, and refer to the original paper~\cite{lanvin2021semantic} for the proof of their properties (e.g., that
$t_1 \, \consist \, t_2$ if and only if there exists
$t_1'$ and $t_2'$ such that $t_1\preccurlyeq t_1'$,
$t_2\preccurlyeq t_2'$, and $t_1'\leq t_2'$.)

The use of consistent subtyping means that the type-checker has failed to type the expression with the static rules $\vds$, and switched to a \emph{gradual mode} of operation, which is described by the rules in Figure~\ref{fig:gradual-rules}. Figure~\ref{fig:gradual-rules} presents only the rules that are specific to the gradual type system, but for every ``static'' rule in Figure~\ref{fig:static-rules} there exists the same rule in the gradual system where each $\vds$ is changed into $\vdg$. When operating in gradual mode, the type-checker works
differently, since the goal is to check whether operations like application or projection \emph{could} succeed at runtime.

\begin{figure}
	\begin{equation*}
		\begin{gathered}
			\TRuleAppDyn
			\\[0mm]
			\TRuleAppStar
			\\[1mm]
			\TRuleProjDyn
			\\[1mm]
			\TRuleProjStar
			\\[1mm]
			\TRuleCaseStar
			\\[1mm]
			\TRulePlusStar
			\\[1mm]
			\TRuleLambdaDyn
			\\[1mm]
			\TRuleLambdaStar
		\end{gathered}
	\end{equation*}
	\caption{Gradual Type System}
	\label{fig:gradual-rules}
\end{figure}

\emph{For projections}, \RuleRef{typing:gradual:proj-dyn}{rule~($\text{proj}_\dyn$)} checks that types can
materialize into correct ones (i.e., $\intTop$ for the index, and a tuple type
for the tuple), in which case all we can deduce for the projection is the type `$\dyn$'.
However, in \RuleRef{typing:gradual:proj-star}{rule~($\text{proj}_\star$)},
if we know that the expression has type $\tuple{t_0,\ldotsTwo,t_n}$ and that
the index materializes into an integer, then it can be typed with
`$\dyn$` intersected with
the union of the tuple contents: $\dyn \land \bigvee_{i=0..n} t_i$.

\emph{For applications}, \RuleRef{typing:gradual:app-dyn}{rule $(\text{app}_?)$} checks whether the (gradual)
types of the argument and of the function can materialize into two static types
such that the materialized argument type is a subtype of the materialized function's
domain.  When this condition holds, the application is assigned type `$\dyn$'.
While this type is imprecise---essentially indicating the application may yield
results of any type---it represents the only sound deduction in this context, albeit
less precise than what \RuleRef{typing:static:app}{rule (app)} would provide if applicable. For instance, it
is unsound to assign type $\intTop$ to the application of the function
$\lam[\{\intTop\rarr\intTop\}]{x}{x}$ to a dynamic argument. Although this
function returns integers when given integer inputs (justifying type
$\intTop\rarr\intTop$),  it would return a Boolean result if passed a Boolean
argument at runtime.

\RuleRef{typing:static:app}{Rule (app)} must not type such applications
either, which is why its domain must be static. For all types $s$ and $t$ we
have $\dyn\land(s\rarr t)\leq(s\lor\dyn)\rarr t$ and $\dyn\leq s\lor\dyn$.
Without the side condition, \RuleRef{typing:static:app}{rule (app)} would then
give type $\intTop$ to the application of an expression of type
$\dyn\land(\intTop\rarr\intTop)$ to a dynamic argument. An example of such an
expression is $\texttt{case}\,\,e\,\,\p{\intTop\rarr\lam[\{\intTop\rarr\intTop\}]{x}{x}}$
with $e$ of type `$\dyn$': \RuleRef{typing:gradual:case-star}{rule (case$_\star$)}
types it with $\dyn\land(\intTop\rarr\intTop)$, and it evaluates to
$\lam[\{\intTop\rarr\intTop\}]{x}{x}$ when $e$ evaluates to an integer. Its
application to an expression $e'$ of type `$\dyn$' would then have type
$\intTop$, yet when $e$ evaluates to an integer and $e'$ to $\tttrue$, the
application evaluates to $\tttrue$, which is not an integer. In
general, under erasure, when an expression of gradual type $t$ is typed in the
gradual system, the only run-time guarantee on its value is that it belongs to
$\dynSup{t}$. Since
$\dynSup{(t_1\rarr t_2)}=\dynInf{t_1}\rarr\dynSup{t_2}$, a function of type
$t_1\rarr t_2$ is guaranteed to return results in $\dynSup{t_2}$ only on
arguments in $\dynInf{t_1}$, while an argument of type $u$ can be any value of
$\dynSup{u}$: the application is safe when $\dynSup{u}\leq\dynInf{t_1}$. For a
static domain $t_1$ this condition is $u\leq t_1$, which is what
\RuleRef{typing:static:app}{rule (app)} checks. For a gradual domain $t$,
\RuleRef{typing:static:app}{rule (app)} applies by subsumption at the static
domain $\dynInf{t}$, since $t\rarr s\leq\dynInf{t}\rarr s$, to every argument of
type $u$ such that $\dynSup{u}\leq\dynInf{t}$ (equivalently, $u\leq\dynInf{t}$):
for instance, to an integer at the domain $\dyn\lor\intTop$, as at the
beginning of this section. When the function is not a $\lambda$-abstraction,
the condition has a cost (for $\lambda$-abstractions, see
\RuleRef{typing:gradual:lambda-dyn}{rule ($\lambda_\dyn$)} below). For instance,
if $f$ is a variable, or an application, of type $\dyn\rarr\intTop$ and $e$ has
type `$\dyn$', then \RuleRef{typing:static:app}{rule (app)} does not apply to
$\app{f}{e}$, which gets type `$\dyn$' by
\RuleRef{typing:gradual:app-dyn}{rule (app$_\dyn$)}. This cannot be avoided:
since $\dyn\land(\intTop\rarr\intTop)\leq\dyn\rarr\intTop$, $f$ may evaluate at
run time to $\lam[\{\intTop\rarr\intTop\}]{x}{x}$, and $e$ to a Boolean.

\emph{For $\lambda$-abstractions}, \RuleRef{typing:gradual:lambda-dyn}{rule
($\lambda_\dyn$)} has the premises of \RuleRef{typing:static:lambda}{rule
($\lambda$)}, but its conclusion replaces each domain $t_i$ of the interface by
its maximal materialization $\dynSup{t_i}$. Since
$\dynSup{t_i}\rarr s_i\leq t_i\rarr s_i$, the type it deduces is more precise
than the one deduced by \RuleRef{typing:static:lambda}{rule ($\lambda$)}. The
rule is sound because the body has been checked under the assumption
$x\,{:}\,t_i$, and the weak system of Figure~\ref{fig:strong-rules}, which
describes what holds at run time, does not distinguish this assumption from
$x\,{:}\,\dynSup{t_i}$: \RuleRef{typing:weak:var-circ}{rule (var$^\circ$)} types
$x$ with $t_i\land\dyn$, which is equivalent to $\dynSup{t_i}\land\dyn$.
\RuleRef{typing:weak:lambda-circ}{Rule ($\lambda^\circ$)} of the weak system
normalizes domains in the same way. \RuleRef{typing:gradual:lambda-dyn}{Rule
($\lambda_\dyn$)} lets \RuleRef{typing:static:app}{rule (app)} apply
$\lambda$-abstractions whose interface has gradual domains. For instance, in
the gradual system, \RuleRef{typing:static:lambda}{rule ($\lambda$)} gives
$\lam[\{\dyn\rarr\intTop\}]{x}{x{+}1}$ the type $\dyn\rarr\intTop$ (its body
has type $\intTop\land\dyn\leq\intTop$ by
\RuleRef{typing:gradual:plus-star}{rule (plus$_\star$)}), which
\RuleRef{typing:static:app}{rule (app)} can use only for arguments of type
$\bott$. \RuleRef{typing:gradual:lambda-dyn}{Rule ($\lambda_\dyn$)} gives it
the type $\topp\rarr\intTop$, so that \RuleRef{typing:static:app}{rule (app)}
gives type $\intTop$ to its application to any argument, including an argument
of type `$\dyn$'. This is sound, since the body either fails or returns an
integer. \RuleRef{typing:gradual:lambda-dyn}{Rule ($\lambda_\dyn$)} is not in
the static system: there, it would type
$\app{(\lam[\{\dyn\rarr\dyn\}]{w}{w})}{3}$ with `$\dyn$', while the result $3$
does not have type `$\dyn$' in $\vds$ (since $3\not\leq\dyn$), which contradicts
Theorem~\ref{thm:static-soundness}.

\emph{For interfaces}, the side condition of
\RuleRef{typing:static:lambda}{rule ($\lambda$)} in Figure~\ref{fig:static-rules} restricts the interfaces in
which `$\dyn$' occurs in a domain. Given a domain $t$, we call $\dynInf{t}$ its
\emph{static part}, which contains the values that the arrow accepts with
certainty, and $\dynSup{t}\setminus\dynInf{t}$ its \emph{dynamic part}, which
contains the values that the arrow accepts only because `$\dyn$' may
materialize into them. An interface $\{t_i\rarr s_i\}_{i\in I}$ is
\emph{separated} if the dynamic part of each of its domains is disjoint from
the static part of every domain, that is, if
$(\dynSup{t_i}\setminus\dynInf{t_i})\land\dynInf{t_j}\leq\bott$ for all
$i,j\in I$. Every static interface is separated, and so is every interface
with a single arrow. Without this condition,
Theorem~\ref{thm:static-soundness} would fail. Consider the interface
$\{\intTop\rarr\intTop;\,\dyn\rarr\dyn\}$ and the type
$t_\circ=(\intTop\rarr\intTop)\land(\dyn\rarr\dyn)$ that
\RuleRef{typing:static:lambda}{rule ($\lambda$)} would deduce for the identity
function annotated with it. Since $\dynInf{t_\circ}\simeq\topp\rarr\bott$ and
$\dynSup{t_\circ}\simeq\intTop\rarr\intTop$, we have
$t_\circ\leq\intTop\rarr(\dyn\land\intTop)$, so
\RuleRef{typing:static:app}{rule (app)} would give type $\dyn\land\intTop$ to
the application of this function to $3$. But the result $3$ does not have type
$\dyn\land\intTop$ in $\vds$, since $3\not\leq\dynInf{(\dyn\land\intTop)}=\bott$.
The integers belong to the static part of the first domain and to the dynamic
part of the second one: on them, $t_\circ$ guarantees results in
$\intTop\land\dynInf{\dyn}=\bott$, which the premise for the first arrow does
not justify. For the programmer, the condition means that a value accepted with
certainty by one arrow of an interface must not be accepted only dynamically by
another arrow. In practice, an arrow whose domain is `$\dyn$', used as a
fallback for the values not handled by the other arrows, must exclude their
static parts: instead of the interface above, the programmer writes
$\{\intTop\rarr\intTop;\,(\dyn\setminus\intTop)\rarr\dyn\}$, that is, in
Elixir syntax, \elix{(integer() -> integer()) and (dynamic(not integer()) -> dynamic())},
and the application of the identity function to $3$ then has type $\intTop$.
An abstraction whose interface is not separated is not rejected: it is typed
by the gradual system, whose \RuleRef{typing:gradual:lambda-dyn}{rule
($\lambda_\dyn$)} has no side condition, and
Theorem~\ref{thm:gradual-soundness} applies to it.

\begin{remark}[Alternative design]\label{rem:separated}\em
Another design leaves programmers free to write any interface, and lets
\RuleRef{typing:static:lambda}{rule ($\lambda$)} deduce for
$\lam[\IFace]{x}{e}$ with $\IFace=\{t_i\rarr s_i\}_{i\in I}$ the type $\bigwedge_{i\in I}(t_i'\rarr s_i)$, where
$t_i'=\dynInf{t_i}\lor(\dyn\land(\dynSup{t_i}\setminus\bigvee_{j\in I}\dynInf{t_j}))$
removes from the dynamic part of $t_i$ the static parts of all domains. Since
$t_i'\leq t_i$, the premises of the rule still hold for $x\,{:}\,t_i'$, and the
interface $\{t_i'\rarr s_i\}_{i\in I}$ is separated; when $\IFace$ is
separated, $t_i'\simeq t_i$ for every $i$. For the interface
$\{\intTop\rarr\intTop;\,\dyn\rarr\dyn\}$ above, this design deduces
$(\intTop\rarr\intTop)\land((\dyn\setminus\intTop)\rarr\dyn)$, the type of the
separated interface that the programmer would write. The two designs are
equivalent in practice: the alternative changes neither the codomains of the
arrows nor the union of their domains, only the way the domain is split among
the arrows. The choice between the two is a design decision that we leave for when type annotations will be added to the compiler (see Section~\ref{sec:elixir-integration}). 
\end{remark}

Assigning type `$\dyn$' to the application of a function to a dynamic argument, as \RuleRef{typing:gradual:app-dyn}{rule (app$_\dyn$)} does, constitutes a significant limitation of adding `$\dyn$' to a type system: `$\dyn$' tends to escape and contaminate the entire system, making type deductions less precise and, thus, less useful. Generally, there is no way to statically determine a concrete return type when a function is applied to a dynamic value, unless runtime type checks are inserted to enforce the function's signature---the approach taken by current sound gradual typing systems. However, this solution is unavailable to us: a core design principle of~\cite{castagna2023design} requires the type system integration to affect only static type-checking without modifying Elixir's runtime behavior. Fortunately, Elixir's VM already performs type checks both implicitly through strong operations such as `$+$', and explicitly via programmer-defined guards. We leverage both mechanisms in the type system by adding to types \emph{strong arrows} that record this information, and by adding to the typing rules the \RuleRef{typing:gradual:app-star}{rule $(\text{app}_\star)$} which uses this information.
Strong arrows are singled out by a $\star$ symbol:
\begingroup
\setlength{\abovedisplayskip}{3pt}
\setlength{\belowdisplayskip}{3pt}
\[ \textbf{Types} \qquad\qquad t\, \bnfeq\, \cdots \bnfor \strongType{\fun{t}{t}} \]
\endgroup
As for arrows, the domain of a strong arrow is a contravariant position and its
codomain a covariant one, so that
$\dynInf{(\strongType{\fun{t}{s}})}=\strongType{\fun{\dynSup{t}}{\dynInf{s}}}$ and
$\dynSup{(\strongType{\fun{t}{s}})}=\strongType{\fun{\dynInf{t}}{\dynSup{s}}}$.
Since the interpretation of strong arrows (Section~\ref{sub:semantic-strong-arrows} and Appendix~\ref{sec:strong-subtyping}) is
contravariant in the domain and covariant in the codomain, $\dynInf{t}\leq\dynSup{t}$ still
holds for every type $t$.
A strong arrow denotes functions that work as usual on their domain,
but when applied to an argument outside their domain they
either \textit{fail on an explicit runtime type check},
or \emph{return a value of their codomain type}, or diverge.
In Section~\ref{sub:strong} (code in line~\ref{complexguard}) we gave the
example of the function \elix{second_strong}, which in our calculus
can be encoded as
\begingroup
\setlength{\abovedisplayskip}{4pt}
\setlength{\belowdisplayskip}{3pt}
\[
	\lam[\{\pc{\TopType,{\intTop},..}\,\rarr\intTop\}]{x}{\texttt{case}\, {x} \, \texttt{(}\pc{\TopType,\intTop} \rarr \proj{1}{x}\texttt{)}} \]
\endgroup
It is a function that returns an integer if its argument is a tuple whose second element is an integer,
and otherwise ``fails'', that is, it reduces to $\omegaCase$.
The notion of strong arrow is not relevant to a standard static type system,
but it matters in a gradual type system where uncertainty is both a problem (modules
are not annotated, and the type-checker must infer types) and a feature
(some programming idioms are inherently dynamic). The purpose of a strong
arrow is then to guarantee that a function, when applied to a dynamic argument,
will return a value of a specific type. This guarantee is used by the \RuleRef{typing:gradual:app-star}{rule $(\text{app}_\star)$} to deduce for the application that specific type. As all the other rules in Figure~\ref{fig:gradual-rules},
the rule is only used when static type-checking fails. As all the rules marked by a $\star$, this rule has to preserve
the flexibility of the typing, as other functions would then struggle to
type-check a fully static return type. Thus, rules annotated by $\star$
introduce `$\dyn$' in their conclusion, in the form of an intersection.
This property, which was described in \S\ref{sub:prop} of the
introduction, is called \emph{dynamic propagation}. Alongside with
`$\dyn$', a static type is propagated to be used by the type-checker
to detect type incompatibilities. If an argument of type
$(\dyn \land \intTop)$ is used where a Boolean is expected, the
type-checker raises a static type error: \RuleRef{typing:gradual:app-dyn}{rule (app$_\dyn$)}
(and, for a strong function, \RuleRef{typing:gradual:app-star}{rule (app$_\star$)})
accepts this application, since $\dyn\land\intTop$ materializes into $\bott$, but the
implementation of these rules rejects the applications that are typable only through such a
materialization (Section~\ref{sec:implem-app-dyn}). And if an argument of type
$\dyn \land (\intTop \lor \boolTop)$ is used where a Boolean is expected, the type-checker in gradual mode will allow it,
by considering that the argument could become a Boolean at runtime (corresponding to the materialization of $\dyn$ into \boolTop).
The  discussion in Section~\ref{subsec:dynamic-propagation-removal}
revisits this design choice for strong applications, and describes a variant
that drops the dynamic component from their result type.

\newcommand{\leftequation}[2]{
	\noindent\makebox[\textwidth][c]{%
		\makebox[0pt][l]{\text{#1}}
		\hfill
		$#2$
		\hfill
		\makebox[0pt][r]{}
	}
}

The introduction \RuleRef{typing:gradual:lambda-star}{rule for strong arrows $(\lambda_\star)$} requires an auxiliary
type-checking judgment $\Gamma \vdw e \gradual t$ defined in Figure~\ref{fig:strong-rules}.
This type system models the type checks performed by the Elixir runtime.
Indeed, if $e$ is closed and $\varnothing \vdw e \gradual t$, then $e$ either diverges, or fails with a known runtime error,
or evaluates to a value to which this system gives no type disjoint from $t$ (as in Theorem~\ref{thm:gradual-soundness} below; see Corollary~\ref{app:cor:gradual-soundness} in the appendix). Therefore, this system must accept expressions such
as {$\texttt{case}\,\, 42 \,\, \p{\boolTop \rarr 5}$}
which directly reduces to $\omegaCase$. This system is similar to the declarative
one of Figure~\ref{fig:static-rules}, but with additional ``escape hatches'' that
make strong operations permissible no matter the type of their
operands. For instance, since
`$+$' is strong (the Elixir VM checks
at runtime that the operands of an
addition are both integers), then \RuleRef{typing:weak:add-circ}{rule (add$^\circ$)}
only asks that its terms are well-typed. If the addition does not fail, then it returns an integer (typed as $\intTop{\land}\dyn$ for dynamic propagation).
Other such operations are tuple projection, pattern-matching, and also function application.
Using this system, we infer strong function types with \RuleRef{typing:gradual:lambda-star}{rule $(\lambda_\star)$} of the gradual type system $\vdg$;
if a function $\lam[\{t_1\rarr t_2\}]{x}{e}$ has type $t_1 \rarr t_2$, then this
type is strong if, with $x$ of type $\dyn$, the body $e$ can be checked to
have type $t_2$ (actually $t_2{\wedge}\dyn$ for dynamic propagation)
using the rules of Figure~\ref{fig:strong-rules}. The rules explicitly allow expressions that are known to fail at compile time.
As another example, consider \RuleRef{typing:weak:case-circ}{rule~(case$^\circ$)} in Figure~\ref{fig:strong-rules}, which does not have an exhaustiveness condition because an
escaping expression will not return a value but fail at
\begin{wraptable}{r}{6.5cm}
	\vspace{2.5mm}\smash{ \(
		\hspace*{-2mm}\typingrule{\RuleDef{typing:weak:case-bot-circ}{(\text{case}_\bott^\circ)}}
		{\Gamma \vdw e \gradual t \qquad t \land \biglor_i \tau_i \simeq \bott}
		{\Gamma \vdw \Case{e} \gradual \bott}
		\)}
\end{wraptable}
runtime.
Note that, in this rule, if no pattern matches, any type can be chosen for the result; thus the \RuleRef{typing:weak:case-bot-circ}{rule~$(\text{case}_\bott^\circ)$} on the right---which types a case expression that always fails---is admissible. This is why the expression $\texttt{case}\,42\,\p{\boolTop \rarr 5}$ we hinted at above is well-typed with type $\bott$. Finally, for technical reasons we need to prove that every closed expression well-typed with type $t$ in the weak system has the dynamic type ``$t\land\dyn$'' (see Corollary~\ref{cor:dynamic-typing} in the appendix): the \RuleRef{typing:weak:cst-circ}{rule (cst$^\circ$)} handles constants (which are always well-typed), \RuleRef{typing:weak:var-circ}{rule (var$^\circ$)} handles variables, \RuleRef{typing:weak:lambda-circ-dyn}{rule $(\lambda_\dyn^\circ)$} handles $\lambda$-abstractions, and \RuleRef{typing:weak:and-circ}{rule (and$^\circ$)} serves, in practice,
to introduce the intersection with the dynamic type.
\RuleRef{typing:weak:lambda-circ-dyn}{Rule $(\lambda_\dyn^\circ)$} gives every
$\lambda$-abstraction the type $\function\land\dyn$, even when its body is
ill-typed, since at run time a $\lambda$-abstraction is a function whose
interface is not checked. Evaluation can produce such abstractions: when
$\lam[\{\intTop\rarr(\intTop\rarr\intTop)\}]{x}{\lam[\{\intTop\rarr\intTop\}]{y}{x}}$
is applied, by \RuleRef{typing:gradual:app-dyn}{rule (app$_\dyn$)}, to an
expression of type `$\dyn$' that evaluates to $\tttrue$, the result is
$\lam[\{\intTop\rarr\intTop\}]{y}{\tttrue}$, whose body does not have type
$\intTop$.

\begin{figure}
	\begin{equation*}
		\begin{gathered}
			\TRuleCstWeak
			\quad
			\TRuleVarWeak
			\quad\,\,
			\TRuleTupleWeak
			\\[1.5mm]
			\TRuleLambdaWeak
			\quad
			\TRuleLambdaDynWeak
			\\[1mm]
			\TRuleLambdaStarWeak
			\quad
			\TRuleAndWeak
			\\[1mm]
			\TRuleAppWeak
			\quad
			\TRuleAppDynWeak
			\\[1mm]
			\TRuleAppStarWeak
			\\[1mm]
			\TRuleProjWeak
			\\[1mm]
			\TRuleProjIntTopWeak
			\quad
			\TRuleProjToppWeak
			\\[1mm]
			\TRuleCaseWeak
			\\[1mm]
			\TRuleAddWeak
			\quad
			\TRuleSubWeak
		\end{gathered}
	\end{equation*}
	\caption{Weak Type System}
	\label{fig:strong-rules}
\end{figure}

To summarize, we have presented in Figures~\ref{fig:static-rules},~\ref{fig:gradual-rules}, and~\ref{fig:strong-rules}
three declarative systems that work together to model different typing disciplines over programs and their guarantees:
Figure~\ref{fig:static-rules} presents a fully static
discipline, where subtyping is used to check compatibility
between types, and function type annotations are enforced. This is what is expected
of a fully annotated program.
Figure~\ref{fig:gradual-rules} only comes into play when the previous type-checking fails. It uses
a more relaxed relation on types, consistent subtyping, to check programs whose types are gradual
(i.e., where `$\dyn$' occurs in them). Its soundness property is therefore an
erasure property: the program is not instrumented with new checks, so uncertainty
introduced by `$\dyn$' may surface as an existing runtime failure rather than as an
inserted contract violation. What cannot happen is that a well-typed expression
returns a value outside the type inferred for it.
Figure~\ref{fig:strong-rules} serves as an auxiliary system to infer strong function types,
but its elaboration mirrors the semantics of the BEAM VM: every syntactically correct program typechecks
with type $\topp$ (Lemma~\ref{lem:weak-totality} in the appendix), since it either diverges, returns a value (necessarily of type $\topp$), or fails
due to VM checks. However, some programs have more precise types which are passed around like information
to be used later.

With this clear distinction, we formulate three type safety results that depend on whether the unsafe
$\omega$-rules or the gradual rules are used to type expressions.


\begin{Theorem}[Soundness]\label{thm:static-soundness}
	For every expression $e$ and type $t$ such that $\JudgmentTypeS{\varnothing}{e}{t}$
	is derived without using any $\omega$ rules,
	either there exists a value $v : t$ such that $e \seqreduces v$, or $e$ diverges.
\end{Theorem}

\begin{Theorem}[$\omega$-Soundness]\label{thm:soundness}
	For every expression $e$ and type $t$ such that
	$\JudgmentTypeS{\varnothing}{e}{t}$ is
	derived using $\omega$-rules,
	either $e$ diverges, or $e \seqreduces v$ with $v : t$, or $e \seqreduces \omegaOutOfRange$.
\end{Theorem}

\begin{Theorem}[Gradual Soundness]\label{thm:gradual-soundness}
	For every closed expression $e$ and type $t$ such that $\JudgmentTypeG{\varnothing}{e}{t}$,
	either there is a value $v$ such that $e \seqreduces v$ and
	$s\land t\not\leq\bott$ for every type $s$ such that $\varnothing \vdw v \tc s$,
	or there exists $p$ such that $e \seqreduces \omega_p$,
	or $e$ diverges.
\end{Theorem}
The first theorem (Theorem~\ref{thm:static-soundness}) states that, in the absence of gradual typing, if no
warning is emitted and no use of `$\dyn$' is involved, then we are in a classic strict
static typing system. If a
warning is raised but gradual typing is still not used, then the second theorem
(Theorem~\ref{thm:soundness})
states that the only possible runtime failure is the out-of-range selection of
a tuple. If gradual typing is used, then the third theorem
(Theorem~\ref{thm:gradual-soundness}) states that any resulting value
will have the shape described by the inferred type, or else execution fails at
one of the runtime checks already present in the program. This shape is
expressed by the condition on $v$: the weak system of
Figure~\ref{fig:strong-rules} deduces for $v$ no type disjoint from $t$. For instance,
if the type $t$ deduced for a given expression is `$\intTop$' then any
value the expression reduces to is necessarily an integer; if it is
`$\dyn$', then the value can be any  value; if it is
`$\dyn\rarr\dyn$', then the value will be a $\lambda$-abstraction. When no
$\lambda$-abstraction occurs in $v$, that is, when $v$ is a constant (recall
that tuples of constants are constants), the condition is equivalent to
$\varnothing\vdw v\tc t$. In general, it cannot be replaced by
$\varnothing\vdw v\tc t$: Appendix~\ref{app:gradual-safety} gives a closed
expression $e$ such that $\JudgmentTypeG{\varnothing}{e}{t}$ whose value is a
pair of abstractions to which the weak system does not give the type $t$. Note that, while the first two
theorems ensure that well-typed expressions produce only well-typed
values of the same type, the third theorem ensures only that any value
produced by the expression will have the expected shape in the
erased runtime:
because of weak-reduction, a gradually-typed expression can return a
$\lambda$-abstraction whose body
is not well-typed---though, it will be (type) safe in every context. Thus, in particular, if
a closed expression $e$ is of type $t_1\to t_2$, then
Theorem~\ref{thm:gradual-soundness} ensures that it can only return
$\lambda$-abstractions, and these can be applied safely. Let $e$ return $v$,
and let $e'$ be a closed expression such that
$\JudgmentTypeG{\varnothing}{e'}{\dynInf{t_1}}$. Since
$t_1\to t_2\leq\dynInf{t_1}\to t_2$ and $\dynInf{t_1}$ is static, subsumption
and \RuleRef{typing:static:app}{rule (app)} give $\app{e}{e'}$ the type $t_2$,
and, since reduction is deterministic, the
reduction of $\app{e}{e'}$ goes through $\app{v}{e'}$. So, by
Theorem~\ref{thm:gradual-soundness} applied to $\app{e}{e'}$, the application
$\app{v}{e'}$ diverges, fails, or returns a value as described by the theorem
for the type $t_2$. The interfaces of these abstractions need not be subtypes
of $t_1\to t_2$ (see the example before
Corollary~\ref{app:cor:gradual-functions} in the appendix).

The proofs of these theorems are given in Appendix~\ref{app:soundness}.


\section{Guard Analysis}\label{sec:guard-analysis}

Section~\ref{sec:safe-erasure} shows how to handle dynamic
types in languages with explicit type tests.
However, our focus is on exploring languages that use patterns and guards rather than relying solely on type tests.
These are more general, as type cases
can be encoded as guard type-tests on capture variables.

In Elixir, patterns have the form of constants (i.e, basic types' constants or tuples thereof) containing (pairwise distinct) capture variables,
whereas guards consist of complex expressions formed by boolean combinations
(\elix{and}, \elix{or}, \elix{not}) from a limited set of expressions
such as type tests (\elix{is_integer}, \elix{is_atom}, \elix{is_tuple},
etc.), equality tests (\elix{==}, \telix{!=}), comparisons (\elix{<},
\elix{<=}, \elix{>}, \elix{>=}), data selection (\elix{elem},
\elix{hd}, \elix{tl}, \elix{!\mbox{\sl{map}}.\mbox{\it{key}}!}), and size functions
(\elix{tuple_size}, \elix{map_size}, \elix{length}). The complete
syntax for patterns and guards for (Featherweight) Elixir can be found
in Section~\ref{sec:fw-elixir}, Figure~\ref{app:elixir-concrete-syntax}. To define our typed guard analysis, we introduce in Core Elixir a simplified syntax for patterns and guards given in
Figure~\ref{fig:patterns}. The revised Core Elixir syntax for case expressions is
$\caseExpr{}$, where $e$ represents the expression being matched and
$\ov{pg\to e}$ denotes a non-empty list of branches. Each branch consists of
a pair $pg$---formed by a pattern $p$ and a guard $g$---and the corresponding expression $e$ to be
executed. Additionally, we introduce a new expression $\texttt{size}\, e$
to calculate the size of a tuple, applicable in both expressions and guards,
enhancing the complexity of guards to gauge the precision of our analysis.
It is typed by the following rules of the static, gradual, and weak systems, respectively (the virtual machine checks that the argument of \texttt{size} is a tuple, as it checks the arguments of an addition):
\begin{gather*}
	\TRuleSize\qquad\TRuleSizeStar
	\\[1mm]
	\TRuleSizeWeak
\end{gather*}

Guards are constructed using three identifiers: guard atoms $a$, representing
simplified expressions, type tests ($a \isof \tau$), and comparisons ($a = a$,
$a \neq a$, $\GuardLt{a}{a}$), which are combined using boolean operators \texttt{and} and \texttt{or}.
Equality $v = v'$ on values is syntactic equality up to $\alpha$-renaming and
up to the interfaces of the $\lambda$-abstractions that $v$ and $v'$ contain,
and the ordering on values $<_\text{term}$ is derived from the total order on values used in Elixir (see in the appendix, Figure~\ref{fig:elixir-ordering} for a formal definition), which ignores interfaces as well.
As on the BEAM, where type annotations do not exist at run time, a program
cannot observe the interfaces of its functions.
The test types $\tau$ are extended to include
union $\tau \lor \tau$ and negation $\neg\tau$, allowing for more expressive
tests. For instance, it is now possible to verify whether a variable $x$
is either an integer or a tuple ($x \isof \intTop \lor \tupleTop$) or to
ascertain that it is not a tuple ($ x \isof \neg\tupleTop$).

This syntax closely resembles Elixir's concrete syntax; the main difference is
that $\elix{not}$ is absent from guards, which is not restrictive: in Section~\ref{sec:fw-elixir} we
outline a translation that encodes the $\elix{not}$ in the guards of Figure~\ref{fig:patterns}. In what follows, to improve readability,
we will sometimes use (\When{$p$\!\!\!}{\!\!\!$g$}) to denote
the pattern-guard pair $pg$.
\begin{figure}
	\centering
	\hspace*{-2.5pt}
	\begin{minipage}{.529\textwidth}
		\begin{grammarfig}
			\textbf{Exprs} \!\! & e \!\!&\bnfeq&\!\!\ldots \bnfor \caseExpr{} \bnfor
			\texttt{size}\, e\\
			\textbf{Patterns} \!\! & p \!\!&\bnfeq&\!\!c \bnfor x  \bnfor \tuple{\ov{p}}
			\\
			\textbf{Guards} \!\!      & g\!\!
			&\bnfeq&\!\! a \isof\tau\bnfor a = a \bnfor a \,\,\texttt{!=}\,\, a
			\\
			&&\bnfor&\!\! \GuardLt{a}{a} \bnfor\GuardAnd{g}{g} \bnfor \GuardOr{g}{g}
			\\
			\textbf{Atoms} \!\! &a\!\!
			&\bnfeq&\!\! c \bnfor x \bnfor \tuple{\ov{a}} \bnfor\proj{a}{a}\bnfor \sizeTup{a}
			\\
			\textbf{Tests} \!\! & \tau\!\!
			&\bnfeq&\!\! c \bnfor b \bnfor \tuple{\ov{\tau}}
			\bnfor \OpenTuple{\ov{\tau}}
			\\
			&&\bnfor&\!\! \tau \lor \tau \bnfor \neg\, \tau
		\end{grammarfig}
		\small(where variables occur at most once in each pattern) \\
		\captionof{figure}{Patterns and Guards}%
		\label{fig:patterns}
	\end{minipage}\hfill %
	\begin{minipage}{.465\textwidth}
		\[
			\smallemath{
				\begin{array}{rlr}
					v/c     = & \!\!\!\! \{\}  & \text{if } v =c    \\
					v/x     = & \!\!\!\! \{x \mapsto v\}        \\
					\{v_1,\! ..,\! v_n\}/\{p_1,\! ..,\! p_n\} =
						 & \!\!\!\! \bigcup_{i=1}^{n} \sigma_i
						 & \!\!\!\! \text{if } v_i/p_i = \sigma_i   \\
		      & \!\!\!\!   & \text{ for all } i =1..n  \\
					v/p      = & \!\!\!\! \fail    & \text{otherwise} \\[2mm]
					v/(\when{p}{g})  =  & \!\!\!\! \sigma
						& \text{if } v/p = \sigma \text{ and } \\  & \!\!\!\!
					& g\,\sigma \seqreducesG \mathsf{true}  \\
					v/(\when{p}{g})  = & \!\!\!\! \fail    & \text{otherwise}
				\end{array}
			}
		\]
		\vspace{5.7mm}
		\captionof{figure}{Matching}%
		\label{fig:vp-vpg}
	\end{minipage}
\end{figure}

The operational semantics defined in Figure~\ref{fig:reductions} is extended with the rules in
Figure~\ref{fig:pattern-matching-semantics} to account for pattern-matching and the size operator. The updated evaluation
contexts and a new guard evaluation contexts are defined as follows:
\begingroup
\setlength{\abovedisplayskip}{5pt}
\setlength{\belowdisplayskip}{5pt}
\[
	\begin{array}{l@{\hspace{1em}}ll}
		\textbf{Context}       & \Context      & \bnfeq \cdots
		\mid \GuardSize\,\,{\Context}
		\mid \CasePat{\Context}                                \\
		\textbf{Guard Context} & \GuardContext & \bnfeq
		\ContextHole \mid \GuardAnd{\GuardContext}{g}
		\mid \GuardOr{\GuardContext}{g} \mid \GuardIsOf{\GuardContext}{t} \mid \GuardLt{\GuardContext}{a} \mid \GuardLt{v}{\,\GuardContext} \\
		&& \mid \GuardEq{\GuardContext}{a} \mid \GuardEq{v}{\GuardContext}
		\mid \GuardNeq{\GuardContext}{a} \mid \GuardNeq{v}{\,\GuardContext}
	\end{array}
\]
\endgroup
This semantics relies on the functions for matching values to patterns $v/p$, and
values to guarded patterns $v/(pg)$, as defined in Figure~\ref{fig:vp-vpg}.
When $v$ is a value and $p$ a pattern, $v/p$ results in an environment
$\sigma$ that assigns the capture variables in $p$ to corresponding matching values occurring
in $v$. For example, $v/x$ returns the environment where $x$ is bound to
$v$, and $\tuple{v_1,v_2}/\tuple{x, y}$ yields ${x \mapsto v_1, y \mapsto
			v_2}$. Similarly, $v/(pg)$ creates such an environment but also verifies that
the guard $g$ evaluates to $\smalltrue$ within the created environment;
if $v$ does not match $p$ or the guard condition fails, the operation
returns the token $\fail$. For instance, $v/(\When{x}{x \isof \intTop})$
results in ${x \mapsto v}$ if $v$ is an integer and $\fail$ otherwise.

An important aspect of the semantics of pattern matching is that within
a specific branch, if a reduction results in an error (i.e., reduces to
an $\omega$), then the entire guard fails, and that branch is
discarded. For instance, consider the guard $(\GuardOr{\texttt{size}\,x =
		2}{x \isof \intTop})$. If $x$ is not a tuple, taking its size
will lead to an error. Consequently, even if $x$ is an integer, the guard
 fails (rules \RuleRef{reduction:guard-context-omega}{\SmallRedRule{Context$_\omega$}} and \RuleRef{reduction:guard-context-fail}{\SmallRedRule{Context$_{\fail}$}} in Figure \ref{fig:pattern-matching-semantics}: an error in an atom makes the whole guard reduce to $\fail$), instead of evaluating to $\GuardTrue$ as could be expected from the disjunction. Every guard thus evaluates to $\GuardTrue$, to $\GuardFalse$, or to $\fail$.

It should also be noted that, in Elixir, both `\texttt{and}' and `\texttt{or}'
operators exhibit short-circuit behavior. Specifically, if the left-hand
side of an \texttt{and}-guard evaluates to a value other than \smalltrue, the right-hand
side is not evaluated and the conjunction evaluates to \smallfalse\ (rule
\RuleRef{reduction:guard-and-false}{\SmallRedRule{And$_\false$}}); this is why the rule is written with the
condition $v\neq\smalltrue$. If the evaluation of the left-hand side errors,
rule \RuleRef{reduction:guard-context-omega}{\SmallRedRule{Context$_\omega$}} makes the whole guard fail. By contrast, the right-hand side of an
\texttt{or}-guard is evaluated only when the left-hand side evaluates exactly
to \smallfalse\ (rule \RuleRef{reduction:guard-or-false}{\SmallRedRule{Or$_\false$}}); if the left-hand side
evaluates to \smalltrue, the disjunction succeeds immediately (rule
\RuleRef{reduction:guard-or-true}{\SmallRedRule{Or$_\true$}}); if it evaluates to any other value, the
disjunction evaluates to \smallfalse\ immediately (rule
\RuleRef{reduction:guard-or-v}{\SmallRedRule{Or$_v$}}), and if evaluation of the left-hand
side errors, rule \RuleRef{reduction:guard-context-omega}{\SmallRedRule{Context$_\omega$}} makes the whole guard fail.

\begin{figure}
	\center
	\begin{tabular}{lrcll}
		\RuleDef{reduction:guard-case}{\RedRule{Case}}
		 & $\caseExprIndex[v]$                                         
		 & $\reduces$
		 & $e_j\, \sigma$
		 & if $v / (\when{p_j}{g_j}) = \sigma$ and                     \\[0mm]
		 &
		 &
		 &
		 & for all $i\!<\!j \quad v / (\when{p_i}{g_i}) = \fail$ \\[0mm]
		\RuleDef{reduction:guard-case-omega}{\RedRule{Case$_\omega$}}
		 & $\caseExprIndex[v]$
		 & $\kern-0.5em\reduces\kern-0.5em$
		 & $\omegaCase$
		 & if $\match{v}{\when{p_i}{g_i}} = \fail$ for all $i\leq n$   \\
		\RuleDef{reduction:guard-size}{\RedRule{Size}}
		 & $\GuardSize\,{\tuple{v_1,\ldotsTwo, v_n}}$
		 & $\kern-0.5em\reduces\kern-0.5em$
		 & $n$
		 &                                                             \\
		\RuleDef{reduction:guard-size-omega}{\RedRule{Size$_\omega$}}
		 & $\GuardSize\,\,v$
		 & $\kern-0.5em\reduces\kern-0.5em$
		 & $\omegaSize$
		 & if $v \neq \{\ov{v}\}$                                      \\[4mm]
	\end{tabular}
	\begin{minipage}{.47\linewidth}
		\hspace*{-3mm}\begin{tabular}{l rcl l}
			\RuleDef{reduction:guard-and-true}{\RedRule{And$_\true$}}
			 & $\GuardAnd{\GuardTrue}{g}$
			 & $\kern-1em\reducesG\kern-1em$
			 & $g$
			 &                               \\
			\RuleDef{reduction:guard-and-false}{\RedRule{And$_\false$}}
			 & $\GuardAnd{v}{g}$
			 & $\kern-1em\reducesG\kern-1em$
			 & $\GuardFalse$
			 & if $v \not = \GuardTrue$      \\
			\RuleDef{reduction:guard-or-true}{\RedRule{Or$_\true$}}
			 & $\GuardOr{\GuardTrue}{g}$
			 & $\kern-1em\reducesG\kern-1em$
			 & $\GuardTrue$
			 &                               \\
			\RuleDef{reduction:guard-or-false}{\RedRule{Or$_\false$}}
			 & $\GuardOr{\GuardFalse}{g}$
			 & $\kern-1em\reducesG\kern-1em$
			 & $g$
			 &                               \\
			\RuleDef{reduction:guard-or-v}{\RedRule{Or$_v$}}
			 & $\GuardOr{v}{g}$
			 & $\kern-1em\reducesG\kern-1em$
			 & $\GuardFalse$
			 & if $v \notin \boolTop$        \\
			\RuleDef{reduction:guard-eq-true}{\RedRule{Eq$_\true$}}
			 & $\GuardEq{v}{v'}$
			 & $\kern-1em\reducesG\kern-1em$
			 & $\GuardTrue$
			 & \text{if } $v = v'$           \\
			\RuleDef{reduction:guard-eq-false}{\RedRule{Eq$_\false$}}
			 & $\GuardEq{v}{v'}$
			 & $\kern-1em\reducesG\kern-1em$
			 & $\GuardFalse$
			 & \text{else }   \\
			 \RuleDef{reduction:guard-lt-true}{\RedRule{Lt$_\true$}}
			 & $\GuardLt{v}{v'}$
			 & $\kern-1em\reducesG\kern-1em$
			 & $\GuardTrue$
			 & \text{if } $v <_{\text{term}} v'$           \\
			\RuleDef{reduction:guard-lt-false}{\RedRule{Lt$_\false$}}
			 & $\GuardLt{v}{v'}$
			 & $\kern-1em\reducesG\kern-1em$
			 & $\GuardFalse$
			 & \text{else }
		\end{tabular}
	\end{minipage}\hspace*{2pt}
	\begin{minipage}{.5\linewidth}
		\begin{tabular}{l  rcl  l}
			\RuleDef{reduction:guard-neq-true}{\RedRule{NotEq$_\true$}}
			 & $\GuardNeq{v}{v'}$
			 & $\kern-1em\reducesG\kern-1em$
			 & $\GuardTrue$
			 & \text{if } $v \neq v'$               \\
			\RuleDef{reduction:guard-neq-false}{\RedRule{NotEq$_\false$}}
			 & $\GuardNeq{v}{v'}$
			 & $\kern-1em\reducesG\kern-1em$
			 & $\GuardFalse$
			 & \text{else }                         \\
			\RuleDef{reduction:guard-oftype-true}{\RedRule{OfType$_\true$}}
			 & $v \isof t$
			 & $\kern-1em\reducesG\kern-1em$
			 & $\GuardTrue$
			 & \text{if } $v \in t$ \\
			\RuleDef{reduction:guard-oftype-false}{\RedRule{OfType$_\false$}}
			 & $v \isof t$
			 & $\kern-1em\reducesG\kern-1em$
			 & $\GuardFalse$
			 & \text{else}                          \\
			\RuleDef{reduction:guard-context-g}{\RedRule{Context$_g$}}
			 & $\GuardContextWith{g}$
			 & $\kern-1em\reducesG\kern-1em$
			 & $\GuardContextWith{g'}$
			 & if $g \reducesG g'$                  \\
			\RuleDef{reduction:guard-context-a}{\RedRule{Context$_a$}}
			 & $\GuardContextWith{a}$
			 & $\kern-1em\reducesG\kern-1em$
			 & $\GuardContextWith{a'}$
			 & if $a \reduces a'$                  \\
			\RuleDef{reduction:guard-context-omega}{\RedRule{Context$_\omega$}}
			 & $\GuardContextWith{a}$
			 & $\kern-1em\reducesG\kern-1em$
			 & $\fail$
			 & if $a \reduces \omega_p$          \\
			\RuleDef{reduction:guard-context-fail}{\RedRule{Context$_{\fail}$}}
			 & $\GuardContextWith{\fail}$
			 & $\kern-1em\reducesG\kern-1em$
			 & $\fail$
			 & if $\GuardContext\neq\ContextHole$
		\end{tabular}
	\end{minipage}
	\caption{Pattern Matching and Guard Reductions}%
	\label{fig:pattern-matching-semantics}
\end{figure}

\subsection{Typing Pattern Matching: Pattern Accepted Types}\label{sec:typing-pattern-matching}

To type the expression $\caseExprI{}$ we aim to precisely type each branch's expression $e_i$
by analyzing the set of values for which  the pattern-guard
pair $p_ig_i$ succeeds, that is, $\{v \in \textbf{Values}\mid \match{v}{p_ig_i}\not=\fail\}$.
The best typing results are obtained when such a set coincides with the set of values of a given type. However, because of the presence of guards, not every such set of values corresponds perfectly to a type. For
example, we have seen in \S\ref{sub:ga} the guard in
line~\ref{test2}---expressed in our formalism by
the pattern-guard pair
$(\When{x}{ \GuardOr {\texttt{(}\proj{0}{x} \isof \boolTop\texttt{)}}
		{\texttt{(}\proj{0}{x} = \proj{1}{x}\texttt{)}} })$---, which
matches all tuples where either the first element is a Boolean, or the first two elements
are identical: yet there does not exist a specific type that denotes exactly the set of all such tuples.
To address this, given a pattern-guard pair $pg$, we approximate the set of values that match $pg$ by defining two types termed as the
\emph{possibly accepted type} $\pAccType{pg}$
(in our example it is $\OpenTuple{\boolTop} \lor \OpenTuple{\topp,\topp}$
since any value accepted by the pattern will belong to this type)
and the \emph{surely accepted type}
$\acceptedType{pg}$ (here it is $\OpenTuple{\boolTop}$ since all tuples starting
with a Boolean are surely accepted). Through these types, we derive an approximating type $t_i$
encompassing all values reaching $e_i$.
When the matched expression is of type $t$, $t_i$ is formulated as
$(t\wedge\pAccType{p_i g_i})\!\smallsetminus\!\bigvee_{j{<}i}\acceptedType{p_j g_j}$.
In other words, the values in $t_i$ are those that may be produced by $e$
(i.e., those in $t$), and may be captured by $p_i g_i$ (i.e., those $\pAccType{p_i g_i}$)
and which are not surely captured by a preceding branch (i.e., minus those in
$\acceptedType{p_j g_j}$ for all $j<i$).
This type $t_i$ can be utilized to generate the type environment under
which $e_i$ is typed. This environment, denoted as $\TypeEnv{t_i}{p_i}$,
assigns the deducible type of each capture variable of the pattern $p_i$,
assuming the pattern matches a value in $t_i$. The definition of this
environment is a standard concept in semantic subtyping and is detailed in
Appendix~\ref{app:guardanalysis}, Figure~\ref{app:fig:typing-environments}.
A first coarse approximation of the typing discipline for pattern-matching expressions is given by the following
rule (given here only for presentation purposes but not included in the system):\\[2mm]
\centerline{\(
	\typingruleExtra{\RuleDef{typing:case:case-omega-coarse}{\text{(case$_\omega${\footnotesize(coarse)})}}}
	{\Gamma \vd e : t
		\quad
		(\forall i{\leq}n)
		\,;(t_{i} \nleq \bott\,\Rightarrow\,\Gamma, \TypeEnv{t_{i}}{p_i} \vd e_i : s)}
	{\Gamma \vd \caseExprI{} : s}
	{\smallemath{\begin{array}{l}t_i=(t\wedge\pAccType{p_ig_i})\!\smallsetminus\!\bigvee_{j{<}i}\acceptedType{p_jg_j}\\t \leq \biglor_{i\leq n} \pAccType{p_ig_i}{}\end{array}}}
	\)}\\[2mm]
\noindent
The rule types a case-expression of $n$ branches. For the $i$-th
branch with pattern $p_i$ and guard $g_i$, it computes $t_i$ and
produces the type environment $\TypeEnv{t_i}{p_i}$. This environment
is used to type $e_i$ only if $t_{i} \nleq \bott$ (i.e., $t_i$ is not empty), thus ensuring that some values may
reach the branch and checking for case redundancy. The side condition
$t \leq \biglor_{i\leq n} \pAccType{p_ig_i}{}$ ensures that every value of type
$t$ (i.e., every possible result of $e$) may \emph{potentially} be captured by some branch, addressing
exhaustiveness. The rule is labeled with $\omega$, indicating the emission of a warning, as the union $\biglor_{i\leq
		n} \pAccType{p_ig_i}{}$ might over-approximate the set
of captured values\cut{,
risking run-time exhaustiveness failures}. However, if $t\leq\biglor_{i\leq
		n} \acceptedType{p_ig_i}{}$ holds, then there's no warning (cf.\ \RuleRef{typing:case:case}{rule (case)} later in this section), since all values in $\biglor_{i\leq
		n} \acceptedType{p_ig_i}{}$ are surely captured by some pattern-guard pair, and so are those in $t$.

The typing rule for case expressions is actually more complicated than
the one above, since it performs a finer-grained analysis
of Elixir guards that is also used to compute their surely/possibly accepted types.
Let us look at it in detail:
\begingroup
\setlength{\abovedisplayskip}{5pt}
\setlength{\belowdisplayskip}{5pt}
\[\begin{array}{l}
	\TRuleCasePgOmegaWeak
	\\[4mm]
	\hspace*{-1mm}\textit{where }
	\Gamma\,;t \vdash ({p_i}{g_i})_{i\leq n}
	\leadsto (t_{ij},\BoolB_{ij})_{i\leq n, j\leq m_i}
	\end{array}\]
\endgroup
\noindent
In contrast to the prior rule, the system now computes for each $\when{p_i}{g_i}$ pair, a list of $m_i$ types
$t_{i1},...,t_{i{m_i}}$  which
together contain every value of type $t$ that reaches $p_ig_i$ and is accepted by it (Lemma~\ref{lem:guard-necessary} below). The rule types each $e_i$ expression $m_i$-times, each with a distinct environment $\TypeEnv{t_{i j}}{p_i}$.
The $t_{ij}$ at issue are derived from an auxiliary deduction system:
$\Gamma;t \vdash ({p_i}{g_i})_{i\leq n}\leadsto
	(t_{ij},\BoolB_{ij})_{i\leq n, j\leq m_i}$. This system, detailed in
the rest of this section, inspects each $g_i$ for OR-clauses,
and for each such clause it generates a pair $(t,\BoolB)$ that indicate the clause's type $t$ and a Boolean flag  $\BoolB$ indicating its
exactness. For example, the guard
$(\GuardOr {\proj{0}{x} \isof \boolTop} {\proj{0}{x} =\proj{1}{x}})$
of our example is formed by two OR-clauses and the analysis produces two pairs $(\OpenTuple{\boolTop},\smalltrue)$ and
$(\OpenTuple{\neg\boolTop,\topp},\smallfalse)$: the first flag is \smalltrue\ since the type for the first clause is exact ($\OpenTuple{\boolTop}$ exactly contains all the values that match $\proj{0}{x} \isof \boolTop$); the second clause is analyzed only for the tuples that the first one does not surely accept, those whose first element is not a Boolean, and its flag is \smallfalse\ since the type is an approximation ($\OpenTuple{\neg\boolTop,\topp}$ strictly contains the values of this kind that match $\proj{0}{x} =\proj{1}{x}$). Likewise, the guard $(\GuardOr {x \isof \intTop}
	{\proj{0}{x} \isof \intTop})$  will only produce exact types, namely $(\intTop,\smalltrue)$ and
$(\OpenTuple{\intTop}, \smalltrue)$.
The analysis processes guards in Elixir's evaluation order and accounts for potential clause failures.
The analysis of guard $g_i$ refines the types of the capture variables of $p_i$. A guard uses only the capture variables of its pattern: a guard that tests a variable $z$ bound outside the pattern is compiled by adding $z$ to the matched expression and to every pattern, that is, $\texttt{case}\,\tuple{e,z}\,(\tuple{p_i,z}\,g_i\rightarrow e_i)_{i\leq n}$ (where the $p_i$ do not bind $z$). Otherwise the refinement of $z$ would be lost when the accepted types are computed, and a clause could be declared exact although it depends on the value of $z$.

Given $\Gamma;t \vdash ({p_i}{g_i})_{i\leq n}\leadsto
	(t_{ij},\BoolB_{ij})_{i\leq n, j\leq m_i}$, the possibly accepted
type for $p_ig_i$ is the union of all $t_{ij}$'s, while the surely
accepted type for $p_ig_i$ is the union of all $t_{ij}$'s for which
$\BoolB_{ij}$ is true. Thus, we have $\pAccType{p_ig_i}=\bigvee_{j\leq
		m_i} t_{ij}$ and $\acceptedType{p_ig_i}=\bigvee_{\{j\leq
		m_i\,|\,\BoolB_{ij}\}} t_{ij}$. In our example, if $g$ is the guard
$(\GuardOr {\proj{0}{x} \isof \intTop} {\proj{0}{x} =\proj{1}{x}})$, then
$\pAccType{\when{x}{g}}= \OpenTuple{\intTop}\lor\OpenTuple{\topp,\topp}$
and $\acceptedType{\when{x}{g}}= \OpenTuple{\intTop}$. Likewise, if $g$
is the guard $(\GuardOr {x \isof \intTop} {\proj{0}{x} \isof \intTop})$,
then the possibly and surely accepted types of $xg$ are the same, both
being $\intTop\lor\OpenTuple{\intTop}$, indicating that the approximation
	is exact. In this second case we can use a rule identical to the \RuleRef{typing:case:case-omega}{\text{(case$_\omega$)} rule} but using a stricter side condition that ensures exhaustiveness and, thus, that does not emit any warning:
\begingroup
\setlength{\abovedisplayskip}{3pt}
\setlength{\belowdisplayskip}{5pt}
\[\begin{array}{l}
		\TRuleCasePg
		\\[4mm]
		\hspace*{-1mm}\textit{where }
		\Gamma\,;t \vdash ({p_i}{g_i})_{i\leq n}
		\leadsto (t_{ij},\BoolB_{ij})_{i\leq n, j\leq m_i}
			\!\text{ and }
			\sAccType{p_i g_i} = \bigvee_{\{j\leq m_i\,\mid\, \BoolB_{i j}\}} t_{i j}
 \end{array}\]
\endgroup
\noindent
The displayed equation for $\sAccType{p_i g_i}$ is the operational use of the
	``surely accepted'' component computed by the analysis: the \RuleRef{typing:case:case}{(case)} side
condition selects exactly those generated pieces whose Boolean flag is
\smalltrue, since only those pieces are sufficient to guarantee that the
corresponding pattern-guard pair will accept the scrutinee.
When using this analysis, type safety depends
on the side conditions used.
\RuleRef{typing:case:case}{Rule (case)} with $t\leq\biglor_{i\leq n} \acceptedType{p_ig_i}{}$
is safe for exhaustiveness, ensuring the same static guarantee as Theorem~\ref{thm:static-soundness}, as stated by the following updated version of the theorem:
\begin{restatable}[Static Soundness]{theorem}{guardStaticSoundness}\label{thm:guard-soundness}
		If $\JudgmentTypeS{\varnothing}{e}{t}$ is derived without using $\omega$-rules and with the \RuleRef{typing:case:case}{(case) rule} with condition $t\leq\biglor_{i\leq n} \acceptedType{p_ig_i}{}$,
	then either $e \seqreduces v$ with $v: t$, or $e$ diverges.
\end{restatable}
\noindent
\RuleRef{typing:case:case-omega}{Rule (case$_\omega$)} will be used whenever the \RuleRef{typing:case:case}{rule (case)} fails---meaning that our guard analysis is not precise---raising a warning and adding $\omegaCase$
to the set of explicit runtime errors in Theorem~\ref{thm:soundness}, which becomes:
\begin{restatable}[$\omega$-Soundness]{theorem}{guardOmegaSoundness}\label{thm:guard-omega-soundness}
	If $\JudgmentTypeS{\varnothing}{e}{t}$
	is derived using $\omega$-rules and the
		\RuleRef{typing:case:case}{(case)} and \RuleRef{typing:case:case-omega}{(case$_\omega$)} rules,
	then either $e \seqreduces v$ with $v: t$, or $e$ diverges, or $e \seqreduces \omegaCase$ or $e \seqreduces \omegaOutOfRange$.
\end{restatable}

To complete the type system we need to add a rule for pattern matching expression in the gradual system:
\[\begin{array}{l}
			\TRuleCasePgStar
			\\[5mm]
			\textit{where }
			\Gamma\,;t \vdash ({p_i}{g_i})_{i\leq n}
			\leadsto (t_{ij},\BoolB_{ij})_{i\leq n, j\leq m_i}
			\!\!\text{ and }\!
			\pAccType{p_i g_i} \!=\! \bigvee_{j\leq m_i} t_{i j}
		\end{array}\]
The gradual \RuleRef{typing:case:case-star}{rule (case$_\star$)} simply checks that the scrutinee may be covered
by some patterns, propagates the dynamic type, and does not modify the type safety of Theorem~\ref{thm:gradual-soundness}.
The weak system of Section~\ref{sec:safe-erasure} is extended likewise, by
rule \RuleRef{typing:case:case-circ}{(case$^\circ$)} of
Appendix~\ref{app:guard-safety}, which has the premises of
rule~(case$_\star$) in the weak system and no side condition.

\begin{restatable}[Gradual Soundness]{theorem}{guardGradualSoundness}\label{app:thm:gradual-guard-soundness}
	If $e$ is closed and $\JudgmentTypeG{\varnothing}{e}{t}$ using rules
		\RuleRef{typing:case:case}{(case)}, \RuleRef{typing:case:case-omega}{(case$_\omega$)}, and \RuleRef{typing:case:case-star}{(case$_\star$)},
	then either there is a value $v$ such that $e \seqreduces v$ and
	$s\land t\not\leq\bott$ for every type $s$ such that $\varnothing \vdw v \tc s$,
	or there exists $p$ such that $e \seqreduces \omega_p$,
	or $e$ diverges.
\end{restatable}


\begin{remark}[Naive Type Narrowing]\label{rem:narrowing}
	{\em
	Let $x$ be a variable of type $\topp$ and consider the following code snippet: }\\[-1mm]
	\centerline{$\texttt{case}\,x\; (\OpenTuple{y,z}\texttt{\,when\,} y\isof\intTop \to \texttt{size}(x)+y,...)$}\smallskip

	\noindent{\em The current typing rules would reject this code since the \texttt{size} operator expects a tuple, but $x$ is of type $\topp$. This despite the fact that $\texttt{size}(x)$ will only be called if the pattern $\OpenTuple{y,z}$, succeeds, implying $x$ being a tuple.  To solve this issue we need to refine the type of the variables occurring in the matched expression. Given an expression $e$ we define its skeleton $\texttt{sk(}e\texttt{)}$ as:}
	\begingroup
	\setlength{\abovedisplayskip}{0pt}
	\setlength{\belowdisplayskip}{2pt}
	\begin{align*}
		\mathtt{sk}\,\texttt{(}x\texttt{)}
		 & = x
		 &                                                                                                                                    \\[-2pt]%
		\mathtt{sk} \, \texttt{(}  \texttt{\{}  e_{{\mathrm{1}}}  {,} \, \, ... \, {,} \,  e_{{n}}  \texttt{\}} \texttt{)}
		 & = \texttt{\{}\mathtt{sk}\,\texttt{(} e_{{\mathrm{1}}}\texttt{)}{,}\,\,...\,{,}\,\mathtt{sk}\,\texttt{(}e_{n}\texttt{)}\texttt{\} } \\[-2pt]
		\mathtt{sk}\,\texttt{(}e\texttt{)}
		 & = \topp
		 & \text{for any other expression}
	\end{align*}
	\endgroup
	\em The skeleton of an expression is thus a pattern that matches the
	structure and variables of that expression while leaving out
	any functional parts. For example, the skeleton of a tuple formed by a variable and an application
	like \emph{$\tuple{x, e  \texttt{(}  e_{{\mathrm{1}}}  {,} \, \, ... \, {,} \,  e_{{n}}  \texttt{)}}$} is $ \tuple{x,\topp} $ which
	is the pattern that captures in $x$ the first projection matches any value in the second projection.\par
		In practice, if $e$ is being matched, its skeleton $\texttt{sk}(e)$
		is added to the patterns of all branches. Thus, any type narrowing that occurs in their guard analysis
		is also applied to the variables of $e$.
		This is possible by adding intersections---noted \texttt{\&}---to patterns:
		a value matches the intersection pattern $p_1 \!\mathrel{\texttt{\&}} p_2$ iff it matches both $p_1$ and $p_2$; now every
		pattern $p_i$ in  $\caseExprI{}$ can be compiled as $\texttt{sk}(e)  \!\mathrel{\texttt{\&}} p_i$, after renaming the capture variables of $p_i$ apart from the variables of $e$, so that no variable occurs on both sides of an intersection pattern.
		Then, we handle dependencies between variables (e.g., if pattern $x  \!\mathrel{\texttt{\&}} \OpenTuple{y, z}$
		is followed by a guard $y \isof \intTop$, then the
		type of $x$ is refined to $\OpenTuple{\intTop, \topp}$),
		using an environment update $\Gamma[x \refine t]_p$
		(see next section)
		that narrows the type of $x$ in $\Gamma$ to $t$, and  uses pattern $p$
		to properly refine the type of other variables in $\Gamma$ that depend on $x$.
		The pattern $p$ is therefore part of the guard analysis judgment. In the
		next section we write this judgment as $\Gamma ; p \vdash g \mapsto
		\mathcal{R}$, for a guard result $\mathcal{R}$ as defined below. The
		pattern component is threaded unchanged through rules that do not inspect
		it, and is used explicitly by rules such as \RuleRef{guard:analysis:var}{$[\textsc{var}]$} and
		\RuleRef{guard:analysis:or}{$[\textsc{or}]$}. This makes the dependency local in the rules, explains
		the notation $\Gamma[x \refine \tau]_p$ when it first appears, and is the
		form used in the soundness proof (see Lemma
		\ref{lem:safe-success-environment}) and in the Elixir type checker.
\end{remark}


\subsection{Guard Analysis System.}\label{sub:guard-analysis-system}

In this section we illustrate how to derive the judgment $\Gamma;t \vdash ({p_i}{g_i})_{i\leq n}\leadsto (t_{ij},\BoolB_{ij})_{i\leq n, j\leq m_i}$ that is central to typing case expressions and to infer function types from their guards. Our guard analysis derives two main judgments:

\begin{itemize}
\item \textit{Guard Analysis Judgment}: $\Gamma ; p \vd g \mapsto \mathcal{R}$ represents the analysis of a single guard $g$ under type environment $\Gamma$ and current pattern $p$, producing a result $\mathcal{R}$ (see below) that describes the conditions under which the guard succeeds or fails.

\item \textit{Pattern-Guard Sequence Judgment}: $\Gamma;t \vdash ({p_i}{g_i})_{i\leq n}\leadsto (t_{ij},\BoolB_{ij})_{i\leq n, j\leq m_i}$ represents the analysis of a sequence of pattern-guard pairs under environment $\Gamma$ with scrutinee type $t$, producing refined types and exactness flags for use in case expression typing.
\end{itemize}

 \noindent The core of our guard analysis is the derivation of the first judgment form $\Gamma ; p \vd g \mapsto \GuardR$, which associates each guard with a current pattern $p$ and a result $\GuardR$ defined as follows:
%
\[
    \begin{array}{l@{\hspace{3em}}rlll}
        \textbf{Guard Results}   & \GuardResult & \bnfeq \Result{\GuardS}{\GuardT_1,\ldotsTwo,\GuardT_m} \bnfor \Result{\GuardS}{\GuardFalse} \bnfor \omega
                                 & (m\geq 1) \\[.5mm]
        \textbf{Environments}    & \GuardS, \GuardT & \bnfeq (\Gamma, \BoolB) \\
    \end{array}
\]

\subsubsection{Guard Analysis Success Results} When a guard $g$ may succeed, the guard analysis produces a \emph{safe environment} $\GuardS$ for the whole guard and a list of \emph{success environments} $\GuardT_1,\ldotsTwo,\GuardT_m$, one for each OR-clause of the guard that may succeed. Each environment consists of a type environment $\Gamma$ paired with a Boolean exactness flag $\BoolB$:

\begin{itemize}
\item The \textit{safe environment} $\GuardS = (\Gamma, \BoolB)$ specifies a necessary condition on the types of the variables of the guard such that the guard does not error (i.e., does not reduce to $\fail$). When the flag $\BoolB$ of the environment is $\smalltrue$, it means that the condition is also sufficient.

\item Each \textit{success environment} $\GuardT = (\Gamma, \BoolB)$ describes a way in which the guard may evaluate to $\smalltrue$: the guard evaluates to $\smalltrue$ only if the types of its variables satisfy one of the success environments, and if they satisfy a success environment whose flag $\BoolB$ is $\smalltrue$, then the guard evaluates to $\smalltrue$.
\end{itemize}

\begin{example}
%
Consider the guard $(\GuardSize{x} = 2)$ formed by a single OR-clause, attached
to the pattern $p=x$ and analyzed under the environment $(x:\topp)$. This guard
produces an error if and only if $x$ is \emph{not} a tuple. Thus, its
\emph{safe environment} is the pair $(x:\tupleTop,\smalltrue)$: the flag is
$\smalltrue$ since $x:\tupleTop$ is a necessary and sufficient condition for
the clause not to error. Likewise, the analysis can determine that this guard
succeeds precisely when $x$ is a tuple of size 2. Thus, its \emph{success
environment} is $(x:\tuple{\topp,\topp}, \smalltrue)$---the type
$\tuple{\topp,\topp}$ exactly captures all tuples of size 2. In summary, we
expect our system to deduce the judgment
$\JudgmentGuard[\,;\,x]{x\boundTo\topp}{\GuardSize{x} = 2}{\{(x:\tupleTop,\smalltrue);(x:\tuple{\topp,\topp}, \smalltrue)\}}$.

In contrast, for the guard $(x \isof \intTop \,\,\texttt{and}\,\, x > y)$
attached to the pattern $p=\tuple{x,y}$ and analyzed under
$(x:\topp,y:\topp)$, we know that if it succeeds, then $x$ must be an integer,
but our type system cannot encode the relational constraint $(x > y)$. Hence,
the success environment is $((x\!:\!\intTop, y\!:\!\topp), \smallfalse)$---we
know that $x$ is an integer, but the type $\intTop$ over-approximates the
actual values that satisfy $x > y$. Note that $y$ is not constrained to be an
integer since Elixir allows comparing any two values ($>$ defines a total order
on all values). Therefore, this guard will never error, whatever values $x$ and
$y$ are bound too. The safe environment for this guard is thus
$((x{:}\topp,y{:}\topp),\smalltrue)$. In summary, our analysis will derive the
judgment
$\JudgmentGuard[\,;\,\tuple{x,y}]{(x\boundTo\topp,y\boundTo\topp)}{x \isof \intTop \,\,\texttt{and}\,\, x > y}{\{((x{:}\topp,y{:}\topp),\smalltrue);((x\!:\!\intTop, y\!:\!\topp), \smallfalse)\}}$.
\end{example}
\noindent
When a guard is formed by several OR-clauses, each success environment describes the way a clause may succeed, taking into account the clauses that precede it. Concretely, the analysis of a disjunction guard $g_1 \,\textsf{or}\, g_2$ produces the success environments of $g_1$, which describe the conditions under which $g_1$ succeeds, plus those of $g_2$, which describe the conditions under which $g_1$ reduces to $\smallfalse$ \emph{without erroring} and $g_2$ succeeds; the safe environment is one for the whole guard, since a guard that errors in any of its clauses errors as a whole.
The analysis can also produce a result $\Result{\GuardS}{\GuardFalse}$, which has no success environment: it corresponds to a guard that never evaluates to $\smalltrue$.

\subsubsection{Guard Analysis Failure Results} The cases for a faulty guard that either always errors (reduces to $\fail$) or never evaluates to $\smalltrue$ are captured by failure results $\GuardFail$.
\begingroup
\setlength{\abovedisplayskip}{3pt}
\setlength{\belowdisplayskip}{5pt}
\[
    \begin{array}{l@{\hspace{3em}}rlll}
        \textbf{Failure Results} & \GuardFail & := &  \Result{\GuardS}{\GuardFalse} \bnfor \omega
    \end{array}
\]
\endgroup
Failure results are a subset of the guard results that, for presentation purposes, we single out by ranging them over with $\GuardFail$.
The result $\omega$ states that the guard always errors, for every value of the types of its variables.

\subsubsection{Guard Analysis Rules}\label{sec:guard-analysis-rules}
Appendix~\ref{app:guardanalysis} gives the complete rules to derive $\Gamma ; p \vd g \mapsto \GuardR$ judgments.
Their typing premises, of the form $\Gamma\vd a : t$, are derived without $\omega$-rules.
Hereafter, we comment only on the most
significant ones by unrolling a series of examples.
Consider the guard (${\When{\tuple{x,y}}{x \isof \tupleTop}}$),
that performs a
type test on a capture \emph{variable} $x$. Here the current pattern is
$p=\tuple{x,y}$. The analysis of the guard $x\isof\tupleTop$ is performed under
the hypothesis that $x:\topp$ and $y:\topp$, since both capture variables do not
	have any further constraint (cf.\ rule \RuleRef{guard:accepted:accept}{\textsc{[accept]}} at the  end of this section).
\begin{wraptable}{r}{7.55cm}
	\vspace{2.5mm}\smash{ \(
		\hspace*{-1mm}  \ruleGuardVar
		\)}
\end{wraptable}
The guard analysis rule that handles this
case is \RuleRef{guard:analysis:var}{\smalltextsc{[Var]}} given here  on the right, where the
notation $\Gamma[x \refine t]_p$ denotes the environment obtained from
$\Gamma$ after refining the typing of $x$ with $t$ in the current
pattern $p$ (here, $\Gamma[x \refine \tupleTop]_{\tuple{x,y}}$), i.e.,
ascribing $x$ the type $\Gamma(x)\land t$; see
Figure~\ref{app:def:environment-updates} in
Appendix~\ref{app:guardanalysis} for the formal definition).
This produces the judgment
\begingroup
\setlength{\abovedisplayskip}{2pt}
\setlength{\belowdisplayskip}{2pt}
\[\JudgmentGuard[\,;\,\tuple{x,y}]
	{(x\boundTo\topp,y\boundTo\topp)}
	{(x \isof \tupleTop)}
	{\Result {\p{(x\boundTo\topp,y\boundTo\topp), \smalltrue}}
		{\p{(x\boundTo\tupleTop,y\boundTo\topp), \smalltrue}}}\]
\endgroup
in which the first element of the result---the safe environment---leaves the type environment for variables unchanged, since
this  guard cannot error (paired with Boolean flag $\smalltrue$, since this analysis is
exact), while the second element---the success environment---contains $(x\boundTo\tupleTop,y\boundTo\topp)$
indicating that the guard will succeed if and only if $x$ is a tuple
(and this condition is also sufficient as indicated again by the Boolean flag $\smalltrue$).

If we refine the previous guard by adding a \emph{conjunction}
\begingroup
\setlength{\abovedisplayskip}{2pt}
\setlength{\belowdisplayskip}{2pt}
\[
	\When{\tuple{x,y}} {\GuardAnd {(x \isof \tupleTop)} {(\GuardSize{x} = 2)}}
\]
\endgroup
then the new guard now specifically matches tuples of size 2. Its analysis is done by the following rule:
\begingroup
\setlength{\abovedisplayskip}{11pt}
\setlength{\belowdisplayskip}{11pt}
\[\GuardAndRuleSingleFlat\]
\endgroup
In this rule, the success environment produced by the analysis of the first
component $x \isof \tupleTop$ of the \textsf{and} (in our case, $\Delta_1 = (x : \tupleTop,y:\topp)$) is used to analyze the second component ($\GuardSize{x} = 2$), which is
then handled by successive uses of the rules
\RuleRef{guard:analysis:eq2}{\smalltextsc{[Eq$_2$]}} and \RuleRef{guard:analysis:size}{\smalltextsc{[Size]}}:
\begingroup
\setlength{\abovedisplayskip}{-3pt}
\setlength{\belowdisplayskip}{3pt}
\[
	\ruleGuardEqTwo\qquad
	\ruleGuardSize
\]
\endgroup
where $\tupleTop^i$ is the type of all the tuples of size $i$. Rule \RuleRef{guard:analysis:eq2}{\smalltextsc{[Eq$_2$]}}
corresponds to the best-case scenario of a guard equality: when one
of the terms has a singleton type ($\Gamma\vd a_2 : c$), a sufficient
condition for both terms to be equal is that the other term gets
this type as well ($\Gamma ; p \vd \GuardIsOf{a_1}{c}$).
In our example, this means doing the analysis
$\Gamma ; \tuple{x,y} \vd \GuardIsOf{\GuardSize{x}}{2}$ with rule \RuleRef{guard:analysis:size}{\smalltextsc{[Size]}}. This rule
asks two questions (i.e., checks two premises): ``can $x$ be a tuple'' (this produces a non-erroring environment), and ``can $x$ be a tuple of size 2?'' (which in our case refines $x$ to be of type $\tuple{\topp,\topp}$). The most general versions of these rules make approximations and can be found in Appendix~\ref{app:guardanalysis} (Figure~\ref{app:fig:guard-analysis-approx-rules}). A disequality $\GuardNeq{a_1}{a_2}$ is analyzed in the same way, with the test $\GuardIsOf{a_1}{\neg c}$ in place of $\GuardIsOf{a_1}{c}$ (rules \RuleRef{guard:analysis:neq1}{\smalltextsc{[Neq$_1$]}} and \RuleRef{guard:analysis:neq2}{\smalltextsc{[Neq$_2$]}} of Appendix~\ref{app:guardanalysis}).

The safe environment of the conjunction is $\Phi_1$, the one of its first component: if $g_1$ evaluates to $\smallfalse$, the conjunction evaluates to $\smallfalse$ without evaluating $g_2$, so the values for which $g_1$ evaluates to $\smallfalse$ do not make the conjunction error, whatever $g_2$ would do on them. This environment is exact if the one of $g_1$ is and if $g_2$ cannot error on the values for which $g_1$ succeeds, which rule \RuleRef{guard:analysis:and}{\smalltextsc{[And]}} checks by requiring the safe environment of $g_2$ to be exact and equal to $\Delta_1$. In our example, $\Phi_2 = \Delta_1 = (x : \tupleTop,y:\topp)$: since $x$ is a tuple, $\GuardSize{x}$ does not error, and the safe environment $(x:\topp,y:\topp)$ of the conjunction is exact.

To go further, we can check that the second element of this
tuple has type $\intTop$, by adding another conjunct to the guard:
\(
\When{\tuple{x,y}}
{ \GuardAnd {\GuardAnd {(x \isof \tupleTop)} {(\GuardSize{x} = 2)}}
	{ {(\GuardProj{1}{x} \isof \Int)} } }
\).
Now, rule \RuleRef{guard:analysis:proj}{\smalltextsc{[Proj]}} applies:\\[1mm]
\centerline{
	\(
	\ruleGuardProj
	\)}\\[1.6mm]
where $\tupleTop^{>i}$ represents tuples of size greater than $i$ (e.g., $\tupleTop^{>1} = \OpenTuple{\topp,\topp}$).
This rule reads from left to right: after checking that the index $a'$ is a
singleton integer $i$ (in our example, ``$1$''),
the non-erroring environment is computed by checking that the tuple
has more than $i$ elements. In our example,
($\GuardEq{\GuardSize{x}}{2}$) has already refined $x$ to be of type
$\tuple{\topp,\topp}$. Finally, the success environment checks that
the tuple is of size greater than $i$ with $t$ in $i$-th position
(in our example, it has type $\OpenTuple{\topp,\intTop}$);
since $x$ was a tuple of size two, the intersection of those two types
is $\tuple{\topp,\intTop}$.

In the case of a disjunction, a guard can succeed if its first component
succeeds, or if the first fails (but does not error) and the second
succeeds (guards being evaluated in a left-to-right order).
Consider
\(
{\When{\tuple{x,y}}
		{ \GuardOr {\GuardAnd {\p{x \isof \tupleTop}} {\p{\GuardSize{x} = 2}}}
			{ {(\GuardIsOf{y}\boolTop)} }} }\label{guard:and}
\)
whose analysis uses rule \RuleRef{guard:analysis:or}{\smalltextsc{[or]}}\\[1.5mm]
\centerline{\( \GuardOrRuleSingle \)}
The first term of the \textsf{or}, that is, $\GuardAnd {\p{x \isof \tupleTop}} {\p{\GuardSize{x} = 2}}$, is analyzed with rule \RuleRef{guard:analysis:and}{\smalltextsc{[And]}} given before, which produces
\scalebox{0.9}{$\Result{\p{(x\boundTo\topp,y\boundTo\topp),\smalltrue}}{\p{(x\boundTo\tuple{\topp,\topp}, y\boundTo\topp),\smalltrue}}$}.
The second term, thus, is analyzed under the environment
$(x\boundTo\neg\tuple{\topp,\topp},y\boundTo\topp)$
which is obtained by subtracting the success environment of the first guard
from its non-erroring one (i.e., because tuples of size two make the first guard
succeed, they will never reach the second guard).
This is done by computing the type
\begingroup
\setlength{\abovedisplayskip}{3pt}
\setlength{\belowdisplayskip}{3pt}
\[
	t = \acc{\tuple{x,y}}{x:\topp,y:\topp}\!\smallsetminus\acc{\tuple{x,y}}{x:\tuple{\topp,\topp}, y:\topp}
	= \tuple{\topp,\topp}\!\smallsetminus\tuple{\tuple{\topp,\topp},\topp}
	= \tuple{\neg\tuple{\topp,\topp},\topp}
\]
\endgroup
where the notation $\acc{p}\Gamma$ (defined in Appendix~\ref{app:guardanalysis}, Figure~\ref{app:def:accepted-types})
denotes the type of values that are accepted by a pattern $p$ \emph{and} which,
when matched against $p$, bind the capture variables of $p$ to types in $\Gamma$ (e.g.,
$\acc{\tuple{x,y}}{x:\intTop,y:\boolTop} = \tuple{\intTop,\boolTop}$).
This choice of $t$ is motivated by the fact that the analysis of the first term
is \emph{exact} (since the Boolean flag is $\smalltrue$), therefore it is safe to assume that the values that make the first guard succeed, never end up in the
second guard. (The rule subtracts the maximal materialization $\dynSup{(\acc{p}{\Delta_1})}$ of the accepted type, which is the accepted type itself when, as here, the types are static; for gradual types it removes all the values that $\Delta_1$ may describe.)
Because this is a disjunction, the two ways that the guard succeeds are not mixed into a single environment, but give two distinct success environments, while the guard has a single safe environment, the one of its first term: a value on which the first term errors makes the whole guard error. The Boolean flags then record which results are exact. The success environment of the second term is exact only if the safe environment of the first term is exact, since otherwise a value described by it might make the first term error; and the safe environment of the disjunction is exact only if, in addition, the second term cannot error on the values that reach it, that is, if its own safe environment is exact and equal to $\Gamma,\TypeEnv{t}{p}$.

So far our guards could not error, but it is a common feature in Elixir that
guards that error short-circuit a branch of a case expression. For example, the guard
\begingroup
\setlength{\abovedisplayskip}{3pt}
\setlength{\belowdisplayskip}{3pt}
\[\When{\tuple{x,y}}{ \GuardOr {(\GuardSize{x} = 2)} {x \isof \boolTop} }\]
\endgroup
only succeeds when the first projection of the matched value is a
tuple of size two, and fails for all other values \emph{including}
when the first projection is a boolean (in which case $\GuardSize{}\!\!$ raises an error).
This is handled by rule \RuleRef{guard:analysis:or}{\smalltextsc{[Or]}} as well, by considering
the non-erroring environment of a guard and using it as a base to analyze the second term of a disjunction. In our example, the non-erroring environment is
$(x\boundTo\tupleTop,y\boundTo\topp)$, and the second term is found instantly
to be false. This will raise a warning, as a clause of a guard that
only evaluates to false is a sign of a possible mistake.

\begin{remark}[Safe environments of conjunctions]\label{rem:safe-and}\em
The safe environment of $\GuardAnd{g_1}{g_2}$ is the one of $g_1$, even when $g_2$ errors on some of the values for which $g_1$ succeeds: rule \RuleRef{guard:analysis:and}{\smalltextsc{[And]}} then only marks it as inexact. A more precise analysis would remove from it the values for which $g_1$ surely succeeds and $g_2$ surely errors. Consider the Elixir guard
\begin{minted}[fontsize=\small]{elixir}
(is_integer(x) or is_tuple(x)) and tuple_size(x) == 2 or is_integer(x)
\end{minted}
\noindent which compiles to a Core guard equivalent to $\GuardOr{\GuardAnd{(x\isof\intTop\lor\tupleTop)}{(\GuardSize{x}=2)}}{(x\isof\intTop)}$, analyzed below with pattern $x$ under $x:\topp$. On an integer, \elix{tuple_size} errors and the whole guard fails, so the last disjunct never succeeds, and the guard accepts exactly the tuples of size two. Our analysis gives the first disjunct the inexact safe environment $(x:\topp,\smallfalse)$, analyzes $x\isof\intTop$ under $x:\neg\tuple{\topp,\topp}$, and returns the success environments $(x:\tuple{\topp,\topp},\smalltrue)$ and $(x:\intTop,\smallfalse)$: the clause may accept integers. Removing the integers from the safe environment of the conjunction would give $(x:\neg\intTop,\smalltrue)$; the last disjunct would then be analyzed under $x:\neg\intTop\setminus\tuple{\topp,\topp}$, found to be always false, and reported by a warning, which points at a probable mistake. We decided not to refine the analysis in this way, for simplicity: the system seemed already complicated enough.
\end{remark}

\subsubsection{Pattern-Guard Sequence Analysis}\label{sec:pattern-guard-sequence-analysis}
In a final processing step, the guard analysis judgment
$\Gamma ; p \vd g \mapsto \GuardResult$ is used to derive the judgment
$\Gamma ; t \vd (p_i g_i)_{i\leq n}
	\leadsto (t_{ij},\BoolB_{ij})_{i\leq n, j\leq m_i}$ used to type a case expression.
Rule \RuleRef{guard:accepted:accept}{\smalltextsc{[accept]}} analyzes a single
pattern-guard pair under its incoming type $t$ and translates the success
environments $(\Delta_i, \BoolC_i)_{i\in I}$ into
$(\dynSup{t}\land\acceptedType{p}_{\Delta_i}, \BoolC_i)_{i\in I}$,
preserving their precision flags. Here $t$ is the current residual type.
The intersection with $\dynSup{t}$ retains restrictions on the incoming value
that may be lost when its type is projected onto the pattern's capture variables.
For example, consider a scrutinee of type $\tuple{\texttt{:count},\intTop}\lor\tuple{\texttt{:flag},\atomTop}$
and the clause $\When{\tuple{\mathit{tag},x}}{ {\mathit{tag} \isof \texttt{:count}}} \to x\,{+}\,1$.
The guard analysis in the premises of \RuleRef{guard:accepted:accept}{\smalltextsc{[accept]}}  produces the success environment $(\mathit{tag}:\texttt{:count},\,x:\intTop\lor\atomTop)$,
yielding the pattern accepted type $\tuple{\texttt{:count},\intTop\lor\atomTop}$ which has lost the correlation between
the two components. Its intersection with $\dynSup{t}$ yields $\tuple{\texttt{:count},\intTop}$,
the type of the values that reach this clause and are accepted by it, so that the body
$x\,{+}\,1$ is typed with $x:\intTop$ and is well typed, whereas it would be rejected with
$x:\intTop\lor\atomTop$.
Guards that always fail are handled by rule \RuleRef{guard:accepted:fail}{\textsc{[fail]}}.
\begingroup
\setlength{\abovedisplayskip}{5pt}
\setlength{\belowdisplayskip}{5pt}
\begin{equation*}
	\begin{gathered}
		\ruleAccept
		\qquad
		\ruleFail
	\end{gathered}
\end{equation*}
\endgroup
The sequence of successive guard-pattern pairs in a case expression
is handled by \RuleRef{guard:accepted:seq}{\smalltextsc{[Seq]}}, which takes care to
refine the possible types as the analysis advances, by subtracting
from the potential type $t$ the surely accepted type $\bigvee_{(s, \textsf{true})\in\mathcal{A}} s$ of the analysis
of the current guard-pattern.
\begingroup
\setlength{\abovedisplayskip}{1pt}
\setlength{\belowdisplayskip}{5pt}
\begin{equation*}
	\ruleSeq
\end{equation*}
\endgroup

\noindent The intersection in rule \RuleRef{guard:accepted:accept}{\smalltextsc{[accept]}}
bounds the generated types by $\dynSup{t}$. It does not change the residual
computed by \RuleRef{guard:accepted:seq}{\smalltextsc{[seq]}}:
subtracting $\dynSup{t}\land\acc{p}{\Delta}$ from $t$ gives the same result
as subtracting $\acc{p}{\Delta}$. Rule \RuleRef{guard:accepted:seq}{\smalltextsc{[seq]}} produces the auxiliary types
$\Gamma\,;t \vdash ({p_i}{g_i})_{i\leq n}
\leadsto (t_{ij},\BoolB_{ij})_{i\leq n, j\leq m_i}$
used in the typing rules for case expressions.

\subsubsection{Guard Analysis Properties} Finally, the guard analysis provides two key properties regarding the types involved in the typing rules for case expressions.

\begin{restatable}[Surely accepted types are sufficient]{lemma}{surelyAcceptedTypesAreSufficient}\label{lem:guard-sufficient}
	Given the guarded pattern sequence analysis $\Gamma\,;t \vdash ({p_i}{g_i})_{i\leq n} \leadsto {(t_{ij},\BoolB_{ij})}_{i\leq n, j\leq m_i}$, for all $i, j$ such that $\BoolB_{i j} = \smalltrue$ and $t_{i j}\not\leq\bott$, for all values $v$,
	\[
		(v : t_{i j}) \,\, \Longrightarrow \,\, \exists (i_0\leq i) \,\text{ s.t. }\, (v/p_{i_0}g_{i_0} \neq \fail)
	\]
\end{restatable}

This lemma expresses the fact that if a value $v$ belongs to a
surely accepted type $t_{i j}$ associated with
a branch $p_i g_i$ of a case expression, then this
value will never reach any branches after $p_i g_i$.
This property is crucial for typing, because
it ensures that surely accepted types are excluded from
subsequent branches, thereby refining the type of the scrutinee.

\begin{restatable}[Possibly accepted types are necessary]{lemma}{possiblyAcceptedTypesAreNecessary}\label{lem:guard-necessary}
	Given the guarded pattern sequence analysis $\Gamma\,;t \vdash ({p_i}{g_i})_{i\leq n} \leadsto {(t_{ij},\BoolB_{ij})}_{i\leq n, j\leq m_i}$,
	for all $i\leq n$ and all values $v$ such that $v : t$ and $v/(p_{h} g_{h}) = \fail$ for every $h<i$,
	\[
		(v/(p_i g_i) \neq \fail) \,\, \Longrightarrow \,\, \exists j \leq m_i
		\text{ s.t. } (v : t_{i j})
	\]
\end{restatable}

This lemma expresses the fact that if a value $v$ of type $t$ that is not accepted by the guarded patterns preceding
$p_i g_i$ is accepted by $p_i g_i$, then its type must
belong to one of the possible types $t_{i j}$ for some $j \leq m_{i}$ that were computed by the guard analysis. This ensures that, during typing, within
a branch, the types of the capture variables depend
directly on the corresponding possibly accepted type.

Both lemmas are implied by a technical lemma (Lemma~\ref{lem:safe-success-environment}) on safe/success environments found in Appendix~\ref{app:guard-analysis-correctness}.

\section{Arity and Strong Arrows}\label{sec:arity}
The examples of the introduction use two extensions of unary
arrow types of semantic subtyping. The first one comes from Elixir and Erlang: function arity is part
of function identity, and it can be tested by guards. This is reflected by the
test $\elix{is_function(f, n)}$, which checks whether \elix{f} is a function of
arity \elix{n}, and whose usage we showed in the definition of the
\elix{curry} function in lines~\ref{curry1}--\ref{curry2}. To support this feature in Core Elixir, we extend the syntax as follows:
\begingroup
\setlength{\abovedisplayskip}{2pt}
\setlength{\belowdisplayskip}{2pt}
\[
  \begin{array}{lrcl}
    \textbf{Expressions} \qquad & e &\bnfeq &\cdots \bnfor \lambda^\IFace{\,\ov{x}\,}.{e}
    \bnfor\app{e}{\,\ov{e}\,}                                               \\
    \textbf{Base Types}        & b &\bnfeq &\cdots \bnfor \function_n       \\
    \textbf{Types}             & t& \bnfeq &\cdots \bnfor \fun{\ov{t}}{t}
  \end{array}
\]
\endgroup
\noindent
Abstractions have lists of parameters $\ov{x}$, applications
have lists of arguments $\ov{e}$, the domain of an arrow type is a list of types
$\ov t$, and the base type $\function_n$ is the type of all functions of arity
$n$ (the type $\function$ of all functions is then the union of all the
$\function_n$). Since base types are test types, it is now possible to test whether a value is a function of arity
$n$, by checking the test type $\function_n$.
The reduction rules of Figure~\ref{fig:reductions} extend in the same way:
evaluation contexts include $\app{\Context}{\ov{e}}$ and
$\app{v}{\ov{v},\Context,\ov{e}}$, which evaluate the function and then its
arguments from left to right; rule \RuleRef{reduction:app}{\textsc{[App]}}
reduces $\app{(\lambda^\IFace\ov{x}.e)}{\ov{v}}$ to $e\Subst{\ov{x}}{\ov{v}}$
when the lists $\ov{x}$ and $\ov{v}$ have the same length, and rule
\RuleRef{reduction:app-omega}{\textsc{[App$_\omega$]}} applies whenever the
function is not an abstraction with as many parameters as arguments.
The typing rules of Section~\ref{sec:safe-erasure} extend to these forms
componentwise. In particular, \RuleRef{typing:static:app}{rule (app)} requires
every type of the domain $\ov t$ to be static, and
\RuleRef{typing:gradual:lambda-dyn}{rules ($\lambda_\dyn$)} and
\RuleRef{typing:weak:lambda-circ}{($\lambda^\circ$)} replace every type of each
domain $\ov t$ of the interface by its maximal materialization, and
\RuleRef{typing:weak:lambda-circ-dyn}{rule ($\lambda^\circ_\dyn$)} types an
abstraction with $n$ parameters with $\function_n\land\dyn$. The side
condition of \RuleRef{typing:static:lambda}{rule ($\lambda$)} reads each domain
$t^1,\ldots,t^n$ as the set of argument lists it describes: its static part is
$\dynInf{(t^1)}\times\cdots\times\dynInf{(t^n)}$ and its dynamic part is
$(\dynSup{(t^1)}\times\cdots\times\dynSup{(t^n)})\setminus(\dynInf{(t^1)}\times\cdots\times\dynInf{(t^n)})$.

The second extension, strong arrows, is specific to our safe erasure approach. Strong arrows record the fact
that the body of a function contains enough checks to guarantee that the function will return only values from its codomain, even when the function is applied in gradual mode. As
for multi-arity arrows, this requires extending the semantic subtyping
framework: the subtyping relation used in the previous sections, namely the one
defined in Lanvin's thesis~\cite{lanvin2021semantic}, works for the type
constructors already present in that theory, but it does not define subtyping
for non-unary arrows or for strong arrows.

We thus need to extend the current theory to define the subtyping relation on these new
types and deduce the algorithm to decide this relation. 
In semantic subtyping this is not
always straightforward: although the techniques to do so are extensively explained in the
literature (e.g., \cite{FCB08,castagna2005gentle,lmcs:6098,castagna2022programming}), it may not be obvious how to adapt them
to specific situations. 
There are two ways to do so: either by defining an encoding of your
custom types into existing types or by extending the semantic interpretation of types to
support them. 

Multi-arity arrow types admit an encoding into the existing theory: one can
represent them using unary arrows and pairs, where one component encodes the arrow type
(with multiple arguments represented as a tuple type), and the other encodes the arity
as the singleton type of the corresponding integer constant. Under this encoding, two
arrow types are comparable only if they share the same arity component (see the direct
interpretation in \S\ref{subsec:multi-arity} below). The set of all functions of arity
$n$ is then encoded as $\tuple{n,(\bott\to\topp)}$, and
$\tuple{\intTop,(\bott\to\topp)}$ is the type of all functions (in the interpretation of Definition~\ref{def:multi-arity} below; for the one of~\citet{lanvin2021semantic}, see Remark~\ref{rem:two-models}). 
Since this encoding is not
efficient for an implementation---even more so when combined with strong arrows---we
have not adopted it; nevertheless, it is instructive to present it, as it shows how
functions of arbitrary arity can be accommodated within the existing theory.

Strong arrows cannot be obtained in the same way. Their interpretation must
distinguish the ordinary domain condition of an arrow from the stronger property
used by safe erasure: outside the domain, a strong function may fail, but if it returns a value, then it belongs to the codomain. We therefore give a direct
set-theoretic interpretation for both extensions and derive from it the corresponding
subtyping algorithms. This consists of two steps:
\begin{enumerate}[left=5pt,parsep=0pt,topsep=2pt,itemsep=0pt]
  \item defining the semantic interpretation of the new type;
  \item deriving from this interpretation the decomposition rules to check subtyping for the new type.
\end{enumerate}
We succinctly describe these steps in the next two sections. In Section~\ref{subsec:multi-arity}, we outline the interpretation of multi-arity function types and how to use it to decide subtyping . In Section~\ref{sub:semantic-strong-arrows}, we present the same development for strong
arrow types. We defer to the appendixes the proofs of
the related properties. We assume knowledge of the basic definitions of semantic
subtyping, which can be found in any of the references cited above (among them,
we suggest~\cite{castagna2022programming} for a simple introduction).

\subsection{Semantic Interpretation of Multi-Arity Function Types}\label{subsec:multi-arity}
Introducing a new type constructor for multi-arity functions
requires an interpretation in the domain of semantic subtyping. In semantic
subtyping, types are interpreted as sets in a domain $\Domain$ whose
elements represent the values of the language
(see~\cite[Definition 2]{castagna2022programming} for the definition of a concrete domain and its explanation). Let
$\DOmega$ denote the set $\Domain\cup\{\Omega\}$, where $\Omega$ is a special
element not belonging to $\Domain$ and representing all type errors; let
$\fparts{S}$ denote the set of finite subsets of a set $S$. We suppose that
$\Domain$ contains the disjoint union
$\biguplus_{n\in\mathbb{N}}\fparts{\product{\Domain^n}{\DOmega}}$: each relation is
thus tagged by its arity $n$, so that relations of different arities, for instance the
empty relations, are distinct elements of $\Domain$. In what follows,
$\fparts{\product{\Domain^n}{\DOmega}}$ denotes the $n$-th summand of this union.
Multi-arity function spaces are then interpreted as follows:
\begin{Definition}\label{def:multi-arity}
  Let $X_1,\ldotsTwo,X_n$, and $Y$ be subsets of the domain $\Domain$. We define
    \begin{align*}
      \fun{\p{X_1,\ldotsTwo,X_n}}{Y} &\eqdef
      \left\{ R \in \fparts{\product{\Domain^n}{\DOmega}} \right. \\
      &\qquad \left. \mid \forall (d_1,\ldotsTwo,d_n,\delta)   \in R.\,\,
        (\forall i\in\{1,...,n\}.\,\, d_i \in X_i)
        \Rightarrow \delta \in Y\right\}
    \end{align*}
  \end{Definition}  
In a nutshell, the space of functions of arity $n$ is defined as the set of finite sets of $n{+}1$-tuples $(d_1,\ldotsTwo,d_n,\delta)$ such that if the first $n$ components are in the domain of the function type, then the last component
$\delta$ is in its codomain. Since $Y$ is a subset of $\Domain$, it implies that if the arguments $d_i$'s are in the right $X_i$'s, then $\delta\not=\Omega$: the application does not raise a type error.\footnote{As explained by~\cite[Section 3.2]{castagna2022programming}, $\Omega$ is a constant  whose
    main purpose is to
    avoid $\topp\to\topp$ to coincide with the set of all functions
    (see~\cite[Section 4.2]{FCB08} for full details).} 
This definition is a natural extension of the definition of function spaces in semantic subtyping, where an arrow type is interpreted as the set of the finite approximations of its functions, seen as graphs (i.e., as relations, or sets of pairs): see~\cite[Definition 4.2]{FCB08}. The Definition~\ref{def:multi-arity} above is then used to define the interpretation function $\sem{.}:\mathcal{T}\to\parts\Domain$ from the types $\mathcal{T}$ of Core Elixir into sets within $\Domain$, using standard semantic subtyping techniques. In particular, a multi-arity function type $(t_1,...t_n)\to t$ is interpreted as $(\sem{t_1},\ldotsTwo,\sem{t_n})\to\sem{t}$, the base type $\function_n$ as the set $\fparts{\product{\Domain^n}{\DOmega}}$ of all relations of arity $n$, and $\function$ as the union of these sets. This semantic interpretation induces the definition of the subtyping relation as the containment of the interpretations of types: $t_1 \leq t_2$ if and only if $\sem{t_1} \subseteq \sem{t_2}$. Using
set-theoretic equivalences, the problem of deciding whether a subtyping relation $t_1 \leq t_2$ holds, can then be simplified
to an emptiness checking problem: $t_1 \leq t_2 \iff \sem{t_1} \subseteq \sem{t_2}\iff\sem{t_1} \smallsetminus \sem{t_2}\subseteq\varnothing\iff \sem{t_1\smallsetminus t_2}\subseteq\varnothing$.
This emptiness check can itself be decomposed over each
disjoint component of a type (i.e., tuples, integers, etc.) and checked by algorithms described
in \cite[Section 4]{lmcs:6098} which are defined for \emph{disjunctive normal forms}
of literals $\ell$, that is, atomic types (i.e., a product, an arrow, ...) and their negations.
For multi-arity functions, this means that we have to decide the emptiness of a disjunctive normal form
\(
\bigvee_{i\in I}\bigwedge_{j\in J} \ell_{i j}
\)
where all  $\ell_{i j}$ have the form of multi-arity arrow types $\ov{t}\rarr t$ or their negations (a literal $\function_n$ fixes the arity of an intersection, $\neg\function_n$ makes the intersections of arity $n$ empty, and $\neg\function_m$ does not affect those of arity $n\neq m$). To simplify
the problem of checking that such a normal form is empty notice that, according to
the interpretation of Definition~\ref{def:multi-arity}, intersections of different arities are empty; therefore we can consider only intersections of arrows of the same arity. Since a union of types/sets is empty if and only if each type/set composing it is empty, then we have to check the emptiness of intersections of arrows of the same arity or their negations. Therefore, the problem
simplifies into checking for every intersection forming the union \(\bigvee_{i\in I}\bigwedge_{j\in J} \ell_{i j}\), that the intersection of positive arrows is contained in the union of the
negated ones:
\begingroup
\setlength{\abovedisplayskip}{3pt}
\setlength{\belowdisplayskip}{1pt}
\begin{equation}\label{eq:multi-arity}
  \bigwedge_{i\in P} (t_i^{(1)},...,t_i^{(n)})\rarr t_i
  \leq \bigvee_{j\in N} (t_j^{(1)},...,t_j^{(n)})\rarr t_j
\end{equation}
\endgroup
(note that if any negative arrow were not of arity $n$, then this arrow would not play any role in the containment and could be removed). Since each literal can be interpreted as a set according to Definition~\ref{def:multi-arity}, we reformulate this problem as a set-containment problem. We then solve it, as stated by Theorem~\ref{thm:set-containment-multi-arity}, using the set manipulation techniques developed by~\citet{frisch2004theorie}. The full details of the semantic interpretation and the proof of the theorem, for this interpretation and for the one of~\citet{lanvin2021semantic} (Remark~\ref{rem:two-models}), are given in Appendix~\ref{app:functions}.

\begin{restatable}[Multi-arity Set-Containment]{theorem}{multiAritySetContainment}\label{thm:set-containment-multi-arity}
  Let $n\in\mathbb{N}$, let $P$ and $N$ be finite sets, and let
  $(X_i^{(1)})_{i\in P}$, \ldots, ${(X_i^{(n)})}_{i\in P}$,
  $(X_i)_{i\in P}$, $(Y_i^{(1)})_{i\in N}$, \ldots, $(Y_i^{(n)})_{i\in N}$,
  $(Y_i)_{i\in N}$ be families of subsets of the domain $\Domain$.
  Then,

\[\begin{array}{l}
 \displaystyle\Intersection{P}{\Function{\Arguments{X_i}}{X_i}}
        \subseteq
        \Union[j]{N}{\Function{\Arguments{Y_j}}{Y_j}}
    \quad \Leftrightarrow\quad  \exists j_0 \in N. \\
    \hspace*{13mm}\forall \iota : P \!\rightarrow\! [ 1,n+1 ].
    \left(\exists k{\in}[1,n].\,\, Y_{j_0}^{(k)} \!\subseteq
        \!\!\!\!\!\!\!\!\!\displaystyle
        \bigcup_{\{i\in P \,\mid\, \iota(i) = k\}}\!\!\!\!\!\! X_i^{(k)}\right)
        \textbf{or }\!\displaystyle
        \left(\bigcap_{\{i\in P \,\mid\, \iota(i) = n+1 \}} \!\!\!\!\!\!\!\!\!X_i \subseteq Y_{j_0}\right)
\end{array}\]
where a union over an empty set is $\varnothing$, an intersection of codomains
over an empty set is $\DOmega$, and the intersection of an empty family of
function spaces of arity $n$ is the set $\fparts{\product{\Domain^n}{\DOmega}}$ of
all relations of arity $n$.
\end{restatable}
Here, $\iota$ ranges over maps from positive-arrow indices to the $n$
argument positions and the codomain position $n+1$. This theorem states that to check whether an intersection of arrows of arity $n$ is contained in a union of arrows of the same arity, we need to find at least one arrow in the union (the ``$\exists j_0 \in N$" in the right-hand side of the formula) for which such a containment holds. The theorem then reduces this containment problem on arrows, to multiple smaller containment checks on their domain and return types. It thus enables
the definition of a recursive algorithm that decides subtyping.  Notice that the multiple smaller problems are related by an ``\textbf{or}" which means that a naive implementation of the algorithm described by Theorem~\ref{thm:set-containment-multi-arity} may have to backtrack and, therefore, unroll all the memoized solutions found in the current run.

Following~\citet{frisch2004theorie}, we can define a backtrack-free algorithm to decide whether an intersection of arrows is contained in a single arrow (the one corresponding to $j_0$). That is an algorithm that for all $n\in\mathbb{N}$ decides 
\begin{equation}\label{algo:multi-arity}
  \textstyle\bigwedge_{f \in P} f \,\,\leq\,\, (t_1,\ldotsTwo,t_n) \rightarrow t
\end{equation}
where $P$ is a finite set of arrows of arity $n$, that is, arrows of the form $(t_1',\ldots,t_n') \rightarrow t'$ (when $P$ is empty, the intersection is $\function_n$).
This algorithm is expressed by the function $\Phi_n$ of $n+2$ arguments defined as:
\begingroup
\setlength{\abovedisplayskip}{5pt}
\setlength{\belowdisplayskip}{5pt}
\[
  \begin{array}{l}
    \Phi_n(t_1,\ldotsTwo,t_n,(b,t),\varnothing) \;=\;
    \left( \exists k\in[1,n].\,\, t_k \leq \bott \right) \textbf{ or }
    \left( b \textbf{ and } t \leq \bott \right)                                                \\[1mm]
    \Phi_n(t_1,\ldotsTwo,t_n,(b,t),\{ (t_1',\ldotsTwo,t_n') \rightarrow t' \} \cup P) \;=\;
    \Phi_n(t_1,\ldotsTwo,t_n, (\tttrue, t \wedge t'), P)\, \textbf{ and }                    \\
    \hspace{4cm}
    \forall k\in[1,n].\,\, \Phi_n(t_1,\ldotsTwo,t_k\!\smallsetminus\!t_k'\,, \ldotsTwo, t_n, (b,t), P)
  \end{array}
\]
\endgroup
The Boolean $b$ records whether the codomain of some arrow of $P$ has been
intersected with $t$: if none has, the condition on the codomain fails, since an
intersection of codomains over an empty set is $\DOmega$, which contains $\Omega$
(Theorem~\ref{thm:set-containment-multi-arity}).
The call $\Phi_n(t_1, \ldotsTwo, t_n, (\ttfalse,\neg t), P)$ decides $\eqref{algo:multi-arity}$
(see Theorem~\ref{thm:arity-decide} in Appendix~\ref{app:functions}: for the interpretation of Definition~\ref{def:multi-arity}, $\Phi_n$ is $\Phi'_n$ without its Booleans for the arguments, which are then always true).
To summarize, since an intersection of arrows of arity $n$ is a subtype of a union of arrows of the same arity
if and only if it is a subtype of one of the arrows in the union
(a corollary of Theorem~\ref{thm:set-containment-multi-arity} corresponding to the ``$\exists j_0{\in}N$'' in the statement),
then the subtyping problem in~\eqref{eq:multi-arity} reduces to
finding one negative arrow ($j_0\in N$) such that $\Phi_n(t_{j_0}^{(1)},\ldotsTwo, t_{j_0}^{(n)}, (\ttfalse,\neg t_{j_0}), P)$ returns true.

\begin{remark}[Lanvin's interpretation]\label{rem:two-models}\em
The subtyping relation of this work, used in the previous sections, is the
one of Lanvin~\citet{lanvin2021semantic}, extended to the new
types. It is induced by a refinement of the interpretation of
Definition~\ref{def:multi-arity}, in which the inputs of relations also range
over a new element $\invOmega$ that belongs to every domain
(Appendix~\ref{app:functions}). We have presented the interpretation of
Definition~\ref{def:multi-arity} because it is simpler. Every result of this
section has a counterpart for Lanvin's interpretation, and
Appendices~\ref{app:functions} and~\ref{sec:strong-subtyping} prove both
at once. The type safety results of Sections~\ref{sec:safe-erasure}
and~\ref{sec:guard-analysis} hold for both relations:
Appendices~\ref{app:soundness} and~\ref{app:guardanalysis} prove them for
both. The two differ only on arrows with an empty domain component (for
gradual types, in their minimal or maximal materializations). With
$\invOmega$, the first clause of
Theorem~\ref{thm:set-containment-multi-arity} requires in addition that some
$i\in P$ have $\iota(i)=k$; $\Phi_n$ needs a Boolean for each argument; and, in
the first equivalence of Section~\ref{sub:semantic-strong-arrows}, the disjunct
$a\leq\bigvee_{j\in J}t_j$ requires in addition $J\neq\varnothing$, and the
rewriting there of an intersection of strong arrows as a single strong arrow
requires a non-empty family. Moreover, $(\bott,\ldotsTwo,\bott)\rarr\topp$ is then the
largest arrow type of arity $n$, but, for $n\geq1$, it is strictly contained in
$\function_n$, and the encoding of multi-arity arrows with pairs described at
the beginning of this section does not hold: it cannot distinguish
$\fun{\p{\bott,\intTop}}{\intTop}$ from $\fun{\p{\bott,\boolTop}}{\intTop}$.
Lanvin's relation is contained in the relation induced by
Definition~\ref{def:multi-arity}, and it is strictly smaller: for instance
$\intTop\rarr\intTop\leq\bott\rarr\boolTop$ holds only for the latter, in
which $\bott\rarr\boolTop$ is equivalent to $\function_1$. Through the minimal
and maximal materializations of Section~\ref{sec:gradual-typing}, the two
relations also differ on arrows whose domain is `$\dyn$': for instance,
$\dyn\rarr\dyn\leq\dyn\rarr\bott$ holds only for the relation induced by
Definition~\ref{def:multi-arity}.
\end{remark}

\noindent

\subsection{Semantic Interpretation of Strong Arrow Types}\label{sub:semantic-strong-arrows}
Integrating the strong arrow types introduced in Section~\ref{sec:gradual-typing} into the semantic subtyping framework requires a similar approach to that used above for multi-arity function types. We must first define how to interpret strong arrow types as sets within the domain $\Domain$. We then use this interpretation to decompose the emptiness problem of a disjunctive normal form of arrow literals into simpler, more tractable problems.

For the sake of simplicity, we present the semantic interpretation of unary strong arrows. Therefore, we suppose that $\Domain$ contains $\fparts{\Domain\times\DOmega}$. The extension to multi-arity is straightforward. In this domain the function space $X\to Y$ is the set of all elements in $\fparts{\Domain\times\DOmega}$ that are only composed by pairs such that if the first projection is in $X$, then the second projection is in $Y$.

The interpretation of a strong arrow type $(X \rarr Y)^\star$ must be a subset (in general, a strict one) of the interpretation of $X \rarr Y$, insofar as it must contain only the relations that represent the strong functions in  $X \rarr Y$. These are functions that for every possible argument (whether in $X$ or not) return either a value in $Y$ or $\Omega$. In other words, the strong arrow type $(X \rarr Y)^\star$ is interpreted as the set of finite sets of pairs $(d,\delta)$, where if $d$ is in the domain of the function then $\delta$ is in its codomain, but with the additional requirement that if $d$ is \emph{not} in the domain of the function, then $\delta$ must be either in $Y$ or be $\Omega$. More compactly, $(X \rarr Y)^\star = (X\rarr Y)\cap\fparts{\Domain\times Y_\Omega}$. The intersection shows that the (interpretation of a) strong arrow type $(X \rarr Y)^\star$ is a subset of the  arrow type $(X \rarr Y)$.

We now use this interpretation to derive an algorithm for deciding subtyping in the presence of strong arrow types. As the interpretation reveals, strong arrow types are subsets of arrow types. This relationship makes it straightforward to extend the existing framework: in the presence of strong arrows, the disjunctive normal form of a type remains a union of intersections of literals on the same constructors. However, for functional spaces, the literals can now be either arrows or strong arrows. Therefore, deciding subtyping of function spaces requires solving emptiness problems of the following form:
\begin{align*}
	 & \bigwedge_{i \in I} (t_i \rarr s_i) \land
	\bigwedge_{j \in P} (t_j \rarr s_j)^{\star} \land
	\bigwedge_{k \in R} \neg (t_k \rarr s_k) \land
	\bigwedge_{l \in Q} \neg (t_l \rarr s_l)^{\star}
	\quad \leq \quad  \mathds{O}
\end{align*}
This containment can be further simplified since $\bigwedge_{j \in P} (t_j \rarr s_j)^{\star} = (\bigvee_{j \in P} t_j\rarr \bigwedge_{j \in P} s_j)^{\star}$ (see Lemma~\ref{lem:strong-intersection} in Appendix~\ref{sec:strong-subtyping}; when $P$ is empty, both sides are the set of all unary relations). Therefore, we can rewrite the previous containment as:
\begin{align*}
  & \bigwedge_{i \in I} (t_i \rarr s_i) \land
  (t \rarr s)^{\star} \land
  \bigwedge_{k \in R} \neg (t_k \rarr s_k) \land
  \bigwedge_{l \in Q} \neg (t_l \rarr s_l)^{\star}
  \quad \leq \quad  \mathds{O}
\end{align*}    
This holds if and only if one of the following formulas holds (cf.\ Lemma~\ref{lem:exists-parts})
\begin{eqnarray}
		  \text{either} &&\exists i_0 \in R. \, \bigwedge_{i \in I} (t_i \rarr s_i) \land
			             (t \rarr s)^{\star} \leq (t_{i_0} \rarr s_{i_0})     \label{cond:first}    \\[2mm]
			\text{or} && \exists i_0 \in Q. \,   \bigwedge_{i \in I} (t_i \rarr s_i) \land
			             (t \rarr s)^{\star} \leq (t_{i_0} \rarr s_{i_0})^{\star} \label{cond:second}
\end{eqnarray}
Finally, the first condition \eqref{cond:first} can be solved by applying this equivalence:
	\begin{equation*}
		\bint_{i\in I} (t_i \rarr s_i) \inter \StrongType{\p{c \rarr d}}
		\leq
		\p{a \rarr b}
		\,\,\iff \,\,
		\begin{cases}
			a \leq \buni\limits_{i\in I} t_i \union c \\[-1mm]
			\forall J \subseteq I. \quad
			\left( a \leq \buni\limits_{j \in J} t_j  \right) \lor
			\left( \bint\limits_{j \in I\setminus J} s_j \inter d \leq b \right)
		\end{cases}
	\end{equation*}
and the second condition \eqref{cond:second} by applying the equivalence
	\begin{equation*}
		\bint_{i\in I} (t_i \rarr s_i) \inter \StrongType{\p{c \rarr d}}
		\leq
		\StrongType{(a \rarr b)}
		\,\, \iff \,\,
		\begin{cases}
			a \leq \buni\limits_{i\in I} t_i \union c \\[-1mm]
			\forall J \subseteq I. \,\,
			\left( \buni\limits_{j \in J} t_j =  \mathds{1}  \right) \lor
			\left( \bint\limits_{j \in I\setminus J} s_j \inter d \leq b \right)
		\end{cases}
	\end{equation*}
Technical details and proofs of the above definitions and results, are given in Appendix~\ref{sec:strong-subtyping}, for this interpretation and for the one of~\citet{lanvin2021semantic} (Remark~\ref{rem:two-models}), for which, in the first equivalence, the disjunct $a\leq\bigvee_{j\in J}t_j$ requires in addition $J\neq\varnothing$, and the simplification above requires $P\neq\varnothing$: the case $P=\varnothing$ is decided without a strong arrow on the left, as explained after Theorem~\ref{thm:algo-strong}.

\begin{remark}
{\em 
	Strong arrows describe functions whose behavior is constant
	outside their domain: they necessarily error, diverge, or return a value
	of their precise codomain type. A function of type $(\intTop \rarr \intTop)^\star$,
	when given Booleans, can either error or \emph{return integers};
	thus it cannot also have type $(\boolTop \rarr \boolTop)^\star$,
  unless it never returns a value, that is, unless it has type $(\intTop\lor\boolTop \rarr \bott)^\star$ (it diverges on integers and Booleans, and fails or diverges on the other arguments). This explains why an intersection of strong arrows is, as we stated earlier, a strong arrow itself.
}
\end{remark}

\section{Inference}\label{sec:inference}
The problem of inference for Elixir consists of finding a type for
functions defined by several pattern-matching clauses, without any type annotation.
Inference appears in this work mainly as a convenience tool: indeed, one
could simply decide that every function must be annotated, and inference
would not be required. In our case, it is both an interesting research question and
a practical one: writing annotations for untyped code is not without effort, and inference
can help by suggesting annotations to the programmer. In the case
of anonymous functions, being able to infer their types means that annotating
can be made optional (e.g., this is convenient when passing short anonymous
functions created on the fly to functions that process enumerable data, as in the code in
line~\ref{bump2}).
To study inference, we thus add  to Core Elixir expressions, \emph{non-annotated} $\lambda$-abstractions formed by a list of pattern-matching clauses:
\begingroup
\setlength{\abovedisplayskip}{3pt}
\setlength{\belowdisplayskip}{3pt}
\[
     e \bnfeq\,\, \cdots \bnfor \lamPat[]{\,\ov{p g \rarr e}\,}
\]
\noindent Operationally, the expression $\lamPat[]{\,\ov{p g \rarr e}\,}$ is equivalent to $\lambda x.\texttt{case\,}x\,(\ov{p g \rarr e})$ where $x$ is fresh and the interface of the $\lambda$-abstraction is omitted. This essentially means that a multi-clause function definition is encoded as a single function whose body is a single case expression (which is exactly what the Elixir compiler does).

\subsection{Inferring Interfaces from Guards}
To infer the type of such functions, we use the guard analysis defined
in Section~\ref{sec:guard-analysis} to infer for each clause a list of accepted types $t_i$
that represent every type potentially accepted by the clause.
We then type the body of the function for each of these types, producing $t_i'$,
and take the intersection of the resulting types $\bigwedge_i (t_i \rarr t_i')$
as the type of the function.
For instance, the analysis of the guard in
\[ \PatLam{ \When{x}{(\GuardOr{x \isof \intTop}{x \isof \boolTop})} \rarr x } \]
\endgroup
produces the two accepted types $\intTop$ and $\boolTop$; type-checking the function
with $\intTop$ as input gives $\intTop$ as result, and likewise for $\boolTop$.
Hence, the inferred type is $(\intTop\rarr\intTop)\land(\boolTop\rarr\boolTop)$.
Formally, the new expression is typed by the \RuleRef{typing:inference:infer}{rule (\textsc{infer})} below
\begingroup
\setlength{\abovedisplayskip}{3pt}
\setlength{\belowdisplayskip}{4pt}
\[
     \typingrule{\RuleDef{typing:inference:infer}{\textsc{(infer)}}}
     {\!\!
          \Gamma\,;\mbox{$\topp$} \vdash ({p_i}{g_i})_{i\leq n}
          \!\!\leadsto\! (t_{i j},\BoolB_{ij})_{i\leq n, j\leq m_i}
          \,
          (\forall i\, \forall j)
          \,;
          (\Gamma, x : t_{i j} \vdg \texttt{case}\,x\,\texttt{do}\,(\,\ov{p g \rarr e}\,) : t_{i j}')\!\!
     }
     { \Gamma \vdg \lamPat[]{\,\ov{p g \rarr e}\,} :
          \bigwedge_{i j} (t_{i j} \rarr t_{i j}') }
\]
\endgroup
where $x$ is a fresh variable, $\topp$ is chosen as its initial type (meaning that the argument could be of any type),
and the Boolean flags $\BoolB_{i j}$ are discarded (the exactness analysis is not required).
The Elixir compiler (since v.1.18 where gradual types were introduced) chooses $\dyn$ instead of $\topp$ as initial type, which is the case of a domain containing `$\dyn$' described below.

The extension of the type system with inference is sound in the following sense. Replacing $\lamPat[]{\,\ov{p g \rarr e}\,}$ by $\lambda^{\mathbb{I}}x.\texttt{case\,}x\,(\ov{p g \rarr e})$, where $\mathbb{I}=\{t_{ij}\rarr t'_{ij}\}_{ij}$ is the inferred interface, turns an instance of \RuleRef{typing:inference:infer}{rule (\textsc{infer})} into an instance of \RuleRef{typing:static:lambda}{rule ($\lambda$)} with the same premises, since the domains $t_{ij}$ are static, hence $\mathbb{I}$ is separated, and its variant for a domain containing `$\dyn$', described below, into an instance of \RuleRef{typing:gradual:lambda-dyn}{rule ($\lambda_\dyn$)}. Remark~\ref{app:rem:inference} in Appendix~\ref{app:guardanalysis} gives the details and the condition under which the soundness theorems follow.

In some cases, the domain inferred by the guard analysis for the function (i.e., $\bigvee_{ij}t_{ij}$) may not be precise. When this occurs, the programmer can assist the inference process by providing the domain type: it then suffices to replace this type for $\topp$ in the \RuleRef{typing:inference:infer}{rule (\textsc{infer})} to take into account this information. If this type contains `$\dyn$', so may the accepted types $t_{ij}$, and the conclusion of the rule then uses, in the gradual system, their maximal materializations $\dynSup{t_{ij}}$ as domains, as \RuleRef{typing:gradual:lambda-dyn}{rule ($\lambda_\dyn$)} does for interfaces. For example, if we swap the order of the or-guards in the first clause of \elix{test} in line~\ref{test1}, then the type inferred for the function would be the one in lines~\ref{testty1}--\ref{testty4} but where the second arrow (line~\ref{testty2}) has domain \elix{{:int, ..}} instead of \elix{{:int, term(), ..}}. Although the type checker would produce a warning (because of the use of \RuleRef{typing:static:proj-omega}{(proj$_\omega$)}), this type would accept as input \elix{{:int}}, which fails. This can be avoided if the programmer provides the input type \elix{{tuple(), tuple(), ..} or {boolean()}} to the inference process. This can be done by using partial annotations, as we proposed in \cite[Section 3.2]{castagna2023design}. These are annotations that do not specify the output type, leaving the system to deduce it. In the specific example we would use the following partial annotation:
\begin{minted}{elixir}
!\tp! {{tuple(), tuple(), ..} or {boolean()}} -> _
\end{minted}
More generally, we can suppose that also the multi-clause functions may be annotated with an \emph{optional} interface:
\setlength{\abovedisplayskip}{0pt}
\setlength{\belowdisplayskip}{3pt}
\[
\begin{array}{lrcl}
     \textbf{Expressions} & e &\bnfeq& \cdots \bnfor \lamPat[\mathbb{I}\!]{\,\ov{p g \rarr e}\,}\\
               \textbf{Interfaces} & \mathbb{I} &\bnfeq& \varepsilon\bnfor\{t_i\to t_i'\}_{i=1..n}
               \end{array}
\]
where $\varepsilon$ denotes the empty string. The previous $\lambda$-abstractions of the form $\lambda^\IFace{x}.{e}$ are then syntactic sugar for single clause functions with one pattern variable and no guards, that is, $\lamPat[\mathbb{I}\!]{\,{x\rarr e}\,}$. The typing rules for multi-clause $\lambda$-abstractions with a non-empty interface $\mathbb{I}$ are the same as the ones without interface, except that the domain of the interface is used instead of $\topp$ and the tested types are the ones in the interface rather than those deduced by the guard analysis; as in \RuleRef{typing:static:lambda}{rule ($\lambda$)}, the interface must be separated. As in \RuleRef{typing:gradual:lambda-dyn}{rule ($\lambda_\dyn$)}, in the gradual system the conclusion may instead replace each domain $t_i$ of the interface by $\dynSup{t_i}$, with no such condition.

\subsection{Dynamic propagation with inferred types.}
Inferring static function types for existing code in a dynamic language can disrupt continuity,
as existing code may rely on invariants that are not captured by types.
Furthermore, in a set-theoretic type system, no property guarantees
that a given inferred type is the most general; consider, for example, that the
successor function could be given types $\intTop \rarr \intTop$ but also any
variation of $(2\rarr 3)\land(3 \rarr 4)\land((\intTop{\setminus}(2{\lor} 3))\rarr\intTop)$
using singleton types. While both types are correct and can be related by
subtyping, it is the role of the programmers to choose the one that corresponds
to their intent and to annotate the function accordingly.

Thus, we need to introduce some flexibility so that inferred static types do not
prematurely enforce this choice.
We achieve this by adding a dynamic arrow intersection that points
the full domain (the union of the $t_i$'s) to $\dyn$.
%
%
\begingroup
\setlength{\abovedisplayskip}{2pt}
\setlength{\belowdisplayskip}{3pt}
\[
     \typingrule{\RuleDef{typing:inference:infer-star}{(\textsc{infer}_{\star})}}
     {\!\!
          \Gamma\,;\mbox{$\topp$} \vdash ({p_i}{g_i})_{i\leq n}
          \!\!\leadsto\! (t_{i j},\BoolB_{ij})_{i\leq n, j\leq m_i}
          \,
          (\forall i\, \forall j)\,;
          (\Gamma, x : t_{i j} \vdg \texttt{case}\,x\,\texttt{do}\,(\,\ov{p g \rarr e}\,) : t_{i j}')\!\!
     }
     { \Gamma \vdg \lamPat[]{\,\ov{p g \rarr e}\,} :
          \bigwedge_{i j} (t_{i j} \rarr t_{i j}')
          \land (\bigvee_{i j} t_{ij} \rarr \dyn) }
\]
\endgroup
Now `$\dyn$' gets automatically intersected with each possible return type during
function application. While using \RuleRef{typing:inference:infer-star}{rule (\textsc{infer}$_\star$)} by default appears necessary
when typing a dynamic language, being able to type-check using only \RuleRef{typing:inference:infer}{rule (\textsc{infer})}, with $\vds$ in place of $\vdg$,
gives a stronger type safety guarantee, as eliminating the use of `$\dyn$' during
type-checking (and thus, of gradual rules) allows controlling for explicit errors
(see Theorems~\ref{thm:static-soundness}, \ref{thm:soundness},
\ref{thm:gradual-soundness}).

\smallskip\noindent
\textbf{Multi-arity.} 
We have presented inference for single-arity functions,
but the same principle straightforwardly applies to multi-arity functions presented in Section~\ref{sec:arity}: anonymous
functions become {$\lamPat[]{\,\ov{\ov{p} g \rarr e}}$}, and the current guard analysis
can be repurposed to produce accepted types for each argument by wrapping
these arguments into a tuple pattern.

\section{Featherweight Elixir}\label{sec:fw-elixir}
Throughout this article, we have used Core Elixir as the language for formal definitions and technical discussions. For the reader convenience we summarize the full syntax of Core Elixir in Figure~\ref{fig:full-core-elixir-syntax} and the complete definition of its operational semantics in Appendix~\ref{app:core-elixir}.
\begin{figure}
	\centering
	\begin{grammarfig}
		\textbf{Expressions} & e &\bnfeq& c \bnfor x\bnfor \lamPat[\mathbb{I}\!]{\,\ov{\ov{p} g \rarr e}\,}
		\bnfor\app{e}{\,\ov{e}\,} \bnfor\tuple{\ov{e}} \bnfor \proj{e}{e}
		\bnfor \caseExpr{}
		\bnfor e + e
		\\
		\textbf{Patterns} & p &\bnfeq& c \bnfor x  \bnfor \tuple{\ov{p}}
		\\
		\textbf{Guards}      & g
		&\bnfeq& a \isof \tau \bnfor a = a \bnfor a \,\texttt{!=}\,\, a \bnfor a \,\,\texttt{<}\,\,a \bnfor \GuardAnd{g}{g} \bnfor \GuardOr{g}{g} \\[0mm]
		\textbf{Guard atoms} &a
		&\bnfeq& c\bnfor x\bnfor\proj{a}{a}\bnfor \sizeTup{a} \bnfor \tuple{\ov{a}}\\[0mm]
		\textbf{Test types} & \tau
		&\bnfeq& b \bnfor c \bnfor
		\pc{\ov{\tau}} \bnfor \tuple{\ov{\tau}}
		\bnfor \tau \lor \tau \bnfor \neg \tau \\[1mm]
		\textbf{Base types} & b
		&\bnfeq& \intTop \bnfor \boolTop \bnfor \function \bnfor \function_n \bnfor \tupleTop  \\
		\textbf{Types}      & t
		&\bnfeq& b \bnfor c \bnfor \fun{\ov{t}}{t} \bnfor \tuple{\ov{t}}
		\bnfor \OpenTuple{\ov{t}} \bnfor t \lor t \bnfor \neg \,t \bnfor \dyn \\
        \textbf{Interfaces} & \mathbb{I} &\bnfeq& \varepsilon\bnfor\{\ov{t_i}\to t_i'\}_{i=1..n}
 	\end{grammarfig}
	\caption{Core Elixir}
	\label{fig:full-core-elixir-syntax}
\end{figure}

To bridge the gap between Core Elixir and the concrete Elixir syntax we used in the examples of Section~\ref{sec:walkthrough}, we define and briefly discuss the syntax of \emph{Featherweight Elixir} (FW-Elixir). FW-Elixir is a strict subset of Elixir that covers all the language features discussed in this work, including tuples, anonymous and multi-arity functions, case-expressions with patterns and guards, and more. Figure~\ref{app:elixir-concrete-syntax} presents the formal syntax for FW-Elixir. All the examples provided in the introduction are written in a syntax that is valid for FW-Elixir and, thus, for Elixir as well.

\begin{figure}
	\begin{grammarfig}
		\textbf{Expressions} & \FWE &\bnfeq&
		\FWL \bnfor x \bnfor \FWA\ \EFn{\FWE} \bnfor \EApp{\FWE}{\FWE_1,..,\FWE_n} \bnfor
		\FWE + \FWE
		\\
		&&\bnfor& \ECase{\FWE} \bnfor
		\ETuple{\FWE_1,..,\FWE_n} \bnfor \EElem{\FWE}{\FWE}
		\\
		\textbf{Singletons} & \FWL &\bnfeq& n \bnfor k \bnfor \tuple{\ov{\FWL}}
		\\
		\textbf{Patterns} & \FWP &\bnfeq& \FWL \bnfor x
		\bnfor \tuple{\FWP_1,..,\FWP_n}
		\\
		\textbf{Guards}
		&\FWG &\bnfeq& \FWD \bnfor \FWC \bnfor \gnot{\FWG} \bnfor \gand{\FWG}{\FWG}
		\bnfor \gor{\FWG}{\FWG} \bnfor \FWD \,\,\texttt{==}\,\, \FWD \bnfor \FWD \,\texttt{!=}\,\, \FWD
		\\
		&&\bnfor& \FWD \,\,\texttt{<}\,\,\FWD \bnfor \FWD \,\,\texttt{<=}\,\, \FWD \bnfor \FWD \,\,\texttt{>}\,\,\FWD \bnfor \FWD \,\,\texttt{>=}\,\,\FWD 
		\\
		\textbf{Selectors}
		&\FWD &\bnfeq& \FWL \bnfor x \bnfor \EElem{\FWD}{\FWD} \bnfor \EtupleSize{\FWD}
		\bnfor \tuple{\ov{\FWD}} \\
		\textbf{Checks}
		&\FWC &\bnfeq& \EisInt{\FWD} \bnfor \EisAtom{\FWD} \bnfor \EisTuple{\FWD}
		\\
		&&\bnfor& \EisFun{\FWD} \bnfor \EisFunction{\FWD}{n}\\[2mm]
		\textbf{Base types} & \FWB
		&\bnfeq& \texttt{integer()} \bnfor \texttt{atom()} \bnfor \texttt{function()} \bnfor \texttt{tuple()}  \\
		\textbf{Types}      & \FWT
		&\bnfeq& \FWB\bnfor \FWL \bnfor \fun{\ov{\FWT}}{\FWT} \bnfor \tuple{\ov{\FWT}}
		\bnfor \OpenTuple{\ov{\FWT}} \bnfor \FWT\;\texttt{or}\;\FWT \bnfor \FWT\;\texttt{and}\;\FWT \bnfor \texttt{not}\,\FWT \\
		&&\bnfor& \texttt{none()} \bnfor \texttt{term()} \bnfor \texttt{dynamic()}\\
		\textbf{Annotations} & \FWA
  &\bnfeq& \texttt{\$}\ \FWT \bnfor \varepsilon  
	\end{grammarfig}
	\small(where $x$ ranges over variables, $n$ ranges over integers, and $k$ ranges over atoms)
	\caption{Featherweight Elixir}
	\label{app:elixir-concrete-syntax}
\end{figure}

If we look at the definitions of the expressions in Figure~\ref{fig:full-core-elixir-syntax} and~\ref{app:elixir-concrete-syntax}, there is a clear 1-1 correspondence between them. So it is easy to see that the  \elix{second_strong} function we defined in the introduction (line~\ref{complexguard}) corresponds to the following Core Elixir expression:
\begingroup
\setlength{\abovedisplayskip}{3pt}
\setlength{\belowdisplayskip}{3pt}
\[\lam[\{\pc{\TopType,\intTop}\rarr\intTop\}]{x}{\texttt{case}\,
		{x} \, \texttt{(}\pc{\TopType,\intTop} \rarr \proj{1}{x}\texttt{)}}\]
\endgroup
It is therefore easy to transpose the typing and reduction rules defined for Core Elixir expressions to the expressions of FW-Elixir. The correspondence for guards and patterns of the two languages is instead more nuanced. In some cases this correspondence is clear: for example, it is clear that the second clause of the \elix{test} function in the introduction (line~\ref{test2}) can be expressed in Core Elixir as a branch of a case-expression with pattern $x$ and guard:
\begingroup
\setlength{\abovedisplayskip}{3pt}
\setlength{\belowdisplayskip}{3pt}
\[\GuardOr
	{\texttt{(}x \isof \boolTop\texttt{)}}{\texttt{(}\proj{0}{x}
		= \proj{1}{x}\texttt{)}}\]
\endgroup

But the link between guards and patterns of the two languages is subtler. While Core Elixir type tests are more general than those that can be performed by FW-Elixir guards, the latter look more expressive than Core Elixir guards, since they can use negation which is not present in Core Elixir. We already gave in Section~\ref{sec:walkthrough} the example of the guard \elix{not(tuple_size(x)==0)} (line \ref{first}) that cannot immediately be transposed into a Core Elixir guard, although the equivalent guard \mintinline[escapeinside=::,bgcolor=snow]{elixir}{tuple_size(x)!=0} can. As a different example, consider the following definition of \elix{negate}:
\begin{minted}{elixir}
def negate(x) when not(is_function(x) or is_tuple(x)), do: not x
\end{minted}
\newcommand{\negsp}{\hspace*{-2.5pt}}
If we assume \elix{integer()} and \elix{boolean()} to be the only basic types used in FW-Elixir, then this definition of \elix{negate}  is equivalent to the second clause of \elix{negate} given in line~\ref{neg2} in the introduction:  the type \elix{boolean()} is the complement of \elix{integer()!\negsp! or !\negsp!function()!\negsp! or !\negsp!tuple()} (where the type \elix{function()} denotes the type of all functions) and all values of type \elix{integer()} are captured by the first clause of \elix{negate}. 

It is not immediately obvious how to systematically express these negated guards in Core Elixir, although it must be possible since there are equivalent definitions that do not use negation.
To adapt the guard and pattern analysis of Core Elixir to FW-Elixir (and thus to Elixir), we must demonstrate that FW-Elixir's apparent expressiveness advantage is illusory—its guards can be compiled into Core Elixir guards. The formal definition of this compilation requires two mutually recursive functions. The compilation pushes all negations of a guard into its leaves. For example, the Elixir guard \elix{not(tuple_size(x)==0)} is compiled into the Core Elixir guard $\texttt{size}(x)\texttt{!=}0$, while \elix{not(is_function(x) or is_tuple(x))} becomes $(x\isof\neg\function)\texttt{ and } (x\isof\neg\tupleTop)$. Likewise, it encodes guard relations present only in FW-Elixir into the relations of Core Elixir (e.g., \elix{x >= y} is encoded as $(y\;\texttt{<}\;x)\;\texttt{or}\;(x\;\texttt{=}\;y)$).

The complete definition of the compilation is given in Appendix~\ref{app:guard-compilation}. As there is a one-to-one correspondence between the expressions of Core Elixir and FW-Elixir, so there is the same correspondence on their values. The compilation preserves the guard semantics, that is, an FW-Elixir value succeeds on a guard if and only if the corresponding Core Elixir value succeeds on the compilation  of that guard. The only delicate case is that of bare selectors: in Elixir, a selector used as a guard, such as \elix{elem(x,0)} in \elix{elem(x,0) or is_tuple(x)}, must evaluate to a Boolean when it is an operand of \elix{and}, \elix{or}, or \elix{not}, and otherwise the whole guard fails (in our example, on every tuple whose first element is not a Boolean); the compilation turns such a selector into a Core Elixir guard that fails when the selector is not a Boolean (see the cases for $\Tg(\FWD)$ and $\Ng(\FWD)$ in Figures~\ref{fig:guard-compilation-1} and \ref{fig:guard-compilation-2} in the appendix). The property is proved by induction on the guard, for a stronger statement that relates the guards that evaluate to $\smalltrue$, those that evaluate to $\smallfalse$, and those that error or yield a non-Boolean value (Appendix~\ref{app:guard-compilation}). This compilation is systematically applied to Elixir guards by our implementation of the Elixir type checker---outlined in the next section---to perform its guard analysis.

\section{Implementation}\label{sec:implem}
\newcommand{\tmin}[1]{\infDyn{#1}}
\newcommand{\tmax}[1]{\supDyn{#1}}
\newcommand{\functionTop}{\function}


As previously mentioned, all features presented in this paper have been implemented in the Elixir compiler at varying degrees of maturity. This integration started with the version 1.17 of Elixir and will continue in the foreseeable future according to a roadmap constantly updated and available in the Elixir documentation page on types~\cite{elixirtypes}. While the previous sections provide a high-level specification of the type system for Elixir, this section offers an overview of some key implementation aspects: how data structures are specifically tailored to represent each Elixir type constructor; how the typing rules from previous sections are realized in practice, sometimes with restrictions to align with Elixir developers' intuitions; and how the type system implementation impacts the broader Elixir development ecosystem.

\subsection{Data Structures}\label{sec:data-structures}
The core of our type system's implementation draws inspiration from the implementation of CDuce~\cite{cduce,BCF03}. 

The data-structure representing a type is called \emph{Descr} (for ``Descriptor''). It is essentially a record, mapping each atomic 
type to its representation. For this presentation, we suppose to have just four atomic types---\texttt{integer}, \texttt{atom}, \texttt{tuple}, and \texttt{function}---for which we use four different representations:\footnote{
In the current implementation of Elixir, there are also \felix{binary()}, \felix{float()}, \felix{pid()}, \felix{port()}, \felix{reference()}, and \felix{list()}, with similar representations.}
\begin{enumerate}
    \item \texttt{integer} is the base type representing all integers. While in
          Featherweight Elixir this type has non-empty subtypes---e.g., the
          singleton type \texttt{42}---in Elixir we currently did not implement
          singleton types of integers, therefore \texttt{integer} is
          \emph{indivisible} in the implemented type language: no non-empty
          proper subtype of \texttt{integer} is represented. Thus, every
          representable type in Elixir either contains all integer values or
          none.
          This is in contrast to the \texttt{atom} type (see next item), which does support singleton subtypes and more granular distinctions.
          The indivisible integer type will therefore be represented by a simple Boolean indicating its presence or not.
    \item \texttt{atom} is the base type representing all atoms. The case for
         atoms is different from the one of integers, since it is a common Elixir
         programming practice to use atoms in tuples as flags that determine the
         content of the rest of the tuple. We have seen an example with the
         \elix{:int} atom in Section~\ref{sub:ga}, but a better example is given
         by \emph{GenServer}---one of the most used modules in Elixir---whose
         options are passed as pairs in which the first element is an atom that indicates
         the option name and determines the type of the second element of the pair:
\begin{minted}{elixir}
!\tp! !\ty! option() = {:debug, debug()} or 
                  {:name, name()} or
                  {:timeout or :hibernate_after, timeout()} or 
                  {:spawn_opt, [Process.spawn_opt()]} 
\end{minted}
        In view of the widespread use of this programming pattern for atoms, our implementation in Elixir included from the very beginning atom singleton types, meaning that \texttt{atom}---unlike \texttt{integer}---is not an indivisible type. However, every non-empty representable subtype of \texttt{atom} is always of one of the following two forms:
        \begin{itemize}
            \item[-] it is a finite union of atoms $\{a_1,\ldots,a_n\}$, or
            \item[-] it is all the atoms minus a (possibly empty) finite set of atoms $\{a_1,\ldots,a_n\}$
        \end{itemize}
        Thus, we can represent any atom-only type expressible in this grammar by a pair in which the first element is a flag that indicates whether the type is a finite or co-finite union of atoms, that is, $(\texttt{\textcolor{purple}{:union}}\;,\,\{a_1,\ldots,a_n\})$ or $(\texttt{\textcolor{purple}{:minus}}\;,\, \{a_1,\ldots,a_n\})$, and where the base type \texttt{atom} corresponds to $(\texttt{\textcolor{purple}{:minus}}\;,\, \{\;\})$.
    \item \texttt{tuple} is the base type representing all tuples. Any representable type containing only tuples (i.e., any subtype of \texttt{tuple}) can always be put in disjunctive normal form. Furthermore,
        it is always possible to eliminate intersections
        of tuple types (by intersecting the components), and of their negations (by distributing them over disjunctions: $\tuple{t_1,t_2}\wedge\neg\tuple{s_1,s_2} =  \tuple{t_1{\wedge} s_1, t_2\wedge\neg s_2}\vee\tuple{t_1{\wedge}\neg s_1,t_2}$, and likewise for larger arities). Therefore, any representable tuple-only type can be represented as a finite union of tuple types. Thus, we can store tuple types as a list
        of tuple atom types (of different arities): 
        $[\tuple{t_{i 1},\ldots,t_{i n_i}}]_{i \in \{1,..,n\}}$
        where $n_i$ is the arity of the $i$-th tuple type, with $n_i{\not=}n_j$ for $i{\not=}j$. Finally, every tuple atom type is equipped with a Boolean flag that indicates whether it is an open tuple type (i.e., of the form $\OpenTuple{\bar t}$) or a closed tuple type (i.e., of the form $\tuple{\bar t}$).
    \item \texttt{function} is the base type representing all function values. In semantic subtyping every representable subtype of \texttt{function} is (equivalent to)
        a propositional logic formula where literals
        are the base types $\function_n$ of Section~\ref{sec:arity}, atom arrow types $(t_1,\ldots,t_n)\rarr t$,
        or their negations. We can store such a formula using a binary decision diagram (BDD), that is a binary tree where each node is labeled
\begin{figure}[t]
\begin{minipage}{.46\textwidth}
\centering
\begin{tikzpicture}
    \node (root) at (0,0) {$(\intTop\rarr\intTop)$};
    \node (bool-node) at (-1,-1) {$(\boolTop\rarr\boolTop)$};
    \node (zero1) at (1,-1) {$0$};
    \node (one) at (-2,-2) {$1$};
    \node (zero2) at (0,-2) {$0$};
    
    \draw[->] (root) -- (zero1) node[midway,right] {$\neg$};
    \draw[->] (root) -- (bool-node) node[midway,left] {};
    \draw[->] (bool-node) -- (one) node[midway,left] {};
    \draw[->] (bool-node) -- (zero2) node[midway,right] {$\neg$};
\end{tikzpicture}
\caption{BDD representation of a function type}%
\label{fig:bdd-representation}
\end{minipage}
\begin{minipage}{.53\textwidth}
\begin{center}
\begin{tabular}{l|r}
$\intTop$ & $\smalltrue$ \\
\midrule
$\atomTop$ & $(\texttt{:union}, \{\texttt{:foo}, \texttt{:bar}\})$ \\
\midrule

$\tupleTop$ & $[\tuple{\intTop, \neg\,\intTop}, \tuple{\neg\,\intTop, \topp}]$ \\
\midrule
$\function$ & as in Figure~\ref{fig:bdd-representation} \\
\end{tabular}
\end{center}
\caption{Example of a Descr}%
\label{tab:descr-example}
\end{minipage}
\end{figure}
 by a type $\function_n$ or an atom arrow type, and the leaves are labeled by either $0$ or $1$. The formula represented by a BDD is the union of all paths from the root to the $1$-labeled leaves, where each path corresponds to the conjunction of the literals along the path, either positively (if taking a left path after the literal), or negatively (if taking a right path after the literal).
        For instance, type $(\intTop\rarr\intTop)\land(\boolTop\rarr\boolTop)$ will be represented by the BDD given in Figure~\ref{fig:bdd-representation} where there is just one path leading to a $1$-labeled leaf.

        Strong arrows do not require a different representation: it suffices to
        tag the arrow literals that label the BDD nodes. Ordinary arrow atoms of the form $(t_1,\ldots,t_n)\rarr t$ are represented by the triple \texttt{(\textcolor{purple}{:weak}\,,\,$[t_1,\ldots,t_n]$\,,\,$t$)}, and their strong version by the triple \texttt{(\textcolor{purple}{:strong}\,,\,$[t_1,\ldots,t_n]$\,,\,$t$)}.  The BDD
        still represents a DNF of such atoms, and a path to a $1$-labeled leaf is
        extracted as a conjunction containing types $\function_n$, ordinary arrows, strong arrows, and
        their negations. The treatment specific to strong arrows happens after this
        extraction. First, all positive strong arrows in the conjunction are
        normalized to a single strong arrow by the simplification
        $\bigwedge_{j \in P} (t_j \rarr s_j)^{\star} =
        (\bigvee_{j \in P} t_j\rarr \bigwedge_{j \in P} s_j)^{\star}$ (cf.\ Lemma~\ref{lem:strong-intersection} in Appendix~\ref{sec:strong-subtyping}) and 
        the resulting mixed weak/strong subtyping checks are then solved using the equivalences of their semantic interpretations (see the
        algorithm of Theorem~\ref{thm:algo-strong}, Theorem~\ref{thm:strong-no-strong-left} when the conjunction has no positive strong arrow, and the equivalences
        of Theorems~\ref{thm:strong-arrow-types} and~\ref{thm:strong-strong}).
        
\end{enumerate}
The type represented by a Descr is the union of all the atomic types present in
its fields. For instance, the type $\intTop \lor (\texttt{:foo} \texttt{ or }
\texttt{:bar}) \lor (\tuple{\topp,\topp}\smallsetminus\tuple{\intTop,\intTop})
\lor ((\intTop\rarr\intTop)\land(\boolTop\rarr\boolTop))$ will be represented by
its four components described in Figure~\ref{tab:descr-example},  where we
distributed the negation to obtain the tuple type. Note also that the negation of any atomic type is represented by a descriptor that saturates the other atomic types: for instance, $\neg\intTop\land\neg\atomTop$ is represented by a descriptor with the \texttt{integer} field set to \texttt{false}, the \texttt{atom} field set to \texttt{(\texttt{:union}, \{\})}, and the \texttt{tuple} and \texttt{fun} fields set to their top elements, which represent the types of all tuples and of all functions.

This structure provides an efficient way to perform complex type operations, including subtype checking and type inference. In particular, since all of these atomic types are mutually
disjoint, the set operations can be performed separately on each atomic type, allowing for a modular implementation. 
This modularity, crucial for the performance of the type-checker, is preserved when we add gradual types. This is done by using the property stated in Section~\ref{sec:gradual-typing} that every gradual type $t$ is equivalent to  $\dynInf{t}\lor (\dyn \land \dynSup{t})$ and thus can be represented by two static types, that is two Descr's as the above, corresponding to $\dynInf{t}$ and $\dynSup{t}$. Notice that the representation of a gradual type by two static types is not unique. For instance, $\intTop\lor(\dyn\land\boolTop)$ is already in the sought form. But by representing this type by its two materialization extrema, that is as $\intTop\lor(\dyn\land(\intTop\lor\boolTop))$, we have that all the type operations can be done modularly on each separate descriptor, since they maintain the invariance that the two descriptors are in an order relation corresponding to minimal and maximal materialization. Static types are then represented by two identical (shared) descriptors.

\newcommand{\gdom}[1]{\texttt{grad\_dom}(#1)}


\subsection{Typing Rules Specificities}
The typing rules presented in the previous sections are implemented in the Elixir compiler, with some specificities that reflect some practical needs. This implies that some rules are implemented with further additional conditions, or by using type operators more refined than those given by the theoretical system.

This is particularly the case for the rules that are used to type function applications which are specialized to take into account the strict evaluation semantics of Elixir. We start by briefly describing how the application \RuleRef{typing:static:app}{rule (app)} is implemented and then examine the consequences on the \RuleRef{typing:gradual:app-dyn}{rule (app$_\dyn$)}.

\subsubsection{Implementation of \RuleRef{typing:static:app}{(app)}}
The declarative system we presented in Figure~\ref{fig:static-rules} does \emph{not} describe a type-checking algorithm. For instance, the \RuleRef{typing:static:app}{(app)} rule assumes that the type of the function is an arrow type whose domain is exactly the type of the argument. But when typing an expression the typing algorithm always returns the most precise type it can deduce for that expression. For functions this rarely is an arrow whose domain is the type of the argument: in general this type is a union of intersections of arrows and their negations, and usually just an intersection.
To implement the \RuleRef{typing:static:app}{(app)} rule, the algorithm must subsume the type of the function (whatever it is) to the \emph{smallest arrow type whose domain is the argument type}: the codomain of this arrow is the type of the application.
 
For instance, when applying a function of type $(\intTop\rarr\intTop)\land(\atom\rarr\atom)\land(\tupleTop\rarr\tupleTop)$ to an argument of type $(\intTop\lor\atom)$, the type algorithm  subsumes the type of the function using the following subtyping relation
\[(\intTop\rarr\intTop)\land(\atom\rarr\atom)\land(\tupleTop\rarr\tupleTop) \leq (\intTop\lor\atom)\rarr(\intTop\lor\atom)\]
and deduces $(\intTop\lor\atom)$ as the type of the application. To make this subsumption possible, two conditions must be satisfied: the function expression must have an arrow type---i.e., a subtype of $\bott\rarr\topp$, the largest arrow type---and the argument type must be a subtype of the domain of the function type. So in a system with semantic subtyping the \RuleRef{typing:static:app}{rule (app)} for an application $\app{e}{e'}$ is implemented in three steps:
\begin{enumerate}
    \item Deduce the type $t$ of the function expression $e$ and check that $t \leq \bott\rarr\topp$.
    \item Deduce the type $s$ of the argument expression $e'$ and check that $s\leq \dom{t}$, where $\dom{t}$ is the domain of the function type $t$.
    \item Compute the codomain of the smallest arrow type supertype of $t$ and whose domain is $s$; this is denoted $t \circ s$ and defined as $\min\{u\bnfor t\leq s\to u\}$.
\end{enumerate}
The effective computation of the domain and result application operators are standard in semantic subtyping and can be found in the literature (see in particular~\cite{lmcs:6098} for a detailed explanation of the application operator). The domain operator has been extended to gradual types by Lanvin~\cite{lanvin2021semantic} (Section~\ref{sec:implem-app-dyn}).

For instance if $t$ is an arrow type, that is, if $t\leq \bott\rarr\topp$, then it is equivalent to a type in the following form, where every clause is non-empty and every $P_i$ is non-empty 
    \(
    t \simeq \bigvee_{i \in I} \left(\bigwedge_{s_i\to t_i \in P_i}\! s_i\to t_i \wedge \bigwedge_{s_i\to t_i \in N_i} \neg(s_i\to t_i)\right)
    \).
In the case of static types the domain of $t$ is then defined as
    $\dom{t} \eqdef \bigwedge_{i \in I} \bigvee_{s_i \to t_i \in P_i} s_i
    $. In a nutshell, the domain of an intersection of arrows is the union of the domains of the arrows (negated arrows do not play any role\footnote{Although natural, this is not immediate. Knowing that a function has type $\intTop \to \boolTop$ is informative for domain computation: it guarantees that the function behaves well on every $\intTop$ argument, yielding a Boolean result. By contrast, knowing that a function has type $\neg(\intTop\to\boolTop)$ contributes nothing to the domain, since this type provides no guarantee about the function's behavior on any specific argument---it merely witnesses the existence of at least one integer 
    for which the function either returns a non-Boolean value or produces the error $\Omega$ (cf.\ Section~\ref{subsec:multi-arity}). This is why negated arrows play no role in domain computation, even when their input type is a singleton: $\neg(\textsf{42}\to\intTop)$ is not a subtype of $\textsf{42} \to \neg\intTop$, since membership in the negated arrow type does not entail that the function yields a non-integer result for $\textsf{42}$---the function may equally well map $\textsf{42}$ to $\Omega$. }), and the domain of a union is the intersection of the domains. 

For gradual types, the domain of \RuleRef{typing:static:app}{rule (app)} must be static (Section~\ref{sec:gradual-typing}). For a static type $d$, we have $s\leq d$ if and only if $\dynSup{s}\leq d$, so the smallest static domain that accepts an argument of type $s$ is its maximal materialization $\dynSup{s}$. Moreover, $t\leq d\rarr u$ holds if and only if $\dynInf{t}\leq d\rarr\dynInf{u}$ and $\dynSup{t}\leq d\rarr\dynSup{u}$. Therefore, step~2 checks that $\dynSup{s}\leq\dom{\dynSup{t}}$, and step~3 computes the least $u$ such that $t\leq\dynSup{s}\rarr u$, that is,
\[
t \circ s = (\dynInf{t} \circ \dynSup{s}) \lor (\dyn \land (\dynSup{t} \circ \dynSup{s}))
\]
where the two applications on the right-hand side are those of static types. This is the application operator of the Elixir compiler~\cite[Section~13.2]{duboc-phd}. The static part of the result, $\dynInf{t}\circ\dynSup{s}$, comes only from the static part $\dynInf{t}$ of the type of the function, applied to all the values that the argument may produce at run time, that is, to $\dynSup{s}$; what comes from $\dynSup{t}$ is intersected with `$\dyn$'. For instance, let
\[
t_0 = \dyn \land (\intTop \rarr \intTop) \land (\boolTop \rarr \boolTop)
\]
so that $\dynInf{t_0}\simeq\bott$ and $\dynSup{t_0}\simeq(\intTop \rarr \intTop) \land (\boolTop \rarr \boolTop)$. The application of a function of type $t_0$ to an argument of type $\intTop$, or of type $\dyn\land\intTop$, has type $\dyn\land\intTop$. Its application to an argument of type `$\dyn$' is not typed by \RuleRef{typing:static:app}{rule (app)}, since $\dynSup{\dyn}=\topp\not\leq\intTop\lor\boolTop$, but by \RuleRef{typing:gradual:app-dyn}{rule (app$_\dyn$)} (Section~\ref{sec:implem-app-dyn}). Without the restriction to static domains, \RuleRef{typing:static:app}{rule (app)} would give it type $\intTop$, since $t_0\leq(\intTop\lor\dyn)\rarr\intTop$ and $\dyn\leq\intTop\lor\dyn$, although a function of type $t_0$ may return a Boolean on such an argument (Section~\ref{sec:gradual-typing}). For the functions whose types the compiler infers, the restriction has no cost: the compiler starts from `$\dyn$' (Section~\ref{sec:inference}), so the accepted types may be gradual, but the domains of the inferred arrows are their maximal materializations, as in \RuleRef{typing:gradual:lambda-dyn}{rule ($\lambda_\dyn$)}, hence static types.

\subsubsection{Implementation of \RuleRef{typing:gradual:app-dyn}{(app$_\dyn$)}}\label{sec:implem-app-dyn}
Consider the application of a function of type $\boolTop\to\boolTop$ to an argument of type $\dyn\land\intTop$. Intuitively, we want this application to be rejected: the function works only with Boolean arguments while the type $\dyn\land\intTop$ indicates that the argument can only evaluate to integer values. The argument type $\dyn\land\intTop$ is not uncommon, since the presence of the dynamic type may result from a dynamic propagation for enhanced flexibility (to allow using such an argument where a strict subtype of \intTop{} is expected). However, in this case the flexibility is excessive: since $\dyn\land\intTop$ can materialize to $\bott$, we can apply \RuleRef{typing:gradual:app-dyn}{rule (app$_\dyn$)} and deduce the type $\dyn$ for the application, instead of rejecting it. The logic of this deduction is that if the argument materializes to $\bott$ (i.e., the argument is a diverging expression), then the application is sound: which is correct, since the application will never return. However, accepting an application as sound solely because its argument—and thus the entire application—\emph{might} diverge is nonsensical: if the argument actually diverges, then we do not care what type-checker deduced. Therefore, in practice we implement the \RuleRef{typing:gradual:app-dyn}{rule (app$_\dyn$)} to reject applications when the only solution is to materialize the argument type to $\bott$. We achieve this by introducing a \emph{compatibility} relation, which is a restricted version of the consistent subtyping relation, and using it in our implementation of \RuleRef{typing:gradual:app-dyn}{(app$_\dyn$)} instead of the usual consistent subtyping relation. This yields a stricter and, consequently, still sound typing of gradual function applications.

Recall that consistent subtyping, denoted by $\consist$, is a relation between two types $t_1$ and $t_2$ such that if $t_1 \consist t_2$, then there exist $t_1', t_2'$ where $t_1 \materialize t'_1$, $t_2 \materialize t_2'$, and $t_1' \leq t_2'$. When we add the restriction that $t_1'$ cannot be empty, we obtain the compatibility relation, denoted by $t_1 \sqsubseteq_c t_2$.
Formally, \emph{compatibility} is defined as follows:
    \begin{equation}\label{eq:consub}
         t_1 \sqsubseteq_c t_2 \text{ iff } (\text{if } \tmin{t_1} \leq \bott \text{ then } (\tmax{t_1}\land \tmax{t_2} \not\leq \bott) \text{ else } \tmin{t_1} \leq \tmax{t_2})
    \text{ or } (t_1 \simeq \bott \land t_2 \simeq \bott) 
    \end{equation}
Compatibility clearly implies consistent subtyping: if $t_1$ is compatible with $t_2$, then either the first case of the conditional holds (in which case $\tmin{t_1}$ is empty, so $\tmin{t_1} \leq \tmax{t_2}$), or the second case holds, which directly corresponds to consistent subtyping. The final or condition in the definition \eqref{eq:consub} ensures that the empty type is compatible with itself, which is necessary for subtyping to imply compatibility. In practice, this means that \RuleRef{typing:gradual:app-dyn}{(app$_\dyn$)} accepts a materialization of the argument type to $\bott$ only if the function explicitly expects arguments of type $\bott$. While defining functions with empty domains is mildly interesting in Elixir, this may be crucial in other dynamic languages. For instance, \texttt{ty}---a type-checker for Python developed by Astral~\cite{ty}---uses a function \texttt{assert\_never()} that expects an argument of type $\bott$ to check exhaustiveness of nested multiple if/elif commands: the checked expression is passed in the last else-branch to \texttt{assert\_never()} to statically verify that its type narrowed to the empty type~\cite{meyer-pc}.

As for \RuleRef{typing:static:app}{rule (app)}, the implementation of \RuleRef{typing:gradual:app-dyn}{rule (app$_\dyn$)} checks the type of the argument against the domain of the type of the function, but with compatibility instead of subtyping, and with the domain of a gradual type, as defined by Lanvin~\cite{lanvin2021semantic} and extended to types whose maximal materialization contains values other than functions~\cite{duboc-phd}:
\[
\dom{t} \eqdef \dom{\tmax{t}\land(\bott\rarr\topp)} \lor (\dyn \land \dom{\tmin{t}})
\]
for every gradual type $t$ such that $\tmin{t}\leq\bott\rarr\topp$. The domains on the right-hand side are those of static types defined in the previous section: in particular, $\dom{\bott}=\topp$, since $\bott$ is the union of no clause, and $\dom{\bott\rarr\topp}=\bott$. The minimal materialization of $\dom{t}$, that is $\dom{\tmax{t}\land(\bott\rarr\topp)}$, contains the arguments accepted by all the functions of type $\tmax{t}$, and its maximal materialization, $\dom{\tmin{t}}$, the arguments accepted by all the functions of type $\tmin{t}$ (since $\tmin{t}\leq\tmax{t}\land(\bott\rarr\topp)$, the former is contained in the latter). For instance, the domain of $\dyn \land (\intTop \rarr \intTop)$ is $\intTop \lor \dyn$: integers are accepted for sure, and other values may be accepted at run time, since `$\dyn$' may materialize into another arrow type, for instance into $\boolTop \rarr \boolTop$. Likewise, the domain of $t_0$ is $\intTop\lor\boolTop\lor\dyn$, and the domain of `$\dyn$' is `$\dyn$'.

The condition implemented for \RuleRef{typing:gradual:app-dyn}{rule (app$_\dyn$)} is that the argument type $t_2$ must be compatible with the domain of the function type $t_1$, that is, $t_2 \sqsubseteq_c \dom{t_1}$. Since the maximal materialization of $\dom{t_1}$ is $\dom{\tmin{t_1}}$, when $\tmin{t_2}$ is not empty this is the side condition $t_1\consist t_2\rarr\topp$ of the rule, that is, $\tmin{t_2}\leq\dom{\tmin{t_1}}$; when $\tmin{t_2}$ is empty, compatibility further requires that $\tmax{t_2}$ and $\dom{\tmin{t_1}}$ intersect, or that $t_2$ and $\dom{t_1}$ be both empty, while the side condition reduces to $\tmin{t_1}\leq\bott\rarr\topp$. So compatibility implies the side condition in every case. Safety is preserved since the application is typed by a supertype of $\dyn$ (the type deduced by the \RuleRef{typing:gradual:app-dyn}{(app$_\dyn$) rule}), which is always safe. The implementation of \RuleRef{typing:gradual:app-star}{rule (app$_\star$)} checks its side condition $t_2\consist t_1$ by compatibility too, that is, $t_2 \sqsubseteq_c t_1$.

\subsection{Elixir Integration Status.}\label{sec:elixir-integration}
The core analyses presented in this work have been progressively implemented in
Elixir since version 1.17. Version 1.17 (June 2024~\cite{elixir117}) introduced set-theoretic types for atoms, maps, structs and the indivisible base types, inferring types from patterns within a single function and using them only to emit compile-time warnings. Version 1.18 (December 2024~\cite{elixir118}) added tuples and lists, made gradual typing available through \elix{dynamic()}, type-checked local and remote calls (except in \elix{for}, \elix{with} and closures), and performed gradual inference of patterns and return types, without yet inferring guards. Version 1.19 (October 2025~\cite{elixir119}) added type inference and checking of anonymous functions and function captures, and type checking of the dispatch and implementation of \emph{protocols} (akin to Haskell type classes). In v1.19 type inference for functions is refined
to take partially into account the types of the operators the functions use.

Elixir v1.20, released on June 3, 2026~\cite{elixir120}, completes the first
milestone of the integration: it performs type inference and gradual type
checking of every Elixir program, without any type annotation. It implements the
full guard analysis described in Section~\ref{sec:guard-analysis}, including
precision information and the computation of possibly and surely accepted types
for pattern/guard pairs, and uses this analysis to infer the intersection types
for multi-clause functions described in Section~\ref{sec:inference}. This
milestone still requires no modification to the syntax of Elixir. As of October
2026, v1.21 is in development and scheduled for November 2026; v1.21 and v1.22
(planned for May 2027) focus on the performance of the type checker and on
recursive and parametric types, which are needed to type standard library
modules such as \elix{Enum}~\cite{elixirroadmap26}. The first modifications of the
syntax will come with the subsequent milestones: explicit annotations for
\emph{structs} (which are named maps with a predefined set of keys, used as data
containers) and then the \elix{!\tp!}-prefixed function type annotations
demonstrated in our examples. The logic that powers annotated functions
is already implemented, and we use a branch of the compiler that enables them
to test the \emph{If-T} benchmark in Section~\ref{if-t-benchmark}.

The next steps of the integration of
the type system into Elixir are detailed in a regularly updated roadmap in the
Elixir documentation~\cite{elixirtypes}. Each release follows an approximately
six-month development cycle.

We used release candidate version 1.18-rc0 in September 2024 to perform a first assessment of the overhead introduced by our typechecker on five substantial and extensively tested codebases: Remote is one of the largest Elixir codebases with over a million lines of code (this is a conservative estimation, only the exact number of modules is known); Credo, Livebook, and Phoenix are among the most popular Elixir packages; and Hex is the package manager for the Elixir ecosystem.

\smallskip
\begin{center}
\begin{tabular}{|l|r|r|r|r|r|r|}
\hline\rowcolor{violet2}
Codebase & \multicolumn{1}{c|}{LoC} & \multicolumn{1}{c|}{Files} & Modules & \makecell{Type Checking\\ Time} & \makecell{Total Compilation\\ Time} & \multicolumn{1}{c|}{\%}\\[-1pt]
\hline
Remote & $>$1,000,000 & $>$10,000 & 18,059 & 11.116 s & 707.598 s  & 1.6\\[-1.5pt]
\rowcolor{violet1}
Livebook & 61093 & 254 & 299 & 0.177 s & 4.112 s&4.3\\[-1.5pt]
Credo & 29181 & 252 & 264 & 0.059 s & 1.305 s &4.5\\[-1.5pt]
\rowcolor{violet1}
Phoenix &  21389& 71& 88 &  0.049 s& 0.525 s&9.3\\[-1.5pt]
Hex & 15632 & 196 & 241 & 0.091 s & 1.339 s  &6.8\\[-1pt]

\hline
\end{tabular}
\end{center}
\smallskip

\noindent
The ``type-checking time'' includes type-checking, checking for deprecated APIs, and verifying function definitions. These overheads are comparable to previously existing ad hoc checks embedded within the Elixir compiler, which have since been replaced by our type system. Note that type-checking essentially consists of checking subtyping relations and that the complexity of subtype checking for arrow types is not greater than that for tuple types or map types (included in v1.18). Thus, although the v1.18 implementation of gradual set-theoretic types did not include arrow types, it already provided a precise assessment of the performance of the monomorphic system presented here, demonstrating that our implementation scales effectively, handling large codebases with minimal performance impact.

These results were confirmed by repeating the evaluation about a year later with version 1.19, released on October 16, 2025~\cite{elixir119}, which adds function types, performs function type inference, includes protocol checking, and performs a more complete analysis of map types. We computed the same metrics but on the codebases as they had evolved (since in some cases with closed source code we were given access only to the latest version). Both the general compilation time and the type-checking time were significantly reduced due to improvements in both the general Elixir compiler and in the implementation of the type system, despite the additional type checks implemented in it.

\smallskip
\begin{center}
\begin{tabular}{|l|r|r|r|r|r|r|}
\hline\rowcolor{violet2}
Codebase & \multicolumn{1}{c|}{LoC} & \multicolumn{1}{c|}{Files} & Modules & \makecell{Type Checking\\ Time} & \makecell{Total Compilation\\ Time} & \multicolumn{1}{c|}{\%}\\[-1pt]
\hline
Remote & $>$1,000,000 & $>$10,000 & 24292 & 19.476 s &  228.801 s & 8.5\\[-1.5pt]
\rowcolor{violet1}
Livebook & 61,170 & 256 & 258 & 0.102 s & 3.051 s & 3.3\\[-1.5pt]
Credo & 28,493 & 251 & 314 & 0.027 s & 1.095 s & 2.4\\[-1.5pt] 
\rowcolor{violet1}
Phoenix & 22,497 & 74 & 110 & 0.029 s & 0.461 s & 6.3\\[-1.5pt]
Hex & 15,476 & 196 & 207 & 0.065 s & 1.232 s & 5.3\\[-1.5pt]

\hline
\end{tabular}
\end{center}

\smallskip

\noindent
The only exception is the Remote codebase, which has grown significantly since the previous evaluation, with a 35\% increase in the number of modules. The type-checking time for Remote has increased by 75\% compared to the previous evaluation, and now it takes nearly 20 seconds to type-check 24 thousand modules. This, combined with the  substantial improvement of the compiler (which reduced the total compilation time by 68\% despite the increased number of modules), explains the much higher percentage of type-checking time relative to total compilation time. Remarkably, the new typing features introduced in v1.19 found a type error in the Remote codebase, that went undetected by the previous versions. Specifically, the type-checker detected a call of the \elix{String.Chars} protocol that may be called with a tuple argument in an error branch of some logger. Since the \elix{String.Chars} protocol does not provide an implementation for tuples, this would result in a runtime failure if the error branch is ever executed. This error was undetected by tests because the calling of an error branch is extremely uncommon, which explains why it slipped through: classic needle in a haystack.

The implementation of our type system has been distributed with Elixir since
version 1.17, released in June 2024. Feedback from developers has been positive
across versions 1.17, 1.18, and 1.19. Version 1.20 and its release candidates
have received similarly positive feedback while exercising a substantially more
complete inference engine. Particularly appreciated was the fact that the type
system does not require any syntactic change to Elixir, and it was able to
uncover previously hidden errors without introducing significant compilation overhead and
without requiring Elixir developers to explicitly add type annotations. For
existing codebases with reasonable test coverage, most type system reports came
from uncovering dead code---code which won't ever be executed---but also a few
actual type mismatch bugs. The type system has already found issues in major
projects like Phoenix, Livebook, Postgrex, and Flame, particularly identifying
unused function clauses and unreachable code paths. Some bugs had persisted
undetected in production code for over two years.
For instance, our type system found a bug in the Phoenix framework---which is used by over 14,600 public websites---where an exception handler directly accessed the \elix{message} field of an exception value. Since not every exception struct defines this field, the error branch could itself fail while reporting the original error; the fix replaced the field access with \elix{Exception.message/1}~\cite{phoenix-bug-fix}. Our system already found several dozens of lines of dead code in Postgrex, Flame, and Phoenix LiveView, as we detail in the next section. Although these few examples may seem negligible with respect to the number of lines examined, it is important to note that Elixir projects in general, and the cited libraries in particular, have extensive test coverage, and that these are bugs found in \emph{production code} that has already been merged and passed through testing, code reviews, sometimes even Quality Assurance (QA) processes---code that has been running in production for months or even years.



\section{Empirical Evaluation}\label{sec:eval}
How precise is our type system? Does it produce types precise enough to be useful? How does it compare with other gradual type systems in terms of precision? Answering these questions is not always straightforward. In our case the strategy of integration into Elixir that we have followed, which is incremental and conservative, brings two further hurdles. First, we are postponing the implementation of type annotations since we want to focus on implementing types (and making developers get used to them) without any syntactic modification to the language. This means that we cannot use type annotations as a way to measure the precision of our type system, either by mutating them to see whether type errors are detected, or by removing dynamics to see how precise types can be: these are two techniques that are commonly used by the small body of work on measuring gradual typing empirically~\cite{TunnellWilsonGreenmanPombrioKrishnamurthi18,LGFD21,LGFD23,RakAmnouykitMcCrevanMilanovaHirzelDolby20,NsoforGreenman25,GuoGreenman25}. Second, our type system is being integrated in a language with hundreds of millions of lines of code used in production, and therefore types are deployed in a very conservative way, using highly dynamic---thus, imprecise---types so that false positives in existing well-tested code are nearly nonexistent. This means that the current integration strategy of our type system yields, by design, not very precise types (in terms of the ``type precision'' relation defined in Section \ref{sec:gradual-typing}) and in any case much less precise than if we had implemented type annotations.

Empirical studies of gradual typing make precision observable by choosing an external reference point. Corpus studies of Python annotations measure how often programmers use static annotations, where they fall back to dynamic types, and which patterns or checker limitations explain those fallbacks~\cite{RakAmnouykitMcCrevanMilanovaHirzelDolby20,NsoforGreenman25}. Benchmark studies instead use curated positive and negative examples: precision is the ability to accept programs justified by runtime tests and reject close variants where the tests do not justify the operation~\cite{GuoGreenman25}. A third family measures programmer-facing consequences of precision and runtime enforcement, either through surveys over alternative gradual semantics or rational-programmer experiments that compare the debugging information produced by erasure, transient, and natural semantics~\cite{TunnellWilsonGreenmanPombrioKrishnamurthi18,LGFD21,LGFD23}. Our current Elixir deployment lacks user-written annotations, so the evaluation below uses observables already available in the compiler: dead-code discoveries, exactness of guard-derived accepted-types, benchmark precision on guarded examples, and the dependence of typability on dynamic propagation.

We therefore distinguish the evidence already produced by the current compiler, namely \emph{the discovery of dead code}, from four quantitative experiments.
\begin{enumerate}
    \item A guard-analysis exactness experiment, which measures how often the static analysis computes exact accepted-types for pattern/guard pairs.
    \item An extension of Guo and Greenman's \emph{If-T} benchmark~\cite{GuoGreenman25}, which measures precision on curated positive and negative examples involving type narrowing.
    \item An arrow-return informativeness experiment, which measures how often inferred function arrows return a type more informative than just \elix{dynamic()}.
    \item A dynamic-propagation removal experiment, which measures the additional warnings produced when dynamic propagation is disabled.
\end{enumerate}
The reproduction artifacts for the dead-code scan and these four quantitative
experiments are archived in a public repository~\cite{CoreElixirExperiments}.
The artifact is organized by numbered experiment directories corresponding to
guard exactness, the \emph{If-T} benchmark, arrow-return informativeness,
dynamic-propagation removal, and the dead-code commit scan.

\subsection{Dead code detection}
Dead code detection is a useful precision check for static analysis.
Theodoridis \emph{et al.}~\citet{Theodoridis22} use dead code elimination
(DCE) as a ``lens'' or ``black-box oracle'' for compiler analyses, and Migeed
\emph{et al.}~\citet{migeed24} apply branch elimination---a form of DCE---to
measure gradual-type-system precision. We use the same check for Elixir: we
searched 7,150 non-merge commits on the default branches of 16 production
Elixir repositories, between May~1, 2024, and May~31, 2026, to include main-branch
type-system testing before v1.17. The search crossed dead-code terms such as
\texttt{dead code}, \texttt{unused}, \texttt{unused clause},
\texttt{unreachable}, and \texttt{redundant} with type/compiler terms such as
\texttt{type system}, \texttt{typesystem}, \texttt{type checker},
\texttt{compiler}, \texttt{warning}, and release names such as
\texttt{Elixir 1.18}. Appendix~\ref{app:dead-code-protocol} gives the complete
keyword protocol and accepted commits.

We report two evidence buckets. \emph{Direct} commits name the type system,
type checker, compiler, or a relevant Elixir release. \emph{Warning-linked}
commits name an Elixir warning or dead-code cleanup, but not the type-system
pass itself. We exclude broader static-cleanup commits unless their metadata
ties the deletion to such a warning.

The scan found fourteen dead-code fixes across nine projects, removing at least
179 lines of code. The direct bucket contains nine commits, accounting for 91
deleted lines, attributed to the type system, compiler, or release-specific
type-aware warnings~\cite{postgrex-bug-fix,flame-bug-fix,live-view-bug-fix,live-view-dead-code-engine,ecto-dead-code-repo-schema,exdoc-dead-code-erlang,livebook-dead-code-type-system,livebook-dead-code-compiler,ash-unused-code-type-checker}.
The warning-linked bucket is less explicit and contains five commits,
accounting for 88 deleted lines~\cite{phoenix-unused-clause-elixir-118,spitfire-elixir-118-warnings,postgrex-unused-clauses,live-view-remove-dead-code,live-view-latest-elixir-warnings}.

For the mixed Ash patch, we count only the five deleted lines in its two
type-checker cleanup hunks. All counted fixes predate the v1.20.0 release (June~3,
2026) and therefore reflect fixes found while the type system was being tested
against Elixir main and release candidates.

\begin{table}[ht]
\centering
\small
\begin{tabular}{|l|r|r|r|}
\hline\rowcolor{violet2}
\textbf{Project} & \textbf{Direct} & \textbf{Warning-linked} & \textbf{Lines} \\[-1pt]
\hline
Postgrex & 1 / \makebox[.8em][r]{15} & 1 / 45 & 60 \\[-1.5pt]
\rowcolor{violet1}
Flame & 1 / \makebox[.8em][r]{9} & -- & 9 \\[-1.5pt]
Phoenix LiveView & 2 / \makebox[.8em][r]{12} & 2 / 15 & 27 \\[-1.5pt]
\rowcolor{violet1}
Ecto & 1 / \makebox[.8em][r]{4} & -- & 4 \\[-1.5pt]
ExDoc & 1 / \makebox[.8em][r]{11} & -- & 11 \\[-1.5pt]
\rowcolor{violet1}
Livebook & 2 / \makebox[.8em][r]{35} & -- & 35 \\[-1.5pt]
Ash & 1 / \makebox[.8em][r]{5} & -- & 5 \\[-1.5pt]
\rowcolor{violet1}
Phoenix & -- & 1 / 14 & 14 \\[-1.5pt]
Spitfire & -- & 1 / 14 & 14 \\[-1.5pt]
\hline
\textbf{Total} & \textbf{9 / 91} & \textbf{5 / 88} & \textbf{179} \\[-1.5pt]
\hline
\end{tabular}
\caption{Dead code removals found in the commit scan.
Entries in the Direct and Warning-linked columns are commits / deleted lines; 
for Ash, only the isolated type-checker cleanup hunks are counted.
}
\label{tab:dead_code}
\end{table}
The count is conservative: broader static-cleanup commits are omitted unless
their metadata ties the deletion to an Elixir warning.

These results give a qualitative lower bound on precision, not a ratio.
Most of the warnings behind the accepted commits report a clause that can
never match, or a function applied outside its domain, because of the type
inferred for some expression. A checker that inferred only \elix{dynamic()}
would report none of them, since \elix{dynamic()} intersects, and is
compatible with (Equation~\eqref{eq:consub}), every non-empty type. Each
such commit therefore shows that, without annotations and in the current
conservative deployment, the checker inferred a type precise enough to rule
out a branch; this type may be gradual, since $\dyn\land t\land\tau$ is
empty exactly when $t\land\tau$ is. The remaining warnings report a clause
already covered by the previous ones, and witness the exactness of guard
analysis (Section~\ref{sec:guard-exactness}). Since the maintainers of these
mature, well-tested libraries removed the code, all these warnings were true
positives. The scan measures neither missed dead code nor false positives;
the experiments that follow address precision quantitatively.

\subsection{Guard-analysis exactness}\label{sec:guard-exactness}
Directly related to dead-code detection is the precision of guard analysis.
Recall from Section~\ref{sec:guard-analysis} that each clause of a case expression or of a multi-clause function is formed by a pattern $p$ and a guard $g$, that is, by a pattern/guard pair $pg$ (written \elix{p when g} in Elixir). Such a pair is \emph{exact} when its possibly accepted type and its surely accepted type (Section~\ref{sec:typing-pattern-matching}) coincide: for instance, the pair in the first clause of \elix{test} (line~\ref{test1}) is exact, while the one in the second clause (line~\ref{test2}) is not (see Section~\ref{sub:ga}). In the current type-checker implementation, this condition is captured by the \texttt{precise?} predicate. When \texttt{precise?} holds, there is no gap between the values the analysis deems possible and the values it proves accepted; therefore, redundant branches identified by the analysis can be classified as dead code.

The unit of observation is one pattern/guard pair analyzed by the type checker. For each codebase, the experiment records the number of analyzed pairs, the number for which \texttt{precise?} holds, and the exactness ratio:\smallskip
\[
    \frac{\text{exact pattern/guard pairs}}{\text{analyzed pattern/guard pairs}}.\smallskip
\]
We report this ratio per project and in aggregate across the corpus.
\begin{table}[ht]
\centering
\begin{tabular}{|l|r|r|r|}
\hline\rowcolor{violet2}
\textbf{Library} & \textbf{Analyzed Pairs} & \textbf{Exact Pairs} & \textbf{Exactness} \\[-1pt]
\hline
Elixir Standard Library & 37,685 & 23,150 & 61.43\% \\[-1.5pt]
\rowcolor{violet1}
Phoenix Web Framework & 3,542 & 3,318 & 93.68\% \\[-1.5pt]
Ecto Database Library & 5,207 & 4,415 & 84.79\% \\[-1.5pt]
\rowcolor{violet1}
LiveView & 5,290 & 4,480 & 84.69\% \\[-1.5pt]
Postgrex & 7,563 & 2,634 & 34.83\% \\[-1.5pt]
\rowcolor{violet1}
Flame & 686 & 648 & 94.46\% \\[-1.5pt]
\hline
\end{tabular}
\caption{Exact pattern/guard pairs in large Elixir codebases}
\label{tab:precise_guards}
\end{table}
Across these six codebases, 38,645 of 59,973 analyzed pattern/guard pairs are exact, for a weighted aggregate exactness ratio of 64.44\%. For the broader set of open-source Elixir projects listed in \hyperref[app:guard-exactness-corpus]{Appendix~\ref*{app:guard-exactness-corpus}}, 141,755 of 164,701 analyzed pairs are exact, for an aggregate exactness ratio of 86.07\%. Postgrex is the main outlier. Its lower exactness ratio comes mostly from generated type-extension dispatch code: Postgrex generates decoder functions that repeatedly pattern match on runtime lists of type descriptors, with clauses such as \mintinline[escapeinside=!!,fontsize=\small,breaklines]{elixir}+[{Extension, mod} | types]+. The current exactness analysis does not treat such list-head patterns as exact, so these generated dispatch cases are conservatively counted as inexact. This is a limitation of the implemented exactness check rather than of the guards themselves; an equivalent dispatch representation based on tuples would be represented precisely. The corresponding guarded-only counts and reproduction metadata are kept with the supplemental guard-exactness data.
This proportion gives an indirect way to compare the precision of our type system with the precision of other sound gradual systems that insert dynamic checks at compilation. When a guard is used to check the parameters of a function, the result of the guard analysis is used to determine the static type of the parameters, which automatically yields a strong type for that function. If the guard analysis is exact, then the strong function type yields the same kind of safety as the one provided by a sound gradual typing system that inserts a dynamic check at the function entry point. In those cases, our sound erasure gradual type system is at least as precise as a non-erasure sound gradual typing approach.

\subsection{Extension of the \emph{If-T} benchmark}\label{if-t-benchmark}
The guard-exactness experiment of Section~\ref{sec:guard-exactness} measures precision on real code, but it does not by itself provide controlled positive and negative examples. To obtain such examples, we port the \emph{If-T} benchmark of Guo and Greenman~\cite{GuoGreenman25}, whose current implementations are maintained in a public repository~\cite{GuoGreenmanIfTBenchmark}, with Elixir cases. Each benchmark item consists of a positive program, in which a pattern, guard, or built-in runtime check correctly narrows the type of a variable that is then used in 
a language operation, and a negative variant which should fail to typecheck. For a detailed account of the benchmark and of the properties tested by each item, we refer the reader to Guo and Greenman~\cite{GuoGreenman25}.

This experiment follows the benchmark methodology of \emph{If-T}, but adapts the examples to Elixir's patterns, guards, and safe-erasure discipline rather than treating the results as a direct cross-language comparison.

For this experiment we used the research branch of the Elixir compiler that supports
asserted type annotations for functions, since Elixir does not support type
annotations as of v1.20. These assertions are used to express
the benchmark's intended function boundaries, and reflect the behavior described
in the typing rules of the system described in the paper. We also compile the
Elixir programs with warnings as errors, so an Elixir type warning is counted as
a rejected program. Appendix~\ref{app:ift-benchmark-repro} gives the
corresponding artifact and adaptation details.

Table~\ref{tab:ift-core-results} gives the result on the thirteen core
benchmark items. The notation follows \emph{If-T}: {\footnotesize$\bigcirc$} means that the
positive program is accepted and the negative program is rejected, while
$\times$ means that one of these expectations fails. When a benchmark item has
several independent negative witnesses, each witness is checked separately; the
item receives {\footnotesize$\bigcirc$} only if all of them are rejected.

\begin{table}[ht]
	\centering
	\resizebox{\textwidth}{!}{%
		\begin{tabular}{lcccccccccccc}
			\toprule
			\textbf{Benchmark} & \textbf{TR} & \textbf{TS} & \textbf{Flow} & \textbf{mypy} & \textbf{Pyright} & \textbf{Sorbet} & \textbf{Luau} & \textbf{MLsem} & \textbf{TClojure} & \textbf{\textsf{ty}} & \textbf{Pyrefly} & \textbf{\color{fireenginered}Elixir} \\
			\midrule
			positive           & {\footnotesize$\bigcirc$}  & {\footnotesize$\bigcirc$}  & {\footnotesize$\bigcirc$}    & {\footnotesize$\bigcirc$}    & {\footnotesize$\bigcirc$}       & {\footnotesize$\bigcirc$}      & {\footnotesize$\bigcirc$}    & {\footnotesize$\bigcirc$}     & {\footnotesize$\bigcirc$}        & {\footnotesize$\bigcirc$}           & {\footnotesize$\bigcirc$}       & {\footnotesize$\bigcirc$}      \\
			negative           & {\footnotesize$\bigcirc$}  & {\footnotesize$\bigcirc$}  & {\footnotesize$\bigcirc$}    & {\footnotesize$\bigcirc$}    & {\footnotesize$\bigcirc$}       & {\footnotesize$\bigcirc$}      & {\footnotesize$\bigcirc$}    & {\footnotesize$\bigcirc$}     & {\footnotesize$\bigcirc$}        & {\footnotesize$\bigcirc$}           & {\footnotesize$\bigcirc$}       & {\footnotesize$\bigcirc$}      \\
			connectives        & {\footnotesize$\bigcirc$}  & {\footnotesize$\bigcirc$}  & {\footnotesize$\bigcirc$}    & {\footnotesize$\bigcirc$}    & {\footnotesize$\bigcirc$}       & {\footnotesize$\bigcirc$}      & {\footnotesize$\bigcirc$}    & $\times$       & {\footnotesize$\bigcirc$}        & {\footnotesize$\bigcirc$}           & {\footnotesize$\bigcirc$}       & {\footnotesize$\bigcirc$}      \\
			nesting\_body      & {\footnotesize$\bigcirc$}  & {\footnotesize$\bigcirc$}  & {\footnotesize$\bigcirc$}    & {\footnotesize$\bigcirc$}    & {\footnotesize$\bigcirc$}       & {\footnotesize$\bigcirc$}      & {\footnotesize$\bigcirc$}    & {\footnotesize$\bigcirc$}     & {\footnotesize$\bigcirc$}        & {\footnotesize$\bigcirc$}           & {\footnotesize$\bigcirc$}       & {\footnotesize$\bigcirc$}      \\
			struct\_fields     & {\footnotesize$\bigcirc$}  & {\footnotesize$\bigcirc$}  & {\footnotesize$\bigcirc$}    & {\footnotesize$\bigcirc$}    & {\footnotesize$\bigcirc$}       & $\times$        & {\footnotesize$\bigcirc$}    & {\footnotesize$\bigcirc$}     & {\footnotesize$\bigcirc$}        & {\footnotesize$\bigcirc$}           & {\footnotesize$\bigcirc$}       & {\footnotesize$\bigcirc$}      \\
			tuple\_elements    & {\footnotesize$\bigcirc$}  & {\footnotesize$\bigcirc$}  & {\footnotesize$\bigcirc$}    & {\footnotesize$\bigcirc$}    & {\footnotesize$\bigcirc$}       & {\footnotesize$\bigcirc$}      & {\footnotesize$\bigcirc$}    & {\footnotesize$\bigcirc$}     & {\footnotesize$\bigcirc$}        & {\footnotesize$\bigcirc$}           & {\footnotesize$\bigcirc$}       & {\footnotesize$\bigcirc$}      \\
			tuple\_length      & $\times$    & {\footnotesize$\bigcirc$}  & {\footnotesize$\bigcirc$}    & {\footnotesize$\bigcirc$}    & {\footnotesize$\bigcirc$}       & $\times$        & $\times$      & {\footnotesize$\bigcirc$}     & {\footnotesize$\bigcirc$}        & $\times$             & {\footnotesize$\bigcirc$}       & {\footnotesize$\bigcirc$}      \\
			alias              & {\footnotesize$\bigcirc$}  & {\footnotesize$\bigcirc$}  & $\times$      & $\times$      & {\footnotesize$\bigcirc$}       & {\footnotesize$\bigcirc$}      & $\times$      & {\footnotesize$\bigcirc$}     & {\footnotesize$\bigcirc$}        & $\times$             & {\footnotesize$\bigcirc$}       & $\times$        \\
			nesting\_condition & {\footnotesize$\bigcirc$}  & $\times$    & $\times$      & $\times$      & $\times$         & {\footnotesize$\bigcirc$}      & $\times$      & {\footnotesize$\bigcirc$}     & {\footnotesize$\bigcirc$}        & $\times$             & $\times$         & {\footnotesize$\bigcirc$}      \\
			merge\_with\_union & {\footnotesize$\bigcirc$}  & {\footnotesize$\bigcirc$}  & {\footnotesize$\bigcirc$}    & $\times$      & {\footnotesize$\bigcirc$}       & {\footnotesize$\bigcirc$}      & $\times$      & {\footnotesize$\bigcirc$}     & {\footnotesize$\bigcirc$}        & {\footnotesize$\bigcirc$}           & {\footnotesize$\bigcirc$}       & {\footnotesize$\bigcirc$}      \\
			predicate\_2way    & {\footnotesize$\bigcirc$}  & {\footnotesize$\bigcirc$}  & {\footnotesize$\bigcirc$}    & {\footnotesize$\bigcirc$}    & {\footnotesize$\bigcirc$}       & $\times$        & $\times$      & {\footnotesize$\bigcirc$}     & {\footnotesize$\bigcirc$}        & {\footnotesize$\bigcirc$}           & {\footnotesize$\bigcirc$}       & {\footnotesize$\bigcirc$}      \\
			predicate\_1way    & {\footnotesize$\bigcirc$}  & $\times$    & {\footnotesize$\bigcirc$}    & {\footnotesize$\bigcirc$}    & {\footnotesize$\bigcirc$}       & $\times$        & $\times$      & {\footnotesize$\bigcirc$}     & {\footnotesize$\bigcirc$}        & {\footnotesize$\bigcirc$}           & {\footnotesize$\bigcirc$}       & {\footnotesize$\bigcirc$}      \\
			predicate\_checked & {\footnotesize$\bigcirc$}  & $\times$    & {\footnotesize$\bigcirc$}    & $\times$      & $\times$         & $\times$        & $\times$      & {\footnotesize$\bigcirc$}     & {\footnotesize$\bigcirc$}        & $\times$             & $\times$         & {\footnotesize$\bigcirc$}      \\
			\bottomrule
		\end{tabular}%
	}
	\caption{Core \emph{If-T} benchmark results. TR abbreviates Typed Racket, TS abbreviates TypeScript, and TClojure abbreviates Typed Clojure.}
	\label{tab:ift-core-results}
\end{table}

The Elixir adaptation satisfies 12 of the 13 core benchmark items. The only
remaining miss is \texttt{alias}, and it is a precision miss on the positive
program rather than an accepted negative. In the Elixir port, the positive
program binds \elix{y = is_binary(x)} and then branches on \elix{y}; the true
branch calls \elix{byte_size(x)}. The checker can type the Boolean value
\elix{y}, but it does not retain the relational fact that \elix{y} being true
implies that \elix{x} is a binary. Thus, in the body of \elix{if y do ...},
\elix{x} remains at type \elix{term()}, and the valid call to
\elix{byte_size(x)} is rejected. The negative alias witnesses are rejected as
expected: one uses \elix{x + 1} under the saved test, and the other rebinds
\elix{y} to \elix{true} before calling \elix{byte_size(x)}. The table entry
therefore records a missing alias-tracking refinement, not a soundness issue or
an accepted negative program. On these examples the Elixir type system detects all the type errors that are present in the code.

\subsection{Return informativeness}
The arrow-return experiment we present in this section measures precision through the function types
inferred by the compiler, complementing the guard, benchmark, and warning-based
experiments of Section~\ref{sec:guard-exactness}. In the current implementation, an inferred function type is
represented as a set of arrows. Each arrow has an inferred domain and an
inferred return type. Since the system is intentionally conservative for
existing unannotated code, all function parameters are initially typed with
\elix{dynamic()}; the question is whether the analysis can nevertheless recover
useful information about the values returned by these functions.

For each inferred arrow, we therefore record a binary score on its return type:
the score is 0 when the return type is exactly \elix{dynamic()}, and 1
otherwise. For instance, \elix{dynamic(integer())} counts as informative.\footnote{\label{fn:dynamic-t}In error messages, Elixir prints the intersection \felix{t and dynamic()} as \felix{dynamic(t)}.} This is natural: this type
still carries the gradual component needed by safe erasure, but it also indicates that
the type system has inferred a static shape since functions with such a return type will only produce results that are integer values.
More generally, every return type other than \elix{dynamic()} is informative
because of how compatibility (Equation~\eqref{eq:consub}) treats the two
parts of a gradual type $t\simeq\dynInf{t}\lor(\dyn\land\dynSup{t})$: its
\emph{static part} $\dynInf{t}$, that is, the part that is not intersected
with $\dyn$, and its dynamic part $\dyn\land\dynSup{t}$. A non-empty static
part is checked as strictly as a static type: all of it must be accepted by
the expected type $s$ (that is, $\dynInf{t}\leq\dynSup{s}$). When the static
part is empty ($\dynInf{t}\simeq\bott$, as for $\dyn\land\intTop$, whose
maximal materialization is $\intTop$), the check is optimistic: it suffices
that some value of $\dynSup{t}$ is accepted
($\dynSup{t}\land\dynSup{s}\not\leq\bott$). For instance, if $t$ is a
non-empty static type, a result of type $\dyn\lor t$ passes only where all
of $t$ is accepted, and a result of type $\dyn\land t$ wherever some value
of $t$ is. Up to equivalence, \elix{dynamic()} is the only type that has an
empty static part and an unbounded dynamic part: it is compatible with
every non-empty type, so passing a result of type \elix{dynamic()} where a
non-empty type is expected is never reported. For every other return type
$t$, there is a non-empty type $s$ with which $t$ is not compatible, so
passing the result where an $s$ is expected is reported as an error.
The resulting ratio is
\smallskip
\[
    \frac{\text{arrows whose return is not exactly \texttt{dynamic()}}}
         {\text{inferred arrows}}.
\smallskip
\]
This coarse measure is not correlated with the precision relation of
Section~\ref{sec:gradual-typing}; rather, it is a simple corpus-level test of
whether inference collapses function results to pure dynamic information or
keeps useful return information.

We ran this measurement on 17 selected open-source Elixir projects, comprising
517,999 lines of Elixir code. Table~\ref{tab:arrow_return_excerpt}
shows a representative excerpt: the two largest projects, three projects near
the low and high ends of the distribution, and the weighted total. The full
per-project table is given in Appendix~\ref{app:arrow-return-corpus}.

\begin{table}[ht]
\centering
\begin{tabular}{|l|r|r|r|r|}
\hline\rowcolor{violet2}
\textbf{Project} & \textbf{LoC} & \textbf{Arrows} & \textbf{Informative} & \textbf{Ratio} \\[-1pt]
\hline
Blockscout & 183,142 & 26,384 & 15,493 & 58.7\% \\[-1.5pt]
\rowcolor{violet1}
Ash & 108,532 & 13,267 & 8,479 & 63.9\% \\[-1.5pt]
Livebook & 55,119 & 7,424 & 5,566 & 75.0\% \\[-1.5pt]
\rowcolor{violet1}
Credo & 23,712 & 4,441 & 2,274 & 51.2\% \\[-1.5pt]
Nerves & 4,831 & 274 & 126 & 46.0\% \\[-1.5pt]
\rowcolor{violet1}
Spitfire & 3,853 & 119 & 93 & 78.2\% \\[-1.5pt]
\hline
\textbf{Total} & \textbf{517,999} & \textbf{69,976} & \textbf{43,158} & \textbf{61.7\%} \\[-1.5pt]
\hline
\end{tabular}
\caption{Excerpt of inferred-arrow return informativeness on selected open-source Elixir projects}
\label{tab:arrow_return_excerpt}
\end{table}

Overall, 43,158 of the 69,976 inferred arrows have a return type more
informative than exactly \elix{dynamic()}, for a weighted ratio of 61.7\%.
This result is useful because it is measured precisely in the setting where the
type system is most constrained: no user annotations are available (therefore all function parameters are initially assumed to be of type \elix{dynamic()}), and the
compiler is deliberately conservative to avoid false positives. The majority of
inferred function arrows still recover return information. The largest projects
also follow this trend: Blockscout and Ash dominate the corpus by size and by
number of inferred arrows, and both have a majority of informative returns
(58.7\% and 63.9\%, respectively). The lower-ratio projects, such as Nerves,
SQL, and Credo, identify codebases where inference more often returns exact
\elix{dynamic()}, while projects such as Spitfire and Livebook show that the
same analysis can recover informative return types at a much higher rate. For
lower-ratio projects such as Nerves, a substantial source of imprecision is
the module-boundary restriction on inferred types. This experiment runs the
compiler in inference mode
on unannotated code; to avoid dependency loops during inference, inferred
function types are not used across module boundaries. Consequently, calls
through remote module interfaces---for instance wrappers around \elix{Mix},
\elix{System}, \elix{File}, \elix{Path}, shell tasks, or dynamic module
dispatch---often return exact \elix{dynamic()} even when the callee's
implementation may infer a more precise result.

\subsection{Dynamic-propagation removal experiment}\label{subsec:dynamic-propagation-removal}
Dynamic propagation is the mechanism that, when a strong function is applied to
dynamically typed arguments, propagates the dynamic component from the argument
to the result, yielding a more permissive use of the result. The Elixir type
checker also has an internal static mode, not exposed as a compiler option, that
disables this propagation: where the default checker assigns a propagated type
of the form $t \land \dyn$, the static variant assigns the static type $t$.
Exposing this mode would give programmers stronger control over the gradual
discipline of the typechecker.

We use this variant to estimate how many warnings are avoided by dynamic
propagation. The experiment runs both modes on the same codebase and counts
only the additional warnings produced by disabling dynamic propagation,
ignoring those already present in the full run.
We run the experiment on the same selected open-source repository set used for
the other corpus measurements. For each project, we compiled first with the full
checker and then with the static variant, counting only type warnings reported
for the project itself, not for its dependencies.
Table~\ref{tab:dynamic_propagation_removal} reports the baseline number of type
warnings, the number produced by the static variant, and the delta.

\begin{table}[ht]
\centering
\small
\begin{tabular}{|l|r|r|r|}
\hline\rowcolor{violet2}
\textbf{Project} & \textbf{Full checker} & \textbf{Static variant} & \textbf{Additional warnings} \\[-1pt]
\hline
Blockscout & 37 & 50 & +13 \\[-1.5pt]
\rowcolor{violet1}
Ash & 33 & 45 & +12 \\[-1.5pt]
Livebook & 4 & 6 & +2 \\[-1.5pt]
\rowcolor{violet1}
HexPm & 12 & 12 & +0 \\[-1.5pt]
Ecto & 0 & 0 & +0 \\[-1.5pt]
\rowcolor{violet1}
Credo & 0 & 6 & +6 \\[-1.5pt]
PhoenixLiveView & 0 & 1 & +1 \\[-1.5pt]
\rowcolor{violet1}
Phoenix & 0 & 0 & +0 \\[-1.5pt]
MixSBOM & 2 & 2 & +0 \\[-1.5pt]
\rowcolor{violet1}
Postgrex & 0 & 12 & +12 \\[-1.5pt]
OpenApiSpex & 2 & 7 & +5 \\[-1.5pt]
\rowcolor{violet1}
ExDoc & 1 & 2 & +1 \\[-1.5pt]
Nerves & 0 & 0 & +0 \\[-1.5pt]
\rowcolor{violet1}
Spitfire & 1 & 2 & +1 \\[-1.5pt]
SQL & 1 & 11 & +10 \\[-1.5pt]
\rowcolor{violet1}
Flame & 0 & 1 & +1 \\[-1.5pt]
AbsintheFederation & 1 & 1 & +0 \\[-1.5pt]
\hline
\textbf{Total} & \textbf{94} & \textbf{158} & \textbf{+64} \\[-1.5pt]
\hline
\end{tabular}
\caption{Type warnings produced with and without dynamic propagation}
\label{tab:dynamic_propagation_removal}
\end{table}

The static variant increases the number of project type warnings from 94 to
158, for 64 additional warnings across the corpus, or roughly a 68\% increase
over the default checker. This increase is relatively small. It suggests that
a programmer who wants the stronger guarantees given by strong functions could
enable this mode and fix the remaining warnings in many cases. The increase is
not uniform: six projects are unchanged, while the largest increases occur in
Blockscout (largest codebase, with close to 200 000 lines of code), Ash,
Postgrex, SQL, Credo, and OpenApiSpex.

A caveat is that fixing these warnings often means adding explicit checks after function applications in
a way that does not really improve the quality of the code. For example, if \elix{f(x)} returns either an integer or a Boolean,
then \elix{f(x) + 1} must be rewritten so that the result of \elix{f(x)} is checked with
\elix{is_integer} before using it in the addition. These checks can be somewhat redundant:  dynamic propagation avoids forcing the programmer
to write such checks in the default mode.

\subsection{Summary}
The experiments in this section measure complementary aspects of precision. The dead-code detection shows that the checker already exposes unreachable code in mature libraries, thereby gauging the precision of the type system. Guard exactness measures how often the guard analysis produces no approximation gap on pattern/guard pairs---an indicator of how frequently strong arrows can be deduced and, with them, how closely the precision of our system approaches that of non-erasure semantics. The \emph{If-T} benchmark compares the precision of our system against other gradually-typed languages on controlled examples of positive and negative narrowing. The arrow-return experiment gauges whether inferred function types retain useful return information in the absence of annotations, that is, when all arguments are initially assumed to be of dynamic type. Finally, the dynamic-propagation removal experiment shows that, on a selected corpus of code, disabling dynamic propagation increases the number of type warnings by 68\%, though the absolute count remains small. Taken together, these measurements paint a concrete picture of a conservative deployment: the type system deliberately preserves dynamic flexibility, yet still recovers enough static information to diagnose dead code, type guarded idioms precisely, and infer non-trivial function result types at scale.

Finally, a less tangible---yet for us equally significant---indicator is the reception
our work has received, both from the Elixir community and beyond. Within the
Elixir community, feedback has been unanimously positive: many developers have
expressed genuine excitement about the new type system and have actively
engaged with the features progressively rolled out in language release
candidates. Beyond Elixir, the most significant sign of adoption is the recent
decision by Astral to incorporate several of the ideas presented here---on
gradual typing and intersection types~\cite{ty}---into \textsf{ty}, an
extremely fast and widely used Python type-checker.\footnote{The Astral
ecosystem, comprising \texttt{Ruff}, \texttt{uv}, and \texttt{ty}, collectively
reaches hundreds of millions of downloads per month, making it arguably the
most widely adopted new developer infrastructure in the Python ecosystem in the
past decade.} This adoption has gone as far as reimplementing the very
optimizations developed for Elixir~\cite{ty-optimization} and described in the
second author's PhD thesis~\cite{duboc-phd}, of which this work is a
substantial excerpt.

\section{Related work}\label{sec:related}%
Two works closely align with ours by using semantic subtyping to establish a type system for Erlang and Elixir (the latter
being a compatible superset of the former with which it shares a common
functional core). The most closely related is our own companion paper~\cite{castagna2023design}, which focuses on the design principles for integrating semantic subtyping into Elixir, omitting the formal technical details that are developed here. We discussed at length in the introduction how our two works complement each other. The other relevant work is
by Schimpf \emph{et al.} \citet{schimpf2023set} who propose a type system for Erlang based on
semantic subtyping, implement it (see~\cite{SWB23}), and provide useful benchmarks regarding its
expressiveness compared both to Dialyzer~\cite{Dialyzer} and Gradualizer~\cite{gradualizer}. The work by Schimpf \emph{et al.} \citet{schimpf2023set} is rather
different from ours, since they adapt the existing theory of semantic subtyping
to Erlang, while the point of our work is to show how to \emph{extend} semantic
subtyping with features motivated by or specific to Elixir: how to add gradual typing without
modifying Elixir's compilation and how to extract the most information from the
expressive guards of Erlang/Elixir. Also, their work extensively uses type reconstruction, while we rely more on explicit type annotations, gradual typing, and the inference of guards accepted types. Concerning these two works we want to signal that both \citet{castagna2023design} and
\citet{schimpf2023set} provide extensive comparison of the semantic
subtyping approach with existing typing efforts for Erlang and Elixir, and we invite the reader to refer to their analyses for more details on this aspect.

\paragraph{\it Parametric polymorphism.}
Parametric polymorphism is outside the scope of this work, but we expect
its addition to our system to be mostly straightforward: it requires no new
ideas, only the re-establishment of some of our results. The theory of
polymorphic semantic subtyping exists and is
standard~\cite{CX11,polyduce1,polyduce2}. The Etylizer
project~\cite{SWB23} has already ported it to the BEAM, as a type checker
for Erlang, but it does not consider gradual typing. The extension of the
polymorphic set-theoretic type theory of~\cite{CX11,polyduce1,polyduce2}
to gradual types is covered by~\cite{castagna2019gradual,lanvin2021semantic},
on which the gradual theory of this work is based. What remains to be done is
to re-establish the results that concern the constructs we introduce.
First, strong and multi-arity arrows must be added to the polymorphic
interpretation of types, and the subtyping and tallying algorithms extended
accordingly. Second, the soundness of our system rests on erasure
(Theorem~\ref{thm:gradual-soundness}), whereas the soundness results
of~\cite{castagna2019gradual,lanvin2021semantic} rely on the insertion of
casts: our soundness proofs must therefore be redone in the presence of
type instantiation, which, as in~\cite{castagna2019gradual}, would be
restricted to static types. Finally, polymorphism yields no new strong
arrows: no runtime check exists for a type variable, so, for instance, the
polymorphic type of \elix{second} given in Section~\ref{sub:strong} is not
strong (see also the end of this section).
The eventual practical integration
of polymorphism into Elixir---which initially will only be achievable through the
systematic use of explicit type annotations without resorting to type
reconstruction---will require a non-trivial adaptation of the tallying and
constraint solving algorithms described in~\cite{castagna2019gradual}, as well
as a deeper understanding of the techniques necessary for producing informative
error messages and for pretty-printing the types resulting from these
algorithms.

\paragraph{\it Maps.}
Two recent works extend the theory of semantic subtyping by specifically
targeting Elixir's map types. The first work~\citet{Cas23records} introduces a
system that seamlessly types maps used as \emph{records}---i.e., maps with a fixed,
statically known set of keys, where accessing an unknown key is a (type)
error---and maps used as \emph{dictionaries}, where the set of keys may vary
dynamically, keys can be computed by expressions, and lookups of undefined keys do not result in an error. This work
builds on the theory of record types for semantic subtyping developed in
Frisch's PhD dissertation~\citet{frisch2004theorie}, and is orthogonal to our
contributions, making it directly compatible with the type system presented
here. The second work~\citet{CP25} extends the theory of semantic subtyping by adding
\emph{row polymorphism}~\cite{Remy89,Wand89}, and motivates its use for typing Elixir's maps.
Row polymorphism is particularly useful in Elixir because maps are a primary
data structure for encoding structured data, including records, configurations,
state representations, and structs. In many idiomatic patterns, maps are
extended or partially updated across function boundaries (e.g., by adding new
fields in pipeline expressions or merging configuration data), and row polymorphism
enables the typing of such patterns without losing precision. As in the previous
case, the system by~\citet{CP25} builds on the record type theory of Frisch and
extends it with techniques borrowed from the polymorphic type system of
CDuce~\cite{polyduce1,polyduce2} to handle row variables. Like~\cite{Cas23records},
this work is orthogonal to ours and can be easily integrated with our system
without requiring changes to its core structure.

\paragraph{\it Other languages using semantic subtyping.}
Elixir and Erlang are among the latest languages to embrace semantic subtyping techniques.
Other languages in this category include CDuce~\cite{cduce} which lacks gradual typing and guards,
but supply the latter with powerful regular expression patterns; Ballerina~\cite{ballerina} which
is a domain-specific language for network-aware applications whose emphasis is on the use of read-only and write-only types and shares with Elixir the typing of records given by~\citet{Cas23records}; Lua\emph{u}~\cite{luau,luausemsub} Roblox's gradually typed dialect of Lua, a dynamic scripting language for games with emphasis on performance, with a type system that switches to semantic subtyping when the original syntactic subtyping fails~\cite{luausemsubrelease};  Julia~\cite{Julia} with a type system that is based on a combination of syntactic and semantic subtyping and supports an advanced type system for modules. Finally, Python's type specification~\cite{pythontyping} has recently started incorporating concepts from gradually-typed semantic subtyping. Although this is not yet reflected in practice, there are promising attempts such as the \texttt{ty} type-checker~\cite{ty} that aim to include union, intersection, and negation types in Python and that we discussed at the end of Section~\ref{sec:eval}. Similarly, TypR~\cite{typr} is an ongoing effort to equip R with a gradual type checker based on set-theoretic types, built on the same semantic-subtyping foundations; it checks R code (and the C code it calls) against types written by the programmer in comments, and assigns the dynamic type to unbound names.
While some of these languages employ gradual typing and/or guards, none share Elixir's emphasis on these features or, consequently, on the typing techniques we developed in this work. Nevertheless, we believe that portions of our work could be adapted to these languages, particularly the techniques for safe erasure gradual typing (strong functions and dynamic propagation) and the extension of semantic subtyping to multi-arity function spaces.

\paragraph{\it Gradual typing in semantic subtyping.}
The thesis by Lanvin~\citet{lanvin2021semantic} defines a semantic subtyping approach
to gradual typing, which forms the basis of the gradual typing aspects of our
system, since we borrow from \citet{lanvin2021semantic} the definitions of
subtyping, precision, and consistent subtyping for gradual types. The main
difference with Lanvin~\cite{lanvin2021semantic} is that he considers that sound
gradual typing is achievable by inserting casts in the compiled code whenever
necessary, while our work shows a way to adapt gradual typing to achieve
soundness while remaining in a full erasure discipline. The relations defined by~\cite{lanvin2021semantic} are also implemented at Meta for the gradual typing of
the (Erlang) code of WhatsApp~\cite{eqwalizer}, and whose differences with the
semantic subtyping approach are detailed in~\cite{castagna2023design}, to which
we defer this discussion.~\citet{lanvin2021semantic} builds on and extends the work by Castagna \emph{et al.~}\cite{castagna2019gradual} who show how to perform ML-like type reconstruction
in a gradual setting with set-theoretic types. As anticipated in Section~\ref{sec:contributions}, this is
one of the limitations of our work. To address it we count on adapting
the results of~\cite{castagna2024inf} on  type inference for dynamic
languages. An alternative option is to utilize the approach by \citet{CPN16} which employs traditional, less computationally demanding type reconstruction techniques than those in~\cite{castagna2024inf}, but lacks the capability to infer intersection types for functions.

\paragraph{\it Industrial solutions for typing dynamic languages.}
Industry systems have adopted gradual typing to facilitate the progressive adoption of typing disciplines. TypeScript~\cite{bierman2014}, built on JavaScript,  is a most prominent example, offering a compile-time-only type system featuring erased annotations, structural types, unions, intersections, and control flow narrowing~\cite{typescript-narrowing}. However, its subtyping relation for unions and intersections is defined syntactically rather than semantically, resulting in ``incomplete'' union and intersection types that do not always align with expectations/properties of the intended set-theoretic interpretation. Moreover, the subtyping relation was intentionally designed to be unsound in certain---quoting~\cite{typescript-unsound}---``carefully considered'' cases, which contrasts with our design philosophy. Despite these acknowledged limitations in soundness and completeness, TypeScript represents a clear success story in terms of both adoption and tooling ecosystem. Our work reproduces these key features while addressing soundness and completeness concerns, and providing additional benefits grounded in a robust theoretical framework.
A particularly illustrative example of such benefits involves JavaScript's heavy reliance on dictionary data structures. As detailed in~\cite[Section 5]{Cas23records}, the syntactic approach employed by TypeScript (but also the one of Flow, see below) produces imprecise and ambiguous typing for these structures, necessitating various restrictions. In contrast, semantic subtyping provides more general and precise typing for dictionary structures while eliminating the need for such restrictions. Another key difference concerns the treatment of dynamic values. TypeScript separates the permissive and unsound \texttt{any} type from the safer \texttt{unknown} type. The former is an escape hatch: in particular, TypeScript lets programmers write user-defined type predicates of the form \texttt{x is T}, which can override the type checker's knowledge without performing a corresponding runtime check, and therefore allow a dynamic value to be treated as having an incorrect static type. The latter is called a top type in the TypeScript documentation~\cite{typescript-unknown} and is closer to our \texttt{term()} than to our \texttt{dynamic()}: both represent all possible values as a static top type. Consequently, such a value must be refined before it can be used by an operator with a more specific domain: TypeScript rejects such uses for \texttt{unknown}, while our strict treatment of static types issues a warning for \texttt{term()}. Our \texttt{dynamic()} occupies a different point. It does not behave like \texttt{any}: whenever a computation depends on uncertain information, this uncertainty is propagated in the inferred type rather than being hidden by an unsound escape hatch. This may reduce precision, but strong functions can recover static information from existing checks. At the same time, \texttt{dynamic()} is less restrictive than \texttt{unknown} or \texttt{term()}, since the system does not require such a refinement before every use. Instead, checks refine dynamic values and preserve the information they discover; for instance, in a branch where \texttt{is\_integer(x)} holds, a variable initially typed \texttt{dynamic()} is refined to \texttt{dynamic(integer())}.

Flow~\cite{Chaudhuri_Vekris_Goldman_Roch_Levi_2017} (built on JavaScript, too) has a similar surface, but with stricter inference (a feature that we only partially integrate into our system, as we
present inference as a way to suggest more strictly enforced
type annotations). It also has a smaller ecosystem.
Hack~\cite{facebookhack} (built on PHP) features a static type checker that is able to introduce
runtime checks via HHVM (HipHop Virtual Machine) type hints, providing operators `is' and `as' to assert and cast types (casts are only supported for \texttt{bool}, \texttt{int}, \texttt{float} and \texttt{string}~\cite{hack-type-assertions}).
It is a pragmatic construction that really compares to our 
leveraging of existing BEAM VM checks and guard mechanisms.
The difference is that we use more expressive structural types, thus there is 
a greater distance, in Elixir than in Hack, between the types that are VM-checkable and those that feature in type annotations.

Sorbet~\cite{sorbet} (built on Ruby) is a gradual type system with an erasure discipline, a dual design similar to TypeScript with \texttt{T.untyped} (unsafe escape hatch) and \texttt{T.anything} (explicit check required),
and it can optionally insert runtime checks for type assertions, similarly to Hack and to our approach. It supports union (\texttt{T.any}) and intersection (\texttt{T.all}) types, and uses the latter for
control flow narrowing in a way very similar to our system. One main difference with our approach is that the Sorbet system
is largely nominal as types are based on classes and modules
rather than structural shapes. Another more substantial difference is that, as for TypeScript or Flow, the subtyping relation is syntactic rather than semantic. While Sorbet---unlike TypeScript---does not push unsoundness, and---like us---it explains containment in terms of a set-theoretic interpretation of types, it still does not fully comply with such an interpretation. For instance, Sorbet permits the intersection of two modules that export the same method with different signatures (i.e., types). Since the actual method depends on the order in which the two modules are imported, then the intersection of the types of these two modules should give this method the union of the two signatures. However, while Sorbet accepts the calls whose arguments are in the intersection of the domains of the signatures (as expected when applying a function with a union type), if the return types are incompatible then the call is rejected (instead of being typed with the union of the return types). Furthermore, if the intersection of the domains is empty, then a code specifying the intersection of the two modules is still accepted, despite the fact that the method in common to the modules can never be called. This shows that the basic subtyping properties of the union of two arrow types are not accounted for. It is a direct consequence of the syntactic nature of the subtyping relation, which does not completely capture the intended semantics of types.

In general, the major difference of
our approach with industry solutions is that it is built on a sound foundation of gradual
set-theoretic types, thus avoiding the pitfall of ad-hoc rules that are added 
later on: for instance, the overload call evaluation of Python (PEP 484) consists of a
six-step algorithm~\cite{python_overload_spec} that performs expansions to attempt to find matches with arguments; steps five and six were recently added to ``determine whether all possible materializations of the argument’s type are assignable to the corresponding parameter type''. By contrast, overloaded (intersected) gradual
function types are a native object of our type algebra. However, these systems
are also not bound to a specific theory and can constantly evolve to match
desired behaviors. While this enhances their flexibility, it also hinders their portability
to other languages and complicates formal reasoning about their properties.

\paragraph{\it Positioning among gradual typing disciplines.}
We discussed above few major examples of industrial type systems for dynamic languages. A larger range of gradual type systems for industrial-grade dynamic languages have explored different trade-offs between soundness, precision, and runtime overhead.  
These have been thoroughly examined by Greenman, Dimoulas, and Felleisen~\cite{Greenman_Dimoulas_Felleisen_2023}, who categorize the implemented semantics into three main disciplines: \emph{erasure}, \emph{transient}, and \emph{natural}. In the \emph{erasure} discipline, types are used only for static analysis and have no influence on the runtime behavior of programs. The other two disciplines change the runtime behavior of programs based on the static type information: the transient discipline inserts shape checks in the code that enforce the correspondence between top-level value constructors and the expected types, ensuring that the \emph{typed portion} of the code cannot ``go wrong''; the natural discipline goes further by inserting wrappers or proxies that enforce for the \emph{entire program} the full type structure at runtime. 

The work on Thorn~\cite{wrigstad2010integrating} is a useful
point of comparison because it also distinguishes a fully checked fragment from
forms of interaction with dynamic code. Thorn introduces \emph{like types} as an
intermediate point between static and dynamic types: uses are checked
statically, but values flowing from dynamic code are checked at run time. Our
static fragment similarly enjoys the usual stronger safety guarantee, but the
gradual part of our system is designed for an erasure semantics rather than for
a check-inserting semantics. Strong functions therefore play a role analogous to
some inserted checks, but only when the needed checks are already present in the
program or in the BEAM.

As noted by~\cite{Greenman_Dimoulas_Felleisen_2023}, the \emph{erasure} discipline is by far the most popular design choice in industry, especially because its predictable behavior and performance. But the current systems using this discipline implement it at the expense of type soundness.
For example, in TypeScript type annotations are completely erased from the emitted JavaScript.  
However, as already pointed out, TypeScript does not guarantee type soundness and requires modifications to the compiler---such as additional static checks~\cite{rastogi2015safe}---to partially recover it.
This issue also affects the already cited Flow and Hack, developed by Meta for JavaScript and PHP, respectively, as well as systems like Mypy~\cite{lehtosalo2017mypy} and the one defined by Rastogi, Chaudhuri, and Hosmer~\cite{Rastogi_Chaudhuri_Hosmer}.

Our work clearly belongs to the class of \emph{erasure} disciplines, and improves its state of the art by introducing a fully erasable gradual type system with an explicit erasure soundness guarantee.
We achieve this by statically identifying and type-checking \emph{strong functions}, which can safely interact with dynamic code in the following sense: if a strong application returns a value, that value is compatible with the inferred type, and otherwise the execution fails at an existing runtime check. The presence of strong functions reduces the gap between the erasure and transient disciplines. Strong functions are, in effect, proxies for the shape checks of the transient discipline: instead of inserting these shape checks, as done by transient systems, our system statically identifies the presence of these checks within the scope of a function application. This entails a different precision trade-off rather than a uniform loss of precision. If no static evidence is available, uncertainty propagates through \texttt{dynamic}; when patterns, guards, or VM-checked operations provide evidence, the analysis can infer very precise domains, for instance as unions of the domains accepted by the clauses of a multi-clause function. A shape check is possible only if there is a corresponding guard that implements it, which is not always the case, thus reducing the number of functions that can be proven strong. Thus, the static detection of a strong function working on composite specifications such as \texttt{list(integer)} may be problematic, since
there is no primitive guard in Elixir to assert that \emph{all} elements of a list have 
a given type. 
The addition of parametric polymorphism to the system, as discussed above, would not change this situation: deducing the strongness of a polymorphic function of type \elix{list(X) -> list(X)} is not easier than deducing the strongness of a monomorphic function of type \elix{list(integer()) -> list(integer())}, since they both require evidence from the analysis of explicit traversals of the result. The only difference is that the application of an argument of type \texttt{list(dynamic())} will give for the polymorphic function a result of type \texttt{list(dynamic())} rather than just \texttt{dynamic()} as in the second case.  
These limitations seem to us acceptable trade-offs to retain both soundness and the advantages of the erasure discipline.

A related way to reduce the cost of sound gradual typing is the work of Richards \emph{et al.}~\cite{richards2017vm}, who observe that modern dynamic-language virtual machines already use speculative shape checks for optimization. Their intrinsic object contracts make gradually-typed contract obligations part of object shapes, so that these obligations can be checked by existing VM shape checks rather than by a separate layer of contracts. We share the broader goal of exploiting checks already available in a production implementation, but at a different layer. Their system changes the VM representation so that dynamic contract checks become cheaper; safe erasure keeps Elixir's runtime behavior unchanged and uses static analysis to type only those interactions whose safety follows from the checks already performed by Elixir operators, pattern matching, and guards.

The soundness property enforced by our system bears resemblance to the notion of \emph{open-world soundness} introduced by Reticulated Python~\cite{Vitousek_Swords_Siek_2017,retphyton}, classified by \cite{Greenman_Dimoulas_Felleisen_2023} as a transient system, which guarantees that well-typed programs, when compiled from a gradually-typed surface language to an untyped target language, can safely interoperate with untyped code. In our case, Elixir's existing runtime behavior already provides sufficient guarantees (via its virtual machine checks and guard mechanisms) to ensure a similar property when endowed with a gradual type system, but \emph{without runtime wrappers or coercions}. Typed Racket~\cite{tobin2008design} (implementing a \emph{natural} discipline) takes a hybrid approach, using occurrence typing for internal precision and contracts at module boundaries to enforce soundness.

\GuillaumeReviews{Other standard criteria for gradual languages are best read as points of comparison rather than as properties our system must satisfy. Complete monitoring~\cite{greenman2019complete} requires every communication path between typed and untyped code to be protected by a run-time check of the boundary type, including for higher-order values. The property does not require these checks to be inserted by a compiler, only to be present. Safe erasure does not satisfy it, because interfaces are trusted without being checked. For instance, the function $\lam[\{\intTop\rarr\intTop\}]{x}{x}$ of Section~\ref{sec:gradual-typing} is well typed but not strong: dynamically typed code may apply it to a Boolean, which then enters a body typed under the assumption $x{:}\intTop$ without passing any check. The system remains sound because \RuleRef{typing:gradual:app-dyn}{rule (app$_\dyn$)} gives such an application the type `$\dyn$': the guarantee is enforced where the result is used, not at the boundary. Checks are present, and exploited, only for strong functions, and even then only for first-order properties: of a functional argument, only the arity can be checked.
In a sense,
our way to deal with untyped code is lazy: we are hoping that, whenever it is used, there will be a strong
function able to recover the needed static type information, and the type system will use that evidence to type an interaction precisely. Without such check, the uncertainty propagates, until the next call which will
perhaps be strong this time.

Vigilance~\cite{gierczak2024vigilant} characterizes a related (actually, stronger) property: whether the combination of translation and dynamic semantics enforces the complete run-time typing history of every value---all types ever associated with it. Safe erasure inserts no translation and relies entirely on the BEAM's pre-existing runtime checks (pattern matching, guards, tag tests). Those checks are not adequate to enforce the full semantic type language: they do not check higher-order function types, they do not enforce composite types like \elix{list(integer())}, and they cannot track the typing history of a value as it flows through the program. So safe erasure is not vigilant in the sense of~\cite{gierczak2024vigilant}. However, the comparison with vigilance is still informative rather than simply negative, and this is what makes it interesting as a lens. Vigilance revealed that Natural and Transient, despite having the same static type system, enforce different adequate type systems at runtime---Natural enforces the full simple type system, Transient enforces only a tag type system. The Elixir situation is analogous but more radical: the ``translation'' is the identity (erasure), and the dynamic semantics is the BEAM's existing operational semantics. The adequate type system that this combination actually enforces is something weaker than the full static type language---roughly, the fragment that the BEAM's tag checks and pattern/guard refinements can witness, captured by the Weak Type System of Figure~\ref{fig:strong-rules}. So the key difference from both complete monitoring and vigilance is that those properties are normative---they tell you what a check-inserting system should \emph{enforce}. Safe erasure is not trying to satisfy either property; instead it accepts what the existing runtime already enforces and tries to capture that faithfully in the static type system itself. The guarantee it offers in exchange is erasure soundness stated in Section~\ref{sec:safe-erasure}: if a well-typed closed expression terminates normally, the returned value is compatible with the inferred type, and any failure that occurs was already present in the untyped program.

Concerning the gradual guarantee~\cite{siek2015refined}, its dynamic part holds
trivially for our system: type annotations are erased, so making them less
precise does not change the behavior of a program. Its static part, instead,
does not hold, and this is a direct consequence of erasure. The function
$\lam[\{\intTop\rarr\intTop\}]{x}{x}$ is well typed, but its less precise
version $\lam[\{\dyn\rarr\intTop\}]{x}{x}$ is not: its body has type `$\dyn$',
which is not a subtype of $\intTop$. Accepting it would be unsound: rule
\RuleRef{typing:static:app}{(app)} would then give type $\intTop$ to its
application to an argument of type `$\dyn$' that evaluates to a Boolean, which
the function returns unchanged. A cast-inserting system accepts the less
precise version by inserting a check. In Elixir this check must be written by
the programmer, for instance as the guard \elix{when is_integer(x)}, which makes
the function well typed again. Again, what we do ensure is the erasure soundness property stated above.} 

Finally, what distinguishes our system from other approaches to gradual typing is the combination of Lanvin's gradual semantic subtyping~\cite{lanvin2021semantic} with strong arrow types---which are pivotal to the precision of our erasure discipline---and dynamic type propagation. It is therefore natural to ask how far these features depend on each other, and in particular whether strong arrows and dynamic type propagation could be used in gradual type systems without semantic subtyping. The answer is yes, but likely at a cost in effectiveness. Strong arrows and dynamic type propagation inherently require intersection types, but this does not mean that intersections must appear explicitly in the type syntax, nor that the system must adopt semantic subtyping. Those intersections could instead be internalized through some form of tagging---so that, for instance, the type $t^{\texttt{?}}$ represents the intersection of $t$ with the dynamic type (this is in fact what Elixir presents to the programmer: see Footnote~\ref{fn:dynamic-t}). The difficulty lies neither in the syntax of strong arrows themselves nor in the characteristics of intersections, but in the analysis that computes the precise type of returned values after the refinements performed inside a function body. In our Elixir setting, those refinements arise from patterns, guards, and strong runtime operations; semantic subtyping provides a uniform way to compute them via unions, intersections, and differences. A more classical constraint-based refinement system might also work, but it would need to recover essentially the same information. Set-theoretic types are therefore not essential to the abstract safe-erasure idea, but they are the natural vehicle for the Elixir instantiation.
They also make explicit a distinction that conventional typed calculi tend to obscure: ordinary semantic arrows are weak specifications, constraining only behavior on their domain—a property that is essential for intersections of arrows and for multi-clause functions. In many conventional type systems, off-domain applications are either ill-typed or rejected by inserted checks, so arrows behave as though they were strong. Our system requires both notions: weak arrows for ordinary set-theoretic function typing, and strong arrows for exploiting existing runtime checks to recover precise gradual results.

To summarize, transient and natural disciplines, such as Typed Racket and Reticulated Python, maintain soundness via runtime instrumentation, which introduces overhead and affects integration for existing untyped code. By contrast, our system provides a sound type system without modifying the runtime, relying instead on Elixir's built-in runtime checks and our guard-aware analysis. Combined with semantic subtyping and dynamic type propagation, our approach occupies a distinct point in the design space: it avoids the unsoundness of TypeScript, Flow, and Hack, and the runtime cost of systems like Reticulated Python and Typed Racket, though this comes with different precision trade-offs and a different treatment of failures from wrapper-based systems. Our \emph{safe erasure} strategy enables Elixir developers to adopt types incrementally, without sacrificing compatibility, performance, or language semantics.

\section{Conclusion}%
\label{sec:conclusion}

We have presented the theoretical foundations of a new type system for Elixir, extending the framework of semantic subtyping with key features to support core idioms of the language: safe-erasure gradual typing, multi-arity function spaces, and guard-aware type analysis. The resulting system is expressive enough to capture idiomatic Elixir code and to provide developers with actionable type information via static warnings and error messages.

Since its initial (and still in progress) integration in Elixir v1.17~\cite{elixir117} the system has received positive feedback from the community and has already proven useful in detecting bugs in large-scale codebases with extensive test coverage.  Beyond immediate usability benefits, the type system lays a foundation for more robust tooling and gradual adoption in existing Elixir projects.

While our design is tailored to Elixir, the underlying theory has broader implications: it extends semantic subtyping in ways that are applicable to other dynamic languages---notably, those with existing large codebases---and offers a general framework for building sound, expressive, and fully erased gradual type systems.

Future work includes completing  the missing parts of the type system in the Elixir compiler according to the roadmaps outlined by~\cite{elixirtypes,castagna2023design}, and continuing to evaluate its
performance and usability in real-world scenarios. From a theoretical standpoint
we aim to extend the type system to include Elixir's first-class module system, to study type reconstruction, improve occurrence typing,
and to devise types to support  concurrency and distribution.
\nocite{FCB02}

\bibliographystyle{alphaurl}
\bibliography{ref}

\fi

\pagebreak

\appendix

\section{Operational Semantics of Core Elixir}\label{app:core-elixir}

The complete syntax of Core Elixir is presented in Figure~\ref{fig:full-core-elixir-syntax} in the main text.
The language has strict reduction semantics defined by the
reduction rules in Figures~\ref{app:fig:standard-and-failure-reductions}. These use the operator $v/(\when{p}{g})$, defined in Figure~\ref{fig:pattern-matching-operation} which returns the result of matching the value $v$ against the pattern $p$ with guard $g$ (i.e., either a substitution for the capture variables in $p$ or a failure), and the guard reductions relation, defined in Figure~\ref{fig:guard-reductions}, which computes the result of a guard (i.e., either a Boolean value or \texttt{fail}).

The semantics is defined in terms of values (ranged over by $v$), evaluation contexts (ranged over by $\Context$), and guard evaluation contexts (ranged over by $\GuardContext$), used to define the semantics of pattern matching.
They are formally defined as follows:
\begingroup
\setlength{\abovedisplayskip}{5pt}
\setlength{\belowdisplayskip}{5pt}
\[
	\begin{array}{llll}
		\textbf{Values}        & v
		                       & \bnfeq            & c \mid \lambda^\mathbb{I}{\ov{x}}.{e} \mid \tuple{\ov{v}}                                       \\
		\textbf{Context}       & \Context          & \bnfeq                                                    & \ContextHole \mid \app{\Context}{\ov{e}}
		\mid \app{v}{\ov{v},\Context,\ov{e}} \mid \tuple{\ov{v}, \Context, \ov{e}}
		\mid \proj{\Context}{e}
		\\
		                       &                   &
		                       & \mid \proj{v}{\Context} \mid \Case{\Context} \mid \Context + e
		\mid v + \Context
		\\
		                       &                   &
		                       & \mid \GuardSize\,\,{\Context} \mid \CasePat{\Context}                                                                   \\
		\textbf{Guard Context} & \GuardContext
		                       & \bnfeq            &
		\ContextHole \mid \GuardAnd{\GuardContext}{g}
		\mid \GuardOr{\GuardContext}{g} \mid \GuardIsOf{\GuardContext}{t}
		\mid \GuardContext = a \mid v = \GuardContext
		\mid \GuardNeq{\GuardContext}{a}
		\mid \GuardNeq{v}{\,\GuardContext} \\
		&&& \mid \GuardLt{\GuardContext}{a} \mid \GuardLt{v}{\,\GuardContext}
	\end{array}
\]
\endgroup

The operational semantics consists of two categories of reduction rules: standard reductions that define successful computation, and failure reductions that handle runtime errors. These rules are presented in Figure~\ref{app:fig:standard-and-failure-reductions}.

\begin{figure}[ht]
	\begin{tabular}{lrclr}
		\RuleDef{reduction:app}{\RedRule{App}}
		 & $\app{(\lambda^\mathbb{I}{\ov{x}}.{e})}{\ov{v}}$
		 & $\kern-0.5em\reduces\kern-0.5em$
		 & $e \Subst{\ov{x}}{\ov{v}}$
		 & if $|\ov{x}|=|\ov{v}|$                                       \\
		\RuleDef{reduction:proj}{\RedRule{Proj}}
		 & $\proj{i}{\tuple{v_0,\ldotsTwo,v_n}}$
		 & $\kern-0.5em\reduces\kern-0.5em$
		 & $v_i$
		 & if $i \in [ 0 \ldotsTwo n ]$                                 \\
		\RuleDef{reduction:plus}{\RedRule{Plus}}
		 & $v + v'$
		 & $\reduces$
		 & $v''$
		 & where $v'' = v + v'$                                         \\[0.5mm]
		 &
		 &
		 &
		 & and $v, v'$ are integers                                     \\
		\RuleDef{reduction:size}{\RedRule{Size}}
		 & $\GuardSize\,{\tuple{v_1,\ldotsTwo, v_n}}$
		 & $\kern-0.5em\reduces\kern-0.5em$
		 & $n$
		 &                                                              \\
		\RuleDef{reduction:guard-case}{\RedRule{Case}}
		 & $\caseExprIndex[v]$                                          
		 & $\reduces$
		 & $e_j\, \sigma$
		 & if $v / (\when{p_j}{g_j}) = \sigma$ and                      \\[0.5mm]
		 &
		 &
		 &
		 & $\forall i\!<\!j.\ v / (\when{p_i}{g_i}) = \fail$ \\
		\RuleDef{reduction:context}{\RedRule{Context}}
		 & $\ContextWith{e}$
		 & $\kern-0.5em\reduces\kern-0.5em$
		 & $\ContextWith{e'}$
		 & if $e \reduces e'$                                           \\[4mm]
		\RuleDef{reduction:app-omega}{\RedRule{App$_\omega$}}
		 & $\app{v}{\ov{v}}$
		 & $\kern-0.5em\reduces\kern-0.5em$
		 & $\omegaApp$
		 & if $v\neq\lambda^\IFace{\ov{x}}.{e}$ with $|\ov{x}|{=}|\ov{v}|$ \\
		\RuleDef{reduction:proj-omega-range}{\RedRule{Proj$_{\omega,\textsc{range}}$}}
		 & $\proj{v}{\tuple{\ov{v}}}$
		 & $\kern-0.5em\reduces\kern-0.5em$
		 & $\omegaOutOfRange$
		 & if $v \neq i$ for $i = 0 \ldotsTwo |\ov{v}|{-}1$              \\
		\RuleDef{reduction:proj-omega-not-tuple}{\RedRule{Proj$_{\omega,\textsc{notTuple}}$}}
		 & $\proj{v'}{v}$
		 & $\kern-0.5em\reduces\kern-0.5em$
		 & $\omegaProjection$
		 & if $v \not = \tuple{\ov{v}}$                                 \\
		\RuleDef{reduction:plus-omega}{\RedRule{Plus$_\omega$}}
		 & $v + v'$
		 & $\kern-0.5em\reduces\kern-0.5em$
		 & $\omegaPlus$
		 & if $v$ or $v'$ not integers                                  \\
		\RuleDef{reduction:size-omega}{\RedRule{Size$_\omega$}}
		 & $\GuardSize\,\,v$
		 & $\kern-0.5em\reduces\kern-0.5em$
		 & $\omegaSize$
		 & if $v \neq \{\ov{v}\}$                                       \\
		\RuleDef{reduction:guard-case-omega}{\RedRule{Case$_\omega$}}
		 & $\caseExprIndex[v]$
		 & $\kern-0.5em\reduces\kern-0.5em$
		 & $\omegaCase$
		 & if $\forall(i\leq n).\match{v}{\when{p_i}{g_i}} = \fail$    \\[1mm]
		\RuleDef{reduction:context-omega}{\RedRule{Context$_\omega$}}
		 & $\ContextWith{e}$
		 & $\kern-0.5em\reduces\kern-0.5em$
		 & $\omega_p$
		 & if $e \reduces \omega_p$                                     \\
	\end{tabular}
	\caption{Standard and Failure Reductions}%
	\label{app:fig:standard-and-failure-reductions}
\end{figure}

\emph{Pattern Matching.}
Since patterns contain capture variables, the reduction of pattern matching implies the creation
of a substitution $\sigma$ that binds the capture variables of the pattern to the values they capture. Figure~\ref{fig:pattern-matching-operation}
defines the pattern matching operation. The form $p_1\&p_2$ used in this
figure is an auxiliary intersection pattern introduced by the guard analysis:
it is not a surface pattern, and it matches a value exactly when both component
patterns match that value. Patterns are linear: no variable occurs twice in a
pattern, and in particular no variable occurs in both $p_1$ and $p_2$ (the
capture variables of a pattern are renamed apart from the variables of the
skeleton that the guard analysis intersects with it), so that $\sigma_1\cup\sigma_2$
below is a substitution.

\begin{figure}[ht]
	\[
		\begin{array}{lll}
			v/c                                     & = \{\}                      & \text{if } v =c                                                          \\
			v/x                                     & = \{x \mapsto v\}                                                                                      \\
			v/(p_1 \&\, p_2)                        & = \sigma_1 \cup \sigma_2    & \text{if } v/p_1 = \sigma_1 \text{ and } v/p_2 = \sigma_2                \\
			\{v_1, \dots, v_n\}/\{p_1, \dots, p_n\} & = \bigcup_{i=1..n} \sigma_i & \text{if } v_i/p_i = \sigma_i \text{ for all } i =1..n                   \\
			v/p                                     & = \fail                     & \text{otherwise}                                                         \\[4mm]
			v/(\when{p}{g})                         & = \sigma                    & \text{if } v/p = \sigma \text{ and } g\,\sigma \seqreducesG \mathsf{true} \\
			v/(\when{p}{g})                         & = \fail                     & \text{otherwise }
		\end{array}
	\]
	where  $\sigma$ denotes substitutions from variables to values
	\caption{Pattern Matching Operation}%
	\label{fig:pattern-matching-operation}
\end{figure}

\emph{Guard Semantics.}
The semantics of pattern matching includes the evaluation of guards.
A pattern-guard branch succeeds if and only if the value matches the pattern and the guard evaluates to true.
Note that a guard can fail (resulting in runtime error $\omega$), in which case the branch is skipped. This does not imply a failure of the whole pattern matching expression since the failure of a guard is part of standard Elixir semantics: for instance consider the following definition:
\begin{minted}{elixir}
def test2(x) when is_integer(elem(x,1)) or elem(x,0) == :int, do: elem(x,1)      !\elabel{test21}!
def test2(x) when is_boolean(elem(x,0)) do: elem(x,0) !\elabel{test22}!
\end{minted}
The application \elix{test2({true})} makes the guard in line \ref{test21} fail, but the application succeeds by returning \elix{true}.

This behavior is formalized by the \RuleRef{reduction:guard-context}{\textsc{Context$_\omega$}}
rule in Figure~\ref{fig:guard-reductions}, while other rules in this figure deal with Boolean operations, type tests, equality checks, and arithmetic operations within guard contexts.
The Boolean rules are intentionally asymmetric. In an \texttt{and}-guard, any
left-hand value different from \smalltrue\ is enough for the conjunction to
evaluate to \smallfalse\ without evaluating the right-hand side. In an \texttt{or}-guard, the
right-hand side is evaluated only when the left-hand side evaluates exactly to
\smallfalse; if the left-hand side succeeds, the disjunction succeeds
immediately; if it evaluates to any other value, the disjunction evaluates to
\smallfalse\ immediately. If the left-hand side of a conjunction or of a disjunction errors, the \RuleRef{reduction:guard-context}{context rule} makes the guard fail. When an atom inside a guard errors, the whole guard reduces to $\fail$, by rule \RuleRef{reduction:guard-context}{\textsc{Context$_\omega$}} applied to the whole guard, or by rule \RuleRef{reduction:guard-ctx}{\textsc{Context$_g$}} applied to a subguard followed by rule \RuleRef{reduction:guard-ctx-fail}{\textsc{Context$_{\fail}$}}. Atoms contain no application, so their reduction terminates; hence every guard $g\sigma$, where $\sigma$ maps the variables of $g$ to values, reduces to exactly one of $\smalltrue$, $\smallfalse$ and $\fail$.
The comparison guard \texttt{<} depends on a total value ordering
relation directly inherited from Elixir which is defined on
Figure~\ref{fig:elixir-ordering}.
In that figure, in the guards Eq and NotEq of Figure~\ref{fig:guard-reductions},
and in the matching of constants of
Figure~\ref{fig:pattern-matching-operation}, $v = v'$ denotes syntactic
equality up to $\alpha$-renaming and up to the interfaces of the
$\lambda$-abstractions that $v$ and $v'$ contain; in
Figure~\ref{fig:elixir-ordering}, $\prec$ is a fixed strict total order on
$\lambda$-abstractions taken up to the same equality. On the BEAM, equality
and order on functions are implementation-defined (two $\alpha$-equivalent
functions created at different places of a program may differ), and nothing in
our development depends on these choices: the guard analysis never decides an
equality or a comparison between two functions (the strict ordering of types of
Section~\ref{sec:strict-ordering-of-types} does not distinguish functions).
Neither $=$ nor $\prec$ depends on interfaces, which do not exist at run time,
so a program cannot observe them.

\newcommand{\lt}{<}

\begin{figure}[ht]
	\begin{tabular}{rrcll}
		\RuleDef{reduction:guard-and-true}{\RedRule{And$_\true$}}
		 & $\GuardAnd{\GuardTrue}{g}$
		 & $\kern-0.5em\reducesG\kern-0.5em$
		 & $g$
		 &                                   \\
		\RuleDef{reduction:guard-and-false}{\RedRule{And$_\false$}}
		 & $\GuardAnd{v}{g}$
		 & $\kern-0.5em\reducesG\kern-0.5em$
		 & $\GuardFalse$
		 & if $v \not = \GuardTrue$          \\
		\RuleDef{reduction:guard-or-true}{\RedRule{Or$_\true$}}
		 & $\GuardOr{\GuardTrue}{g}$
		 & $\kern-0.5em\reducesG\kern-0.5em$
		 & $\GuardTrue$
		 &                                   \\
		\RuleDef{reduction:guard-or-false}{\RedRule{Or$_\false$}}
		 & $\GuardOr{\GuardFalse}{g}$
		 & $\kern-0.5em\reducesG\kern-0.5em$
		 & $g$
		 &                                   \\
		\RuleDef{reduction:guard-or-v}{\RedRule{Or$_v$}}
		 & $\GuardOr{v}{g}$
		 & $\kern-0.5em\reducesG\kern-0.5em$
		 & $\GuardFalse$
		 & if $v \notin \BoolSet$           \\
		 \RuleDef{reduction:guard-lt-true}{\RedRule{Lt$_\true$}}
		 & $\GuardLt{v}{v'}$
		 & $\kern-0.5em\reducesG\kern-0.5em$
		 & $\GuardTrue$
		 & \text{if } $v \mathrel{\lt_{\mathrm{term}}} v'$ \\
		 \RuleDef{reduction:guard-lt-false}{\RedRule{Lt$_\false$}}
		 & $\GuardLt{v}{v'}$
		 & $\kern-0.5em\reducesG\kern-0.5em$
		 & $\GuardFalse$
		 & \text{else } \\
		\RuleDef{reduction:guard-eq-true}{\RedRule{Eq$_\true$}}
		 & $\GuardEq{v}{v'}$
		 & $\kern-0.5em\reducesG\kern-0.5em$
		 & $\GuardTrue$
		 & \text{if } $v = v'$               \\
		\RuleDef{reduction:guard-eq-false}{\RedRule{Eq$_\false$}}
		 & $\GuardEq{v}{v'}$
		 & $\kern-0.5em\reducesG\kern-0.5em$
		 & $\GuardFalse$
		 & \text{else }                      \\
		\RuleDef{reduction:guard-neq-true}{\RedRule{NotEq$_\true$}}
		 & $\GuardNeq{v}{v'}$
		 & $\kern-0.5em\reducesG\kern-0.5em$
		 & $\GuardTrue$
		 & \text{if } $v \neq v'$            \\
		\RuleDef{reduction:guard-neq-false}{\RedRule{NotEq$_\false$}}
		 & $\GuardNeq{v}{v'}$
		 & $\kern-0.5em\reducesG\kern-0.5em$
		 & $\GuardFalse$
		 & \text{else }                   \\
		\RuleDef{reduction:guard-oftype-true}{\RedRule{OfType$_\true$}}
		 & $v \isof t$
		 & $\kern-0.5em\reducesG\kern-0.5em$
		 & $\GuardTrue$
		 & \text{if } $v \in t$              \\
		\RuleDef{reduction:guard-oftype-false}{\RedRule{OfType$_\false$}}
		 & $v \isof t$
		 & $\kern-0.5em\reducesG\kern-0.5em$
		 & $\GuardFalse$
		 & \text{else}                       \\
		\RuleDef{reduction:guard-ctx}{\RedRule{Context$_g$}}
		 & $\GuardContextWith{g}$
		 & $\kern-0.5em\reducesG\kern-0.5em$
		 & $\GuardContextWith{g'}$
		 & if $g \reducesG g'$               \\
		\RuleDef{reduction:guard-ctx-atom}{\RedRule{Context$_a$}}
		 & $\GuardContextWith{a}$
		 & $\kern-0.5em\reducesG\kern-0.5em$
		 & $\GuardContextWith{a'}$
		 & if $a \reduces a'$               \\
		\RuleDef{reduction:guard-context}{\RedRule{Context$_\omega$}}
		 & $\GuardContextWith{a}$
		 & $\kern-0.5em\reducesG\kern-0.5em$
		 & $\fail$
		 & if $a \reduces \omega_p$          \\
		\RuleDef{reduction:guard-ctx-fail}{\RedRule{Context$_{\fail}$}}
		 & $\GuardContextWith{\fail}$
		 & $\kern-0.5em\reducesG\kern-0.5em$
		 & $\fail$
		 & if $\GuardContext\neq\ContextHole$
	\end{tabular}
	\caption{Guard Reductions}%
	\label{fig:guard-reductions}
\end{figure}

\newcommand{\rank}{\mathop{\mathrm{rank}}}

\begin{figure}[ht]
\[
\rank(v)=\begin{cases}
0 & v\;\text{is an integer}\\
1 & v\;\text{is an atom}\\
2 & v\;\text{is a function value}\\
3 & v\;\text{is a tuple}
\end{cases}
\]

\[
\begin{array}{rlcl}
\text{Integers:}   & n \mathrel{\lt_{\mathrm{intra}}} n' &\iff& n<n' \\[2pt]
\text{Atoms:}      & a \mathrel{\lt_{\mathrm{intra}}} a' &\iff& \mathrm{str}(a)
												   \;\text{is lexicographically before}\; \mathrm{str}(a')\\[2pt]
\text{Functions:}  & f \mathrel{\lt_{\mathrm{intra}}} f' &\iff& f \prec f'
												   \\[2pt]
\text{Tuples:} & (v_1,..,v_m) \mathrel{\lt_{\mathrm{intra}}} (v'_1,..,v'_{m'})
				 &\iff& m<m' \;\lor\; \\
				 & & & (m=m' \land \exists i.\,
						v_i \mathrel{\lt_{\mathrm{term}}} v'_i \\
				 & & & \qquad\land \forall j<i.\, v_j = v'_j)
\end{array}
\]

\[
v \mathrel{\lt_{\mathrm{term}}} v'
   \;\;\;\text{iff}\;\;\;
     \rank(v) < \rank(v')
     \;\lor\;
     (\rank(v)=\rank(v') \land v \mathrel{\lt_{\mathrm{intra}}} v')
\]
\caption{Elixir-style value ordering}%
\label{fig:elixir-ordering}
\end{figure}

\noindent
The reduction of guard atoms $a$ comes from standard and failure
reductions defined in Figure~\ref{app:fig:standard-and-failure-reductions}.


\section{Type Safety for Safe Erasure Gradual Typing}\label{app:soundness}

\GuillaumeReviews{This appendix proves the safety property for the erased
gradual system of Section~\ref{sec:safe-erasure}. The proof is organized around
the dependency that makes strong functions sound. The judgment $\Gamma \vdg e:t$
is the gradual type system used to type source programs: it contains the rules
that propagate `$\dyn$' and the rules that introduce or use strong arrows. The
introduction of a strong arrow, however, depends on an auxiliary property of the
function body: when the argument is only known dynamically, every normal return
must still have the announced codomain. This property is expressed by the weak
judgment $\Gamma \vdw e \tc t$.

For this reason the preservation argument is proved first for an extension of
the weak judgment (Definition~\ref{def:enriched-weak}), and the result is then
transferred back to the gradual system by
Lemma~\ref{lem:gradual-to-strong}. The gradual system is the typing discipline
for programs; the weak system records the invariant that justifies strong
function types in an erased semantics.}

Throughout this appendix and Appendix~\ref{app:guardanalysis}, $\leq$ is either of two subtyping relations. On
static types, both are induced by the interpretation of
Appendices~\ref{app:functions} and~\ref{sec:strong-subtyping}, which depends
on a parameter $\OO$: for $\OO=\{\invOmega\}$, it is the interpretation
of~\citet{lanvin2021semantic}, extended to multi-arity functions and strong
arrows, which induces the relation of this work
(Remark~\ref{rem:two-models}); for $\OO=\varnothing$, it is the interpretation
of Section~\ref{sec:arity}. Both relations are extended to gradual types by
the characterization of Section~\ref{sec:gradual-typing}. Every result of these
two appendices holds for both relations. The proofs are written once, for both: they
use the interpretation only through properties that
Appendices~\ref{app:functions} and~\ref{sec:strong-subtyping} prove for both
values of $\OO$, and where the two relations require different arguments, the
proofs say so.

	The terms constituting the source language of Section~\ref{sec:safe-erasure} are defined by the following grammar:
	\[
		\begin{array}{l@{\hspace{1em}}lr}
			\textbf{Terms}      & e ::= x \mid c \mid \lambda^\mathbb{I} x. e \mid e\, e
			\mid \tuple{\ov{e}} \mid \proj{e}{e} \mid e + e
			\mid \texttt{case}\,\,e\,\, (\tau_i \rightarrow e_i)_{i\in I}                \\[1mm]
			\textbf{Values}     & v ::= c \mid \lambda^\mathbb{I} x. e \mid \tuple{\ov{v}}                   \\[1mm]
			\textbf{Interfaces} & \mathbb{I} ::= \{ t_i \to s_i \mid i \in I  \}         \\
		\end{array}
	\]
	Rule~\RuleRef{typing:gradual:lambda}{($\lambda$)} of the static system types only
	abstractions whose interface is separated (Section~\ref{sec:gradual-typing}):
	the case of $\beta$-reduction in the proof of Lemma~\ref{lem:preservation}
	uses this condition.

	The tests of a case expression may overlap: the case rules of
	Figures~\ref{app:fig:gradual-rules} and~\ref{app:fig:strong-rules} type only
	the branches $\tau_i\rarr e_i$ whose first-match region
	$(t\land\tau_i)\setminus\bigvee_{j<i}\tau_j$ is not empty, where $t$ is the
	type of the matched expression.

	\bigskip

\noindent
\GuillaumeReviews{The typing rules for Core Elixir are given in a declarative
style and grouped into three figures. Figure~\ref{app:fig:gradual-rules}
contains the gradual rules for source programs, including the rules that
propagate dynamic information and the rule that introduces strong arrow types.
Figure~\ref{app:fig:tuple-rules} gives the tuple and projection rules.
Figure~\ref{app:fig:strong-rules} gives the weak system used to justify strong
function types. The safety proof below is stated on an extension of this weak
system (Definition~\ref{def:enriched-weak}), which, unlike the weak system
itself, is stable under reduction; the gradual theorem follows from the
inclusion of gradual typing into weak typing.}

\RuleRef{typing:gradual:lambda}{Rule ($\lambda$)} of Figure~\ref{app:fig:gradual-rules}
is admissible in the gradual system: \RuleRef{typing:gradual:lambda-dyn}{rule
($\lambda_\dyn$)} has the same premises and no side condition, and its
conclusion is a subtype of that of rule ($\lambda$). We list rule ($\lambda$)
because it belongs to the static system defined next.

Type-checking a program using only the static rules of
Figures~\ref{app:fig:gradual-rules} and~\ref{app:fig:tuple-rules}, which
are those not annotated with any subscript $\star$ or $\dyn$, is denoted by
$\Gamma \vds e : t$.
Among these static rules, the ones marked $\omega$ deliberately admit
potentially unsafe projections and correspond to compiler warnings. A
derivation that avoids all $\omega$-marked rules gives the strongest safety
guarantee, since it prevents all runtime errors defined in the operational
semantics.

Type-checking with the full gradual system ensures
that a well-typed terminating program evaluates only to a value of the expected type, but admits
various runtime errors.

\begin{figure}
    \begin{equation*}
        \begin{gathered}
            \typingrule{\RuleDef{typing:gradual:cst}{\text{(cst)}}}
            {}{\Gamma \vdg c : c}
            \quad
            \typingrule{\RuleDef{typing:gradual:var}{\text{(var)}}}
            {\Gamma(x) = t}{\Gamma \vdg x : t}
            \\[2mm]
            \typingrule{\RuleDef{typing:gradual:lambda}{(\lambda)}}
            {\mathbb{I}\neq\emptyset \quad \forall (t_i \rarr s_i) \in \mathbb{I}\quad (\Gamma, x : t_i \vdg e : s_i)}
            {\Gamma \vdg \lambda^\mathbb{I} x\,.\, e : \bigwedge_{i \in I} (t_i \rightarrow s_i)}
            \, (\mathbb{I} \text{ separated})
            \\[2mm]
            \TRuleLambdaDyn
            \\[2mm]
            \typingrule{\RuleDef{typing:gradual:lambda-star}{(\lambda_\star)}}
            {\Gamma \vdg \lambda^\mathbb{I} x. e : t_1 \rarr t_2 \qquad
                \Gamma, x : \dyn \vdw e \tc t_2 \land \dyn}
            {\Gamma \vdg \lambda^\mathbb{I} x. e : \strongType{t_1 \rarr t_2}}
            \\[2mm]
            \typingrule{\RuleDef{typing:gradual:app}{\text{(app)}}\!}
            {\Gamma \vdg e_1 : {t_1} \rightarrow {t_2} \quad \Gamma \vdg e_2 : t_1}
            {\Gamma \vdg \app{e_1}{e_2} : t_2}
            \, (t_1 \text{ static})
            \\[2mm]
            \typingrule{\RuleDef{typing:gradual:app-star}{(\text{app}_\star)}}
            {\Gamma \vdg e_1 : \strongType{t_1 \rightarrow t}  \quad \Gamma \vdg e_2 : t_2}
            {\Gamma \vdg \app{e_1}{e_2} : t \land \dyn}
            (t_2 \consist t_1)
            \\[2mm]
            \TRuleAppDyn
            \\[2mm]
            \typingrule{\RuleDef{typing:gradual:case}{(\text{case})}}
            {\Gamma \vdg e : t \quad\,\, (\forall i \in I) \,;
                ((t \land \tau_i) \!\smallsetminus\! (\bigvee_{j<i}\tau_j) \not\leq \mathds{O} \implies \Gamma \vdg e_i : t')}
            {\Gamma \vdg \texttt{case}\, e\, (\tau_i \rightarrow e_i)_i : t'}
            \, (t \leq \lor_i \tau_i)
            \\[2mm]
            \typingrule{\RuleDef{typing:gradual:case-dyn}{(\text{case}_\star)}}
            {\Gamma \vdg e : t \quad\,\, (\forall i \in I) \,;
                ((t \land \tau_i) \!\smallsetminus\! (\bigvee_{j<i}\tau_j) \not\leq \mathds{O} \implies \Gamma \vdg e_i : t')}
            {\Gamma \vdg \texttt{case}\, e\, (\tau _i \rightarrow e_i)_i : t' \land \dyn}
            \, (t \consist \lor_i \tau_i)
            \\[2mm]
            \typingrule{\RuleDef{typing:gradual:plus}{(\text{add})}}
            {\Gamma \vdg e_1 : \intTop \quad \Gamma \vdg e_2 : \intTop}
            {\Gamma \vdg e_1 + e_2 : \intTop}
            \quad
            \typingrule{\RuleDef{typing:gradual:plus-dyn}{(\text{plus}_\star)}}
            {\Gamma \vdg e_1 : t_1 \quad \Gamma \vdg e_2 : t_2}
            {\Gamma \vdg e_1 + e_2 : \texttt{int} \land \dyn}
            \begin{cases} t_1 \consist \texttt{int} \\ t_2 \consist \texttt{int} \end{cases}
            \\
            \typingrule{\RuleDef{typing:gradual:sub}{(\text{sub})}}{\Gamma \vdg e : t_1 \quad t_1 \leq t_2}{\Gamma \vdg e : t_2}
        \end{gathered}
    \end{equation*}
    \caption{Typing rules for the gradual system}%
    \label{app:fig:gradual-rules}
\end{figure}

\begin{figure}
	\begin{equation*}
        \begin{gathered}
            \typingrule{\RuleDef{typing:gradual:tuple}{\text{(tuple)}}}
            {(\forall i = 1..n)\,;\, (\Gamma \vdg e_i : t_i)}
            {\Gamma \vdg \tuple{e_1\texttt{,}\mathtt{...}\texttt{,}e_n} : \tuple{t_1\texttt{,}\mathtt{...}\texttt{,}t_n}}
            \\[2mm]
            \typingrule{\RuleDef{typing:gradual:proj}{\text{(proj)}}}
            {\Gamma \vdg e'\!: \biglor_{i \in K} i \quad \Gamma \vdg e : \OpenTuple{t_0\texttt{,}\mathtt{...}\texttt{,}t_n}}
            {\Gamma \vdg \proj{e'}{e} : \biglor_{i\in K} t_i}
            \,
            \scalebox{0.9}{$K \subseteq [0,n]$}
            \\[2mm]
            \typingrule{\RuleDef{typing:gradual:proj-omega}{\text{(proj$_\omega$)}}}
            {\Gamma \vdg e' : \intTop \quad \Gamma \vdg e :  \tuple{t_0\texttt{,}\mathtt{...}\texttt{,}t_n}}
            {\Gamma \vdg \proj{e'}{e} : \biglor_{i\leq n} t_i}
            \qquad
            \typingrule{\RuleDef{typing:gradual:proj-top-omega}{\text{(proj$^\topp_\omega$)}}}
            {\Gamma \vdg e' : \intTop \quad \Gamma \vdg e : \tupleTop}
            {\Gamma \vdg \proj{e'}{e} : \topp}
            \\[2mm]
            \TRuleProjDyn
            \\[2mm]
            \TRuleProjStar
        \end{gathered}
    \end{equation*}
	\caption{Typing rules for tuples}%
	\label{app:fig:tuple-rules}
\end{figure}

\subsection{Static safety}\label{app:static-safety}

Subject reduction fails for the static system when abstractions are passed as
arguments. Let $A=\intTop\rarr\intTop$,
$F=\lambda^{\{A\rarr\tuple{A,A};\;\neg A\rarr\tuple{\neg A,\neg A}\}}x.\tuple{x,x}$,
and $v=\lambda^{\{\boolTop\rarr\boolTop\}}y.y$. Then
$\JudgmentTypeS{\varnothing}{\app{F}{v}}{\tuple{A,A}\lor\tuple{\neg A,\neg A}}$,
since the type of $F$ is a subtype of
$\topp\rarr(\tuple{A,A}\lor\tuple{\neg A,\neg A})$, but $\app{F}{v}$ reduces to
$\tuple{v,v}$, which has no such type: every type that the static system
deduces for $v$ is a supertype of $\boolTop\rarr\boolTop$, which is a subtype
neither of $A$ nor of $\neg A$. Like the type system of~\citet{FCB08}, whose
rule for abstractions also deduces negations of arrow types, we therefore use
a system that types more values, but only in the proofs. In the presence of
interfaces with `$\dyn$', the abstractions of a program must not receive
negated arrow types (Remark~\ref{rem:marks}):
only the abstractions that an application substitutes for its parameters,
alone or inside tuples, do, and we mark them.

\medskip
\noindent
An \emph{arrow atom} is a weak arrow type $\fun{\p{t^1,\ldotsTwo,t^n}}{s}$, of
arity $n$ (written $t\rarr s$ when $n=1$), or a strong arrow type
$\StrongType{(t\rarr s)}$, of arity $1$; it is static if `$\dyn$' does not
occur in it. Recall that relations are tagged by their arity
(Appendix~\ref{app:functions}).

\begin{lem}[Primality]\label{lem:arrow-primality}
	Let $l$ be the type $\function_n$ or a finite non-empty intersection of
	static arrow atoms of the same arity $n$, and let $N$ be a finite set of
	static arrow atoms and of types
	$\function$ and $\function_m$ ($m\in\mathbb{N}$). Then
	$l\leq\bigvee_{a\in N}a$ if and only if $l\leq a$ for some $a\in N$. Hence
	$l\land\bigwedge_{a\in N}\neg a\not\simeq\bott$ if and only if
	$l\not\leq a$ for every $a\in N$.
\end{lem}

\begin{Proof}
	The ``if'' direction is immediate. By
	Lemma~\ref{lemma:function-interpretation} and
	Definition~\ref{app:def:strong-arrow}, the interpretation of an arrow atom
	$a$ of arity $k$ is the set $\fparts{Z_a}$ of the finite subsets of some set
	$Z_a\subseteq\DO^k\times\DOmega$, taken as relations of arity $k$, and the
	interpretation of $\function_k$ is $\fparts{\DO^k\times\DOmega}$. So the
	interpretation of $l$ is $\fparts{Z}$, where $Z$ is $\DO^n\times\DOmega$ if
	$l$ is $\function_n$, and the intersection of the sets $Z_a$ of the atoms
	$a$ of $l$ otherwise. Suppose that $l\not\leq a$ for every
	$a\in N$. Then $N$ contains neither $\function$ nor $\function_n$, and for
	every $a\in N$ of arity $n$ there is a pair $z_a\in Z\setminus Z_a$. The
	relation $R=\{z_a\mid a\in N \text{ of arity } n\}$ of arity $n$ belongs to
	the interpretation of $l$; it does not belong to the interpretation of any
	$a\in N$ of arity $n$, which does not contain $z_a$, nor to that of any other
	member of $N$, which contains only relations of another arity. So
	$l\not\leq\bigvee_{a\in N}a$. The second statement follows, since
	$l\land\bigwedge_{a\in N}\neg a\simeq\bott$ if and only if
	$l\leq\bigvee_{a\in N}a$.
\end{Proof}

\noindent
For $\OO=\varnothing$ and unary weak arrows, the lemma follows from the
characterization, by~\citet{FCB08}, of the containment of an intersection of
arrows in a union of arrows.

\begin{Definition}[Marked expressions]\label{def:marked}
	Marked expressions are the expressions of
	Sections~\ref{sec:safe-erasure}--\ref{sec:arity} in which abstractions may be \emph{marked}, written
	$\mlambda^\IFace\ov{x}.e$; values include marked abstractions. The
	\emph{marking} $\mk{v}$ of a value $v$ marks the abstractions of $v$ that
	are not inside an abstraction: $\mk{c}=c$; if $v$ is an abstraction
	$\lambda^\IFace\ov{x}.e$ or $\mlambda^\IFace\ov{x}.e$, then $\mk{v}$ is
	$\mlambda^\IFace\ov{x}.e$; and if $v=\tuple{v_1,\ldotsTwo,v_n}$ is not a
	constant, then $\mk{v}=\tuple{\mk{v_1},\ldotsTwo,\mk{v_n}}$. The
	\emph{erasure} $\erase{e}$ of a marked expression $e$ removes its marks,
	and a \emph{marking} of an expression $e$ is any marked expression
	$\hat{e}$ such that $\erase{\hat{e}}=e$. Marked expressions reduce by the
	rules of Figure~\ref{fig:reductions}, their extension of
	Section~\ref{sec:arity} and, for the expressions of
	Section~\ref{sec:guard-analysis}, the rules recalled in
	Appendix~\ref{app:guard-safety}, where a marked abstraction is an abstraction (in
	particular, it belongs to the same test types), except that the rule for
	applications marks the arguments:
	\[
		\app{(\lambda^\IFace\ov{x}.e)}{\ov{v}}\reduces e\Subst{\ov{x}}{\ov{\mk{v}}}
		\qquad\qquad
		\app{(\mlambda^\IFace\ov{x}.e)}{\ov{v}}\reduces e\Subst{\ov{x}}{\ov{\mk{v}}}
	\]
\end{Definition}

\noindent
No reduction rule depends on the marks, and neither does the test $v\in\tau$.
So, for every marked expression $e$, if $e\reduces e'$ then
$\erase{e}\reduces\erase{e'}$, and if $e\reduces\omega_p$ then
$\erase{e}\reduces\omega_p$; conversely, if $\erase{e}\reduces e''$ then
$e\reduces e'$ for some $e'$ with $\erase{e'}=e''$, and if
$\erase{e}\reduces\omega_p$ then $e\reduces\omega_p$. Therefore an expression
$e$ diverges, reduces to a value $v$, or reduces to a failure $\omega_p$ if and
only if, as a marked expression, it diverges, reduces to a marking of $v$, or
reduces to $\omega_p$.

For an abstraction $v$, marked or not, with interface
$\IFace=\{\fun{\p{t_i^1,\ldotsTwo,t_i^n}}{s_i}\mid i\in I\}$, we write $I_v$
for $\bigwedge_{i\in I}\fun{\p{t_i^1,\ldotsTwo,t_i^n}}{s_i}$. When
$\IFace\neq\emptyset$, as rules ($\lambda$) and ($\lambda_\neg$) below require,
$\dynSup{I_v}=\bigwedge_{i\in I}\fun{\p{\dynInf{(t_i^1)},\ldotsTwo,\dynInf{(t_i^n)}}}{\dynSup{s_i}}$
is a non-empty static intersection of arrow atoms of arity $n$.

\begin{Definition}[Enriched static system]\label{def:enriched-static}
	The judgment $\JudgmentTypeSN{\Gamma}{e}{t}$, where $e$ is a marked
	expression, is derived by the static rules (the rules of
	Figures~\ref{app:fig:gradual-rules} and~\ref{app:fig:tuple-rules} not
	annotated with $\star$ or $\dyn$, their extension of
	Section~\ref{sec:arity} and, for the expressions of
	Section~\ref{sec:guard-analysis}, the static rules of
	Appendix~\ref{app:guard-safety}), with $\vdsn$ in place of $\vdg$, where
	\RuleRef{typing:gradual:lambda}{rule ($\lambda$)} applies only to unmarked
	abstractions, by the following rule for marked abstractions:
	\[
		\typingrule{\RuleDef{typing:enriched:lambda-neg}{(\lambda_\neg)}}
		{\IFace\neq\emptyset \quad \forall (t_i\rarr s_i)\in\IFace\,\,\p{\JudgmentTypeSN{\Gamma, x:t_i}{e}{s_i}}}
		{\JudgmentTypeSN{\Gamma}{\mlambda^\IFace x.e}{\dynSup{I_v}\land\textstyle\bigwedge_{h\in H}\neg a_h}}
		\,\,\begin{cases}\IFace\text{ separated}\\ \dynSup{I_v}\not\leq a_h\ \ (h\in H)\end{cases}
	\]
	where $v=\mlambda^\IFace x.e$, $H$ is finite (possibly empty), and every
	$a_h$ is a static arrow atom (the rule extends to several parameters as
	rule~($\lambda$) does), and by the intersection rule:
	\[
		\typingrule{\RuleDef{typing:enriched:and}{(\land)}}
		{\JudgmentTypeSN{\Gamma}{e}{t_1} \quad \JudgmentTypeSN{\Gamma}{e}{t_2}}
		{\JudgmentTypeSN{\Gamma}{e}{t_1\land t_2}}
	\]
\end{Definition}

\noindent
Every rule of the static system is a rule of $\vdsn$, so
$\JudgmentTypeS{\Gamma}{e}{t}$ implies $\JudgmentTypeSN{\Gamma}{e}{t}$, by
the same derivation; the proofs use only this inclusion. Rule
\RuleRef{typing:enriched:and}{($\land$)} mirrors rule
\RuleRef{typing:weak:and-circ}{(and$^\circ$)} of the weak system; the proof of
preservation uses it to type the body of an applied abstraction
(Lemma~\ref{lem:preservation}).

\begin{remark}\label{rem:marks}\em
	The conclusion of rule \RuleRef{typing:enriched:lambda-neg}{($\lambda_\neg$)}
	is static. With $I_v\land\bigwedge_{h\in H}\neg a_h$ instead, its minimal
	materialization could be empty: for
	$f=\lambda^{\{\intTop\rarr(\intTop\lor\dyn)\}}x.x{+}0$ and $A=\intTop\rarr\intTop$,
	the type $I_f\land\neg A$ has minimal materialization $A\land\neg A\simeq\bott$,
	so it is a subtype of $\intTop\rarr(\boolTop\lor\dyn)$, and the application
	$\app{\mk{f}}{3}$, which reduces to $3$, would have type
	$\boolTop\lor\dyn$. Rules \RuleRef{typing:gradual:lambda}{($\lambda$)} and
	\RuleRef{typing:enriched:lambda-neg}{($\lambda_\neg$)} must not type the same
	abstraction either. Otherwise $f$ would have the types $I_f$ and
	$\dynSup{I_f}\land\neg A$, and, by rule
	\RuleRef{typing:enriched:and}{($\land$)}, their intersection, which is
	equivalent to $I_f\land\neg A$ since $I_f\leq\dynSup{I_f}$: the application
	$\app{f}{3}$ would have type $\boolTop\lor\dyn$ as well. With the marks, the abstractions of a program are
	typed by rule~($\lambda$), and those that an application substitutes for its
	parameters, alone or inside tuples, are typed by rule~($\lambda_\neg$).
\end{remark}

\noindent
The types of a value $v$ are described by its \emph{generators}, the types of
the set $\gen(v)$ defined by induction on $v$:
\begin{itemize}
	\item $\gen(c)=\{c\}$;
	\item $\gen(v)=\{I_v\}$ if $v$ is an unmarked abstraction;
	\item if $v$ is a marked abstraction, $\gen(v)$ is the set of the types
	      $\dynSup{I_v}\land\bigwedge_{a\in N}\neg a$, where $N$ is a finite set
	      of static arrow atoms such that $\dynSup{I_v}\not\leq a$ for every
	      $a\in N$;
	\item $\gen(\tuple{v_1,\ldotsTwo,v_n})=\{\tuple{g_1,\ldotsTwo,g_n}\mid g_k\in\gen(v_k)\}$
	      if $\tuple{v_1,\ldotsTwo,v_n}$ is not a constant.
\end{itemize}

\begin{lem}[Generators]\label{lem:generators}
	Let $v$ be a value such that $\JudgmentTypeSN{\Gamma}{v}{s}$ for some type $s$.
	\begin{enumerate}
		\item $\JudgmentTypeSN{\Gamma}{v}{t}$ if and only if $g\leq t$ for some
		      $g\in\gen(v)$. If $v$ is a tuple, its components have types.
		\item $\dynInf{g}\not\simeq\bott$ for every $g\in\gen(v)$.
		\item For all $g_1,g_2\in\gen(v)$ there is $g\in\gen(v)$ such that
		      $g\leq g_1\land g_2$. Hence, if $\JudgmentTypeSN{\Gamma}{v}{t_1}$ and
		      $\JudgmentTypeSN{\Gamma}{v}{t_2}$, then
		      $\JudgmentTypeSN{\Gamma}{v}{t_1\land t_2}$.
		\item For every test type $\tau$ and every $g\in\gen(v)$, $g\leq\tau$ if
		      $v\in\tau$, and $g\land\tau\simeq\bott$ if $v\notin\tau$. Hence
		      $\JudgmentTypeSN{\Gamma}{v}{\tau}$ if and only if $v\in\tau$, and
		      $\JudgmentTypeSN{\Gamma}{v}{\neg\tau}$ if and only if $v\notin\tau$.
		\item If $t$ is static and $\JudgmentTypeSN{\Gamma}{v}{t}$, then
		      $\JudgmentTypeSN{\Gamma}{\mk{v}}{t}$.
		\item If $\JudgmentTypeS{\Gamma}{v}{s'}$ for some type $s'$, in
		      particular if $v$ is a constant, then
		      $\JudgmentTypeSN{\Gamma}{v}{t}$ if and only if
		      $\JudgmentTypeS{\Gamma}{v}{t}$; for a constant, both hold if and
		      only if $v\leq t$.
	\end{enumerate}
\end{lem}

\begin{Proof}
	\begin{enumerate}
		\item ``If'': every generator is deduced for $v$: by
		      \RuleRef{typing:static:cst}{(cst)} for a constant (or by
		      \RuleRef{typing:static:tuple}{(tuple)}, with the same type, for a
		      tuple of constants), and by \RuleRef{typing:gradual:lambda}{($\lambda$)},
		      respectively \RuleRef{typing:enriched:lambda-neg}{($\lambda_\neg$)}
		      with $H=N$, for an unmarked, respectively marked, abstraction, whose
		      premises and side condition on $\IFace$ hold; for a tuple, by
		      induction and \RuleRef{typing:static:tuple}{(tuple)}. Rule
		      \RuleRef{typing:static:sub}{(sub)} concludes. These premises hold,
		      and the components of a tuple have types (by
		      \RuleRef{typing:static:cst}{(cst)} for a tuple of constants), since the derivation of
		      $\JudgmentTypeSN{\Gamma}{v}{s}$, followed upwards through instances
		      of \RuleRef{typing:static:sub}{(sub)} and
		      \RuleRef{typing:enriched:and}{($\land$)}, which keep the expression,
		      reaches an instance of the rule for the outermost construct of $v$:
		      for an abstraction, of the same rule, and for a tuple that is not a
		      constant, of \RuleRef{typing:static:tuple}{(tuple)}, whose premises
		      type the components. ``Only if'': by
		      induction on the derivation; rule \RuleRef{typing:static:sub}{(sub)}
		      is handled by transitivity, rules \RuleRef{typing:static:cst}{(cst)},
		      ($\lambda$) and ($\lambda_\neg$) deduce a generator, rule
		      \RuleRef{typing:static:tuple}{(tuple)} is handled by induction on the
		      components, and rule \RuleRef{typing:enriched:and}{($\land$)} by the
		      first statement of item~3, whose proof does not use item~1: if
		      $g_1\leq t_1$ and $g_2\leq t_2$, some $g\in\gen(v)$ is a subtype of
		      $g_1\land g_2\leq t_1\land t_2$.
		\item A constant is not empty. An intersection of arrow atoms of arity
		      $n$ contains the empty relation of arity $n$, which belongs to every
		      arrow atom of arity $n$ (Definitions~\ref{app:def:multi-arity}
		      and~\ref{app:def:strong-arrow}): so
		      $\dynInf{(I_v)}\not\simeq\bott$. The generators of a marked
		      abstraction are static and not empty, by
		      Lemma~\ref{lem:arrow-primality}. For tuples, componentwise.
		\item The generators of constants and of unmarked abstractions are
		      unique. For a marked abstraction, if $g_1$ and $g_2$ are defined by
		      the sets $N_1$ and $N_2$, take the generator defined by
		      $N_1\cup N_2$. For tuples, componentwise. The consequence is an
		      instance of rule \RuleRef{typing:enriched:and}{($\land$)}.
		\item By induction on $\tau$ (Figure~\ref{app:inductive-def-test-type}).
		      If $\tau=\tau_1\lor\tau_2$ and $v\in\tau$, then $v\in\tau_i$ for some
		      $i$, and $g\leq\tau_i\leq\tau$ by the induction hypothesis; if
		      $v\notin\tau$, then $g\land\tau_1\simeq\bott\simeq g\land\tau_2$, so
		      $g\land\tau\simeq\bott$. If $\tau=\neg\tau_1$ and $v\in\tau$, then
		      $v\notin\tau_1$, so $g\land\tau_1\simeq\bott$, that is, $g\leq\tau$;
		      if $v\notin\tau$, then $v\in\tau_1$, so $g\leq\tau_1$ and
		      $g\land\tau\simeq\bott$. Otherwise $\tau$ is a base type, a
		      constant, $\function_n$, or a tuple of test types, closed or open.
		      If $v$ is a constant $c$ that is not a tuple, then $g=c$ is a
		      singleton type, so $g\leq\tau$ or $g\land\tau\simeq\bott$, and
		      $c\in\tau$ if and only if $c\leq\tau$ (for a base type, since $B(c)$
		      is the least base type that contains $c$).
		      If $v$ is an abstraction with $n$ parameters, marked or not, then
		      $v$ belongs to the test types $\function$ and $\function_n$ and to
		      no other one of these forms, while
		      $g\leq\function_n\leq\function$, since the arrow atoms of arity $n$
		      are subtypes of $\function_n$, and $g\land\tau\simeq\bott$ for every
		      other such test type $\tau$, which is a base type disjoint from
		      $\function_n$, a constant, $\function_m$ with $m\neq n$, or a tuple type. If $v=\tuple{v_1,\ldotsTwo,v_n}$,
		      then $g=\tuple{g_1,\ldotsTwo,g_n}$ with
		      $g_k\in\gen(v_k)$ (for a tuple of constants, $g$ is the product of
		      the singleton types of its components); $v$ and $g$ belong to $\tupleTop$ and are disjoint
		      from the other base types and from the constants that are not tuples; $v$ belongs to a
		      tuple constant $c'$ if and only if $v=c'$, and then $g=c'$, while otherwise
		      $g\land c'\simeq\bott$, since the lengths differ or a component of $g$ is
		      disjoint from the corresponding component of $c'$; and $v$ belongs to
		      $\tuple{\tau_1,\ldotsTwo,\tau_m}$ (respectively
		      $\OpenTuple{\tau_1,\ldotsTwo,\tau_m}$) if and only if $m=n$
		      (respectively $m\leq n$) and $v_k\in\tau_k$ for every $k\leq m$; by
		      the induction hypothesis on the $\tau_k$, $g$ is then a subtype of this test type, and otherwise
		      it is disjoint from it, since the lengths do not match or
		      $g_k\land\tau_k\simeq\bott$ for some $k\leq m$. The consequences
		      follow from items~1 and~2.
		\item Let $g\in\gen(v)$ such that $g\leq t$. Since $t$ is static,
		      $\dynSup{g}\leq t$. Moreover $\dynSup{g}\in\gen(\mk{v})$: it is
		      obtained from $g$ by replacing the generator $I_u$ of every unmarked
		      abstraction $u$ of $v$ that is not inside an abstraction by
		      $\dynSup{I_u}$, which is the generator of $\mk{u}$ defined by
		      $N=\varnothing$; the premises and the side condition of
		      rule~($\lambda_\neg$) for $\mk{u}$ are those of rule~($\lambda$) for
		      $u$. Item~1, for $\mk{v}$, concludes.
		\item A constant is typed by \RuleRef{typing:static:cst}{(cst)} in both
		      systems. The system $\vds$ types only unmarked expressions, so $v$
		      contains no marked abstraction, and $\gen(v)$ is a singleton
		      $\{g\}$, with $g=v$ for a constant. The proof of item~1 applies to
		      $\vds$, which has neither rule ($\lambda_\neg$) nor rule ($\land$),
		      with the derivation of $\JudgmentTypeS{\Gamma}{v}{s'}$ in place of
		      that of $\JudgmentTypeSN{\Gamma}{v}{s}$: so
		      $\JudgmentTypeS{\Gamma}{v}{t}$ if and only if $g\leq t$, that is, by
		      item~1, if and only if $\JudgmentTypeSN{\Gamma}{v}{t}$.\qedhere
	\end{enumerate}
\end{Proof}

\begin{lem}[Dichotomy]\label{lem:dichotomy}
	Let $v$ be a value such that $\mk{v}=v$ and $\JudgmentTypeSN{\Gamma}{v}{s}$
	for some type $s$. For every static type $t$, either
	$\JudgmentTypeSN{\Gamma}{v}{t}$ or $\JudgmentTypeSN{\Gamma}{v}{\neg t}$.
\end{lem}

\begin{Proof}
	By Lemma~\ref{lem:generators}(1), it suffices to find a generator of $v$
	that is a subtype of $t$ or of $\neg t$. By induction on $v$.
	\begin{description}
		\item[Constants.] A constant $c$ is a singleton type: $c\leq t$ or $c\leq\neg t$.
		\item[Marked abstractions.] Let $v$ have $n$ parameters, and let
		      $l=\dynSup{I_v}$. Every generator of $v$ is a subtype of
		      $\function_n$, so only $t\land\function_n$ matters. Write it as a
		      union $\bigvee_{k\in K}(\function_n\land\bigwedge_{m\in M_k}a_{k,m}\land\bigwedge_{h\in H_k}\neg b_{k,h})$,
		      where the $a_{k,m}$ and the $b_{k,h}$ are static arrow atoms of arity
		      $n$ (in a disjunctive normal form of $t$, the other literals either
		      contain $\function_n$, and can be dropped, or are disjoint from it,
		      and remove their disjunct). If some $k$ has $l\leq a_{k,m}$ for every
		      $m\in M_k$ and $l\not\leq b_{k,h}$ for every $h\in H_k$, then
		      $l\land\bigwedge_{h\in H_k}\neg b_{k,h}$ is a generator of $v$ and a
		      subtype of $t$. Otherwise, for every $k$, either some $a_{k,m}$ has
		      $l\not\leq a_{k,m}$, and we choose one and put it in a set $N$, or
		      some $b_{k,h}$ has $l\leq b_{k,h}$. Then
		      $g=l\land\bigwedge_{a\in N}\neg a$ is a generator of $v$, and it is
		      disjoint from every disjunct: from those of the first kind since
		      $g\leq\neg a_{k,m}$, from the others since $g\leq l\leq b_{k,h}$. So
		      $g\leq\function_n\land\neg(t\land\function_n)\leq\neg t$.
		\item[Tuples.] Let $v=\tuple{v_1,\ldotsTwo,v_n}$ not be a constant;
		      then $\mk{v_j}=v_j$ for every $j$. Every generator of $v$ is a subtype
		      of $\tuple{\topp,\ldotsTwo,\topp}$ (with $n$ components), and
		      $t\land\tuple{\topp,\ldotsTwo,\topp}$ is a finite union
		      $\bigvee_{k\in K}\tuple{t_{k,1},\ldotsTwo,t_{k,n}}$ of products of
		      static types, since the intersection of two products of length $n$ is
		      a product and their difference is a finite union of products. By the
		      induction hypothesis, which applies to the components since they
		      have types (Lemma~\ref{lem:generators}(1)), for all $k$ and $j$,
		      $\JudgmentTypeSN{\Gamma}{v_j}{t_{k,j}}$ or
		      $\JudgmentTypeSN{\Gamma}{v_j}{\neg t_{k,j}}$. If some $k$ has
		      $\JudgmentTypeSN{\Gamma}{v_j}{t_{k,j}}$ for every $j$, then
		      Lemma~\ref{lem:generators}(1) gives generators $g_j\leq t_{k,j}$, and
		      $\tuple{g_1,\ldotsTwo,g_n}\leq t$. Otherwise, choose for every $k$ a
		      position $j(k)$ such that
		      $\JudgmentTypeSN{\Gamma}{v_{j(k)}}{\neg t_{k,j(k)}}$. By
		      Lemma~\ref{lem:generators}(1,3), every $v_j$ has a generator $g_j$
		      that is a subtype of $\neg t_{k,j}$ for every $k$ such that $j(k)=j$.
		      Then $\tuple{g_1,\ldotsTwo,g_n}$ is disjoint from the product of
		      index $k$, at position $j(k)$, for every $k$, so it is a subtype of
		      $\neg t$.\qedhere
	\end{description}
\end{Proof}

\begin{cor}[Selection]\label{cor:selection}
	Let $v$ be a value such that $\mk{v}=v$ and
	$\JudgmentTypeSN{\Gamma}{v}{\bigvee_{j\in K}t_j}$, where $K$ is finite and
	every $t_j$ is static. Then $K\neq\varnothing$, and
	$\JudgmentTypeSN{\Gamma}{v}{t_j}$ for some $j\in K$.
\end{cor}

\begin{Proof}
	If $K=\varnothing$, then $v$ has type $\bott$, which contradicts
	Lemma~\ref{lem:generators}(1,2). Otherwise, by induction on the cardinality
	of $K$: let $j_0\in K$; by Lemma~\ref{lem:dichotomy},
	$\JudgmentTypeSN{\Gamma}{v}{t_{j_0}}$, or
	$\JudgmentTypeSN{\Gamma}{v}{\neg t_{j_0}}$, and then, by
	Lemma~\ref{lem:generators}(3), $v$ has type
	$(\bigvee_{j\in K}t_j)\land\neg t_{j_0}$, a subtype of
	$\bigvee_{j\in K\setminus\{j_0\}}t_j$, to which the induction hypothesis
	applies.
\end{Proof}

\begin{lem}[Canonical forms]\label{lem:canonical-forms}
	Let $v$ be a value such that $\JudgmentTypeSN{\Gamma}{v}{t}$.
	\begin{enumerate}
		\item If $t=\fun{\p{t^1,\ldotsTwo,t^n}}{t_0}$ with $t^1,\ldotsTwo,t^n$
		      static, then $v$ is an abstraction with $n$ parameters, marked or not,
		      and $I_v\leq t$.
		\item If $t=\OpenTuple{t_0,\ldotsTwo,t_n}$ (respectively
		      $t=\tuple{t_0,\ldotsTwo,t_n}$), then $v=\tuple{v_0,\ldotsTwo,v_m}$ with
		      $m\geq n$ (respectively $m=n$) and
		      $\JudgmentTypeSN{\Gamma}{v_j}{t_j}$ for every $j\leq n$.
	\end{enumerate}
\end{lem}

\begin{Proof}
	Let $g\in\gen(v)$ such that $g\leq t$ (Lemma~\ref{lem:generators}(1)); $g$
	is not empty (Lemma~\ref{lem:generators}(2)).
	\begin{enumerate}
		\item Since $t\leq\function_n$, $v$ is neither a constant nor a tuple,
		      whose generators are disjoint from $\function$, and it is not an
		      abstraction with $m\neq n$ parameters, whose generators are subtypes
		      of $\function_m$, which is disjoint from $\function_n$. If $v$ is
		      unmarked, then $g=I_v$. If $v$ is marked, then
		      $g=\dynSup{I_v}\land\bigwedge_{a\in N}\neg a$ is static, so
		      $g\leq\dynInf{t}=\fun{\p{t^1,\ldotsTwo,t^n}}{\dynInf{t_0}}$; by
		      Lemma~\ref{lem:arrow-primality}, and since
		      $\dynSup{I_v}\not\leq a$ for every $a\in N$,
		      $\dynSup{I_v}\leq\fun{\p{t^1,\ldotsTwo,t^n}}{\dynInf{t_0}}$. Then
		      $\dynInf{(I_v)}\leq\dynSup{I_v}\leq\dynInf{t}$ and
		      $\dynSup{I_v}\leq\dynSup{t}$, since arrows are monotone in their
		      codomain and $\dynInf{t_0}\leq\dynSup{t_0}$: so $I_v\leq t$.
		\item The generators of abstractions and of constants that are not
		      tuples are not empty and are disjoint from every tuple type. So $v$
		      is a constant that is a tuple, whose singleton type is a product of
		      singletons, or a tuple $\tuple{v_0,\ldotsTwo,v_m}$
		      whose generator is $g=\tuple{g_0,\ldotsTwo,g_m}$ with
		      $g_j\in\gen(v_j)$; the components of $g$ have non-empty minimal
		      materializations (Lemma~\ref{lem:generators}(2)). So $g\leq t$
		      requires $m\geq n$ (respectively $m=n$) and, on both materializations,
		      $g_j\leq t_j$ for $j\leq n$, and Lemma~\ref{lem:generators}(1),
		      applied to the components, which have types, concludes.\qedhere
	\end{enumerate}
\end{Proof}

\noindent
The next two lemmas hold for $\vdsn$ and, with the same proofs, for $\vds$. In
their proofs, rule
\RuleRef{typing:enriched:lambda-neg}{($\lambda_\neg$)} is handled as rule
\RuleRef{typing:static:lambda}{($\lambda$)}: it has the same premises, and its
side conditions depend only on the interface. An abstraction with several
parameters, which are pairwise distinct, is handled as an abstraction with one
parameter.

\begin{lem}[Context agreement]\label{app:lemma:term_context_agreement}
	Let $e$ be a marked expression, $t$ a type, and $\Gamma,\Gamma'$ environments.
	If $\Gamma(z)=\Gamma'(z)$ for every $z\in\fv{e}$, then
	\[
		\JudgmentTypeSN{\Gamma}{e}{t}
		\;\Longleftrightarrow\;
		\JudgmentTypeSN{\Gamma'}{e}{t}.
	\]
\end{lem}

\proof
	We prove the left-to-right implication by induction on the derivation of
	$\JudgmentTypeSN{\Gamma}{e}{t}$ and case analysis on its last rule; the
	converse implication follows by symmetry.
	\begin{description}
		\item[\RuleRef{typing:static:cst}{(cst)}] Immediate: constants do not depend on the environment.
		\item[\RuleRef{typing:static:var}{(var)}] $e=x$. Since $x\in\fv{e}$, we have $\Gamma(x)=\Gamma'(x)$.
		      Thus rule \RuleRef{typing:static:var}{(var)} derives $\JudgmentTypeSN{\Gamma'}{x}{t}$ as well.
		\item[\RuleRef{typing:static:lambda}{($\lambda$)}, \RuleRef{typing:enriched:lambda-neg}{($\lambda_\neg$)}] Suppose that $e$ is $\lambda^{\mathbb{I}} y.\,e'$ or $\mlambda^{\mathbb{I}} y.\,e'$.
		      By inversion, $\JudgmentTypeSN{\Gamma,\,y:t_i}{e'}{s_i}$ for each
		      $(t_i\!\to s_i)\in\mathbb{I}$.
		      The environments $\Gamma,\,y:t_i$ and $\Gamma',\,y:t_i$ agree on
		      every variable of $\fv{e'}$: if $z=y$ they both map it to $t_i$,
		      and otherwise $z\in\fv{e}$ so $\Gamma(z)=\Gamma'(z)$.
		      By the induction hypothesis,
		      $\JudgmentTypeSN{\Gamma',\,y:t_i}{e'}{s_i}$ for all $i$.
		      Re-applying the same rule, with the same set $H$ for ($\lambda_\neg$), yields the result.
		\item[Structural rules] Rules \RuleRef{typing:static:tuple}{(tuple)}, \RuleRef{typing:static:app}{(app)}, \RuleRef{typing:static:case}{(case)}, \RuleRef{typing:static:proj}{(proj)}, \RuleRef{typing:static:proj-omega}{(proj$_\omega$)},
		      \RuleRef{typing:static:proj-top-omega}{(proj$^\topp_\omega$)}, \RuleRef{typing:gradual:plus}{(add)}, \RuleRef{typing:static:sub}{(sub)} and \RuleRef{typing:enriched:and}{($\land$)} pass
		      the same environment to premises whose free variables are included
		      in those of the conclusion, so the induction hypothesis applies
		      directly to each premise. \qed
	\end{description}

\begin{lem}[Static Substitution]\label{app:static_substitution}
	Let $e,e_1$ be marked expressions, $t,t_1$ types, and $x\notin\dom\Gamma$.
	\[
		(\JudgmentTypeSN{\Gamma, x:t_1}{e}{t}) \land (\Gamma \vdsn e_1 : t_1)
		\Longrightarrow (\Gamma \vdsn {e \Subst{x}{e_1}} : {t})
	\]
\end{lem}

\proof\textit{of Lemma~\ref{app:static_substitution}.}
	By induction on the size of the derivation tree and case analysis on the
	last typing rule used to derive $\JudgmentTypeSN{\Gamma, x:t_1}{e}{t}$.
	\begin{description}
		\item[\RuleRef{typing:static:cst}{(cst)}] Immediate since $e$ is a constant and does not depend on $x$.
		\item[\RuleRef{typing:static:var}{(var)}] $e = y$. There are two cases:
		      \begin{itemize}
			      \item $y = x$. Then $e \Subst{x}{e_1} = e_1$
			            so by assumption $\JudgmentTypeSN{\Gamma, x:t_1}{x}{t}$ and $x\notin\dom{\Gamma}$.
			            The last rule is \RuleRef{typing:static:var}{(var)}, so $t=t_1$,
			            and $\Gamma \vdsn e_1 : t_1$ concludes.
			      \item $y \neq x$. Then $e \Subst{x}{e_1} = y$. Since
			            $\Gamma,x:t_1$ and $\Gamma$ agree on every free variable
			            of $y$, Lemma~\ref{app:lemma:term_context_agreement}
			            gives $\JudgmentTypeSN{\Gamma}{y}{t}$.
		      \end{itemize}
		\item[\RuleRef{typing:static:lambda}{($\lambda$)}, \RuleRef{typing:enriched:lambda-neg}{($\lambda_\neg$)}] $e$ is $\lambda^\IFace{y}.{e'}$ or $\mlambda^\IFace{y}.{e'}$.
		      Since substitution is capture-free, we may assume by $\alpha$-renaming that
		      $y \neq x$ and $y \notin \dom{\Gamma}\cup\fv{e_1}$.
		      By inversion, $\JudgmentTypeSN{\Gamma,x:t_1,y:t_i}{e'}{s_i}$
		      for all $(t_i\rarr s_i)\in\IFace$. Since environments are finite
		      maps and $x\neq y$ are both fresh for $\Gamma$, this is the same
		      judgment as $\JudgmentTypeSN{\Gamma,y:t_i,x:t_1}{e'}{s_i}$.
		      Since $y\notin\fv{e_1}$,
		      Lemma~\ref{app:lemma:term_context_agreement} gives
		      $\JudgmentTypeSN{\Gamma,y:t_i}{e_1}{t_1}$, and applying the
		      induction hypothesis with base environment $\Gamma,y:t_i$ yields
		      $\JudgmentTypeSN{\Gamma,y:t_i}{e'\Subst{x}{e_1}}{s_i}$ for all $i\in I$.
		      This concludes by re-applying the same rule, with the same set $H$ for ($\lambda_\neg$), whose side conditions depend only on the interface.
		\item[Structural rules] Rules \RuleRef{typing:static:tuple}{(tuple)}, \RuleRef{typing:static:app}{(app)}, \RuleRef{typing:static:case}{(case)}, \RuleRef{typing:static:proj}{(proj)}, \RuleRef{typing:static:proj-omega}{(proj$_\omega$)}, \RuleRef{typing:static:proj-top-omega}{(proj$^\topp_\omega$)}, \RuleRef{typing:gradual:plus}{(add)}, \RuleRef{typing:static:sub}{(sub)} and \RuleRef{typing:enriched:and}{($\land$)}
		      maintain the same environment in the conclusion and premises,
		      and involve the expression or its sub-expressions
		      in the premises. Hence, they are handled by directly applying the
		      induction hypothesis to their premises.
			  \qed
	\end{description}

\noindent
Lemmas~\ref{lem:progress} and~\ref{lem:preservation} and
Theorems~\ref{app:thm:static-soundness}
and~\ref{app:thm:omega-static-soundness} concern the expressions of
Sections~\ref{sec:safe-erasure} and~\ref{sec:arity}.
Appendix~\ref{app:guard-safety} extends them to the expressions of
Section~\ref{sec:guard-analysis} (Lemma~\ref{lem:guard-static},
Theorems~\ref{thm:guard-soundness} and~\ref{thm:guard-omega-soundness}), which
may also reduce to $\omegaCase$ when rule
\RuleRef{typing:case:case-omega}{(case$_\omega$)} is used.

\begin{lem}[Static Progress]\label{lem:progress}
	Let $e$ be a marked expression such that $\JudgmentTypeSN{\varnothing}{e}{t}$.
	Then either $e$ is a value, or $e\reduces e'$ for some expression $e'$, or
	$e\reduces\omegaOutOfRange$; the last case is impossible if the derivation
	does not use any $\omega$-marked rule.
\end{lem}

\begin{Proof}
	By induction on $e$, every expression is a value, or has the form
	$\ContextWith{x}$ for a variable $x$, or the form $\ContextWith{r}$ where
	$r$ has one of the forms $\app{v}{\ov{v'}}$, $\proj{v}{v'}$,
	$\Case{v}$, and $v+v'$. In a derivation for $\ContextWith{r}$ or
	$\ContextWith{x}$, the expression in the hole is typed under the same
	environment by a subderivation, since evaluation contexts bind no variable
	and every rule types the subexpressions in evaluation position under the
	environment of its conclusion (the branches of a case expression, which a
	rule may leave untyped, are not in evaluation position). No variable has a
	type under $\varnothing$, so $e$ is a value or has the form
	$\ContextWith{r}$. Every such $r$ reduces, to an expression or to a failure,
	and then so does $\ContextWith{r}$
	(rules~\RuleRef{reduction:context}{\textsc{[Context]}}
	and~\RuleRef{reduction:context-omega}{\textsc{[Context$_\omega$]}}). It
	remains to show that $r$ reduces to an expression, or to
	$\omegaOutOfRange$ when $\omega$-marked rules are used. Follow the
	subderivation $\JudgmentTypeSN{\varnothing}{r}{u}$ upwards through instances
	of \RuleRef{typing:static:sub}{(sub)} and
	\RuleRef{typing:enriched:and}{($\land$)}, which keep the expression, to a
	derivation for $r$ that ends with another rule, and consider this rule.
	\begin{description}
		\item[$r=\app{v}{\ov{v'}}$] The rule is \RuleRef{typing:static:app}{(app)},
		      so $v$ has an arrow type of arity $|\ov{v'}|$ whose domain is
		      static. By Lemma~\ref{lem:canonical-forms}(1), $v$ is an abstraction
		      with $|\ov{v'}|$ parameters, and $r$ reduces by the rule for
		      applications.
		\item[$r=\proj{v}{v'}$] If the rule is \RuleRef{typing:static:proj}{(proj)},
		      then $v$ has type $\bigvee_{i\in K}i$ and $v'$ has type
		      $\OpenTuple{t_0,\ldotsTwo,t_n}$, with $K\subseteq[0,n]$. By
		      Lemma~\ref{lem:generators}(1), $v$ has a generator below
		      $\bigvee_{i\in K}i$; it is a constant, since the generators of
		      abstractions and of tuples are not empty
		      (Lemma~\ref{lem:generators}(2)) and are disjoint from $\intTop$; so
		      $v=j$ for some $j\in K$; by
		      Lemma~\ref{lem:canonical-forms}(2), $v'$ is a tuple with at least
		      $n+1$ components, and $r$ reduces by
		      \RuleRef{reduction:proj}{\textsc{[Proj]}}. If the rule is
		      \RuleRef{typing:static:proj-omega}{(proj$_\omega$)} or
		      \RuleRef{typing:static:proj-top-omega}{(proj$^\topp_\omega$)}, then
		      $v$ has type $\intTop$ and $v'$ type $\tupleTop$, so $v$ is an
		      integer and $v'$ a tuple (Lemma~\ref{lem:generators}(4)), and $r$
		      reduces by \RuleRef{reduction:proj}{\textsc{[Proj]}}, or to
		      $\omegaOutOfRange$.
		\item[$r=\Case{v}$] The rule is \RuleRef{typing:static:case}{(case)}, so
		      $v$ has a type $u'\leq\bigvee_{i\in I}\tau_i$. If $v\notin\tau_i$ for
		      every $i\in I$, then, by Lemma~\ref{lem:generators}(1,4), some
		      generator of $v$ would be a subtype of $\bigvee_{i\in I}\tau_i$ and
		      disjoint from it, hence empty, which contradicts
		      Lemma~\ref{lem:generators}(2). So $v$ belongs to some $\tau_i$, and
		      $r$ reduces by \RuleRef{reduction:match}{\textsc{[Match]}}.
		\item[$r=v+v'$] The rule is \RuleRef{typing:gradual:plus}{(add)}. Both
		      $v$ and $v'$ have type $\intTop$, so both are integers, by
		      Lemma~\ref{lem:generators}(4), and $r$ reduces by
		      \RuleRef{reduction:plus}{\textsc{[Plus]}}.\qedhere
	\end{description}
\end{Proof}

\noindent
Since the free variables of an expression typed under $\varnothing$ occur only
in branches that the derivation leaves untyped, the lemma requires neither that
$e$ be closed nor that every branch be attainable.

\begin{lem}[Static Preservation]\label{lem:preservation}
	If $\JudgmentTypeSN{\varnothing}{e}{t}$ and $e\reduces e'$, where $e'$ is an
	expression, then $\JudgmentTypeSN{\varnothing}{e'}{t}$; if the derivation for
	$e$ does not use any $\omega$-marked rule, there is one for $e'$ that does
	not use any.
\end{lem}

\begin{Proof}
	By induction on the derivation of $\JudgmentTypeSN{\varnothing}{e}{t}$. If
	it ends with \RuleRef{typing:static:sub}{(sub)} or
	\RuleRef{typing:enriched:and}{($\land$)}, the induction hypothesis on its
	premises and the same rule conclude. Otherwise it
	ends with the rule for the outermost construct of $e$, which is not a
	constant, a variable or an abstraction, since these do not reduce. If $e$
	reduces by rule~\RuleRef{reduction:context}{\textsc{[Context]}}, then
	$e=\ContextWith{e_0}$ with $\Context\neq\ContextHole$ and $e_0\reduces e_0'$;
	the subexpression of $e$ that contains $e_0$ in evaluation position is typed,
	under $\varnothing$, by a premise of the rule, and the other premises and the
	side conditions depend on it only through its type (for
	\RuleRef{typing:static:case}{(case)}, the matched expression keeps its type,
	so the same branches are typed). The induction hypothesis on this premise and
	the same rule conclude. The other reductions are the following.
	\begin{description}
		\item[Applications] $e=\app{v_f}{u}$, where $v_f$ is an abstraction
		      $\lambda^\IFace x.e_1$ or $\mlambda^\IFace x.e_1$ with
		      $\IFace=\{t_i\rarr s_i\}_{i\in I}$, and $e'=e_1\Subst{x}{\mk{u}}$.
		      The rule is \RuleRef{typing:static:app}{(app)}, with premises
		      $\JudgmentTypeSN{\varnothing}{v_f}{t_1\rarr t}$ and
		      $\JudgmentTypeSN{\varnothing}{u}{t_1}$, where $t_1$ is static. By
		      Lemma~\ref{lem:canonical-forms}(1), $I_{v_f}\leq t_1\rarr t$; by
		      Lemma~\ref{lem:generators}(5),
		      $\JudgmentTypeSN{\varnothing}{\mk{u}}{t_1}$. The derivation for
		      $v_f$, followed upwards through instances of
		      \RuleRef{typing:static:sub}{(sub)} and
		      \RuleRef{typing:enriched:and}{($\land$)}, reaches an instance of
		      ($\lambda$) or ($\lambda_\neg$); its premises are
		      $\JudgmentTypeSN{x:t_i}{e_1}{s_i}$ for $i\in I$, and $\IFace$ is
		      separated. Let
		      $Q=\{i\in I\mid \JudgmentTypeSN{\varnothing}{\mk{u}}{\dynInf{t_i}}\}$
		      and
		      $P=\{i\in I\mid \JudgmentTypeSN{\varnothing}{\mk{u}}{\dynSup{t_i}}\}$.

		      \emph{The codomains.} By the characterization of subtyping of
		      Section~\ref{sec:gradual-typing}, $I_{v_f}\leq t_1\rarr t$ is the
		      conjunction of
		      $\bigwedge_{i\in I}(\dynInf{t_i}\rarr\dynSup{s_i})\leq t_1\rarr\dynSup{t}$
		      and
		      $\bigwedge_{i\in I}(\dynSup{t_i}\rarr\dynInf{s_i})\leq t_1\rarr\dynInf{t}$.
		      Apply Theorem~\ref{app:thm:set-containment-multi-arity}, with $n=1$
		      and a single arrow on the right, to the first inclusion and to the
		      map that sends the indices of $I\setminus Q$ to the domain and those
		      of $Q$ to the codomain. Its first clause would give
		      $t_1\leq\bigvee_{i\in I\setminus Q}\dynInf{t_i}$, hence
		      $\JudgmentTypeSN{\varnothing}{\mk{u}}{\bigvee_{i\in I\setminus Q}\dynInf{t_i}}$,
		      and then, by Corollary~\ref{cor:selection},
		      $\JudgmentTypeSN{\varnothing}{\mk{u}}{\dynInf{t_i}}$ for some
		      $i\in I\setminus Q$, which contradicts the definition of $Q$. So the
		      second clause holds: $Q\neq\varnothing$, since an intersection of
		      codomains over an empty set is $\DOmega$, and
		      $\bigwedge_{i\in Q}\dynSup{s_i}\leq\dynSup{t}$. In the same way,
		      the second inclusion, with $P$ and the types $\dynSup{t_i}$, gives
		      $\bigwedge_{i\in P}\dynInf{s_i}\leq\dynInf{t}$.

		      \emph{Separation.} $Q\subseteq P$, since
		      $\dynInf{t_i}\leq\dynSup{t_i}$. Conversely, let $i\in P$ and
		      $j\in Q$: if $i\notin Q$, then, by Lemma~\ref{lem:dichotomy},
		      $\JudgmentTypeSN{\varnothing}{\mk{u}}{\neg\dynInf{t_i}}$, and, by
		      Lemma~\ref{lem:generators}(3), $\mk{u}$ has type
		      $(\dynSup{t_i}\setminus\dynInf{t_i})\land\dynInf{t_j}$, which is
		      empty since $\IFace$ is separated; this contradicts
		      Lemma~\ref{lem:generators}(1,2). So $P=Q$.

		      \emph{The body.} For every $i\in Q$,
		      $\JudgmentTypeSN{\varnothing}{\mk{u}}{\dynInf{t_i}}$ and
		      $\dynInf{t_i}\leq t_i$, so the Substitution
		      Lemma~\ref{app:static_substitution} applied to the premise for $i$
		      gives $\JudgmentTypeSN{\varnothing}{e_1\Subst{x}{\mk{u}}}{s_i}$. Since
		      $Q\neq\varnothing$, these judgments and rule
		      \RuleRef{typing:enriched:and}{($\land$)}, applied $|Q|-1$ times, give
		      $\JudgmentTypeSN{\varnothing}{e_1\Subst{x}{\mk{u}}}{\bigwedge_{i\in Q}s_i}$.
		      The minimal and maximal materializations of
		      $\bigwedge_{i\in Q}s_i$ are
		      $\bigwedge_{i\in Q}\dynInf{s_i}=\bigwedge_{i\in P}\dynInf{s_i}\leq\dynInf{t}$
		      and $\bigwedge_{i\in Q}\dynSup{s_i}\leq\dynSup{t}$, so
		      $\bigwedge_{i\in Q}s_i\leq t$, and
		      \RuleRef{typing:static:sub}{(sub)} concludes. The derivations of
		      $\mk{u}$ given by Lemma~\ref{lem:generators} use, besides the rules
		      deducing generators, only the premises of the derivation for $u$,
		      and the Substitution Lemma re-applies the rules of the derivations
		      it is given: no new $\omega$-marked rule is used.

		      \emph{Several parameters (Section~\ref{sec:arity}).} Let
		      $\IFace=\{\fun{\p{t_i^1,\ldotsTwo,t_i^n}}{s_i}\}_{i\in I}$, let the
		      arguments be $u_1,\ldotsTwo,u_n$, typed with the static types
		      $t^1,\ldotsTwo,t^n$ of the domain, and let $Q$ (respectively $P$) be
		      the set of the $i$ such that $\mk{u_k}$ has type
		      $\dynInf{(t_i^k)}$ (respectively $\dynSup{(t_i^k)}$) for every $k$.
		      For the maximal materializations, apply
		      Theorem~\ref{app:thm:set-containment-multi-arity} to a map $\iota$
		      that sends every $i\in Q$ to the codomain and every $i\notin Q$ to a
		      position $k$ such that $\mk{u_k}$ has type $\neg\dynInf{(t_i^k)}$
		      (Lemma~\ref{lem:dichotomy}). If the first clause held for a position
		      $k$, then $t^k\leq\bigvee_{\iota(i)=k}\dynInf{(t_i^k)}$, and, by
		      Corollary~\ref{cor:selection}, $\mk{u_k}$ would have type
		      $\dynInf{(t_i^k)}$ for some $i$ with $\iota(i)=k$, which contradicts
		      the choice of $\iota$ (Lemma~\ref{lem:generators}(1--3)). So
		      $Q\neq\varnothing$ and $\bigwedge_{i\in Q}\dynSup{s_i}\leq\dynSup{t}$;
		      likewise $\bigwedge_{i\in P}\dynInf{s_i}\leq\dynInf{t}$. If
		      $i\in P\setminus Q$ and $j\in Q$, choose $k_0$ such that $\mk{u_{k_0}}$
		      has type $\neg\dynInf{(t_i^{k_0})}$; by
		      Lemma~\ref{lem:generators}(1,3), every $\mk{u_k}$ has a generator
		      $g_k$ that is a subtype of $\dynSup{(t_i^k)}\land\dynInf{(t_j^k)}$,
		      and of $\neg\dynInf{(t_i^k)}$ if $k=k_0$; since the $g_k$ are not
		      empty, the product of their interpretations contains an argument
		      list that belongs to the dynamic part of the domain of index $i$ and
		      to the static part of the domain of index $j$, which separation
		      excludes (Section~\ref{sec:arity}). So $P=Q$, and the rest of the
		      proof is the same as for one parameter, with the arguments
		      substituted one at a time (up to $\alpha$-renaming, the parameters
		      are pairwise distinct and not free in the arguments, so that
		      Lemma~\ref{app:lemma:term_context_agreement} types each argument
		      under the parameters not yet substituted).
		\item[\RuleRef{reduction:match}{\textsc{[Match]}}] $e=\Case{v}$ and
		      $e'=e_j$, with $v\in\tau_j$ and $v\notin\tau_i$ for $i<j$. The rule
		      is \RuleRef{typing:static:case}{(case)}, with
		      $\JudgmentTypeSN{\varnothing}{v}{u}$ and
		      $\JudgmentTypeSN{\varnothing}{e_i}{t}$ for every $i$ whose
		      first-match region $(u\land\tau_i)\setminus\bigvee_{k<i}\tau_k$ is
		      not empty. Let $g\in\gen(v)$ with $g\leq u$
		      (Lemma~\ref{lem:generators}(1)). By Lemma~\ref{lem:generators}(4),
		      $g\leq\tau_j$ and $g\land\tau_i\simeq\bott$ for every $i<j$, so $g$
		      is a subtype of the first-match region of $j$, which is therefore
		      not empty (Lemma~\ref{lem:generators}(2)). So
		      $\JudgmentTypeSN{\varnothing}{e_j}{t}$.
		\item[\RuleRef{reduction:proj}{\textsc{[Proj]}}]
		      $e=\proj{j}{\tuple{v_0,\ldotsTwo,v_n}}$ with $j\in[0,n]$, and
		      $e'=v_j$. If the rule is \RuleRef{typing:static:proj}{(proj)}, the
		      index has type $\bigvee_{i\in K}i$, so $j\in K$, and the tuple has a
		      type $\OpenTuple{t_0,\ldotsTwo,t_m}$ with $K\subseteq[0,m]$: by
		      Lemma~\ref{lem:canonical-forms}(2),
		      $\JudgmentTypeSN{\varnothing}{v_j}{t_j}$, and $t_j\leq\bigvee_{i\in K}t_i$.
		      If the rule is \RuleRef{typing:static:proj-omega}{(proj$_\omega$)},
		      the tuple has a type $\tuple{t_0,\ldotsTwo,t_m}$, so $m=n$ and
		      $\JudgmentTypeSN{\varnothing}{v_j}{t_j}$, with
		      $t_j\leq\bigvee_{i\leq n}t_i$. If it is
		      \RuleRef{typing:static:proj-top-omega}{(proj$^\topp_\omega$)}, the
		      type is $\topp$, and $v_j$ has a type, hence type $\topp$: the
		      derivation for the tuple, followed upwards through instances of
		      \RuleRef{typing:static:sub}{(sub)} and
		      \RuleRef{typing:enriched:and}{($\land$)}, reaches
		      \RuleRef{typing:static:tuple}{(tuple)}, whose premises type $v_j$,
		      or \RuleRef{typing:static:cst}{(cst)}, and then $v_j$ is a
		      constant.
		\item[\RuleRef{reduction:plus}{\textsc{[Plus]}}] The rule is
		      \RuleRef{typing:gradual:plus}{(add)}, so $t=\intTop$, and $e'$ is an
		      integer.\qedhere
	\end{description}
\end{Proof}

\begin{Theorem}[Static Type Safety]\label{app:thm:static-soundness}
	If $\varnothing \vds e : t$ has a derivation that does not use any
	$\omega$-marked rule, then either:
	\begin{itemize}
		\item $e \seqreduces v$ and $\JudgmentTypeSN{\varnothing}{\hat{v}}{t}$
		      for a marking $\hat{v}$ of $v$, which is $\varnothing \vds v : t$ if
		      $v$ is a constant;
		\item $e$ diverges.
	\end{itemize}
\end{Theorem}

\begin{Proof}
	Since $e$ is unmarked, $\JudgmentTypeSN{\varnothing}{e}{t}$ by the same
	derivation. Consider the reduction of $e$ as a marked expression. By
	Lemmas~\ref{lem:progress} and~\ref{lem:preservation}, every expression of
	this reduction has type $t$ in $\vdsn$, by a derivation that does not use any
	$\omega$-marked rule, and it is a value or it reduces to an expression. So
	the reduction is infinite, or it ends with a value $\hat{v}$ such that
	$\JudgmentTypeSN{\varnothing}{\hat{v}}{t}$. Since erasing the marks maps
	this reduction onto the reduction of $e$, either $e$ diverges, or
	$e\seqreduces v$ with $v=\erase{\hat{v}}$. If $v$ is a constant, then
	$\hat{v}=v$, and Lemma~\ref{lem:generators}(6) gives
	$\varnothing \vds v : t$.
\end{Proof}

\begin{Theorem}[$\omega$-Static Type Safety]\label{app:thm:omega-static-soundness}
	If $\varnothing \vds e : t$ has a derivation that may use
	$\omega$-marked rules, then either:
	\begin{itemize}
		\item $e \seqreduces v$ and $\JudgmentTypeSN{\varnothing}{\hat{v}}{t}$
		      for a marking $\hat{v}$ of $v$, which is $\varnothing \vds v : t$ if
		      $v$ is a constant;
		\item $e$ diverges;
		\item $e \seqreduces \omegaOutOfRange$.
	\end{itemize}
\end{Theorem}

\begin{Proof}
	As for Theorem~\ref{app:thm:static-soundness}. By
	Lemma~\ref{lem:progress}, the only failure that an expression typed in
	$\vdsn$ can reach is an out-of-range projection: the $\omega$-marked rules
	\RuleRef{typing:static:proj-omega}{(proj$_\omega$)} and
	\RuleRef{typing:static:proj-top-omega}{(proj$^\topp_\omega$)} let the index
	have type $\intTop$ without proving that it is within the bounds of the
	tuple. The other failures are excluded as in the proof of
	Lemma~\ref{lem:progress}: cases are exhaustive, applications have
	abstractions as callees, projections have tuple arguments, and additions have
	integer operands.
\end{Proof}

\begin{remark}[Theorems~\ref{thm:static-soundness} and~\ref{thm:soundness}]\label{app:rem:static-theorems}\em
	Theorems~\ref{thm:static-soundness} and~\ref{thm:soundness} of
	Section~\ref{sec:safe-erasure} are
	Theorems~\ref{app:thm:static-soundness}
	and~\ref{app:thm:omega-static-soundness}: a value $v:t$ in their statements
	is a value $v$ such that $\JudgmentTypeSN{\varnothing}{\hat{v}}{t}$ for a
	marking $\hat{v}$ of $v$, that is, such that $\varnothing\vds v:t$ when $v$
	is a constant. For a value in which abstractions occur, the static system
	does not suffice: with $A$, $F$ and $v$ as at the beginning of this
	subsection, the static system deduces the type
	$\tuple{A,A}\lor\tuple{\neg A,\neg A}$ for $\app{F}{v}$ without
	$\omega$-marked rules, but not for its value $\tuple{v,v}$.
\end{remark}

\label{safety:nongradual-no-omega}



\subsection{Safety of the Gradual System}\label{app:gradual-safety}

The safety of the gradual type system is established through the weak system
of Figure~\ref{app:fig:strong-rules}. Subject reduction fails for the weak
system for the same reason as for the static system: with $A$, $F$ and $v$ as
at the beginning of Appendix~\ref{app:static-safety},
$\JudgmentTypeG{\varnothing}{\app{F}{v}}{\tuple{A,A}\lor\tuple{\neg A,\neg A}}$
by the static rules, hence
$\varnothing\vdw\app{F}{v}\tc\tuple{A,A}\lor\tuple{\neg A,\neg A}$
(Lemma~\ref{lem:gradual-to-strong}), but
$\app{F}{v}$ reduces to $\tuple{v,v}$, and every type that the weak system
deduces for $v$ has a maximal materialization that is a supertype of
$\boolTop\rarr\boolTop$, which is a subtype neither of $A$ nor of $\neg A$. We
therefore prove progress and preservation for an extension $\vdwn$ of the
weak system, which deduces negated arrow types for closed abstractions.
Appendix~\ref{sec:strong-subtyping} interprets strong arrows for functions with
one parameter, and the proofs below consider only those: rules \RuleRef{typing:gradual:lambda-star}{($\lambda_\star$)},
\RuleRef{typing:weak:lambda-circ-star}{($\lambda^\circ_\star$)},
\RuleRef{typing:gradual:app-star}{(app$_\star$)} and
\RuleRef{typing:weak:app-circ-star}{(app$^\circ_\star$)} apply to abstractions
with one parameter and to applications to one argument.

\medskip
\noindent
We write $|e|$ for the size of an expression $e$, the number of nodes of its
syntax tree. The arrows of the interface of an abstraction with $n$ parameters
have arity $n$. Let $v=\lambda^\IFace\ov{x}.e$ be a closed abstraction with
$n$ parameters and interface
$\IFace=\{\fun{\p{t_i^1,\ldotsTwo,t_i^n}}{s_i}\mid i\in I\}$, and let $\vdash$
be a typing judgment. We define:
\begin{itemize}
	\item $I^\circ_v$ is
	      $\bigwedge_{i\in I}\fun{\p{\dynSup{(t_i^1)},\ldotsTwo,\dynSup{(t_i^n)}}}{s_i}$
	      if $\ov{x}:\ov{t_i}\vdash e\tc s_i$ for every $i\in I$, that is, if the
	      premises of rule \RuleRef{typing:weak:lambda-circ}{($\lambda^\circ$)}
	      hold for $v$, and $\function_n$ otherwise; as in
	      Theorem~\ref{app:thm:set-containment-multi-arity}, the intersection of an
	      empty family of arrows of arity $n$ is $\function_n$;
	\item $\Str(v)$, the set of the \emph{strong arrows} of $v$, is empty if
	      $n\neq1$; if $n=1$ and $x$ is the parameter of $v$, it is
	      $\bigcup_{k\in\mathbb{N}}\Str_k(v)$, where $\Str_0(v)=\varnothing$ and
	      $\Str_{k+1}(v)$ is the set of the static strong arrows
	      $\StrongType{(c\rarr d)}$ such that
	      $\dynSup{(I^\circ_v\land\bigwedge_{S\in K}S)}\leq c\rarr d$ for some
	      finite set $K\subseteq\Str_k(v)$, and $x:\dyn\vdash e\tc d\land\dyn$; by
	      induction on $k$, $\Str_k(v)\subseteq\Str_{k+1}(v)$;
	\item a static arrow atom $a$ is \emph{admissible} for $v$ if
	      $\dynSup{(I^\circ_v\land\bigwedge_{S\in K}S)}\not\leq a$ for every
	      finite set $K\subseteq\Str(v)$.
\end{itemize}
Since the maximal materialization of
$\fun{\p{\dynSup{(t_i^1)},\ldotsTwo,\dynSup{(t_i^n)}}}{s_i}$ is
$\fun{\p{\dynSup{(t_i^1)},\ldotsTwo,\dynSup{(t_i^n)}}}{\dynSup{s_i}}$, the type
$\dynSup{(I^\circ_v\land\bigwedge_{S\in K}S)}$ is equivalent to $\function_n$ or
to a finite non-empty intersection of static arrow atoms of arity $n$, to which
Lemma~\ref{lem:arrow-primality} applies.

\begin{Definition}[Enriched weak system]\label{def:enriched-weak}
	For every $N\in\mathbb{N}$, the judgment $\Gamma\vdwnN{N}e\tc t$ is derived by
	the rules of Figure~\ref{app:fig:strong-rules}, their extension of
	Section~\ref{sec:arity} and, for the expressions of
	Section~\ref{sec:guard-analysis}, the weak rules of
	Appendix~\ref{app:guard-safety}, with $\vdwnN{N}$ in place of $\vdw$, and by
	the following rule:
	\[
		\typingrule{\RuleDef{typing:enriched:lambda-circ-neg}{(\lambda^\circ_\neg)}}
		{}
		{\Gamma\vdwnN{N}v\tc I^\circ_v\land\textstyle\bigwedge_{h\in H}\neg a_h}
		\,\,\begin{cases}v\text{ a closed abstraction, }|v|<N\\ a_h\text{ admissible for }v\ \ (h\in H)\end{cases}
	\]
	where $H$ is finite (possibly empty), and $I^\circ_v$ and admissibility are
	defined with $\vdwnN{|v|}$ for $\vdash$. We write $\Gamma\vdwn e\tc t$ if
	$\Gamma\vdwnN{N}e\tc t$ for some $N>|e|$.
\end{Definition}

\noindent
The definition is by induction on $N$: the side conditions of rule
\RuleRef{typing:enriched:lambda-circ-neg}{($\lambda^\circ_\neg$)} for $v$ refer
to the judgment $\vdwnN{|v|}$, and $|v|<N$. Every rule types an expression from
premises on the same expression or on its immediate subexpressions, so a
derivation for $e$ applies rule ($\lambda^\circ_\neg$) only to closed
abstractions that occur in $e$. Hence, for all $N,M>|e|$, the judgments
$\vdwnN{N}$ and $\vdwnN{M}$ derive the same judgments on $e$. In particular,
$\vdwn$ and $\vdwnN{|v|}$ derive the same judgments on the body of $v$, so that
$I^\circ_v$, $\Str(v)$ and admissibility are the same for both, and below we
take them for $\vdwn$. The judgment $\vdwn$ is closed under the rules of the
definition: the premises of a rule, derived in $\vdwn$, are derived at a common
level $N$ larger than the size of the expression typed by its conclusion, where
the rule applies. Every rule of the weak system is a rule of $\vdwn$, so
$\Gamma\vdw e\tc t$ implies $\Gamma\vdwn e\tc t$.

\begin{remark}\label{rem:weak-negations}\em
	Admissibility depends on the strong arrows of an abstraction, and not only on
	$I^\circ_v$. Let
	$v_3=\lambda^{\{\intTop\rarr\intTop;\;\boolTop\rarr(\intTop\lor\boolTop)\}}x.\texttt{case}\,x\,\p{\intTop\rarr x{+}1,\,\boolTop\rarr 42}$.
	Its body has type $\intTop\land\dyn$ under $x:\dyn$, so
	$\StrongType{(\intTop\rarr\intTop)}\in\Str(v_3)$, and
	$I^\circ_{v_3}\land\StrongType{(\intTop\rarr\intTop)}\leq\boolTop\rarr\intTop$:
	on Booleans, $v_3$ returns integers. So $\boolTop\rarr\intTop$ is not
	admissible for $v_3$, although $I^\circ_{v_3}\not\leq\boolTop\rarr\intTop$.
	With the side condition $I^\circ_v\not\leq a_h$ instead, $v_3$ would have the
	type
	$I^\circ_{v_3}\land\neg(\boolTop\rarr\intTop)\land\StrongType{(\intTop\rarr\intTop)}$,
	whose maximal materialization is empty. Rule
	\RuleRef{typing:enriched:lambda-circ-neg}{($\lambda^\circ_\neg$)} applies only
	to closed abstractions. Under $y:\intTop\lor\boolTop$, the body of the
	abstraction $u$ obtained from $v_3$ by replacing $42$ with $y$ has a type
	$d\land\dyn$ under $x:\dyn$ only if $\intTop\lor\boolTop\leq d$. So the same
	definitions, with judgments under $y:\intTop\lor\boolTop$, would make
	$\boolTop\rarr\intTop$ admissible for $u$, while $u\Subst{y}{42}=v_3$, and
	substitution would not preserve the types. The rule also types the abstractions that have no
	\RuleRef{typing:weak:lambda-circ}{($\lambda^\circ$)} derivation, with
	$I^\circ_v=\function_n$, and they need it. Let
	$h=\lambda^{\{\dyn\rarr\dyn\}}w.w$ and
	$P=\lambda^{\{\intTop\rarr(\intTop\rarr\intTop)\}}x.\lambda^{\{\intTop\rarr\intTop\}}y.x$.
	The program $\app{F}{\app{P}{\app{h}{\tttrue}}}$ has type
	$\tuple{A,A}\lor\tuple{\neg A,\neg A}$ in the gradual system, and it reduces
	to $\tuple{L,L}$ with $L=\lambda^{\{\intTop\rarr\intTop\}}y.\tttrue$, whose body
	is not an integer: $L$ needs the type $\neg A$, which rule
	($\lambda^\circ_\neg$) deduces, since $A$ is admissible for $L$. Finally, the
	negated arrow types are a device of the proofs, and they do not describe what
	an abstraction does: for $\OO=\{\invOmega\}$, the abstraction
	$\lambda^{\{\intTop\rarr(\intTop\lor\dyn)\}}x.x{+}0$, which maps integers to
	integers, has the type $\neg(\intTop\rarr\intTop)$.
\end{remark}

\begin{lem}[Progress]\label{lem:gradual-progress}
	If $e$ is closed, then either $e$ is a value, or $e\reduces e'$ for some
	expression $e'$, or $e\reduces\omega_p$ for some $p$.
\end{lem}
\begin{Proof}
	By induction on $e$, a closed expression that is not a value has the form
	$\ContextWith{r}$, where $r$ is an application, a projection, a case
	expression or an addition whose subexpressions in evaluation position are
	values. Every such $r$ reduces, by a rule of Figure~\ref{fig:reductions} or of
	its extension of Section~\ref{sec:arity}, to an expression or to a failure,
	and then so does $\ContextWith{r}$.
\end{Proof}

\noindent
Lemmas~\ref{lem:context-agreement-weak},
\ref{lem:static-typing-imply-dynamic-typing}, \ref{lem:weak-substitution}
and~\ref{lem:well-typedness-after-substitution} and
Corollary~\ref{cor:dynamic-typing} below hold for $\vdwn$, and so for $\vdw$,
whose derivations are those of $\vdwn$ that do not use rule
\RuleRef{typing:enriched:lambda-circ-neg}{($\lambda^\circ_\neg$)}. In their
proofs this rule is an immediate case: it types closed abstractions under every
environment, its side conditions depend only on the abstraction, and a
substitution does not change a closed abstraction. As for the static system, an
abstraction with several parameters, which are pairwise distinct, is handled as
an abstraction with one parameter.

\begin{lem}[Context agreement]\label{lem:context-agreement-weak}
	Let $e$ be an expression, $t$ a type, and $\Gamma,\Gamma'$ environments.
	If $\Gamma(z)=\Gamma'(z)$ for every $z\in\fv{e}$, then
	\[
		\Gamma \vdwn e \tc t
		\;\Longleftrightarrow\;
		\Gamma' \vdwn e \tc t.
	\]
\end{lem}
\begin{Proof}
	We prove the left-to-right implication by induction on the derivation of
	$\Gamma \vdwn e \tc t$ and case analysis on its last rule; the converse
	implication follows by symmetry.
	\begin{description}
		\item[\RuleRef{typing:weak:cst-circ}{(cst$^\circ$)}] Immediate: constants do not depend on the environment.
		\item[\RuleRef{typing:weak:var-circ}{(var$^\circ$)}] $e=x$. Since $x\in\fv{e}$, we have $\Gamma(x)=\Gamma'(x)$.
		      Thus rule \RuleRef{typing:weak:var-circ}{(var$^\circ$)} derives $\Gamma' \vdwn x \tc t$ as well.
		\item[\RuleRef{typing:weak:lambda-circ}{($\lambda^\circ$)}] Suppose $e=\lambda^{\mathbb{I}} y.\,e_0$.
		      By inversion, $\Gamma,\,y:t_i \vdwn e_0 \tc s_i$ for each
		      $(t_i\!\to s_i)\in\mathbb{I}$.
		      The environments $\Gamma,\,y:t_i$ and $\Gamma',\,y:t_i$ agree on every
		      variable of $\fv{e_0}$, for each $i\in I$: if $z=y$ they both map it
		      to $t_i$, and otherwise $z\in\fv{e}$ so $\Gamma(z)=\Gamma'(z)$.
		      By the induction hypothesis, $\Gamma',\,y:t_i \vdwn e_0 \tc s_i$
		      for all $i$.
		      Re-applying rule \RuleRef{typing:weak:lambda-circ}{($\lambda^\circ$)} yields the result.
		\item[\RuleRef{typing:weak:lambda-circ-dyn}{($\lambda^\circ_{\dyn}$)}, \RuleRef{typing:enriched:lambda-circ-neg}{($\lambda^\circ_\neg$)}] Immediate: these rules have no premise, and their conclusions and side conditions do not depend on the environment.
		\item[\RuleRef{typing:weak:lambda-circ-star}{($\lambda^\circ_\star$)}] Suppose $e=\lambda^{\mathbb{I}} y.\,e_0$.
		      By inversion, $\Gamma \vdwn e \tc t_1 \rarr t_2$ and
		      $\Gamma,\,y:\dyn \vdwn e_0 \tc t_2 \land \dyn$.
		      By induction hypothesis on the first premise,
		      $\Gamma' \vdwn e \tc t_1 \rarr t_2$.
		      The environments $\Gamma,\,y:\dyn$ and $\Gamma',\,y:\dyn$ agree on
		      every variable of $\fv{e_0}$, so the induction hypothesis also gives
		      $\Gamma',\,y:\dyn \vdwn e_0 \tc t_2 \land \dyn$.
		      Re-applying rule \RuleRef{typing:weak:lambda-circ-star}{($\lambda^\circ_\star$)} yields the result.
		\item[Other rules] The remaining rules pass the same environment to
		      premises whose free variables are included in those of the
		      conclusion, so the induction hypothesis applies directly to each
		      premise.\qedhere
	\end{description}
\end{Proof}

In $\vdwn$, and so in the weak system, every well-typed expression can be typed with `$\dyn$'.

\begin{lem}[Static typing implies dynamic typing]\label{lem:static-typing-imply-dynamic-typing}
	If $\Gamma \vdwn e \tc t$ then $\Gamma \vdwn e \tc \dyn$
\end{lem}

	\begin{Proof}[Proof]
			We proceed by induction on the size of the derivation tree, and
		case analysis on $\Gamma \vdwn e \tc t$.
		\begin{description}
			\item[\RuleRef{typing:weak:cst-circ}{(cst$^\circ$)}] By subsumption since $c\, \land \dyn\, \leq\, \dyn$.

				\item[\RuleRef{typing:weak:var-circ}{(var$^\circ$)}] By subsumption.

			\item[\RuleRef{typing:weak:and-circ}{(and$^\circ$)}] By induction hypothesis on either of the premises.

			\item[\RuleRef{typing:weak:tuple-circ}{(tuple$^\circ$)}] By induction hypothesis, all components can be
			      typed with $\dyn$. Re-applying \RuleRef{typing:weak:tuple-circ}{(tuple$^\circ$)} and subsumption
			      gives the result, since $\tuple{\dyn,\ldotsTwo,\dyn}\leq\dyn$ when the tuple has at least one component (its minimal materialization is empty). The empty tuple is a constant (Section~\ref{sec:safe-erasure}), typed by \RuleRef{typing:weak:cst-circ}{(cst$^\circ$)}.

			\item[\RuleRef{typing:weak:lambda-circ}{($\lambda^\circ$)}, \RuleRef{typing:weak:lambda-circ-star}{($\lambda^\circ_\star$)}, \RuleRef{typing:weak:lambda-circ-dyn}{($\lambda^\circ_\dyn$)}, \RuleRef{typing:enriched:lambda-circ-neg}{($\lambda^\circ_\neg$)}] Rule \RuleRef{typing:weak:lambda-circ-dyn}{($\lambda^\circ_{\dyn}$)} gives $e$ the type $\function\land\dyn$, and subsumption concludes, since $\function\land\dyn\leq\dyn$.

			\item[\RuleRef{typing:weak:app-circ}{(app$^\circ$)}, \RuleRef{typing:weak:app-circ-star}{(app$^\circ_\star$)}] Replace the use of (app) with \RuleRef{typing:weak:app-circ-dyn}{(app$^\circ_\dyn$)}.

			\item[\RuleRef{typing:weak:app-circ-dyn}{(app$^\circ_\dyn$)}] Immediate since the conclusion is $\dyn$.

			\item[\RuleRef{typing:weak:proj-circ}{(proj$^\circ$)}, \RuleRef{typing:weak:proj-circ-inttop}{(proj$^\circ_\intTop$)}, \RuleRef{typing:weak:proj-circ-topp}{(proj$^\circ_\topp$)}]
			      Replace the projection rule with \RuleRef{typing:weak:proj-circ-topp}{(proj$^\circ_\topp$)}, using
			      subsumption on the premises when necessary.

			\item[\RuleRef{typing:weak:case-circ-dyn}{(case$^\circ$)}] By induction hypothesis, all branches can be typed with $\dyn$.
			      Re-apply the rule with $s = \dyn$.

		\item[\RuleRef{typing:weak:add-circ}{(add$^\circ$)}] Immediate by subsumption since $\texttt{int} \land \dyn \leq \dyn$.

		\item[\RuleRef{typing:weak:sub-circ}{(sub$^\circ$)}] Immediate by induction hypothesis on the single premise.\qedhere
	\end{description}
\end{Proof}

\begin{cor}\label{cor:dynamic-typing}
	If $\Gamma \vdwn e \tc t$, then $\Gamma \vdwn e \tc t \land \dyn$.
\end{cor}

\begin{Proof}
	By application of Lemma~\ref{lem:static-typing-imply-dynamic-typing} and
	\RuleRef{typing:weak:and-circ}{rule (and$^\circ$)}.
\end{Proof}

\noindent
Every expression is well-typed in the weak system, provided that its free
variables are in the environment.

\begin{lem}[Totality]\label{lem:weak-totality}
	If $\fv{e}\subseteq\dom{\Gamma}$, then $\Gamma \vdw e \tc \topp$, and so
	$\Gamma\vdwn e\tc\topp$.
\end{lem}

\begin{Proof}
	By induction on $e$. If $e$ is a constant, a variable (hence in
	$\dom{\Gamma}$), or a $\lambda$-abstraction, then rule
	\RuleRef{typing:weak:cst-circ}{(cst$^\circ$)},
	\RuleRef{typing:weak:var-circ}{(var$^\circ$)}, or
	\RuleRef{typing:weak:lambda-circ-dyn}{($\lambda^\circ_\dyn$)} types it, and
	subsumption concludes. Otherwise, the free variables of every immediate
	subexpression of $e$ are free in $e$, since only $\lambda$-abstractions bind
	variables, so the induction hypothesis types every immediate subexpression
	with $\topp$. Rule \RuleRef{typing:weak:app-circ-dyn}{(app$^\circ_\dyn$)},
	\RuleRef{typing:weak:tuple-circ}{(tuple$^\circ$)},
	\RuleRef{typing:weak:proj-circ-topp}{(proj$^\circ_\topp$)},
	\RuleRef{typing:weak:add-circ}{(add$^\circ$)}, or
	\RuleRef{typing:weak:case-circ-dyn}{(case$^\circ$)} (with $s=\topp$ and every
	branch typed with $\topp$), followed by subsumption, concludes.
\end{Proof}

\begin{figure}
	\begin{equation*}
	        \begin{gathered}
				\TRuleCstWeak
	            \quad
				\TRuleVarWeak
				\quad
				\TRuleAndWeak
	            \\[2mm]
				\TRuleTupleWeak
	            \\[2mm]
	            \TRuleLambdaWeak
	            \qquad
	            \TRuleLambdaDynWeak
	            \\[2mm]
	            \TRuleLambdaStarWeak
            \\[2mm]
	            \TRuleAppWeak
	            \qquad
	            \TRuleAppDynWeak
	            \\[2mm]
			    \TRuleAppStarWeak
	            \\[2mm]
	            \TRuleProjWeak
	            \\[2mm]
	            \TRuleProjIntTopWeak
	            \quad
	            \TRuleProjToppWeak
	            \\[2mm]
	            \TRuleCaseDynWeak
	            \\[2mm]
	            \TRuleAddWeak
	            \quad
            \TRuleSubWeak
        \end{gathered}
    \end{equation*}
	\caption{Typing rules for the weak system}
	\label{app:fig:strong-rules}
\end{figure}

\begin{lem}[Substitution]\label{lem:weak-substitution}
	If $\Gamma, x : s \vdwn e \tc t$ then, for all $e'$ such that $\Gamma \vdwn e' \tc s$ holds, we have $\Gamma \vdwn e [e'/x] \tc t$.
\end{lem}

\proof
	By induction, and case analysis on $\Gamma, x : s \vdwn e \tc t$.
	\begin{description}
		\item[\RuleRef{typing:weak:cst-circ}{(cst$^\circ$)}] Immediate since $e[e'/x]=e$, and constants are typed independently of the environment.

		\item[\RuleRef{typing:weak:var-circ}{(var$^\circ$)}] $e = y$.
		\begin{itemize}
			\item If $y = x$, then by inversion on rule \RuleRef{typing:weak:var-circ}{(var$^\circ$)} we have $(t = s \land \dyn)$.
			Also, $e[e'/x] = e'$ and $\Gamma \vdwn e' \tc s$.
			By Corollary~\ref{cor:dynamic-typing}, we have $\Gamma \vdwn e' \tc s \land \dyn$, which concludes.

			\item If $y \neq x$, then $e[e'/x]=y$. Since $\Gamma,x:s$ and
			      $\Gamma$ agree on every free variable of $y$,
			      Lemma~\ref{lem:context-agreement-weak} gives
			      $\Gamma \vdwn y \tc t$, hence $\Gamma \vdwn e[e'/x] \tc t$.
		  \end{itemize}

			\item[\RuleRef{typing:weak:and-circ}{(and$^\circ$)}] Immediate by IH on both premises, and reapplying rule \RuleRef{typing:weak:and-circ}{(and$^\circ$)}.

			\item[\RuleRef{typing:weak:tuple-circ}{(tuple$^\circ$)}] Immediate by IH on all premises, and
			      reapplying rule \RuleRef{typing:weak:tuple-circ}{(tuple$^\circ$)}.

			\item[\RuleRef{typing:weak:app-circ}{(app$^\circ$)}]
			      We have:
		      $e = e_1\, e_2$,
		      $\Gamma, x:s \vdwn e_1 \tc t_1 \rightarrow t_2$,
		      $\Gamma, x:s \vdwn e_2 \tc t_1$.

		      By IH:
		      $\Gamma \vdwn e_1[e'/x] \tc t_1 \rightarrow t_2$ and
		      $\Gamma \vdwn e_2[e'/x] \tc t_1$.

		      Applying rule \RuleRef{typing:weak:app-circ}{(app$^\circ$)} gives:
		      $\Gamma \vdwn (e_1\, e_2)[e'/x] \tc t_2$

		\item[\RuleRef{typing:weak:app-circ-star}{(app$^\circ_\star$)}] Same as above, ending with applying 	rule \RuleRef{typing:weak:app-circ-star}{(app$^\circ_\star$)}.

		\item[\RuleRef{typing:weak:app-circ-dyn}{(app$^\circ_\dyn$)}]
		      We have $e = e_1\,e_2$, $\Gamma, x : s \vdwn e_1 \tc t_1$, $\Gamma, x : s \vdwn e_2 \tc t_2$.

		      By IH: $\Gamma \vdwn e_1[e'/x] \tc t_1$ and $\Gamma \vdwn e_2[e'/x] \tc t_2$, where obviously $t_1 \leq \topp$ and $t_2 \leq \topp$.

		      Thus, applying \RuleRef{typing:weak:app-circ-dyn}{(app$^\circ_\dyn$)} gives $\Gamma \vdwn (e_1\,e_2)[e'/x] \tc \dyn$.

		\item[\RuleRef{typing:weak:lambda-circ}{($\lambda^\circ$)}]
		      $e = \lambda^\mathbb{I} y. e_0$. By inversion,
		      $(\forall (t_i \rightarrow s_i) \in \mathbb{I}, \Gamma, x:s, y:t_i \vdwn e_0 \tc s_i)$.
		      Since substitution is capture-free, we may assume by $\alpha$-renaming that
		      $y \neq x$ and $y \notin \dom{\Gamma}\cup\fv{e'}$.
		      Since environments are finite maps and $x \neq y$, the above judgments are the same as
		      $(\forall (t_i \rightarrow s_i) \in \mathbb{I}, \Gamma, y:t_i, x:s \vdwn e_0 \tc s_i)$.
		      Since $y \notin \dom{\Gamma}\cup\fv{e'}$, the environments
		      $\Gamma$ and $\Gamma,y:t_i$ agree on every free variable of $e'$,
		      for each $i\in I$.
		      By Lemma~\ref{lem:context-agreement-weak},
		      $\Gamma,y:t_i \vdwn e' \tc s$ for all $i$.
		      By IH,
		      $\forall (t_i \rightarrow s_i) \in \mathbb{I}, (\Gamma, y:t_i \vdwn e_0[e'/x] \tc s_i)$.

		      Applying rule \RuleRef{typing:weak:lambda-circ}{($\lambda^\circ$)} gives:
		      $\Gamma \vdwn (\lambda^\mathbb{I} y. e_0)[e'/x] \tc \bigwedge_{i \in I} (\dynSup{t_i} \rightarrow s_i)$

		\item[\RuleRef{typing:weak:lambda-circ-star}{($\lambda^\circ_\star$)}]
		      Here $e=\lambda^\mathbb{I} y. e_0$.
		      Since substitution is capture-free, we may assume by $\alpha$-renaming that
		      $y \neq x$ and $y \notin \dom{\Gamma}\cup\fv{e'}$.
		      By inversion $\Gamma, x : s \vdwn \lambda^\mathbb{I} y. e_0 \tc t_1 \rarr t_2$ and
		      $\Gamma, x : s, y : \dyn \vdwn e_0 \tc t_2\land\dyn$.
		      Since environments are finite maps and $x \neq y$, the second premise
		      is the same judgment as $\Gamma, y : \dyn, x:s \vdwn e_0 \tc t_2\land\dyn$.
		      Since $y \notin \dom{\Gamma}\cup\fv{e'}$, the environments
		      $\Gamma$ and $\Gamma,y:\dyn$ agree on every free variable of $e'$.
		      By Lemma~\ref{lem:context-agreement-weak}, $\Gamma,y:\dyn \vdwn e' \tc s$.
		      By IH on this premise, $\Gamma, y : \dyn \vdwn e_0[e'/x] \tc t_2\land\dyn$.

		      By IH on the first premise, $\Gamma \vdwn \lambda^\mathbb{I} y. e_0[e'/x] \tc t_1 \rarr t_2$.

		      We can then reapply rule \RuleRef{typing:weak:lambda-circ-star}{($\lambda^\circ_\star$)} to conclude.

			\item[\RuleRef{typing:weak:lambda-circ-dyn}{($\lambda^\circ_\dyn$)}] Immediate: $e[e'/x]$ is a $\lambda$-abstraction, typed by rule \RuleRef{typing:weak:lambda-circ-dyn}{($\lambda^\circ_\dyn$)}.

			\item[\RuleRef{typing:enriched:lambda-circ-neg}{($\lambda^\circ_\neg$)}] Immediate: $e$ is a closed abstraction, so $e[e'/x]=e$, and the rule applies to it under $\Gamma$.

			\item[\RuleRef{typing:weak:proj-circ}{(proj$^\circ$)}, \RuleRef{typing:weak:proj-circ-inttop}{(proj$^\circ_\intTop$)}, \RuleRef{typing:weak:proj-circ-topp}{(proj$^\circ_\topp$)}]
			      Immediate by IH on the premises, and reapplying the same
			      projection rule.

			\item[\RuleRef{typing:weak:case-circ-dyn}{(case$^\circ$)}] $e = \texttt{case}\, e_0\, (\tau_i \rightarrow e_i)_i$ is given type $s'\land \dyn$ (the rule uses $s$, but it is already used in this lemma's statement).

		      By inversion, $\Gamma, x:s \vdwn e_0 \tc u$ for some type $u$ and, for all $i$, if $(u \land \tau_i)\setminus\bigvee_{j<i}\tau_j \not\leq \mathds{O}$, then $\Gamma, x:s \vdwn e_i \tc s'$.

		      By IH, $\Gamma \vdwn e_0[e'/x] \tc u$ and for all $i$ such that $(u \land \tau_i)\setminus\bigvee_{j<i}\tau_j \not\leq \mathds{O}$, we have $\Gamma \vdwn e_i[e'/x] \tc s'$.

		      Applying rule \RuleRef{typing:weak:case-circ-dyn}{(case$^\circ$)} gives $\Gamma \vdwn \texttt{case}\, e_0[e'/x]\, (\tau_i \rightarrow e_i[e'/x])_i \tc s'\land \dyn$.

		\item[\RuleRef{typing:weak:add-circ}{(add$^\circ$)}] Immediate by IH on both premises, and reapplying rule \RuleRef{typing:weak:add-circ}{(add$^\circ$)}.

		\item[\RuleRef{typing:weak:sub-circ}{(sub$^\circ$)}] Immediate by IH and reapplying rule \RuleRef{typing:weak:sub-circ}{(sub$^\circ$)}. \qed
	\end{description}

\begin{lem}[Substitution by value]\label{lem:well-typedness-after-substitution}
	If $\Gamma, x : \dyn \vdwn e \tc t$ and $\Gamma \vdwn v \tc t'$ for a value
	$v$, then $\Gamma \vdwn e[v/x] \tc t$.
\end{lem}

\begin{Proof}
	By Lemma~\ref{lem:static-typing-imply-dynamic-typing},
	$\Gamma\vdwn v\tc\dyn$, and Lemma~\ref{lem:weak-substitution}, with $s=\dyn$,
	concludes.
\end{Proof}

\noindent
The types of a closed value $v$ in $\vdwn$ are described by its \emph{weak
generators}, the types of the set $\genw(v)$ defined by induction on $v$:
\begin{itemize}
	\item $\genw(c)=\{c\land\dyn\}$;
	\item if $v$ is an abstraction, $\genw(v)$ is the set of the types
	      $I^\circ_v\land\bigwedge_{S\in K}S\land\bigwedge_{a\in N}\neg a\land\dyn$,
	      where $K$ is a finite subset of $\Str(v)$ and $N$ a finite set of static
	      arrow atoms admissible for $v$;
	\item $\genw(\tuple{v_1,\ldotsTwo,v_n})=\{\tuple{g_1,\ldotsTwo,g_n}\mid g_k\in\genw(v_k)\}$
	      if $\tuple{v_1,\ldotsTwo,v_n}$ is not a constant.
\end{itemize}
The minimal materialization of every weak generator $g$ is empty, so $g\leq t$
if and only if $\dynSup{g}\leq\dynSup{t}$.

\begin{lem}[Weak generators]\label{lem:weak-generators}
	Let $v$ be a closed value and $\Gamma$ an environment.
	\begin{enumerate}
		\item $\Gamma\vdwn v\tc t$ if and only if $g\leq t$ for some
		      $g\in\genw(v)$.
		\item $\dynSup{g}\not\simeq\bott$ for every $g\in\genw(v)$. Hence, if
		      $\Gamma\vdwn v\tc t$, then $\dynSup{t}\not\simeq\bott$.
		\item For every test type $\tau$ and every $g\in\genw(v)$,
		      $\dynSup{g}\leq\tau$ if $v\in\tau$, and
		      $\dynSup{g}\land\tau\simeq\bott$ if $v\notin\tau$. Hence
		      $\Gamma\vdwn v\tc\tau$ if and only if $v\in\tau$, and
		      $\Gamma\vdwn v\tc\neg\tau$ if and only if $v\notin\tau$.
		\item If $v$ is a constant, then $\Gamma\vdwn v\tc t$ if and only if
		      $v\leq\dynSup{t}$, if and only if $\Gamma\vdw v\tc t$.
	\end{enumerate}
\end{lem}

\begin{Proof}
	\begin{enumerate}
		\item ``If'': every weak generator is deduced for $v$, and rule
		      \RuleRef{typing:weak:sub-circ}{(sub$^\circ$)} concludes. For a
		      constant $c$, rule \RuleRef{typing:weak:cst-circ}{(cst$^\circ$)}
		      deduces $c\land\dyn$. For an abstraction $v$, rule
		      \RuleRef{typing:enriched:lambda-circ-neg}{($\lambda^\circ_\neg$)}, with
		      $H=N$, deduces $I^\circ_v\land\bigwedge_{a\in N}\neg a$, and every
		      $S=\StrongType{(c\rarr d)}\in\Str(v)$ is deduced by rule
		      \RuleRef{typing:weak:lambda-circ-star}{($\lambda^\circ_\star$)}, by
		      induction on the least $k$ such that $S\in\Str_k(v)$: if $K'$ is the
		      finite subset of $\Str_{k-1}(v)$ that puts $S$ in $\Str_k(v)$, then $v$
		      has the type $I^\circ_v\land\bigwedge_{S'\in K'}S'$, by rules
		      ($\lambda^\circ_\neg$) and \RuleRef{typing:weak:and-circ}{(and$^\circ$)},
		      hence the type $c\rarr d$ by (sub$^\circ$), since $c\rarr d$ is static
		      and a supertype of the maximal materialization of that type; and the
		      second premise of ($\lambda^\circ_\star$),
		      $\Gamma,x:\dyn\vdwn e\tc d\land\dyn$ for the parameter $x$ and the body
		      $e$ of $v$, holds by definition of $\Str(v)$ and by
		      Lemma~\ref{lem:context-agreement-weak}, since $\fv{e}\subseteq\{x\}$. Rule (and$^\circ$) and
		      Corollary~\ref{cor:dynamic-typing} then deduce the generator. For a
		      tuple, by induction and rule
		      \RuleRef{typing:weak:tuple-circ}{(tuple$^\circ$)}.

		      ``Only if'': by induction on the derivation. Rule (sub$^\circ$) is
		      handled by transitivity. Rule (and$^\circ$) is handled by the union of
		      the sets $K$ and of the sets $N$ of the two generators, componentwise
		      for tuples. Rule (cst$^\circ$) deduces the generator of a constant,
		      and rule (tuple$^\circ$) is handled by induction on the components,
		      since $\tuple{c_1\land\dyn,\ldotsTwo,c_n\land\dyn}$ is a supertype of
		      $\tuple{c_1,\ldotsTwo,c_n}\land\dyn$ for a tuple of constants. For an abstraction $v$ with $n$ parameters,
		      the other rules are the following.
		      \begin{itemize}
			      \item \RuleRef{typing:weak:lambda-circ}{($\lambda^\circ$)}: its
			            premises hold under $\Gamma$, hence without $\Gamma$, by
			            Lemma~\ref{lem:context-agreement-weak}, since $v$ is closed; so
			            its conclusion is a supertype of $I^\circ_v$ (equal to it if
			            $\IFace\neq\emptyset$), hence of the generator defined by
			            $K=N=\varnothing$.
			      \item \RuleRef{typing:weak:lambda-circ-dyn}{($\lambda^\circ_\dyn$)}:
			            its conclusion $\function\land\dyn$ (or $\function_n\land\dyn$,
			            Section~\ref{sec:arity}) is a supertype of $I^\circ_v\land\dyn$,
			            since every arrow atom of arity $n$ is a subtype of
			            $\function_n$.
			      \item ($\lambda^\circ_\neg$): its conclusion is a supertype of the
			            generator defined by $K=\varnothing$ and $N=H$.
			      \item ($\lambda^\circ_\star$): then $n=1$, and the premises are
			            $\Gamma\vdwn v\tc t_1\rarr t_2$ and
			            $\Gamma,x:\dyn\vdwn e\tc t_2\land\dyn$, where $x$ and $e$ are the
			            parameter and the body of $v$. By the induction hypothesis, some
			            generator
			            $I^\circ_v\land\bigwedge_{S\in K}S\land\bigwedge_{a\in N}\neg a\land\dyn$
			            is a subtype of $t_1\rarr t_2$, so, on maximal materializations,
			            $l\land\bigwedge_{a\in N}\neg a\leq\dynInf{t_1}\rarr\dynSup{t_2}$,
			            where $l=\dynSup{(I^\circ_v\land\bigwedge_{S\in K}S)}$. By
			            Lemma~\ref{lem:arrow-primality}, and since no $a\in N$ is a
			            supertype of $l$, $l\leq\dynInf{t_1}\rarr\dynSup{t_2}$. Since
			            $\fv{e}\subseteq\{x\}$ and $t_2\land\dyn\leq\dynSup{t_2}\land\dyn$,
			            Lemma~\ref{lem:context-agreement-weak} and rule (sub$^\circ$) give
			            $x:\dyn\vdwn e\tc\dynSup{t_2}\land\dyn$. If $K\subseteq\Str_k(v)$,
			            then $S_0=\StrongType{(\dynInf{t_1}\rarr\dynSup{t_2})}$ belongs to
			            $\Str_{k+1}(v)$, and the generator defined by $K\cup\{S_0\}$ and
			            $N=\varnothing$ has a maximal materialization that is a subtype
			            of $S_0$, which is the maximal materialization of
			            $\StrongType{(t_1\rarr t_2)}$
			            (Section~\ref{sec:gradual-typing}).
		      \end{itemize}
		\item A constant $c$ is not empty, and $\dynSup{(c\land\dyn)}=c$. For an
		      abstraction $v$ with $n$ parameters, the type
		      $l=\dynSup{(I^\circ_v\land\bigwedge_{S\in K}S)}$ is equivalent to
		      $\function_n$ or to a finite non-empty intersection of static arrow
		      atoms of arity $n$, so, by Lemma~\ref{lem:arrow-primality},
		      $l\land\bigwedge_{a\in N}\neg a\not\simeq\bott$, since no $a\in N$ is a
		      supertype of $l$. For tuples, componentwise. The consequence follows
		      from item~1, since $g\leq t$ implies $\dynSup{g}\leq\dynSup{t}$.
		\item By induction on $\tau$, as in the proof of
		      Lemma~\ref{lem:generators}(4), with $\dynSup{g}$ in place of $g$. The
		      cases $\tau=\tau_1\lor\tau_2$ and $\tau=\neg\tau_1$ are the same.
		      Otherwise $\tau$ is a base type, a constant, $\function_n$, or a tuple
		      of test types, closed or open. If $v$ is a constant $c$ that is not a
		      tuple, then $\dynSup{g}=c$, and the argument for constants applies.
		      If $v$ is an abstraction with $n$ parameters, then
		      $\dynSup{g}\leq l$, where $l$, as in the proof of item~2, is
		      equivalent to $\function_n$ or to a finite non-empty intersection of
		      static arrow atoms of arity $n$; so $\dynSup{g}\leq\function_n\leq\function$, and
		      $\dynSup{g}\land\tau\simeq\bott$ for every other test type $\tau$ of
		      these forms, while $v$ belongs to $\function$ and $\function_n$ and to
		      no other one. If $v=\tuple{v_1,\ldotsTwo,v_n}$, then
		      $\dynSup{g}=\tuple{\dynSup{g_1},\ldotsTwo,\dynSup{g_n}}$ with
		      $g_k\in\genw(v_k)$ (for a tuple of constants, $\dynSup{g}$ is the
		      product of the singleton types of its components), and the argument
		      for tuples applies, with the induction hypothesis on the $\tau_k$
		      for the closed values $v_k$. For the consequences, $\tau$ and
		      $\neg\tau$ are static, so $g\leq\tau$ if and only if
		      $\dynSup{g}\leq\tau$, and $g\leq\neg\tau$ if and only if
		      $\dynSup{g}\leq\neg\tau$, the minimal materialization of $g$ being
		      empty; items~1 and~2 conclude, since $\neg\tau$ is a test type and
		      $v\in\neg\tau$ if and only if $v\notin\tau$.
		\item Then $\genw(v)=\{v\land\dyn\}$, and $v\land\dyn\leq t$ if and only if
		      $v\leq\dynSup{t}$. The proof of item~1 applies to $\vdw$, since it uses
		      rule ($\lambda^\circ_\neg$) only for abstractions.\qedhere
	\end{enumerate}
\end{Proof}

\begin{lem}[Weak dichotomy]\label{lem:weak-dichotomy}
	Let $v$ be a closed value and $t$ a static type. Then
	$\varnothing\vdwn v\tc t$ or $\varnothing\vdwn v\tc\neg t$.
\end{lem}

\begin{Proof}
	By Lemma~\ref{lem:weak-generators}(1), it suffices to find a weak generator
	$g$ of $v$ such that $\dynSup{g}\leq t$ or $\dynSup{g}\leq\neg t$. By
	induction on $v$.
	\begin{description}
		\item[Constants.] A constant $c$ is a singleton type: $c\leq t$ or
		      $c\leq\neg t$.
		\item[Abstractions.] Let $v$ have $n$ parameters. Call a static arrow atom
		      \emph{positive} for $v$ if it is not admissible, that is, if it is a
		      supertype of $\dynSup{(I^\circ_v\land\bigwedge_{S\in K}S)}$ for some
		      finite $K\subseteq\Str(v)$; finitely many positive atoms are supertypes
		      of the same such type, with $K$ the union of their sets. The maximal
		      materialization of every generator of $v$ is a subtype of
		      $\function_n$, so only $t\land\function_n$ matters. As in the proof of
		      Lemma~\ref{lem:dichotomy}, write it as a union
		      $\bigvee_{k\in D}(\function_n\land\bigwedge_{m\in M_k}a_{k,m}\land\bigwedge_{h\in H_k}\neg b_{k,h})$,
		      where the $a_{k,m}$ and the $b_{k,h}$ are static arrow atoms of arity
		      $n$. If some $k$ has every $a_{k,m}$ positive and every $b_{k,h}$
		      admissible, let $K$ be such that every $a_{k,m}$ is a supertype of
		      $\dynSup{(I^\circ_v\land\bigwedge_{S\in K}S)}$: the maximal
		      materialization of the generator
		      $I^\circ_v\land\bigwedge_{S\in K}S\land\bigwedge_{h\in H_k}\neg b_{k,h}\land\dyn$
		      is a subtype of the disjunct of index $k$, hence of $t$. Otherwise,
		      for every $k$, some $a_{k,m}$ is admissible, and we put it in a set
		      $N$, or some $b_{k,h}$ is positive, and we add to a set $K$ the
		      elements of a finite subset of $\Str(v)$ that makes it positive. The
		      maximal materialization of the generator
		      $g=I^\circ_v\land\bigwedge_{S\in K}S\land\bigwedge_{a\in N}\neg a\land\dyn$
		      is disjoint from every disjunct: from those of the first kind since
		      $\dynSup{g}\leq\neg a_{k,m}$, from the others since
		      $\dynSup{g}\leq\dynSup{(I^\circ_v\land\bigwedge_{S\in K}S)}\leq b_{k,h}$.
		      So $\dynSup{g}\leq\function_n\land\neg(t\land\function_n)\leq\neg t$.
		\item[Tuples.] As in the proof of Lemma~\ref{lem:dichotomy}, with
		      Lemma~\ref{lem:weak-generators}(1) and rule
		      \RuleRef{typing:weak:and-circ}{(and$^\circ$)} in place of
		      Lemma~\ref{lem:generators}(1,3).\qedhere
	\end{description}
\end{Proof}

\begin{lem}[Subject Reduction]\label{lem:subject-reduction}
	If $e$ is closed, $\varnothing\vdwn e\tc t$, and $e\reduces e'$, then
	$\varnothing\vdwn e'\tc t$.
\end{lem}

\begin{proof}[Proof of Subject Reduction]
	By induction on the derivation of $\varnothing\vdwn e\tc t$. If it ends with
	\RuleRef{typing:weak:sub-circ}{(sub$^\circ$)} or
	\RuleRef{typing:weak:and-circ}{(and$^\circ$)}, the induction hypothesis on its
	premises and the same rule conclude. Otherwise it ends with the rule for the
	outermost construct of $e$, which is not a constant, a variable or an
	abstraction, since these do not reduce. Since $e$ is closed, so are $e'$ and
	the subexpressions of $e$ in evaluation position.
	\begin{description}
		\item[Reduction \RuleRef{reduction:context}{\textsc{[Context]}}]
		      $e=\ContextWith{e_0}$ with $e_0\reduces e_0'$ and
		      $\Context\neq\ContextHole$. The subexpression of $e$ that contains $e_0$
		      in evaluation position is typed under $\varnothing$ by a premise of the
		      rule, since evaluation contexts bind no variable, and the other premises
		      depend on it only through its type (for
		      \RuleRef{typing:weak:case-circ-dyn}{(case$^\circ$)}, the matched
		      expression keeps its type, so the same branches are typed). The
		      induction hypothesis on this premise and the same rule conclude.

		\item[Reduction \RuleRef{reduction:app}{\textsc{[App]}}]
		      $e=\app{v}{v_2}$, where $v=\lambda^\IFace x.e_1$ with
		      $\IFace=\{t_i\rarr s_i\}_{i\in I}$, and $e'=e_1\Subst{x}{v_2}$. Let
		      $I^\circ$ be $I$ if the premises of
		      \RuleRef{typing:weak:lambda-circ}{($\lambda^\circ$)} hold for $v$, and
		      $\varnothing$ otherwise, so that
		      $\dynSup{(I^\circ_v)}=\bigwedge_{i\in I^\circ}(\dynSup{t_i}\rarr\dynSup{s_i})$,
		      and let
		      $J=\{j\in I^\circ\mid\varnothing\vdwn v_2\tc\neg\dynSup{t_j}\}$. For every
		      $i\in I^\circ\setminus J$, Lemma~\ref{lem:weak-dichotomy} gives
		      $\varnothing\vdwn v_2\tc\dynSup{t_i}$, hence
		      $\varnothing\vdwn v_2\tc t_i$ by Corollary~\ref{cor:dynamic-typing} and
		      rule (sub$^\circ$), since $\dynSup{t_i}\land\dyn\leq t_i$; so the premise
		      $x:t_i\vdwn e_1\tc s_i$ and Lemma~\ref{lem:weak-substitution} give
		      $\varnothing\vdwn e'\tc s_i$. For every
		      $\StrongType{(c\rarr d)}\in\Str(v)$, the definition of $\Str(v)$ gives
		      $x:\dyn\vdwn e_1\tc d\land\dyn$, and
		      Lemma~\ref{lem:well-typedness-after-substitution}, since $v_2$ has a type
		      by Lemma~\ref{lem:weak-totality}, gives
		      $\varnothing\vdwn e'\tc d\land\dyn$. Therefore, for every finite set
		      $K\subseteq\Str(v)$, rule (and$^\circ$) and
		      Corollary~\ref{cor:dynamic-typing}, with Lemma~\ref{lem:weak-totality}
		      when $B_K$ has no other conjunct than `$\dyn$', give
		      \[
		      	\varnothing\vdwn e'\tc B_K, \qquad\text{where }
		      	B_K=\dyn\land\textstyle\bigwedge_{i\in I^\circ\setminus J}s_i\land\bigwedge_{\StrongType{(c\rarr d)}\in K}d.
		      \]
		      The minimal materialization of $B_K$ is empty, so $B_K\leq t$ if and
		      only if $\dynSup{(B_K)}\leq\dynSup{t}$, and it suffices to show the
		      latter for some $K$: rule (sub$^\circ$) then concludes. Consider the
		      last rule of the derivation.
		      \begin{description}
			      \item[\RuleRef{typing:weak:app-circ}{(app$^\circ$)}] Its premises
			            are $\varnothing\vdwn v\tc t_1\rarr t$ and
			            $\varnothing\vdwn v_2\tc t_1$, with $t_1$ static. By
			            Lemma~\ref{lem:weak-generators}(1), some generator
			            $I^\circ_v\land\bigwedge_{S\in K}S\land\bigwedge_{a\in N}\neg a\land\dyn$
			            of $v$ is a subtype of $t_1\rarr t$, so
			            $l\land\bigwedge_{a\in N}\neg a\leq t_1\rarr\dynSup{t}$, where
			            $l=\dynSup{(I^\circ_v\land\bigwedge_{S\in K}S)}$, and, by
			            Lemma~\ref{lem:arrow-primality}, since no $a\in N$ is a supertype
			            of $l$, $l\leq t_1\rarr\dynSup{t}$. If $K=\varnothing$, apply
			            Theorem~\ref{app:thm:set-containment-multi-arity}, with $n=1$,
			            $P=I^\circ$ and a single arrow on the right, to this inclusion and
			            to the map that sends the indices of $J$ to the domain and those
			            of $I^\circ\setminus J$ to the codomain. If $K\neq\varnothing$,
			            write $\bigwedge_{S\in K}S$ as $\StrongType{(c\rarr d)}$, where
			            $d$ is the intersection of the codomains of the elements of $K$,
			            by Lemma~\ref{lem:strong-intersection}, and apply
			            Theorem~\ref{thm:strong-arrow-types} with the same set $J$. In
			            both cases, the first disjunct implies
			            $t_1\leq\bigvee_{j\in J}\dynSup{t_j}$: then, by rule
			            (and$^\circ$), $v_2$ would have the type
			            $t_1\land\bigwedge_{j\in J}\neg\dynSup{t_j}$, whose maximal
			            materialization is empty, which contradicts
			            Lemma~\ref{lem:weak-generators}(2). So the second disjunct holds,
			            which implies $\dynSup{(B_K)}\leq\dynSup{t}$.
			      \item[\RuleRef{typing:weak:app-circ-dyn}{(app$^\circ_\dyn$)}] Then
			            $t=\dyn$, and, since $e'$ is closed,
			            Lemmas~\ref{lem:weak-totality}
			            and~\ref{lem:static-typing-imply-dynamic-typing} give
			            $\varnothing\vdwn e'\tc\dyn$.
			      \item[\RuleRef{typing:weak:app-circ-star}{(app$^\circ_\star$)}] Its
			            premises are $\varnothing\vdwn v\tc\StrongType{(t_1\rarr t_2)}$
			            and $\varnothing\vdwn v_2\tc\topp$, and $t=t_2\land\dyn$. As for
			            (app$^\circ$), some generator of $v$ gives
			            $l\leq\dynSup{(\StrongType{(t_1\rarr t_2)})}=\StrongType{(\dynInf{t_1}\rarr\dynSup{t_2})}$.
			            If $K\neq\varnothing$, write $\bigwedge_{S\in K}S$ as
			            $\StrongType{(c\rarr d)}$ as above and apply
			            Theorem~\ref{thm:strong-strong}; if $K=\varnothing$, apply
			            Theorem~\ref{thm:strong-no-strong-left}. Both give, for the same
			            set $J$, $\bigvee_{j\in J}\dynSup{t_j}\simeq\topp$, or
			            $\dynSup{(B_K)}\leq\dynSup{t_2}=\dynSup{t}$. The first is
			            impossible, since $v_2$ would have the type
			            $\bigwedge_{j\in J}\neg\dynSup{t_j}\simeq\bott$.
		      \end{description}
		      \emph{Several parameters (Section~\ref{sec:arity}).} Let
		      $v=\lambda^\IFace\ov{x}.e_1$, with $n\neq1$ parameters and
		      $\IFace=\{\fun{\p{t_i^1,\ldotsTwo,t_i^n}}{s_i}\}_{i\in I}$, be applied to
		      $v_1,\ldotsTwo,v_n$ by rule (app$^\circ$), with the static domain
		      $t^1,\ldotsTwo,t^n$, and let $I^\circ$ be as above. Then
		      $\Str(v)=\varnothing$, and, as above,
		      $\dynSup{(I^\circ_v)}\leq\fun{\p{t^1,\ldotsTwo,t^n}}{\dynSup{t}}$. Let $Q$
		      be the set of the $i\in I^\circ$ such that $v_k$ has type
		      $\dynSup{(t_i^k)}$ for every $k$, and apply
		      Theorem~\ref{app:thm:set-containment-multi-arity} to the map $\iota$
		      that sends every $i\in Q$ to the codomain and every
		      $i\in I^\circ\setminus Q$ to a position $k$ such that $v_k$ has type
		      $\neg\dynSup{(t_i^k)}$ (Lemma~\ref{lem:weak-dichotomy}). If the first
		      clause held for a position $k$, then
		      $t^k\leq\bigvee_{\iota(i)=k}\dynSup{(t_i^k)}$, and, by rule
		      (and$^\circ$), $v_k$ would have a type whose maximal materialization is
		      empty. So the second clause holds: $Q\neq\varnothing$, since an
		      intersection of codomains over an empty set is $\DOmega$, and
		      $\bigwedge_{i\in Q}\dynSup{s_i}\leq\dynSup{t}$. For $i\in Q$,
		      every $v_k$ has type $t_i^k$, as above, and the premise
		      $\ov{x}:\ov{t_i}\vdwn e_1\tc s_i$, with the arguments substituted one at
		      a time by Lemma~\ref{lem:weak-substitution} (up to $\alpha$-renaming, the
		      parameters are pairwise distinct, and the arguments are closed), gives
		      $\varnothing\vdwn e'\tc s_i$. Rule (and$^\circ$),
		      Corollary~\ref{cor:dynamic-typing} and rule (sub$^\circ$) conclude.
		      Rule (app$^\circ_\dyn$) is handled as above, and rule
		      (app$^\circ_\star$) concerns applications to one argument.

		\item[Reduction \RuleRef{reduction:proj}{\textsc{[Proj]}}]
		      $e=\proj{j}{v}$, where $v=\tuple{v_0,\ldotsTwo,v_m}$ and $j\in[0,m]$, and
		      $e'=v_j$. If the rule is
		      \RuleRef{typing:weak:proj-circ}{(proj$^\circ$)}, the index has a type
		      $\bigvee_{i\in L}i$ and $v$ a type $\OpenTuple{t_0,\ldotsTwo,t_n}$, with
		      $L\subseteq[0,n]$. By Lemma~\ref{lem:weak-generators}(4),
		      $j\leq\bigvee_{i\in L}i$, so $j\in L$. By
		      Lemma~\ref{lem:weak-generators}(1), $v$ has a weak generator that is a
		      subtype of $\OpenTuple{t_0,\ldotsTwo,t_n}$: $v\land\dyn$ if $v$ is a
		      constant, and $\tuple{g_0,\ldotsTwo,g_m}$ with $g_i\in\genw(v_i)$
		      otherwise. Since the components of its maximal materialization are not
		      empty (Lemma~\ref{lem:weak-generators}(2)), $m\geq n$, and the
		      component of index $j$ is a subtype of $\dynSup{t_j}$; so
		      $\varnothing\vdwn v_j\tc t_j$, by Lemma~\ref{lem:weak-generators}(4) or
		      (1), and $t_j\leq\bigvee_{i\in L}t_i$. If the rule is
		      \RuleRef{typing:weak:proj-circ-inttop}{(proj$^\circ_\intTop$)}, the tuple
		      has a type $\tuple{t_0,\ldotsTwo,t_n}$, so $m=n$, and in the same way
		      $\varnothing\vdwn v_j\tc t_j$, with $t_j\leq\bigvee_{i\leq n}t_i$. If it
		      is \RuleRef{typing:weak:proj-circ-topp}{(proj$^\circ_\topp$)}, then
		      $t=\dyn$, and Lemmas~\ref{lem:weak-totality}
		      and~\ref{lem:static-typing-imply-dynamic-typing} conclude.

		\item[Reduction \RuleRef{reduction:plus}{\textsc{[Plus]}}] The rule is
		      \RuleRef{typing:weak:add-circ}{(add$^\circ$)}, so $t=\intTop\land\dyn$,
		      and $e'$ is an integer $c$, of type $c\land\dyn\leq\intTop\land\dyn$.

		\item[Reduction \RuleRef{reduction:match}{\textsc{[Match]}}]
		      $e=\texttt{case}\,v\,(\tau_i\rarr e_i)_{i\in I}$ and $e'=e_j$, where
		      $v\in\tau_j$ and $v\notin\tau_i$ for $i<j$. The rule is
		      \RuleRef{typing:weak:case-circ-dyn}{(case$^\circ$)}, so $t=s\land\dyn$,
		      $\varnothing\vdwn v\tc u$ for some type $u$, and
		      $\varnothing\vdwn e_i\tc s$ for every $i\in I$ whose first-match region
		      $(u\land\tau_i)\setminus\bigvee_{k<i}\tau_k$ is not empty. Let
		      $g\in\genw(v)$ with $g\leq u$ (Lemma~\ref{lem:weak-generators}(1)). By
		      Lemma~\ref{lem:weak-generators}(3), $\dynSup{g}\leq\tau_j$ and
		      $\dynSup{g}\land\tau_i\simeq\bott$ for $i<j$, so $\dynSup{g}$ is a
		      subtype of the maximal materialization of the first-match region of
		      index $j$, which is therefore not empty
		      (Lemma~\ref{lem:weak-generators}(2)); hence the region is not empty
		      either. So $\varnothing\vdwn e_j\tc s$, and
		      Corollary~\ref{cor:dynamic-typing} gives
		      $\varnothing\vdwn e_j\tc s\land\dyn$.\qedhere
	\end{description}
\end{proof}

\noindent
To link back to the gradual system, we use the fact that every expression well-typed in the former
is well-typed in the latter.

\begin{lem}[Gradual typing implies strong typing]\label{lem:gradual-to-strong}
	If $\Gamma \vdg e : t$ then $\Gamma \vdw e \tc t$.
\end{lem}

	\begin{Proof}[Proof]
		By induction on the derivation of $\Gamma \vdg e : t$. For every rule of
		Figures~\ref{app:fig:gradual-rules} and~\ref{app:fig:tuple-rules}, the
		induction hypothesis gives the premises of a rule of
		Figure~\ref{app:fig:strong-rules} (by subsumption when the latter requires
		the type $\topp$) whose conclusion is a subtype of the conclusion of the
		gradual rule, and subsumption concludes. This rule of
		Figure~\ref{app:fig:strong-rules} is the one with the same name and the
		superscript $\circ$, except in the following cases:
		\begin{itemize}
			\item for \RuleRef{typing:gradual:lambda}{($\lambda$)} and
			      \RuleRef{typing:gradual:lambda-dyn}{($\lambda_\dyn$)}, it is
			      \RuleRef{typing:weak:lambda-circ}{($\lambda^\circ$)}, which has
			      their premises and concludes
			      $\bigwedge_{i\in I}(\dynSup{t_i}\rarr s_i)\leq\bigwedge_{i\in I}(t_i\rarr s_i)$;
			\item for \RuleRef{typing:gradual:plus}{(add)} and
			      \RuleRef{typing:gradual:plus-dyn}{(plus$_\star$)}, it is
			      \RuleRef{typing:weak:add-circ}{(add$^\circ$)}, and for
			      \RuleRef{typing:gradual:case}{(case)} and
			      \RuleRef{typing:gradual:case-dyn}{(case$_\star$)}, it is
			      \RuleRef{typing:weak:case-circ-dyn}{(case$^\circ$)};
			\item for \RuleRef{typing:gradual:proj-omega}{(proj$_\omega$)}, it is
			      \RuleRef{typing:weak:proj-circ-inttop}{(proj$^\circ_\intTop$)}, and for
			      \RuleRef{typing:gradual:proj-top-omega}{(proj$^\topp_\omega$)} and
			      \RuleRef{typing:gradual:proj-dyn}{(proj$_\dyn$)}, it is
			      \RuleRef{typing:weak:proj-circ-topp}{(proj$^\circ_\topp$)};
			\item for \RuleRef{typing:gradual:proj-star}{(proj$_\star$)}, rules
			      \RuleRef{typing:weak:proj-circ-inttop}{(proj$^\circ_\intTop$)} and
			      \RuleRef{typing:weak:proj-circ-topp}{(proj$^\circ_\topp$)} give the
			      types $\bigvee_{i\leq n}t_i$ and $\dyn$, which rule
			      \RuleRef{typing:weak:and-circ}{(and$^\circ$)} intersects.\qedhere
		\end{itemize}
	\end{Proof}

\noindent
Now, the type safety in the gradual system is ensured by the safety of the
enriched weak system $\vdwn$.

\begin{Theorem}\label{app:thm:gradual-soundness} If $e$ is closed and $\varnothing \vdg e : t$, then either $e$ diverges, or $e$ crashes on a
	runtime error $\omega_p$, or $e$ evaluates to a value $v$ such that
	$\varnothing \vdwn v \tc t$, which is $\varnothing\vdw v\tc t$ if $v$ is a
	constant.
\end{Theorem}

\begin{Proof}[Proof of Type safety]
	By Lemma~\ref{lem:gradual-to-strong}, $\varnothing \vdw e \tc t$, hence
	$\varnothing\vdwn e\tc t$. Reduction creates no free variable, so every
	reduct of $e$ is closed, and the subject reduction~\ref{lem:subject-reduction}
	and progress~\ref{lem:gradual-progress} lemmas apply at every step: the
	reduction of $e$ is infinite, or it ends with a failure, or it ends with a
	value $v$ such that $\varnothing\vdwn v\tc t$. If $v$ is a constant,
	Lemma~\ref{lem:weak-generators}(4) gives $\varnothing\vdw v\tc t$.
\end{Proof}

\noindent
Theorem~\ref{thm:gradual-soundness} of Section~\ref{sec:safe-erasure} is the
following consequence of Theorem~\ref{app:thm:gradual-soundness}; its
statement uses only the weak system.

\begin{cor}\label{app:cor:gradual-soundness}
	Let $e$ be a closed expression such that $\varnothing\vdg e:t$ or
	$\varnothing\vdw e\tc t$. Then either $e$ diverges, or
	$e\seqreduces\omega_p$ for some $p$, or $e\seqreduces v$ for a value $v$
	such that $s\land t\not\leq\bott$ for every type $s$ such that
	$\varnothing\vdw v\tc s$. If no $\lambda$-abstraction occurs in $v$, that
	is, if $v$ is a constant (tuples of constants are constants), the last
	condition is equivalent to $\varnothing\vdw v\tc t$.
\end{cor}

\begin{Proof}
	The proof of Theorem~\ref{app:thm:gradual-soundness} starts from
	$\varnothing\vdw e\tc t$, so its conclusion holds under either hypothesis:
	if $e$ evaluates to a value $v$, then $\varnothing\vdwn v\tc t$. Let $s$ be
	such that $\varnothing\vdw v\tc s$. Then $\varnothing\vdwn v\tc s$, and rule
	\RuleRef{typing:weak:and-circ}{(and$^\circ$)} gives
	$\varnothing\vdwn v\tc s\land t$, so
	$\dynSup{(s\land t)}\not\simeq\bott$ by
	Lemma~\ref{lem:weak-generators}(2). By the characterization of subtyping of
	Section~\ref{sec:gradual-typing}, $s\land t\leq\bott$ would imply
	$\dynSup{(s\land t)}\leq\bott$; hence $s\land t\not\leq\bott$.

	Let $v$ be a constant. If $\varnothing\vdw v\tc t$, the argument above,
	with $\varnothing\vdw v\tc t$ in place of $\varnothing\vdwn v\tc t$, gives
	the condition. Conversely, rule \RuleRef{typing:weak:cst-circ}{(cst$^\circ$)}
	gives $\varnothing\vdw v\tc v\land\dyn$, so the condition gives
	$(v\land\dyn)\land t\not\leq\bott$. By the characterization of subtyping,
	and since the minimal materialization of a type is a subtype of its maximal
	one, $\dynSup{((v\land\dyn)\land t)}\simeq v\land\dynSup{t}\not\simeq\bott$.
	Since $v$ is a singleton type, $v\leq\dynSup{t}$, which is
	$\varnothing\vdw v\tc t$ by Lemma~\ref{lem:weak-generators}(4).
\end{Proof}

\noindent
For a result of arrow type, Theorem~\ref{app:thm:gradual-soundness} gives the
following guarantee. It does
not say that the interface of the result is a subtype of that type: the
abstraction $v_3$ of Remark~\ref{rem:weak-negations} has the type
$\boolTop\rarr\intTop$ in the weak system, by rule
\RuleRef{typing:weak:and-circ}{(and$^\circ$)} from $I^\circ_{v_3}$ and
$\StrongType{(\intTop\rarr\intTop)}$, while its interface is not a subtype of
$\boolTop\rarr\intTop$. A gradually typed program can return it with this type:
with $h=\lambda^{\{\dyn\rarr\dyn\}}w.w$ and
$G=\lambda^{\{\bott\rarr(\boolTop\rarr\intTop)\}}z.\texttt{case}\,z\,\p{\intTop\rarr v_3}$,
rule \RuleRef{typing:gradual:lambda}{($\lambda$)} gives $G$ the type
$\bott\rarr(\boolTop\rarr\intTop)$, since no branch is attainable under
$z:\bott$, and rule \RuleRef{typing:gradual:lambda-star}{($\lambda_\star$)}
the type $\StrongType{(\bott\rarr(\boolTop\rarr\intTop))}$, since the body has
type $(\boolTop\rarr\intTop)\land\dyn$ in the weak system under $z:\dyn$. Since
$\app{h}{3}$ has type `$\dyn$' and $\dyn\consist\bott$, rule
\RuleRef{typing:gradual:app-star}{(app$_\star$)} gives $\app{G}{\app{h}{3}}$ the
type $\dyn\land(\boolTop\rarr\intTop)$, and $\app{G}{\app{h}{3}}$ reduces to
$v_3$.

\begin{cor}\label{app:cor:gradual-functions}
	Let $v$ be a closed value such that $\varnothing\vdwn v\tc t_1\rarr t_2$.
	Then $v$ is an abstraction with one parameter, and, for every closed value
	$v'$ such that $\varnothing\vdwn v'\tc\dynInf{t_1}$, the application
	$\app{v}{v'}$ diverges, or reduces to a failure $\omega_p$, or reduces to a
	value $v''$ such that $\varnothing\vdwn v''\tc t_2$.
\end{cor}

\begin{Proof}
	By Lemma~\ref{lem:weak-generators}(1,2), $v$ has a generator whose maximal
	materialization is not empty and is a subtype of
	$\dynSup{(t_1\rarr t_2)}\leq\function_1$. The maximal materializations of
	the generators of constants, of tuples, and of abstractions with $n\neq1$
	parameters are disjoint from $\function_1$, so $v$ is an abstraction with one
	parameter. Since arrows are antitone in their domain and
	$\dynInf{t_1}\leq\dynSup{t_1}$, $t_1\rarr t_2\leq\dynInf{t_1}\rarr t_2$, so
	rule \RuleRef{typing:weak:app-circ}{(app$^\circ$)}, at the static domain
	$\dynInf{t_1}$, gives $\varnothing\vdwn\app{v}{v'}\tc t_2$.
	Lemmas~\ref{lem:subject-reduction} and~\ref{lem:gradual-progress} conclude,
	as in the proof of Theorem~\ref{app:thm:gradual-soundness}.
\end{Proof}




\section{Dynamic type tests}\label{app:dyn}

\begin{figure}[H]
	$B(c)$ maps each constant onto the least base type that contains it (e.g. integers onto $\intTop$, and $\tttrue$ and $\ttfalse$ onto $\boolTop$). The relation $v\in\tau$ is defined by induction on the test type $\tau$ (test types are those of Sections~\ref{sec:safe-erasure}, \ref{sec:guard-analysis} and~\ref{sec:arity}):
	\[
		\begin{array}{l@{\hspace{2em}}l}
			v \in b                   & \text{iff } v \text{ is a constant } c \text{ that is not a tuple and } B(c)\leq b\text{,}\\
			                          & \text{or } v \text{ is an abstraction and } b=\function\text{, or } v \text{ is a tuple and } b=\tupleTop \\
			v \in c                   & \text{iff } v = c \\
			v \in \function_n         & \text{iff } v \text{ is an abstraction with } n \text{ parameters (Section~\ref{sec:arity})} \\
			v \in \tuple{\tau_1,\ldotsTwo,\tau_m} & \text{iff } v=\tuple{v_1,\ldotsTwo,v_m} \text{ and } v_i\in\tau_i \text{ for every } i\leq m \\
			v \in \OpenTuple{\tau_1,\ldotsTwo,\tau_m} & \text{iff } v=\tuple{v_1,\ldotsTwo,v_n} \text{ with } n\geq m \text{ and } v_i\in\tau_i \text{ for every } i\leq m \\
			v \in \tau_1\lor\tau_2    & \text{iff } v\in\tau_1 \text{ or } v\in\tau_2 \\
			v \in \neg\tau            & \text{iff } v\notin\tau
		\end{array}
	\]
	where an abstraction may be marked or not, and tuples of constants are tuples, whose components are their constants.
	\caption{Definition of $v \in \tau$ (Sections~\ref{sec:safe-erasure} and~\ref{sec:guard-analysis})}%
	\label{app:inductive-def-test-type}
\end{figure}

\section{Guard Analysis}%
\label{app:guardanalysis}

We present the formal framework for guard analysis through a series of judgment forms and their associated rules. The guard analysis system operates on two main levels: pattern matching analysis that determines which types are accepted by guarded patterns, and guard analysis that evaluates individual guard conditions.

\begin{figure}[ht]
	\begin{equation*}
		\begin{array}{llr}
			\textbf{Pattern Matching Analysis}\hfill
			 & {\Gamma ; t \vd\ov{\when{p}{g}}\,\,\leadsto\,\,\ov{\AccTypes}}
			\\[1mm]
			\textbf{Guard Analysis}
			 & {\Gamma ; p \vd g \mapsto \mathcal{R}}
		\end{array}
	\end{equation*}
	\begin{grammarfig}
		\textbf{Accepted Types} &\AccTypes &\bnfeq& \ov{(t, \BoolB)} \\[1mm]
		\textbf{Results}
		&\GuardResult &\bnfeq& \{\mathscr{S}\,;\,\mathscr{T}_1,\ldotsTwo,\mathscr{T}_m\} \bnfor \GuardFail \qquad (m\geq 1)\\[1mm]
		\textbf{Environments}
		&\mathscr{S},\mathscr{T} &\bnfeq& (\Gamma, \BoolB) \\[1mm]
		\textbf{Failure Results}
		&\GuardFail &\bnfeq& \omega \bnfor \Result{\mathscr{S}}{\GuardFalse}
	\end{grammarfig}
	\caption{Guard Judgments with $\BoolB,\BoolC \in \{\smalltrue,\smallfalse\}$}
\end{figure}

The meaning of the judgements has been explained in Section~\ref{sub:guard-analysis-system}: just recall that the result of the analysis of a guard that may succeed consists of a safe environment $\mathscr{S}$, which provides a necessary condition for the guard not to error, and of success environments $\mathscr{T}_1,\ldotsTwo,\mathscr{T}_m$, one for each OR-clause that may succeed, at least one of which is satisfied when the guard succeeds; if the Boolean in $\mathscr{S}$ is true, then the condition is also sufficient, and if the Boolean in $\mathscr{T}_k$ is true, then the guard succeeds when $\mathscr{T}_k$ is satisfied (Lemma~\ref{lem:safe-success-environment} gives the precise statements). We write $\Result{\mathscr{S}}{(\Delta_i,\BoolB_i)_{i\in I}}$ for a result, with the convention that it stands for $\Result{\mathscr{S}}{\GuardFalse}$ when $I=\varnothing$. For a result $\GuardResult$ and an environment $\Gamma'$, $\SuccR{\GuardResult}$ is the family of its success environments ($\SuccR{\omega}$ and $\SuccR{\Result{\mathscr{S}}{\GuardFalse}}$ are empty), and $\SafeR{\Gamma'}{\GuardResult}$ is $\BoolB\etet(\Phi\simeq\Gamma')$ if $\GuardResult$ has the safe environment $(\Phi,\BoolB)$, and $\smallfalse$ if $\GuardResult=\omega$, where $\Phi\simeq\Gamma'$ means
that $\Phi$ and $\Gamma'$ have the same domain and $\Phi(x)\simeq\Gamma'(x)$
for every $x$ in it; for a flag $\BoolC$, $\FlagR{\BoolC}{\GuardResult}$ is $\omega$ if $\GuardResult=\omega$, and otherwise the result obtained from $\GuardResult$ by replacing every flag $\BoolB'$ with $\BoolC\etet\BoolB'$. On flags, $\etet$ and $\bigwedge$ denote conjunction. The Guard Analysis judgement explicitly carries the pattern $p$ to which the guard is attached. Rules that do not inspect $p$ simply thread it unchanged; rules such as \RuleRef{guard:analysis:var}{[\textsc{Var}]} and \RuleRef{guard:analysis:or}{[\textsc{Or}]} use it to refine dependent variables through environment updates of the form $\Gamma[x \refine \tau]_p$.

The syntax of guards is presented in Figure~\ref{app:fig:guard-syntax} below, showing the different forms of guard expressions that can be used in pattern matching. Guards can test type membership, equality, and combine conditions using Boolean operators.

\begin{figure}[ht]
	\setlength{\belowcaptionskip}{-10pt}
	\begin{grammarfig}
		\textbf{Guards}      & g
		&\bnfeq& a \isof \tau \bnfor a = a \bnfor a \,\,\texttt{!=}\,\, a\bnfor a \,\,\texttt{<}\,\, a
		\bnfor \GuardAnd{g}{g} \bnfor \GuardOr{g}{g} \\
		\textbf{Guard atoms} &a
		&\bnfeq& c\bnfor x\bnfor\proj{a}{a}\bnfor\sizeTup{a} \bnfor\tuple{\ov{a}}\\
		\textbf{Test types} & \tau
		&\bnfeq& b \bnfor c \bnfor \function_n \bnfor \pc{\ov{\tau}} \bnfor \OpenTuple{\ov{\tau}}
		\bnfor \tau \lor \tau \bnfor \neg \tau
	\end{grammarfig}
	\caption{Guard Syntax. In a guarded pattern $pg$, the free variables of $g$ are capture variables of $p$.}
	\label{app:fig:guard-syntax}
\end{figure}
\noindent

\subsection{Guard Compilation}\label{app:guard-compilation}
Note that in Figure~\ref{app:fig:guard-syntax} there is no negation on guards.
Indeed, the first thing we do is eliminate all negations from guards by pushing
them on the terminal guards, e.g., $\GuardNot{(a==a)}$ becomes $a\texttt{ != } a$. Guard compilation, defined in Figures~\ref{fig:guard-compilation-1} and~\ref{fig:guard-compilation-2}, is given by defining two mutually recursive functions from the set $\mathcal{G}_{\text{Elixir}}$ of Elixir concrete guards of FW-Elixir to the set $\mathcal{G}_{\text{Core}}$ of Core Elixir guards
(syntax in Figure~\ref{fig:patterns}).
Precisely, the
$\Tg: \mathcal{G}_{\text{Elixir}} \rightarrow \mathcal{G}_{\text{Core}}$ function
compiles a concrete guard into a core guard, and the
$\Ng:  \mathcal{G}_{\text{Elixir}} \rightarrow \mathcal{G}_{\text{Core}}$
does so as well, but also pushes down a logical negation into the guard, which means
that, say, a type-check of $\intTop$ becomes a type-check of $\neg \, \intTop$,
and that conjunctions and disjunctions are swapped using De Morgan rules.
Selectors are compiled into guard atoms by $\Td$, and comparisons, whose operands are selectors, into comparisons of atoms.
Bare selectors $\FWD$ can be used as guards. In Elixir, the left operand of \texttt{and} and \texttt{or} and the operand of \texttt{not} must evaluate to a Boolean, otherwise the guard fails, while the right operand of \texttt{and} and \texttt{or} is returned as it is. So $\Tg$ translates a selector into a check that it has singleton type $\tttrue$, and $\Ng$ into a check that it has singleton type $\ttfalse$, both followed by a guard that fails when the selector is not a Boolean; there, $\GuardErr$ stands for the guard $\sizeTup{0}\isof\intTop$, which always fails. A failure and the result $\smallfalse$ have the same effect on a whole guard, so the translation can use this check in every position. Notice that the compilation also defines the semantic of relations present only in FW-Elixir, such as `$\texttt{>=}$', which is compiled as the union of `$\texttt{<}$' and `$\texttt{=}$'. In particular the negation of  $\FWD_1\,\texttt{<}\,\,\FWD_2\,$---i.e., $\Ng(\FWD_1\,\texttt{<}\,\,\FWD_2)$---is compiled  as $(\GuardLt{\Td(\FWD_2)}{\Td(\FWD_1)}) \texttt{ or } (\GuardEq{\Td(\FWD_1)}{\Td(\FWD_2)})$ which is correct since all (FW-)Elixir values are totally ordered (cf. Figure~\ref{fig:elixir-ordering}).
\begin{figure}[t]
\hspace*{-3.5mm}\begin{minipage}[t]{.525\linewidth}
			\small
			\begin{equation*}
\begin{aligned}
\Tg(\FWD) & = (\Td(\FWD) \isof \tttrue) \texttt{ or } \\[-1mm]
          & \phantom{{}={}} ((\Td(\FWD)\isof\neg\boolTop) \texttt{ and } \GuardErr) \\[-1mm]
\Tg(\FWC) & = \Tc(\FWC) \\[-1mm]
\Tg(\GuardAnd{\FWG_1}{\FWG_2}) & = \GuardAnd{\Tg(\FWG_1)}{\Tg(\FWG_2)} \\[-1mm]
\Tg(\GuardOr{\FWG_1}{\FWG_2}) & = \GuardOr{\Tg(\FWG_1)}{\Tg(\FWG_2)} \\[-1mm]
\Tg(\GuardNot{\FWG})& = \Ng(\FWG) \\[-1mm]
\Tg({\FWD_1}\,\texttt{==}\,{\FWD_2})& = \GuardEq{\Td(\FWD_1)}{\Td(\FWD_2)}\\[-1mm]
\Tg({\FWD_1}\,\texttt{!=}\,\,{\FWD_2}) & = \GuardNeq{\Td(\FWD_1)}{\Td(\FWD_2)}  \\[-1mm]
\Tg({\FWD_1}\,\texttt{<}\,\,{\FWD_2}) & = \GuardLt{\Td(\FWD_1)}{\Td(\FWD_2)}\\[-1mm]
\Tg({\FWD_1}\,\texttt{>}\,\,{\FWD_2}) & = \GuardLt{\Td(\FWD_2)}{\Td(\FWD_1)}\\[-1mm]
\Tg({\FWD_1}\,\texttt{<=}\,\,{\FWD_2}) & = \GuardLt{\Td(\FWD_1)}{\Td(\FWD_2)} \texttt{ or } \GuardEq{\Td(\FWD_1)}{\Td(\FWD_2)}  \\[-1mm]
\Tg({\FWD_1}\,\texttt{>=}\,\,{\FWD_2}) & = \GuardLt{\Td(\FWD_2)}{\Td(\FWD_1)} \texttt{ or } \GuardEq{\Td(\FWD_1)}{\Td(\FWD_2)} \\[1mm]
\end{aligned}
\end{equation*}
\end{minipage}
\hspace*{-8mm}
\begin{minipage}[t]{.48\linewidth}
\begin{equation*}
\begin{aligned}
\Td(\EElem{\FWD_1}{\FWD_2}) & = \GuardProj{\Td(\FWD_2)}{\Td(\FWD_1)} \\[-1mm]
\Td(\EtupleSize{\FWD}) & = \GuardSize{\Td(\FWD)}                \\[-1mm]
\Td(\tuple{\FWD_1,..,\FWD_n}) & = \tuple{\Td(\FWD_1),..,\Td(\FWD_n)}   \\[2mm]
\Tc(\EisInt{\FWD}) & = \Td(\FWD) \isof \intTop              \\[-1mm]
\Tc(\EisAtom{\FWD}) & = \Td(\FWD) \isof \atomTop             \\[-1mm]
\Tc(\EisTuple{\FWD}) & = \Td(\FWD) \isof \tupleTop            \\[-1mm]
\Tc(\EisFun{\FWD}) & = \Td(\FWD) \isof \function            \\[-1mm]
\Tc(\EisFunction{\FWD}{n}) & = \Td(\FWD) \isof \function_n
\end{aligned}
\end{equation*}
\end{minipage}
\caption{Guard Compilation: $\Tg$ and auxiliary functions}%
\label{fig:guard-compilation-1}
\end{figure}

\begin{figure}[t]
\centering
\small
\begin{minipage}[t]{.48\linewidth}
\begin{equation*}
\begin{aligned}
	\Ng(\FWD) & = (\Td(\FWD) \isof \ttfalse) \texttt{ or } \\[-1mm]
	          & \phantom{{}={}} ((\Td(\FWD)\isof\neg\boolTop) \texttt{ and } \GuardErr)  \\[-1mm]
	\Ng(\FWC) & = \Nc(\FWC) \\[-1mm]
	\Ng(\GuardAnd{\FWG_1}{\FWG_2}) & = \GuardOr{\Ng(\FWG_1)}{\Ng(\FWG_2)}   \\[-1mm]
	\Ng(\GuardOr{\FWG_1}{\FWG_2})& = \GuardAnd{\Ng(\FWG_1)}{\Ng(\FWG_2)}  \\[-1mm]
	\Ng(\GuardNot{\FWG}) & = {\Tg(\FWG)}  \\[-1mm]
	\Ng({\FWD_1}\,\texttt{==}\,{\FWD_2}) & = \Td(\FWD_1)\,\texttt{!=}\,\,{\Td(\FWD_2)} \\[-1mm]
	\Ng(\FWD_1\,\texttt{!=}\,\,\FWD_2)& = \GuardEq{\Td(\FWD_1)}{\Td(\FWD_2)}        \\[-1mm]
	\Ng(\FWD_1\,\texttt{<}\,\,\FWD_2) & =\GuardLt{\Td(\FWD_2)}{\Td(\FWD_1)} \texttt{ or } \GuardEq{\Td(\FWD_1)}{\Td(\FWD_2)}  \\[-1mm]
	\Ng(\FWD_1\,\texttt{>}\,\,\FWD_2) & =\GuardLt{\Td(\FWD_1)}{\Td(\FWD_2)} \texttt{ or } \GuardEq{\Td(\FWD_1)}{\Td(\FWD_2)} \\[-1mm]
	\Ng(\FWD_1\,\texttt{<=}\,\,\FWD_2) & =\GuardLt{\Td(\FWD_2)}{\Td(\FWD_1)} \\[-1mm]
	\Ng(\FWD_1\,\texttt{>=}\,\,\FWD_2) & =\GuardLt{\Td(\FWD_1)}{\Td(\FWD_2)}
\end{aligned}
\end{equation*}
\end{minipage}
\hfill
\begin{minipage}[t]{.48\linewidth}
\begin{equation*}
\begin{aligned}
\Nc(\EisInt{\FWD})  & = \Td(\FWD) \isof (\neg\intTop      )       \\[-1mm]
\Nc(\EisAtom{\FWD})& = \Td(\FWD) \isof (\neg\atomTop     )       \\[-1mm]
\Nc(\EisTuple{\FWD}) & = \Td(\FWD) \isof (\neg\tupleTop    )      \\[-1mm]
\Nc(\EisFun{\FWD}) & = \Td(\FWD) \isof (\neg\function    ) \\[-1mm]
\Nc(\EisFunction{\FWD}{n})   & = \Td(\FWD) \isof (\neg\function_n  )\\[-1mm]
\end{aligned}
\end{equation*}
\end{minipage}
\caption{Negated Guard Compilation: $\Ng$ and auxiliary functions}%
\label{fig:guard-compilation-2}
\end{figure}

In Elixir, a guard is evaluated as follows: a type check or a comparison evaluates its selectors, and errors if one of them errors; \texttt{G$_1$ and G$_2$} evaluates $\FWG_1$, errors if the result is not a Boolean, returns $\ttfalse$ if it is $\ttfalse$, and the result of $\FWG_2$ otherwise; \texttt{G$_1$ or G$_2$} is symmetric, with $\tttrue$ in place of $\ttfalse$; \texttt{not G} errors if the result of $\FWG$ is not a Boolean, and negates it otherwise; a guard succeeds if and only if it evaluates to $\tttrue$ (an error makes the guard fail).

\begin{lem}[Guard compilation]\label{lem:guard-compilation}
	Let $\FWG$ be an FW-Elixir guard and $\sigma$ a substitution of values for its variables. If $\FWG\sigma$ evaluates to $\tttrue$, then $\Tg(\FWG)\sigma\seqreducesG\smalltrue$ and $\Ng(\FWG)\sigma\seqreducesG\smallfalse$; if it evaluates to $\ttfalse$, then $\Tg(\FWG)\sigma\seqreducesG\smallfalse$ and $\Ng(\FWG)\sigma\seqreducesG\smalltrue$; otherwise, that is, if it errors or evaluates to a value that is not a Boolean, then $\Tg(\FWG)\sigma\seqreducesG\fail$ and $\Ng(\FWG)\sigma\seqreducesG\fail$. In particular, $\FWG\sigma$ succeeds if and only if $\Tg(\FWG)\sigma\seqreducesG\smalltrue$.
\end{lem}

\begin{Proof}
	By induction on $\FWG$. The atom $\Td(\FWD)\sigma$ errors if and only if $\FWD\sigma$ does, and otherwise both evaluate to the same value.
	\begin{description}
		\item[Selectors] Let $\FWD\sigma$ evaluate to $v$, or error. If it errors, both translations fail. If $v=\tttrue$, then $\Td(\FWD)\sigma\isof\tttrue$ is $\smalltrue$, so $\Tg(\FWD)\sigma$ is $\smalltrue$; and $\Td(\FWD)\sigma\isof\ttfalse$ and $\Td(\FWD)\sigma\isof\neg\boolTop$ are $\smallfalse$, so $\Ng(\FWD)\sigma$ is $\smallfalse$. The case $v=\ttfalse$ is symmetric. If $v$ is not a Boolean, the first disjunct of both translations is $\smallfalse$, the test $\Td(\FWD)\sigma\isof\neg\boolTop$ is $\smalltrue$, and $\GuardErr$ fails.
		\item[Checks] $\Tc(\FWC)\sigma$ and $\Nc(\FWC)\sigma$ are $\Td(\FWD)\sigma\isof\tau$ and $\Td(\FWD)\sigma\isof\neg\tau$, where $\tau$ is the type tested by $\FWC$: they fail if $\FWD\sigma$ errors, and otherwise $v\in\neg\tau$ if and only if $v\notin\tau$ (Figure~\ref{app:inductive-def-test-type}).
		\item[Comparisons] Their translations evaluate the same selectors, and fail if one of them errors; otherwise, equality is the same, and, since the order on values is total, $v_1<v_2$ is false if and only if $v_2<v_1$ or $v_1=v_2$.
		\item[$\FWG_1$ \texttt{and} $\FWG_2$] If $\FWG_1\sigma$ evaluates to $\tttrue$, the conjunction behaves as $\FWG_2\sigma$, and so do $\Tg(\FWG_1)\sigma\,\ttand\allowbreak\,\Tg(\FWG_2)\sigma$, since $\Tg(\FWG_1)\sigma\seqreducesG\smalltrue$, and $\Ng(\FWG_1)\sigma\,\ttor\allowbreak\,\Ng(\FWG_2)\sigma$, since $\Ng(\FWG_1)\sigma\seqreducesG\smallfalse$. If it evaluates to $\ttfalse$, the conjunction evaluates to $\ttfalse$, $\Tg(\FWG_1)\sigma\seqreducesG\smallfalse$ makes the first translation $\smallfalse$, and $\Ng(\FWG_1)\sigma\seqreducesG\smalltrue$ makes the second $\smalltrue$. Otherwise the conjunction errors, and both translations fail, since their first operands fail.
		\item[$\FWG_1$ \texttt{or} $\FWG_2$] Symmetric.
		\item[\texttt{not} $\FWG$] Its translations are those of $\FWG$, exchanged, and \texttt{not} exchanges $\tttrue$ and $\ttfalse$ and errors on the other results.\qedhere
	\end{description}
\end{Proof}

\subsection{Strict Ordering of Types}%
\label{sec:strict-ordering-of-types}

To handle comparisons in guards $(a_1 < a_2)$, we can take advantage of
the static type of each guard (if it is available) to decide statically the result of the guard. Here, we define the strict ordering on types $<_{\mathrm{type}}$ that allows us to do this. This is only a detail in the
larger guard analysis system that will be defined in the next subsection~\ref{sec:computation-of-accepted-types}.

From the formalized order on terms $<_{\mathrm{term}}$ (see Figure~\ref{fig:elixir-ordering}) lifted from Elixir, we define a strict ordering on types $<_{\mathrm{type}}$, with the property that
$t_1 <_{\mathrm{type}} t_2$ implies $\forall v_1 \in t_1, \forall v_2 \in t_2, v_1 <_{\mathrm{term}} v_2$ (the converse does not hold: for instance, $1 \not<_{\mathrm{type}} 2$, since the bounds of both types are $r_{\mathsf{int}}$). To do so, we define, for each component type $c$
(integer, atom, tuple, function), a $\min_c$ and $\max_c$ function. Then, we
define a global min and max for each type (as each type is a union of these components).
The ordering of two types $t_1 <_{\mathrm{type}} t_2$ will then be defined
as $t_1 <_{\mathrm{type}} t_2$ iff $\max(t_1) <_{\mathrm{type}} \min(t_2)$. Notice that the $<_{\mathrm{type}}$ ordering and the subtyping relation $\leq$ are unrelated: for instance, $\intTop <_{\mathrm{type}} \atomTop$ but $\intTop \not\leq \atomTop$, and $t_1 \leq t_2$  implies $t_1 \not<_{\mathrm{type}} t_2$ (unless $t_1$ is empty).

The relation $<_{\mathrm{type}}$ will allow us to decide statically the result of
a comparison guard: if $a_1$ is a of type $t_1$ and $a_2$ is of type $t_2$, then $t_1 <_{\mathrm{type}} t_2$ implies that the guard $a_1 < a_2$ will always succeed. Conversely, if $t_2 <_{\mathrm{type}} t_1$, then the guard $a_1 < a_2$ will always fail. If neither of these is true, then the guard
may succeed or fail depending on what happens at runtime.
For clarity, in the rules, we will write $\mathsf{alwaysLess}(t_1, t_2)$ for $t_1 \mathrel{<_{\mathrm{type}}} t_2$, $\mathsf{neverLess}(t_1, t_2)$ for $t_2 \mathrel{<_{\mathrm{type}}} t_1$, and $\mathsf{maybeLess}(t_1, t_2)$ for $\neg\mathsf{alwaysLess}(t_1, t_2) \land \neg\mathsf{neverLess}(t_1, t_2)$. These relations are defined for static types, and in rules \RuleRef{guard:analysis:lt}{(lt)}, \RuleRef{guard:analysis:lt-maybe}{(ltMaybe)} and \RuleRef{guard:analysis:lt-false}{(ltFalse)} the types $t_0$ and $t_1$ are static. This is no loss: if $\Gamma\vds a:t$, then $\Gamma\vds a:\dynSup{t}$, since $t\leq\dynSup{t}$.

\begin{Definition}[Per-component min/max]
	For $c\in\{\mathsf{int}, \mathsf{atom}, \mathsf{tuple}, \mathsf{fun}\}$, define four candidate pairs \((\min_c(t),\max_c(t))\) as follows; $\epsilon$ means “no candidate”, so the global min/max will ignore this
	component. We introduce four constants $r_{\mathsf{int}}, r_{\mathsf{atom}}, r_{\mathsf{tuple}}, r_{\mathsf{fun}}$ to represent the rank token of each component type, such that $r_{\mathsf{int}} < r_{\mathsf{atom}} < r_{\mathsf{fun}} < r_{\mathsf{tuple}}$, as in Figure~\ref{fig:elixir-ordering}.
	\[\hspace*{-4mm}
\begin{array}{l @{\qquad} rlr}
&
\displaystyle (\min_{\mathsf{int}}(t),\max_{\mathsf{int}}(t))=&
\begin{cases}
(r_{\mathsf{int}},r_{\mathsf{int}}) & \text{if } t\land\intTop\not\le\bott \\
(\epsilon,\epsilon) & \text{otherwise}
\end{cases}
\\[1em]
 &
\displaystyle (\min_{\mathsf{atom}}(t),\max_{\mathsf{atom}}(t))=&
\begin{cases}
(\min A,\max A) & \text{if } A=(t\land\atomTop)\ \text{is a non-empty finite set}\\
(r_{\mathsf{atom}},r_{\mathsf{atom}}) & \text{if } (t\land\atomTop) \text{ is cofinite} \\
(\epsilon,\epsilon) & \text{otherwise}
\end{cases}
\\[2em]
 &
\displaystyle (\min_{\mathsf{tuple}}(t),\max_{\mathsf{tuple}}(t))=&
\begin{cases}
(n_{\min}, n_{\max}) & \text{if } t \land \tupleTop\not\le\bott \text{ and its least and}\\
 & \text{greatest arities are } n_{\min} \text{ and } n_{\max}\\
(r_{\mathsf{tuple}},r_{\mathsf{tuple}}) & \text{otherwise and } t \land \tupleTop\not\le\bott \\
(\epsilon,\epsilon) & \text{otherwise}
\end{cases}
\\[2em]
 &
\displaystyle (\min_{\mathsf{fun}}(t),\max_{\mathsf{fun}}(t))=&
\begin{cases}
(r_{\mathsf{fun}},r_{\mathsf{fun}}) & \text{if } t\land\function\not\le\bott \\
(\epsilon,\epsilon) & \text{otherwise}
\end{cases}
\end{array}
\]
In the definitions for $\min_{\mathsf{atom}}$ and $\max_{\mathsf{tuple}}$, we use the fact that it is possible to decide if an atom type is
finite or not (trivial given the representation as a finite or cofinite in
Section~\ref{sec:data-structures}), and if a tuple type has a bounded arity (also trivial given the representation as a list of tuples in
Section~\ref{sec:data-structures}).
\end{Definition}

Now that we have defined a min/max for each component, we define
a global min (respectively, max) by picking the least (respectively, greatest) defined component (following the
ranking order $r_{\mathsf{int}} < r_{\mathsf{atom}} < r_{\mathsf{fun}} < r_{\mathsf{tuple}}$). For instance, if $\max_{\mathsf{tuple}}(t) \neq \epsilon$, then
it should always be that $\max(t) = \max_{\mathsf{tuple}}(t)$.
The functions $\min$ and $\max$ have their images included in
$\{r_{\mathsf{int}}, r_{\mathsf{atom}}, r_{\mathsf{tuple}}, r_{\mathsf{fun}}\}
\cup \mathbb{N} \cup \mathsf{Atoms}$ where $\mathbb{N}$ denotes arity
of tuples, and $\mathsf{Atoms}$ denotes the set of Elixir atoms.
Two such bounds are compared first by their components, in the ranking order;
within the same component, arities are compared as integers, atoms
lexicographically, and a comparison that involves the rank token of that component never holds.
To compare, if $\max(t_1) = 10$ (in $t_1$, tuples of arity 10 are the largest
values with regard to $<_{\mathrm{term}}$) and $\min(t_2) = r_{\mathsf{tuple}}$ (the smallest values in $t_2$ are tuples), then we have that $\neg(t_1 <_{\mathrm{type}} t_2)$. But if $\max(t_1) = r_{\mathsf{fun}}$ (the largest values in $t_1$ are functions) and $\min(t_2) = r_{\mathsf{tuple}}$, then we do have $t_1 <_{\mathrm{type}} t_2$, since every function is smaller than every tuple.

\begin{Definition}[Global min/max]
Let
\begin{enumerate}[nosep,label=\roman*.]
\item
$C_{\min}(t)=\{\min_{\mathsf{int}}(t),\min_{\mathsf{atom}}(t),\min_{\mathsf{tuple}}(t),\min_{\mathsf{fun}}(t)\}\setminus\{\epsilon\}$,
\item
$C_{\max}(t)=\{\max_{\mathsf{int}}(t),\max_{\mathsf{atom}}(t),\max_{\mathsf{tuple}}(t),\max_{\mathsf{fun}}(t)\}\setminus\{\epsilon\}$.
\end{enumerate}
We define
\begin{enumerate}[nosep,label=\roman*.]
	\item
	$\min(t)=\text{the least element of }C_{\min}(t)$, by component order;
	\item $\max(t)=\text{the greatest element of }C_{\max}(t)$, by component order.
\end{enumerate}
\end{Definition}
This order will be used in Figures~\ref{app:fig:guard-analysis-rules} and~\ref{app:fig:guard-analysis-false-failure-rules} of the Guard Analysis rules.

\subsection{Computation of Accepted Types}%
\label{sec:computation-of-accepted-types}
The production rules of Figure~\ref{fig:accepted-types-prod} define how accepted types are computed for guarded patterns. As a quick recap,
an accepted type $(t, \BoolB)$ for a guarded pattern $p g$ pairs a type with a Boolean flag.
The generated types together contain every incoming value accepted by $p g$.
If $\BoolB$ is $\smalltrue$, every value in $t$ is accepted by $p g$;
otherwise, this is not guaranteed.

These rules, which have been explained in Section~\ref{sec:pattern-guard-sequence-analysis}, depend on the more specific guard analysis that will be defined next.

\begingroup
\setlength{\abovedisplayskip}{2pt}
\setlength{\belowdisplayskip}{2pt}
\begin{figure}[ht]
	\setlength{\belowcaptionskip}{-7pt}
	\begin{equation*}
		\begin{gathered}
			\ruleAccept
			\,\,
			\ruleFail
			\\[2mm]
			\ruleSeq
		\end{gathered}
	\end{equation*}
	\caption{Accepted Types Productions}
	\label{fig:accepted-types-prod}
\end{figure}
\endgroup

\subsection{Guard Analysis}

\begingroup
\addtolength{\jot}{0.5em}
\setlength{\abovedisplayskip}{2pt}
\setlength{\belowdisplayskip}{2pt}
\begin{figure}[ht]
  \begin{equation*}
    \begin{gathered}
	  \ruleGuardTrue
	  \quad
	  \ruleGuardFalse
	  \\
      \ruleGuardVar
      \quad
      \ruleGuardSize
      \\
      \ruleGuardProj
      \\
      \ruleGuardEqOne
      \qquad
      \ruleGuardEqTwo
	  \\
      \ruleGuardNeqOne
      \qquad
      \ruleGuardNeqTwo
	  \\
	  \ruleGuardLt
	  \\
	  \ruleGuardLtMaybe
    \end{gathered}
  \end{equation*}
  \caption{Guard Analysis: Main Rules}
  \label{app:fig:guard-analysis-rules}
\end{figure}
\endgroup

Most of the guard analysis rules have been explained in Section~\ref{sub:guard-analysis-system}, and the workings of the other rules can be easily derived from them. In every rule, the typing premises $\JudgmentTypeS{\Gamma}{a}{s}$ are derived without $\omega$-marked rules. Here we summarize the rules.

\begin{description}
\item[Figure~\ref{app:fig:guard-analysis-rules}] defines the main rules for guard analysis. These rules define how type information flows through complex guard expressions.

\item[Figure~\ref{app:fig:guard-analysis-boolean-rules}] defines the boolean rules for guard analysis, which handle the logical combination of guards using AND and OR operators.

\item[Figure~\ref{app:fig:guard-analysis-approx-rules}] defines the approximation rules for guard analysis, which handle more complex guard expressions that involve projection operations and tuple size checks

\item[Figure~\ref{app:fig:guard-analysis-false-failure-rules}] defines the failure and false rules for guard analysis, which define cases where guards definitely fail or when guard evaluation results in false outcomes.

\end{description}
In Table~\ref{tab:guard-rules-numbering} we number these rules for easy reference in the text. The numbering is consistent with the order of the rules in Figures~\ref{app:fig:guard-analysis-rules}, \ref{app:fig:guard-analysis-boolean-rules}, \ref{app:fig:guard-analysis-approx-rules}, and \ref{app:fig:guard-analysis-false-failure-rules}.

\begin{figure}[ht]
  \setlength{\belowcaptionskip}{-7pt}
  \begin{equation*}
    \begin{gathered}
      \ruleGuardAnd
      \\
      \ruleGuardOr
      \\[-3mm]
    \end{gathered}
  \end{equation*}
  \caption{Guard Analysis: Boolean Rules}
  \label{app:fig:guard-analysis-boolean-rules}
\end{figure}

\begin{figure}[ht]
  \begin{equation*}
    \begin{gathered}
      \ruleGuardProjApprox
      \\[2mm]
      \ruleGuardEqApprox
      \\[2mm]
      \ruleGuardSizeApprox
      \\[2mm]
      \ruleGuardLtApprox
      \\[2mm]
      \ruleGuardApprox
    \end{gathered}
  \end{equation*}
  \caption{Guard Analysis: Approximation Rules}
  \label{app:fig:guard-analysis-approx-rules}
\end{figure}

\begin{figure}[ht]
  \setlength{\belowcaptionskip}{-3pt}
  \begin{equation*}
    \begin{gathered}
      \ruleGuardSizeOmega
      \quad
      \ruleGuardEqOmega
      \\
      \ruleGuardProjOmegaTuple
      \quad
      \ruleGuardProjOmegaInt
      \\
      \ruleGuardBoundOmega
      \quad
      \ruleGuardLtOmega
      \\
      \ruleGuardOrF
      \quad
      \ruleGuardOrOmega
      \\
      \ruleGuardAndFFail
      \quad
	  \ruleGuardLtFalse
    \end{gathered}
  \end{equation*}
  \caption{Guard Analysis: False/Failure Rules}
  \label{app:fig:guard-analysis-false-failure-rules}
\end{figure}

\begin{table}[htbp]
\centering
\caption{Guard Analysis Rules Numbering}
\label{tab:guard-rules-numbering}
\begin{tabular}{|c|l|l|}
\hline
\textbf{Rule \#} & \textbf{Rule Name} & \textbf{Description} \\
\hline
1 & \RuleRef{guard:analysis:true}{true} & Basic type check success \\ 
2 & \RuleRef{guard:analysis:false}{false} & Type intersection is empty \\ 
3 & \RuleRef{guard:analysis:var}{var} & Variable type refinement \\ 
4 & \RuleRef{guard:analysis:size}{size} & Tuple size analysis \\ 
5 & \RuleRef{guard:analysis:proj}{proj} & Tuple projection \\ 
6 & \RuleRef{guard:analysis:eq1}{eq$_1$} & Equality check (first form) \\ 
7 & \RuleRef{guard:analysis:eq2}{eq$_2$} & Equality check (second form) \\ 
8 & \RuleRef{guard:analysis:neq1}{neq$_1$} & Disequality check (first form) \\ 
9 & \RuleRef{guard:analysis:neq2}{neq$_2$} & Disequality check (second form) \\ 
10 & \RuleRef{guard:analysis:lt}{lt} & Less than check \\ 
11 & \RuleRef{guard:analysis:lt-maybe}{ltMaybe} & Less than check (maybe) \\ 
\hline
\multicolumn{3}{|c|}{\textbf{Boolean Operators}} \\
\hline
12 & \RuleRef{guard:analysis:and}{and} & Boolean AND operation \\ 
13 & \RuleRef{guard:analysis:or}{or} & Boolean OR operation \\ 
\hline
\multicolumn{3}{|c|}{\textbf{Approximation Rules}} \\
\hline
14 & \RuleRef{guard:analysis:proj-a}{proj$_a$} & Approximated projection \\ 
15 & \RuleRef{guard:analysis:eq-a}{eq$_a$} & Approximated equality or disequality \\ 
16 & \RuleRef{guard:analysis:size-a}{size$_a$} & Approximated size check \\ 
17 & \RuleRef{guard:analysis:lt-a}{lt$_a$} & Approximated less than check \\ 
18 & \RuleRef{guard:analysis:approx}{approx} & Any guard, no information \\ 
\hline
\multicolumn{3}{|c|}{\textbf{False/Failure Rules}} \\
\hline
19 & \RuleRef{guard:analysis:size-omega}{size$_\omega$} & Size operation failure \\ 
20 & \RuleRef{guard:analysis:eq-omega}{eq$_\omega$} & Equality or disequality failure \\ 
21 & \RuleRef{guard:analysis:proj-omega-t}{proj$^t_\omega$} & Projection failure (not a tuple) \\ 
22 & \RuleRef{guard:analysis:proj-omega-i}{proj$^i_\omega$} & Projection failure (index not int) \\ 
23 & \RuleRef{guard:analysis:bound-omega}{bound$_\omega$} & Projection failure (out of bounds) \\ 
24 & \RuleRef{guard:analysis:lt-omega}{lt$_\omega$} & Less than check failure \\ 
25 & \RuleRef{guard:analysis:or-f}{orF} & OR with false branch \\ 
26 & \RuleRef{guard:analysis:or-omega}{or$_\omega$} & OR with error branch \\ 
27 & \RuleRef{guard:analysis:and-fail-l}{and$_\GuardFail$-L} & AND with left branch failure \\ 
28 & \RuleRef{guard:analysis:lt-false}{ltFalse} & Less than check always false \\ 
\hline
\end{tabular}
\end{table}

\subsection{Auxiliary Definitions}

Here are the formal definitions of several notions used in the main text (\S\ref{sec:typing-pattern-matching}):
\begin{itemize}
	\item Accepted types (Figure~\ref{app:def:accepted-types}) determine which types are accepted by pattern matching constructs, providing the foundation for type refinement.
	\item
	Typing environments (Figure~\ref{app:fig:typing-environments}) map variables in patterns to their corresponding types, enabling precise type tracking through pattern matching.
	\item
Environment updates (Figure~\ref{app:def:environment-updates}) define how type environments are refined when guards succeed, providing more precise type information for subsequent analysis.
\end{itemize}

\begin{figure}[t]
	\setlength{\belowcaptionskip}{-7pt}
	\begin{align*}
		\acc{x}{\Gamma} &= \Gamma(x) \text{ if } x \in \dom{\Gamma} \\
		\acc{x}{\Gamma} &= \topp \text{ if } x \notin \dom{\Gamma} \\
		\acc{\tuple{p_1,\ldotsTwo,p_n}}{\Gamma}
			      &= \tuple{\acc{p_1}{\Gamma},\ldotsTwo,\acc{p_n}{\Gamma}} \\
		\acc{c}{\Gamma} &= c \\
		\acc{p_1\,\&\,p_2}{\Gamma}
			      &= \acc{p_1}{\Gamma} \land \acc{p_2}{\Gamma}
	\end{align*}
	\caption{Accepted types}
	\label{app:def:accepted-types}
\end{figure}

\begin{figure}[t]
	If $t \leq \acc{p}{\varnothing}$ then
	$\TypeEnv{t}{p}$ is a map from the variables of $p$ to types:
	\[
		\begin{array}{lll}
			\TypeEnv{t}{x}(x)                      & =  t                                                          \\
			\TypeEnv{t}{\tuple{p_1,\ldots,p_n}}(x) & =  \TypeEnv{\pi_i(t)}{p_i}(x)
			                                       & \text{where } \exists i \text{ unique s.t. } x \in \vars{p_i} \\
			\TypeEnv{t}{p_1\,\&\,p_2}(x)           & =  \TypeEnv{t}{p_1}(x)
			                                       & \text{if } x \in \Vars{p_1}                                   \\
			\TypeEnv{t}{p_1\,\&\,p_2}(x)           & =  \TypeEnv{t}{p_2}(x)
			                                       & \text{if } x \not\in \Vars{p_1} \text{ and } x\in\Vars{p_2}
		\end{array}
	\]
	where $\pi_i(t)$ is the $i$-th projection of the tuple type $t$: for a static $t\leq\tuple{\topp,\ldotsTwo,\topp}$, written as a union $\bigvee_{k\in K}\tuple{t_{k,1},\ldotsTwo,t_{k,n}}$ of products, $\pi_i(t)=\bigvee\{t_{k,i}\mid k\in K,\ t_{k,j}\not\simeq\bott\text{ for every }j\}$, the least type $s$ such that $t\leq\tuple{\topp,\ldotsTwo,s,\ldotsTwo,\topp}$ ($s$ at position $i$); for a gradual $t$, $\pi_i(t)=\pi_i(\dynInf{t})\lor(\dyn\land\pi_i(\dynSup{t}))$, so that $\dynInf{(\pi_i(t))}=\pi_i(\dynInf{t})$ and $\dynSup{(\pi_i(t))}=\pi_i(\dynSup{t})$. Hence $\dynInf{(\TypeEnv{t}{p})}=\TypeEnv{\dynInf{t}}{p}$ and $\dynSup{(\TypeEnv{t}{p})}=\TypeEnv{\dynSup{t}}{p}$, componentwise. For any other type $t$, $\TypeEnv{t}{p}$ stands for $\TypeEnv{(t\land\acc{p}{\varnothing})}{p}$; rules \RuleRef{guard:accepted:accept}{[accept]} and \RuleRef{guard:accepted:fail}{[fail]} use it for types $t$ that need not be below $\acc{p}{\varnothing}$. Since $\acc{p}{\varnothing}$ is static, the two equations above hold for every $t$.
	\caption{Typing Environments}
	\label{app:fig:typing-environments}
\end{figure}

\begin{figure}[t]
	\[
		\begin{array}{lll}
			\forall y\in\dom{\Gamma},\,\,\Gamma[x \refine t](y) & =
			\begin{cases} \Gamma(y)         & \text{if } y \neq x \\
              \Gamma(x) \land t & \text{if } y = x
			\end{cases}                                                \\[4mm]
			\Gamma[x \refine t]_p                               & =
			\left(\Gamma[x \refine t]\,,\, t'/p\right)
			                                                    & \text{where } t' = \acc{p}{\Gamma[x\refine t]}
		\end{array}
	\]
	\caption{Environment Updates}\label{app:def:environment-updates}
\end{figure}

\subsection{Correctness of Guard Analysis}\label{app:guard-analysis-correctness}

The safety analysis of Appendix~\ref{app:soundness} is carried out for a
calculus whose case expressions use type tests $\tau$ instead of guarded
patterns $p g$. To extend it to the case expressions of
Section~\ref{sec:guard-analysis}, whose typing rules are recalled in
Figure~\ref{app:fig:case-typing-rule}, we relate the guard analysis to the
semantics of guards. The proofs of safety use two typing relations for values,
the enriched static system $\vdsn$ (Appendix~\ref{app:static-safety}) and the
enriched weak system $\vdwn$ (Appendix~\ref{app:gradual-safety}), and the
results of this subsection hold for both. For a value $v$ and a type $t$ we
write $v\Vdash t$ for $\JudgmentTypeSN{\varnothing}{v}{t}$ in the
\emph{static instance}, and, when $v$ is closed, for $\varnothing\vdwn v\tc t$
in the \emph{weak instance}. In the static instance, $v$ may contain marked and
unmarked abstractions, and it need not be closed, as in
Lemma~\ref{lem:progress}. Apart from items~5 and~6 of
Lemma~\ref{lem:guard-patterns} and Lemma~\ref{lem:guard-atoms}, which are
proved in each instance, the proofs use only the following properties, which hold in both
instances:
\begin{description}
	\item[(V1)] if $v\Vdash t$ and $t\leq t'$, then $v\Vdash t'$; in particular,
	      if $v\Vdash t$ then $v\Vdash\dynSup{t}$, since $t\leq\dynSup{t}$
	      (rules \RuleRef{typing:static:sub}{(sub)} and
	      \RuleRef{typing:weak:sub-circ}{(sub$^\circ$)});
	\item[(V2)] if $v\Vdash t_1$ and $v\Vdash t_2$, then $v\Vdash t_1\land t_2$
	      (rules \RuleRef{typing:enriched:and}{($\land$)} and
	      \RuleRef{typing:weak:and-circ}{(and$^\circ$)});
	\item[(V3)] if $v\Vdash t$ for some type $t$, then, for every test type
	      $\tau$, $v\Vdash\tau$ if and only if $v\in\tau$, and $v\Vdash\neg\tau$ if
	      and only if $v\notin\tau$ (Lemmas~\ref{lem:generators}(4)
	      and~\ref{lem:weak-generators}(3));
	\item[(V4)] $v\Vdash\bott$ never holds (Lemmas~\ref{lem:generators}(1,2)
	      and~\ref{lem:weak-generators}(2)).
\end{description}
For a type $t$, $\dynSup{(\neg\dynSup{t})}=\neg\dynSup{t}\leq\neg\dynInf{t}=\dynSup{(\neg t)}$
and $\dynInf{(\neg\dynSup{t})}=\neg\dynSup{t}=\dynInf{(\neg t)}$, so
$\neg\dynSup{t}\leq\neg t$, and also $\neg\dynSup{t}\leq\neg\dynInf{t}$. A
pattern $p$ is used with the substitution $\sigma=v/p$; when $v/p\neq\fail$,
$\sigma$ maps every capture variable of $p$, hence every variable of a guard
attached to $p$, to a value, which is closed if $v$ is. For an environment $\Theta$ whose types are
test types, the type $\acc{p}{\Theta}$ is a test type (by
Figure~\ref{app:def:accepted-types}, it is built from test types by products
and intersections, and from $\topp$ for the variables that $\Theta$ does not
map). In particular, for $x\in\vars{p}$ and a test type $\tau$,
$\acc{p}{x:\tau}$ is the test type of the values that match $p$ and whose
part captured by $x$ belongs to $\tau$.

\begin{lem}[Patterns]\label{lem:guard-patterns}
	Let $p$ be a (linear) pattern.
	\begin{enumerate}
		\item For an environment $\Theta$ of test types and a value $v$,
		      $v\in\acc{p}{\Theta}$ if and only if $v/p=\sigma\neq\fail$ and
		      $\sigma(x)\in\Theta(x)$ for every $x\in\vars{p}\cap\dom{\Theta}$.
		      In particular, if $v\Vdash\acc{p}{\Gamma}$ or
		      $v\Vdash\dynSup{(\acc{p}{\Gamma})}$ for some $\Gamma$, then
		      $v/p\neq\fail$.
		\item If $\Gamma'(x)\leq\Gamma(x)$ for every $x\in\vars{p}$, where an
		      environment maps the variables outside its domain to $\topp$, as in
		      the definition of $\acc{p}{\cdot}$, then
		      $\acc{p}{\Gamma'}\leq\acc{p}{\Gamma}$. Moreover
		      $\dynSup{(\acc{p}{\Gamma})}=\acc{p}{\dynSup{\Gamma}}$ and
		      $\dynInf{(\acc{p}{\Gamma})}=\acc{p}{\dynInf{\Gamma}}$, where
		      $\dynSup{\Gamma}$ and $\dynInf{\Gamma}$ are taken componentwise.
		\item If $t\leq\acc{p}{\Gamma}$, then $t\leq\acc{p}{\TypeEnv{t}{p}}$ and
		      $(\TypeEnv{t}{p})(x)\leq\Gamma(x)$ for every $x\in\vars{p}$.
		\item For $x\in\vars{p}$ and a test type $\tau$, the environment update
		      of Figure~\ref{app:def:environment-updates} satisfies
		      $\Gamma[x\refine\tau]_p(y)\leq\Gamma(y)$ for every $y\in\vars{p}$, and
		      $\acc{p}{\Gamma[x\refine\tau]_p}\simeq\acc{p}{\Gamma}\land\acc{p}{x:\tau}$.
		\item If $v\Vdash\tuple{t_1,\ldotsTwo,t_n}$ (respectively
		      $v\Vdash\OpenTuple{t_1,\ldotsTwo,t_n}$), then
		      $v=\tuple{v_1,\ldotsTwo,v_m}$ with $m=n$ (respectively $m\geq n$) and
		      $v_i\Vdash t_i$ for $i\leq n$.
		\item If $t\leq\acc{p}{\varnothing}$, $v\Vdash t$ and $v/p=\sigma\neq\fail$, then
		      $\sigma(x)\Vdash(\TypeEnv{t}{p})(x)$ for every $x\in\vars{p}$.
	\end{enumerate}
\end{lem}

\begin{Proof}
	Items 1 and 2 are proved by induction on $p$, from the definitions of
	matching (Figure~\ref{fig:pattern-matching-operation}), of membership
	(Figure~\ref{app:inductive-def-test-type}) and of accepted types: an
	intersection pattern is linear, so its two components constrain distinct
	variables; materializations commute with products and intersections. The
	second part of item 1 follows from the first and (V1), (V3), since
	$\dynSup{(\acc{p}{\Gamma})}\leq\acc{p}{\varnothing}$. Item 3 is the
	definition of $\TypeEnv{t}{p}$ (Figure~\ref{app:fig:typing-environments}),
	by induction on $p$: for static types, $t$ is contained in the product of
	its projections, and the projections of $t$ are contained in those of
	$\acc{p}{\Gamma}$; for gradual types, apply this to the two
	materializations, which commute with $\TypeEnv{\cdot}{p}$ and with
	$\acc{p}{\cdot}$. Item 4: let
	$\Gamma_1=\Gamma[x\refine\tau]$; since $p$ is linear,
	$t'=\acc{p}{\Gamma_1}\simeq\acc{p}{\Gamma}\land\acc{p}{x:\tau}$ (on both
	materializations, intersections distribute over products), and
	$\Gamma[x\refine\tau]_p$ is $\TypeEnv{t'}{p}$ on $\vars{p}$; by item 3,
	$\TypeEnv{t'}{p}\leq\Gamma_1\leq\Gamma$ on $\vars{p}$, so
	$\acc{p}{\TypeEnv{t'}{p}}\leq t'$ by item 2, and $t'\leq\acc{p}{\TypeEnv{t'}{p}}$
	by item 3. Item 5: in the static instance, by Lemmas~\ref{lem:canonical-forms}(2);
	in the weak instance, as in the case \RuleRef{reduction:proj}{\textsc{[Proj]}}
	of the proof of Lemma~\ref{lem:subject-reduction}: $v$ has a weak generator
	below the tuple type; its maximal materialization is a product
	$\tuple{\dynSup{g_1},\ldotsTwo,\dynSup{g_m}}$ of non-empty types, so $m$ is as
	stated and $\dynSup{g_i}\leq\dynSup{t_i}$, hence $v_i\Vdash t_i$. Item 6, by
	induction on $p$. If $p=x$, then $\sigma(x)=v$ and $(\TypeEnv{t}{p})(x)=t$.
	If $p$ is a constant, it has no variable. If $p=p_1\mathrel{\&}p_2$, then
	$\sigma=\sigma_1\cup\sigma_2$ with $\sigma_i=v/p_i$, and
	$\TypeEnv{t}{p}$ coincides with $\TypeEnv{t}{p_i}$ on $\vars{p_i}$. If
	$p=\tuple{p_1,\ldotsTwo,p_n}$, then $v=\tuple{v_1,\ldotsTwo,v_n}$, and it
	suffices to show $v_i\Vdash\pi_i(t)$, since
	$(\TypeEnv{t}{p})(x)=(\TypeEnv{\pi_i(t)}{p_i})(x)$ for $x\in\vars{p_i}$, and
	$\pi_i(t)\leq\acc{p_i}{\varnothing}$. Since $t\leq\tuple{\topp,\ldotsTwo,\topp}$,
	$v$ has a generator (a weak generator in the weak
	instance) $g$ below $t$, with $g=\tuple{g_1,\ldotsTwo,g_n}$ and $g_i$ a
	generator (a weak generator) of $v_i$, except in the weak instance when $v$
	is a tuple constant $\tuple{c_1,\ldotsTwo,c_n}$: then
	$g=\tuple{c_1,\ldotsTwo,c_n}\land\dyn$, and
	$\dynSup{g}=\tuple{\dynSup{g_1},\ldotsTwo,\dynSup{g_n}}$ for the weak
	generators $g_i=c_i\land\dyn$ of the components. The $g_i$ have
	non-empty maximal materializations, and, in the static instance, non-empty
	minimal materializations, where the $v_i$ have types
	(Lemma~\ref{lem:generators}(1)). The projection of a product of non-empty types is
	its component, and projections are monotone, so $g_i\leq\pi_i(t)$ on the
	maximal materializations and, in the static instance, on the minimal ones.
	Hence $v_i\Vdash\pi_i(t)$.
\end{Proof}

\noindent
The typing premises of the guard analysis concern guard atoms, whose
evaluation is described by the following lemma.

\begin{lem}[Atoms]\label{lem:guard-atoms}
	Let $\Gamma\vds a:s$ be derived without $\omega$-marked rules, and let
	$\sigma$ map the free variables of the atom $a$ to values such that
	$\sigma(y)\Vdash\Gamma(y)$ (respectively $\sigma(y)\Vdash\dynSup{\Gamma(y)}$)
	for every $y\in\fv{a}$. Then $a\sigma$ reduces, without failure, to a value
	$w$ such that $w\Vdash s$ (respectively $w\Vdash\dynSup{s}$).
\end{lem}

\begin{Proof}
	The second statement follows from the first applied to $\dynSup{\Gamma}$,
	since $\dynSup{\Gamma}\vds a:\dynSup{s}$, by induction on the derivation:
	the rules that type atoms, \RuleRef{typing:static:cst}{(cst)},
	\RuleRef{typing:static:var}{(var)}, \RuleRef{typing:static:tuple}{(tuple)},
	\RuleRef{typing:static:proj}{(proj)}, \RuleRef{typing:static:size}{(size)}
	and \RuleRef{typing:static:sub}{(sub)}, still apply when every type is
	replaced with its maximal materialization (the index type of
	\RuleRef{typing:static:proj}{(proj)} and the types $\tupleTop$ and $\intTop$
	are static). The first statement is proved by induction on the derivation.
	Rule \RuleRef{typing:static:sub}{(sub)} is handled by (V1). For
	\RuleRef{typing:static:cst}{(cst)} and \RuleRef{typing:static:var}{(var)},
	$a\sigma$ is a value of the deduced type. For
	\RuleRef{typing:static:tuple}{(tuple)}, the components reduce to values of
	their types, and $\tuple{w_1,\ldotsTwo,w_n}\Vdash\tuple{s_1,\ldotsTwo,s_n}$
	by rule \RuleRef{typing:static:tuple}{(tuple)} or
	\RuleRef{typing:weak:tuple-circ}{(tuple$^\circ$)}. For \RuleRef{typing:static:proj}{(proj)}, the index reduces
	to a value $w'\Vdash\bigvee_{i\in K}i$, which is an integer $j\in K$ by (V3),
	and the tuple to $w\Vdash\OpenTuple{t_0,\ldotsTwo,t_n}$ with
	$K\subseteq[0,n]$; by Lemma~\ref{lem:guard-patterns}(5),
	$w=\tuple{w_0,\ldotsTwo,w_m}$ with $m\geq n$ and $w_j\Vdash t_j$, so
	$\proj{j}{w}$ reduces to $w_j$, and
	$w_j\Vdash\bigvee_{i\in K}t_i$. For \RuleRef{typing:static:size}{(size)},
	the argument reduces to a value $w\Vdash\tupleTop$, which is a tuple by
	(V3), so $\GuardSize{w}$ reduces to an integer $n$, and $n\Vdash\intTop$
	(rule \RuleRef{typing:static:cst}{(cst)}, or
	\RuleRef{typing:weak:cst-circ}{(cst$^\circ$)} and
	\RuleRef{typing:weak:sub-circ}{(sub$^\circ$)}, since $n\leq\intTop$).
\end{Proof}

\noindent
The environments that the analysis produces are refinements of the environment
of the analysis by test types, at least on their maximal materializations.

\begin{lem}[Hulls]\label{lem:guard-hulls}
	Let $p$ be a pattern.
	\begin{enumerate}
		\item Let $\Gamma$ be an environment of static types for the variables
		      of $p$, $\tau$ a test type, and
		      $t=\acc{p}{\Gamma}\land\tau$. Then
		      $\acc{p}{\TypeEnv{t}{p}}\simeq\acc{p}{\Gamma}\land\tau'$ for some test
		      type $\tau'$.
		\item Let $\JudgmentGuard{\Gamma}{g}{\GuardResult}$. For every environment
		      $E$ of $\GuardResult$, safe or success, $E(y)\leq\Gamma(y)$ for every
		      $y\in\vars{p}$, and
		      $\dynSup{(\acc{p}{E})}\simeq\dynSup{(\acc{p}{\Gamma})}\land\tau_E$
		      for some test type $\tau_E$. Every success environment $\Delta$ of
		      $\GuardResult$ satisfies $\Delta(y)\leq\Phi(y)$ for $y\in\vars{p}$,
		      where $\Phi$ is the safe environment of $\GuardResult$. We say that the
		      environments of $\GuardResult$ have \emph{hulls} relative to $\Gamma$.
	\end{enumerate}
\end{lem}

\begin{Proof}
	\begin{enumerate}
		\item By induction on the size of $p$. If $p$ is a variable $x$, then
		      $\acc{p}{\TypeEnv{t}{p}}=t$. If $p$ is a constant $c$, then
		      $\acc{p}{\cdot}=c$, and $\tau'=\topp$. If
		      $p=\tuple{p_1,\ldotsTwo,p_n}$, let $A_i=\acc{p_i}{\Gamma}$. The
		      intersection of a test type with $\tuple{\topp,\ldotsTwo,\topp}$ is a
		      finite union $\bigvee_{k\in K}\tuple{\tau_{k,1},\ldotsTwo,\tau_{k,n}}$
		      of products of test types, so
		      $t=\bigvee_{k\in K}\tuple{A_1\land\tau_{k,1},\ldotsTwo,A_n\land\tau_{k,n}}$
		      and, by definition of $\pi_i$, $\pi_i(t)=A_i\land\tau_i''$, where
		      $\tau_i''=\bigvee\{\tau_{k,i}\mid k\in K,\ A_j\land\tau_{k,j}\not\simeq\bott\text{ for every }j\}$.
		      Since $\TypeEnv{t}{p}$ is $\TypeEnv{\pi_i(t)}{p_i}$ on $\vars{p_i}$, the
		      induction hypothesis on $p_i$ gives
		      $\acc{p_i}{\TypeEnv{t}{p}}\simeq A_i\land\tau'_i$, so
		      $\acc{p}{\TypeEnv{t}{p}}\simeq\acc{p}{\Gamma}\land\tuple{\tau'_1,\ldotsTwo,\tau'_n}$.
		      Otherwise $p$ is an intersection of $k\geq2$ patterns
		      $q_1,\ldotsTwo,q_k$ that are not intersections, with pairwise
		      disjoint variables, and
		      $\acc{p}{\Theta}=\bigwedge_{h}\acc{q_h}{\Theta}$ for every $\Theta$.
		      If some $q_h$ is a variable, then $\acc{q_h}{\TypeEnv{t}{p}}=t$ and,
		      by Lemma~\ref{lem:guard-patterns}(3), the other components are
		      supertypes of $t$, so $\acc{p}{\TypeEnv{t}{p}}=t$. If some $q_h$ is a
		      constant $c$, then $t=\acc{q}{\Gamma}\land(c\land\tau)$, where $q$ is
		      the intersection of the other components, and the induction
		      hypothesis on $q$, with the test type $c\land\tau$, concludes. If all
		      are tuples of the same length $n$, the matching, the accepted types
		      and the environments $\TypeEnv{\cdot}{p}$ of $p$ are those of the
		      tuple pattern whose $i$-th component is the intersection of the
		      $i$-th components of the $q_h$, and the case of tuples applies. If
		      their lengths differ, then $\acc{p}{\Gamma}\simeq\bott$, so
		      $t\simeq\bott$, every projection of $t$ is empty, and
		      $\acc{p}{\TypeEnv{t}{p}}\simeq\bott$.
		\item By induction on the derivation. If two environments have hulls,
		      $\dynSup{(\acc{p}{E'})}\simeq\dynSup{(\acc{p}{\Gamma})}\land\tau_1$ and
		      $\dynSup{(\acc{p}{E})}\simeq\dynSup{(\acc{p}{E'})}\land\tau_2$, then
		      $\dynSup{(\acc{p}{E})}\simeq\dynSup{(\acc{p}{\Gamma})}\land\tau_1\land\tau_2$.
		      So the rules that analyze a premise under an environment of the
		      result of another premise ((size), (proj), the approximation rules,
		      (and), (orF)) preserve hulls, and so do rules (eq$_1$), (eq$_2$), (neq$_1$), (neq$_2$),
		      by the induction hypothesis. Rule (var) refines $\Gamma$ by
		      $\Gamma[x\refine\tau]_p$, whose accepted type is
		      $\acc{p}{\Gamma}\land\acc{p}{x:\tau}$
		      (Lemma~\ref{lem:guard-patterns}(4)); the other atomic rules and
		      rule (approx) return $\Gamma$. In rule (or), the type
		      $t=\acc{p}{\Phi}\setminus\bigvee_{\{i\mid\BoolC_i\}}\dynSup{(\acc{p}{\Delta_i})}$
		      subtracts static types, so, by the induction hypothesis on the first
		      premise,
		      $\dynSup{t}=\dynSup{(\acc{p}{\Phi})}\land\neg\bigvee_i\dynSup{(\acc{p}{\Delta_i})}\simeq\dynSup{(\acc{p}{\Gamma})}\land\tau_\Phi\land\neg\bigvee_i\tau_{\Delta_i}$.
		      By item~1 applied to $\dynSup{\Gamma}$ (Lemma~\ref{lem:guard-patterns}(2)),
		      and since $\TypeEnv{t}{p}$ commutes with $\dynSup{(\cdot)}$, the
		      environment $(\Gamma,\TypeEnv{t}{p})$ of the second premise has a hull
		      relative to $\Gamma$, and so do the success environments of the
		      second premise. The inequalities between environments follow from
		      Lemma~\ref{lem:guard-patterns}(3,4) and the induction hypothesis; in
		      rule (or), $(\TypeEnv{t}{p})(y)\leq\Phi(y)$ since
		      $t\leq\acc{p}{\Phi}$.\qedhere
	\end{enumerate}
\end{Proof}

\begin{cor}[Relative dichotomy]\label{cor:guard-dichotomy}
	Let the environments of $\GuardResult$ have hulls relative to $\Gamma$, let
	$E$ be one of them, with
	$\dynSup{(\acc{p}{E})}\simeq\dynSup{(\acc{p}{\Gamma})}\land\tau_E$, and let
	$v\Vdash\dynSup{(\acc{p}{\Gamma})}$. Then $v\Vdash\dynSup{(\acc{p}{E})}$ if
	and only if $v\in\tau_E$; otherwise $v\Vdash\neg\dynSup{(\acc{p}{E})}$, and
	hence $v\Vdash\neg\acc{p}{E}$ and $v\Vdash\neg\dynInf{(\acc{p}{E})}$.
\end{cor}

\begin{Proof}
	If $v\in\tau_E$, then $v\Vdash\tau_E$ by (V3), and (V2) gives
	$v\Vdash\dynSup{(\acc{p}{E})}$. Conversely, if
	$v\Vdash\dynSup{(\acc{p}{E})}$, then $v\Vdash\tau_E$ by (V1), and
	$v\in\tau_E$ by (V3). If $v\notin\tau_E$, then $v\Vdash\neg\tau_E$ by (V3),
	and $\neg\tau_E\leq\neg\dynSup{(\acc{p}{E})}$, which is a subtype of
	$\neg\acc{p}{E}$ and of $\neg\dynInf{(\acc{p}{E})}$.
\end{Proof}

\noindent
We can now state the correctness of the analysis. Its statements on the safe
and the success environments are the meaning given to them in
Section~\ref{sub:guard-analysis-system}; for exact environments, the
hypothesis is on maximal materializations, which is what the exclusion of
surely accepted values needs (in rule \RuleRef{guard:analysis:or}{(or)} and in
rule \RuleRef{guard:accepted:seq}{[seq]}).

\begin{lem}[Safe/Success environments]\label{lem:safe-success-environment}
	Let $\JudgmentGuard{\Gamma}{g}{\GuardResult}$ be derivable. Its environments
	have hulls relative to $\Gamma$ (Lemma~\ref{lem:guard-hulls}). Let $v$ be a
	value such that $v\Vdash\dynSup{(\acc{p}{\Gamma})}$, and
	$\sigma=v/p$ (Lemma~\ref{lem:guard-patterns}(1)). Write $\Phi$ for the safe
	environment of $\GuardResult$ when $\GuardResult\neq\omega$. Let $\diamond$
	stand either for nothing or for $\dynSup{}$, that is, let $t^\diamond$ be
	either $t$ or $\dynSup{t}$, the same in hypothesis and conclusion.
	\begin{enumerate}
		\item[(G1)] If $v\Vdash(\acc{p}{\Gamma})^\diamond$ and $g\sigma$ does not
		      reduce to $\fail$, then $\GuardResult\neq\omega$ and
		      $v\Vdash(\acc{p}{\Phi})^\diamond$.
		\item[(G2)] If the safe environment of $\GuardResult$ is
		      $(\Phi,\smalltrue)$ and $v\Vdash\dynSup{(\acc{p}{\Phi})}$, then
		      $g\sigma$ does not reduce to $\fail$.
		\item[(G3)] If $v\Vdash(\acc{p}{\Gamma})^\diamond$ and
		      $g\sigma\seqreducesG\smalltrue$, then
		      $v\Vdash(\acc{p}{\Delta})^\diamond$ for some success environment
		      $(\Delta,\BoolC)$ of $\GuardResult$.
		\item[(G4)] If $(\Delta,\smalltrue)$ is a success environment of
		      $\GuardResult$ and $v\Vdash\dynSup{(\acc{p}{\Delta})}$, then
		      $g\sigma\seqreducesG\smalltrue$.
	\end{enumerate}
	In particular, by (G1), if $\GuardResult=\omega$, then $g\sigma$ reduces to
	$\fail$, and by (G3), if $\GuardResult$ has no success environment, then
	$g\sigma$ does not reduce to $\smalltrue$.
\end{lem}

\begin{Proof}
	By induction on the derivation, with case analysis on its last rule. Every
	hypothesis of the lemma implies $v\Vdash\dynSup{(\acc{p}{\Gamma})}$, by (V1),
	since the environments of $\GuardResult$ are below $\Gamma$; so $\sigma$ is
	defined, and, by Lemma~\ref{lem:guard-patterns}(6) and~(3),
	$\sigma(y)\Vdash\dynSup{\Gamma(y)}$ for every $y\in\vars{p}$, and
	$\sigma(y)\Vdash\Gamma(y)$ if $v\Vdash\acc{p}{\Gamma}$. By
	Section~\ref{sec:guard-analysis}, the free variables of $g$ are capture
	variables of $p$. We use Lemma~\ref{lem:guard-atoms} for the typing premises
	$\Gamma\vds a:s$ of the rules, which are derived without $\omega$-marked
	rules: under $v\Vdash\dynSup{(\acc{p}{\Gamma})}$, $a\sigma$ reduces without
	failure to a value $w\Vdash\dynSup{s}$, so $w\in s$ when $s$ is a test type
	(V3). We write ``by IH(G$k$)'' for the induction hypothesis on a premise.
	\begin{description}
		\item[\RuleRef{guard:analysis:true}{(true)}] $g=a\isof\tau$ with
		      $\Gamma\vds a:\tau$, and
		      $\GuardResult=\Result{(\Gamma,\smalltrue)}{(\Gamma,\smalltrue)}$.
		      (G1) and (G3) hold since $\Phi=\Delta=\Gamma$. For (G2) and (G4),
		      $a\sigma$ reduces to $w\in\tau$, so $g\sigma\seqreducesG\smalltrue$.
		\item[\RuleRef{guard:analysis:false}{(false)}] $g=a\isof\tau$ with
		      $\Gamma\vds a:s$ and $s\land\tau\simeq\bott$, and
		      $\GuardResult=\Result{(\Gamma,\smalltrue)}{\GuardFalse}$. (G1) holds
		      since $\Phi=\Gamma$, and (G4) is vacuous. $a\sigma$ reduces to
		      $w\Vdash\dynSup{s}$, so $g\sigma$ does not fail (G2), and
		      $w\notin\tau$: otherwise
		      $w\Vdash\dynSup{s}\land\tau\simeq\dynSup{(s\land\tau)}\simeq\bott$, by
		      (V2) and (V3), which contradicts (V4); so $g\sigma$ is
		      $\smallfalse$ (G3).
		\item[\RuleRef{guard:analysis:var}{(var)}] $g=x\isof\tau$, and
		      $\GuardResult=\Result{(\Gamma,\smalltrue)}{(\Delta,\smalltrue)}$ with
		      $\Delta=\Gamma[x\refine\tau]_p$, so that
		      $\acc{p}{\Delta}\simeq\acc{p}{\Gamma}\land\acc{p}{x:\tau}$
		      (Lemma~\ref{lem:guard-patterns}(4)). (G1) holds since $\Phi=\Gamma$,
		      and $g\sigma$ never fails (G2). If $g\sigma\seqreducesG\smalltrue$,
		      then $\sigma(x)\in\tau$, so $v\in\acc{p}{x:\tau}$
		      (Lemma~\ref{lem:guard-patterns}(1)), $v\Vdash\acc{p}{x:\tau}$ (V3), and
		      (V2) gives $v\Vdash(\acc{p}{\Delta})^\diamond$, since
		      $(\acc{p}{\Delta})^\diamond\simeq(\acc{p}{\Gamma})^\diamond\land\acc{p}{x:\tau}$
		      (G3). If $v\Vdash\dynSup{(\acc{p}{\Delta})}$, then
		      $v\Vdash\acc{p}{x:\tau}$ by (V1), so $v\in\acc{p}{x:\tau}$ and
		      $\sigma(x)\in\tau$ (G4).
		\item[\RuleRef{guard:analysis:size}{(size)}] $g=\sizeTup{a}\isof i$; the
		      premises are
		      $\JudgmentGuard{\Gamma}{a\isof\tupleTop}{\GuardResult_1}$, with the
		      single success environment $(\Phi,\BoolB)$, and
		      $\JudgmentGuard{\Phi}{a\isof\tupleTop^i}{\GuardResult_2}$, whose success
		      environment, if any, is the one of $\GuardResult$. (G1): if $g\sigma$
		      does not fail, then $a\sigma$ reduces to a tuple, so
		      $(a\isof\tupleTop)\sigma\seqreducesG\smalltrue$, and IH(G3) gives
		      $v\Vdash(\acc{p}{\Phi})^\diamond$. (G2): if $\BoolB=\smalltrue$ and
		      $v\Vdash\dynSup{(\acc{p}{\Phi})}$, then IH(G4) gives
		      $(a\isof\tupleTop)\sigma\seqreducesG\smalltrue$, so $a\sigma$ reduces to
		      a tuple and $g\sigma$ does not fail. (G3): if
		      $g\sigma\seqreducesG\smalltrue$, then $a\sigma$ reduces to a tuple of
		      size $i$; as for (G1), $v\Vdash(\acc{p}{\Phi})^\diamond$, and IH(G3) on
		      the second premise concludes. (G4): IH(G4) on the second premise gives
		      that $a\sigma$ reduces to a tuple of size $i$, so
		      $g\sigma\seqreducesG\smalltrue$.
		\item[\RuleRef{guard:analysis:proj}{(proj)}] $g=\proj{a'}{a}\isof\tau$;
		      the premises are $\Gamma\vds a':i$,
		      $\JudgmentGuard{\Gamma}{a\isof\tupleTop^{>i}}{\GuardResult_1}$ with the
		      single success environment $(\Phi,\BoolB)$, and
		      $\JudgmentGuard{\Phi}{a\isof\OpenTuple{\topp,\ldotsTwo,\topp,\tau}}{\GuardResult_2}$
		      ($i$ components $\topp$). The index $a'\sigma$ reduces to $i$. As for
		      (size): $g\sigma$ does not fail if and only if $a\sigma$ reduces to a
		      tuple with more than $i$ components, that is, if and only if
		      $(a\isof\tupleTop^{>i})\sigma\seqreducesG\smalltrue$; and
		      $g\sigma\seqreducesG\smalltrue$ if and only if
		      $(a\isof\OpenTuple{\topp,\ldotsTwo,\topp,\tau})\sigma\seqreducesG\smalltrue$.
		      IH(G3), IH(G4) on the two premises conclude as for (size).
		\item[\RuleRef{guard:analysis:eq1}{(eq$_1$)},
		\RuleRef{guard:analysis:eq2}{(eq$_2$)}] $g=a_1=a_2$, with
		      $\Gamma\vds a_1:c$ and the premise
		      $\JudgmentGuard{\Gamma}{a_2\isof c}{\GuardResult}$ (for (eq$_2$),
		      exchange the roles). $a_1\sigma$ reduces without failure to $c$, so
		      $g\sigma$ fails if and only if $a_2\sigma$ fails, and it is
		      $\smalltrue$ if and only if $a_2\sigma$ reduces to $c$, that is, if and
		      only if $(a_2\isof c)\sigma\seqreducesG\smalltrue$. The induction
		      hypothesis concludes.
		\item[\RuleRef{guard:analysis:neq1}{(neq$_1$)},
		\RuleRef{guard:analysis:neq2}{(neq$_2$)}] $g=\GuardNeq{a_1}{a_2}$, with
		      $\Gamma\vds a_1:c$ and the premise
		      $\JudgmentGuard{\Gamma}{a_2\isof\neg c}{\GuardResult}$ (for (neq$_2$),
		      exchange the roles). As for (eq$_1$), $g\sigma$ fails if and only if
		      $a_2\sigma$ fails, and it is $\smalltrue$ if and only if $a_2\sigma$
		      reduces to a value other than $c$, that is, if and only if
		      $(a_2\isof\neg c)\sigma\seqreducesG\smalltrue$. The induction
		      hypothesis concludes.
		\item[\RuleRef{guard:analysis:lt}{(lt)},
		\RuleRef{guard:analysis:lt-maybe}{(ltMaybe)},
		\RuleRef{guard:analysis:lt-false}{(ltFalse)}] $g=a_0<a_1$, with
		      $\Gamma\vds a_0:t_0$ and $\Gamma\vds a_1:t_1$, $t_0$ and $t_1$
		      static. Both atoms reduce without failure to values $w_0\Vdash t_0$
		      and $w_1\Vdash t_1$, so $g\sigma$ does not fail; the safe environment is
		      $(\Gamma,\smalltrue)$ and the success environment, if any, is
		      $\Gamma$, which gives (G1)--(G3). The bounds of
		      Section~\ref{sec:strict-ordering-of-types} depend only on the test
		      types $\intTop$, $\atomTop$, the constants, $\function$, $\tupleTop$ and
		      $\tupleTop^n$ that meet a type, and a value $w\Vdash t_0$ belongs only
		      to test types that meet $t_0$ (by (V2)--(V4)); so, if
		      $\mathsf{alwaysLess}(t_0,t_1)$, then $w_0<w_1$ (G4), and if
		      $\mathsf{neverLess}(t_0,t_1)$, then $w_0\not<w_1$ and $g\sigma$ is
		      $\smallfalse$ (G3).
		\item[Approximation rules] For
		      \RuleRef{guard:analysis:eq-a}{(eq$_a$)}, for both operators, and
		      \RuleRef{guard:analysis:lt-a}{(lt$_a$)}, $g\sigma$ does not fail if
		      and only if both atoms reduce to values, that is, if and only if the
		      tests $a_0\isof\topp$ and $a_1\isof\topp$ of the premises are
		      $\smalltrue$. For \RuleRef{guard:analysis:proj-a}{(proj$_a$)}, if
		      $g\sigma$ does not fail, then $a'\sigma$ reduces to an integer and
		      $a\sigma$ to a tuple, so the tests $a'\isof\intTop$ and
		      $a\isof\tupleTop$ of the premises are $\smalltrue$. By IH(G3) on the two premises,
		      the second analyzed under the success environment of the first, the
		      result $\Delta$ satisfies (G1) and (G3), since $g\sigma$ is
		      $\smalltrue$ only if it does not fail. For (G2), in
		      (eq$_a$) and (lt$_a$) with $\BoolB\etet\BoolC=\smalltrue$: IH(G4) on
		      the second premise, and then on the first, whose success environment
		      $\Phi$ is above $\Delta$, shows that both atoms reduce to values. Rule
		      \RuleRef{guard:analysis:size-a}{(size$_a$)} is similar: $g$ does not
		      fail if and only if $(a\isof\tupleTop)\sigma\seqreducesG\smalltrue$.
		      Their success environments are not exact (G4).
		\item[\RuleRef{guard:analysis:approx}{(approx)}]
		      $\GuardResult=\Result{(\Gamma,\smallfalse)}{(\Gamma,\smallfalse)}$:
		      (G1) and (G3) hold since the environments are $\Gamma$, and (G2) and
		      (G4) are vacuous.
		\item[$\omega$-rules] \RuleRef{guard:analysis:size-omega}{(size$_\omega$)},
		      \RuleRef{guard:analysis:eq-omega}{(eq$_\omega$)} (both operators),
		      \RuleRef{guard:analysis:proj-omega-t}{(proj$^t_\omega$)},
		      \RuleRef{guard:analysis:proj-omega-i}{(proj$^i_\omega$)},
		      \RuleRef{guard:analysis:bound-omega}{(bound$_\omega$)},
		      \RuleRef{guard:analysis:lt-omega}{(lt$_\omega$)} conclude $\omega$
		      from a failure result $\GuardFail$ for a test that $g\sigma$ makes
		      $\smalltrue$ whenever $g\sigma$ does not fail ($a\isof\tupleTop$,
		      $a_i\isof\topp$, $a'\isof\intTop$, $a\isof\tupleTop^{>i}$, the latter
		      when the index reduces to $i$ by $\Gamma\vds a':i$). By IH(G1) and
		      IH(G3), that test does not reduce to $\smalltrue$ under
		      $v\Vdash\dynSup{(\acc{p}{\Gamma})}$, so $g\sigma$ fails (G1).
		\item[\RuleRef{guard:analysis:or-omega}{(or$_\omega$)}] $g_1$ is analyzed
		      with result $\omega$, so $g_1\sigma$ fails (IH(G1)), and so does
		      $g\sigma$.
		\item[\RuleRef{guard:analysis:and-fail-l}{(and$_\GuardFail$-L)}] $g_1$
		      and $g$ have the same failure result $\GuardFail$. If
		      $\GuardFail=\omega$, then $g_1\sigma$ fails (IH(G1)), and so does
		      $g\sigma$. Otherwise $\GuardFail=\Result{(\Phi,\BoolB)}{\GuardFalse}$,
		      and $g_1\sigma$ does not reduce to $\smalltrue$ (IH(G3)), so $g\sigma$
		      fails if $g_1\sigma$ fails, and is $\smallfalse$ otherwise: (G1) and
		      (G2) follow from IH(G1) and IH(G2), and (G3) and (G4) are vacuous.
		\item[\RuleRef{guard:analysis:and}{(and)}] $g=\GuardAnd{g_1}{g_2}$, with
		      premises
		      $\JudgmentGuard{\Gamma}{g_1}{\Result{(\Phi,\BoolB)}{(\Delta_i,\BoolC_i)_{i\in I}}}$
		      and $\JudgmentGuard{\Delta_i}{g_2}{\GuardResult_i}$ for $i\in I$.
		      (G1): if $g\sigma$ does not fail, neither does $g_1\sigma$, and IH(G1)
		      concludes. (G2): suppose $\BoolB=\smalltrue$, every $\GuardResult_i$
		      has the safe environment $(\Phi_i,\smalltrue)$ with
		      $\Phi_i\simeq\Delta_i$, and $v\Vdash\dynSup{(\acc{p}{\Phi})}$. By
		      IH(G2), $g_1\sigma$ does not fail. If it is $\smallfalse$, so is
		      $g\sigma$. If it is $\smalltrue$, IH(G3) with $\diamond=\dynSup{}$
		      gives $i$ such that $v\Vdash\dynSup{(\acc{p}{\Delta_i})}=\dynSup{(\acc{p}{\Phi_i})}$,
		      and IH(G2) on the premise for $i$ shows that $g_2\sigma$ does not fail.
		      (G3): if $g\sigma\seqreducesG\smalltrue$, then $g_1\sigma$ and
		      $g_2\sigma$ reduce to $\smalltrue$; IH(G3) gives $i$ such that
		      $v\Vdash(\acc{p}{\Delta_i})^\diamond$; then $\GuardResult_i\neq\omega$,
		      by IH(G1) on the premise for $i$, and IH(G3) on it gives a success
		      environment $(\Delta,\BoolC)$ of $\GuardResult_i$ with
		      $v\Vdash(\acc{p}{\Delta})^\diamond$, and $(\Delta,\BoolC_i\etet\BoolC)$
		      is a success environment of $\GuardResult$. (G4): let
		      $\BoolC_i=\BoolC=\smalltrue$ and $v\Vdash\dynSup{(\acc{p}{\Delta})}$; by
		      Lemma~\ref{lem:guard-hulls}, $\Delta\leq\Delta_i$ on $\vars{p}$, so
		      $v\Vdash\dynSup{(\acc{p}{\Delta_i})}$ (Lemma~\ref{lem:guard-patterns}(2),
		      (V1)), and IH(G4) on both premises gives
		      $g_1\sigma\seqreducesG\smalltrue$ and $g_2\sigma\seqreducesG\smalltrue$.
		\item[\RuleRef{guard:analysis:or}{(or)}] $g=\GuardOr{g_1}{g_2}$, with
		      premises
		      $\JudgmentGuard{\Gamma}{g_1}{\Result{(\Phi,\BoolB)}{(\Delta_i,\BoolC_i)_{i\in I}}}$
		      and $\JudgmentGuard{\Gamma'}{g_2}{\GuardResult'}$, where
		      $\Gamma'=\Gamma,\TypeEnv{t}{p}$ and
		      $t=\acc{p}{\Phi}\setminus\bigvee_{\{i\mid\BoolC_i\}}\dynSup{(\acc{p}{\Delta_i})}$.
		      We first show: \emph{if $g_1\sigma\seqreducesG\smallfalse$ and
		      $v\Vdash(\acc{p}{\Phi})^\diamond$, then
		      $v\Vdash(\acc{p}{\Gamma'})^\diamond$.} For every $i$ with
		      $\BoolC_i=\smalltrue$, $v\not\Vdash\dynSup{(\acc{p}{\Delta_i})}$, by
		      IH(G4), so $v\Vdash\neg\dynSup{(\acc{p}{\Delta_i})}$ by
		      Corollary~\ref{cor:guard-dichotomy}. By (V2),
		      $v\Vdash(\acc{p}{\Phi})^\diamond\land\bigwedge_i\neg\dynSup{(\acc{p}{\Delta_i})}$,
		      which is $t^\diamond$ (the subtracted types are static), and
		      $t\leq\acc{p}{\TypeEnv{t}{p}}=\acc{p}{\Gamma'}$
		      (Lemma~\ref{lem:guard-patterns}(3)), so (V1) concludes.
		      (G1): if $g\sigma$ does not fail, neither does $g_1\sigma$, and IH(G1)
		      concludes. (G2): suppose $\BoolB=\smalltrue$, the safe environment of
		      $\GuardResult'$ is $(\Phi',\smalltrue)$ with $\Phi'\simeq\Gamma'$, and
		      $v\Vdash\dynSup{(\acc{p}{\Phi})}$. By IH(G2), $g_1\sigma$ does not
		      fail; if it is $\smalltrue$, so is $g\sigma$; if it is $\smallfalse$,
		      then $v\Vdash\dynSup{(\acc{p}{\Gamma'})}=\dynSup{(\acc{p}{\Phi'})}$ by
		      the claim, and $g_2\sigma$ does not fail, by IH(G2).
		      (G3): if $g\sigma\seqreducesG\smalltrue$, either
		      $g_1\sigma\seqreducesG\smalltrue$, and IH(G3) gives some $\Delta_i$;
		      or $g_1\sigma\seqreducesG\smallfalse$ and
		      $g_2\sigma\seqreducesG\smalltrue$; then
		      $v\Vdash(\acc{p}{\Phi})^\diamond$ by IH(G1), so
		      $v\Vdash(\acc{p}{\Gamma'})^\diamond$ by the claim,
		      $\GuardResult'\neq\omega$ by IH(G1), and IH(G3) gives a success
		      environment $(\Delta,\BoolC)$ of $\GuardResult'$, which is in
		      $\GuardResult$ as $(\Delta,\BoolC\etet\BoolB)$.
		      (G4): for $(\Delta_i,\smalltrue)$, IH(G4) gives
		      $g_1\sigma\seqreducesG\smalltrue$. For $(\Delta,\BoolC\etet\BoolB)$
		      with $\BoolC=\BoolB=\smalltrue$ and
		      $v\Vdash\dynSup{(\acc{p}{\Delta})}$: by Lemma~\ref{lem:guard-hulls},
		      $\Delta\leq\Gamma'\leq\Phi$ on $\vars{p}$, so
		      $v\Vdash\dynSup{(\acc{p}{\Phi})}$, and $g_1\sigma$ does not fail, by
		      IH(G2); if it is $\smalltrue$, so is $g\sigma$; otherwise IH(G4) on the
		      second premise gives $g_2\sigma\seqreducesG\smalltrue$.
		\item[\RuleRef{guard:analysis:or-f}{(orF)}] $g=\GuardOr{g_1}{g_2}$, with
		      premises
		      $\JudgmentGuard{\Gamma}{g_1}{\Result{(\Phi,\BoolB)}{\GuardFalse}}$ and
		      $\JudgmentGuard{\Phi}{g_2}{\GuardResult'}$, and
		      $\GuardResult=\FlagR{\BoolB}{\GuardResult'}$. Since $g_1$ has no
		      success environment, $g_1\sigma$ does not reduce to $\smalltrue$, by
		      IH(G3) with $\diamond=\dynSup{}$; so $g\sigma$ is $g_2\sigma$ when
		      $g_1\sigma$ does not fail. (G1): if $g\sigma$ does not fail, then
		      $g_1\sigma$ is $\smallfalse$, $v\Vdash(\acc{p}{\Phi})^\diamond$ by
		      IH(G1), and IH(G1) on the second premise gives
		      $\GuardResult'\neq\omega$ and the conclusion for its safe environment,
		      which is that of $\GuardResult$. (G2) and (G4): the hypothesis implies
		      $v\Vdash\dynSup{(\acc{p}{\Phi})}$, since the environments of
		      $\GuardResult'$ are below $\Phi$, and $\BoolB=\smalltrue$; by IH(G2),
		      $g_1\sigma$ is $\smallfalse$, and IH(G2), respectively IH(G4), on the
		      second premise concludes. (G3): $g_1\sigma$ is $\smallfalse$ and
		      $g_2\sigma\seqreducesG\smalltrue$; by IH(G1),
		      $v\Vdash(\acc{p}{\Phi})^\diamond$, and IH(G3) on the second premise
		      concludes.\qedhere
	\end{description}
\end{Proof}

\noindent
The guarded pattern sequence analysis
$\Gamma\,;t\vdash(p_ig_i)_{i\leq n}\leadsto(t_{ij},\BoolB_{ij})_{i\leq n,j\leq m_i}$
is derived by rule \RuleRef{guard:accepted:seq}{[seq]}: the pair $p_ig_i$ is
analyzed under $\Gamma;t_i$, where $t_1=t$ and
$t_{i+1}=t_i\setminus\bigvee_{\{j\mid\BoolB_{ij}\}}t_{ij}$, by rule
\RuleRef{guard:accepted:accept}{[accept]}, from
$\Gamma,\TypeEnv{t_i}{p_i}\,;p_i\vdash g_i\mapsto\Result{\_}{(\Delta_{ij},\BoolB_{ij})_{j\leq m_i}}$
with $t_{ij}=t_i^{\Uparrow}\land\acc{p_i}{\Delta_{ij}}$, or by rule
\RuleRef{guard:accepted:fail}{[fail]}, which gives $m_i=1$ and
$(t_{i1},\BoolB_{i1})=(\bott,\smalltrue)$. Write $\Gamma_i$ for
$\Gamma,\TypeEnv{t_i}{p_i}$.

\begin{lem}[Clauses]\label{lem:guard-clauses}
	Let $v\Vdash t_i$. If $v/p_i\neq\fail$, then $v\Vdash\acc{p_i}{\Gamma_i}$. If
	$v/p_i=\fail$, then $v\Vdash\neg\dynSup{t_{ij}}$ for every $j\leq m_i$.
\end{lem}

\begin{Proof}
	By the definition of $\TypeEnv{t_i}{p_i}$
	(Figure~\ref{app:fig:typing-environments}) and
	Lemma~\ref{lem:guard-patterns}(3), applied to $t_i\land\acc{p_i}{\varnothing}$,
	$t_i\land\acc{p_i}{\varnothing}\leq\acc{p_i}{\Gamma_i}$. If $v/p_i\neq\fail$,
	then $v\in\acc{p_i}{\varnothing}$ (Lemma~\ref{lem:guard-patterns}(1)), so
	$v\Vdash\acc{p_i}{\Gamma_i}$ by (V3), (V2) and (V1). If $v/p_i=\fail$, then
	$v\Vdash\neg\acc{p_i}{\varnothing}$ by Lemma~\ref{lem:guard-patterns}(1) and
	(V3). In the case of rule \RuleRef{guard:accepted:accept}{[accept]},
	$\dynSup{t_{ij}}\leq\dynSup{(\acc{p_i}{\Delta_{ij}})}
	\leq\acc{p_i}{\varnothing}$ by Lemma~\ref{lem:guard-patterns}(2).
	The same bound holds trivially for rule
	\RuleRef{guard:accepted:fail}{[fail]}, where $t_{ij}=\bott$.
	Hence $v\Vdash\neg\dynSup{t_{ij}}$ by (V1).
\end{Proof}

\surelyAcceptedTypesAreSufficient*

\begin{Proof}
	If $\JudgmentTypeS{\varnothing}{v}{t_{ij}}$, then $v\Vdash t_{ij}$ in the
	static instance, since $\vds$ is included in $\vdsn$; if $v$ is closed and
	$\varnothing\vdw v\tc t_{ij}$, then $v\Vdash t_{ij}$ in the weak instance. In
	both cases $v\Vdash\dynSup{t_{ij}}$ (V1), so it suffices to show that $v\Vdash\dynSup{t_{ij}}$ and $\BoolB_{ij}=\smalltrue$
	imply $v/(p_ig_i)\neq\fail$. The pair $(t_{ij},\BoolB_{ij})$ is not the one
	of rule \RuleRef{guard:accepted:fail}{[fail]}, since no value has type
	$\bott$ (V4). So $t_{ij}=\dynSup{t_i}\land\acc{p_i}{\Delta_{ij}}$ for an
	exact success environment $\Delta_{ij}$ of the analysis of $g_i$. Thus
	$v\Vdash\dynSup{(\acc{p_i}{\Delta_{ij}})}$ by (V1).
	By Lemma~\ref{lem:guard-patterns}(1), $v/p_i=\sigma\neq\fail$, and
	Lemmas~\ref{lem:guard-hulls}(2) and~\ref{lem:safe-success-environment}(G4) give
	$g_i\sigma\seqreducesG\smalltrue$.
\end{Proof}

\begin{lem}[Exclusion]\label{lem:guard-exclusion}
	Let $v\Vdash t$, and let no pair $p_hg_h$ with $h<i$ accept $v$. Then
	$v\Vdash t_i$. If, moreover, $p_ig_i$ does not accept $v$, then
	$v\Vdash\neg\dynSup{t_{ij}}$ for every $j$ such that $\BoolB_{ij}=\smalltrue$.
\end{lem}

\begin{Proof}
	By induction on $i$, $v\Vdash t_i$ holds for $i=1$; let it hold for $i$, and
	let $p_ig_i$ not accept $v$. If $v/p_i=\fail$, then
	$v\Vdash\neg\dynSup{t_{ij}}$ for every $j$ by Lemma~\ref{lem:guard-clauses}.
	Otherwise, $v\Vdash\acc{p_i}{\Gamma_i}$ by Lemma~\ref{lem:guard-clauses},
	and for every $j$ with $\BoolB_{ij}=\smalltrue$, if
	$t_{ij}=\bott$ then $v\Vdash\neg\bott$ by (V1); otherwise
	$t_{ij}=\dynSup{t_i}\land A_{ij}$, where
	$A_{ij}=\acc{p_i}{\Delta_{ij}}$ for an exact success environment
	$\Delta_{ij}$ of the analysis of $g_i$.
	This environment has a hull relative to $\Gamma_i$
	(Lemma~\ref{lem:guard-hulls}), and
	$v\Vdash\dynSup{(\acc{p_i}{\Gamma_i})}$ by (V1).
	By Lemma~\ref{lem:safe-success-environment}(G4),
	$v\not\Vdash\dynSup{A_{ij}}$, since $p_ig_i$ does not accept $v$.
	Corollary~\ref{cor:guard-dichotomy} therefore gives
	$v\Vdash\neg\dynSup{A_{ij}}$.
	Since $\dynSup{t_{ij}}\leq\dynSup{A_{ij}}$, we have
	$\neg\dynSup{A_{ij}}\leq\neg\dynSup{t_{ij}}$, so
	$v\Vdash\neg\dynSup{t_{ij}}$ by (V1).
	In both cases, by (V2) and (V1), and since
	$\neg\dynSup{s}\leq\neg s$,
	$v\Vdash t_i\land\bigwedge_{\{j\mid\BoolB_{ij}\}}\neg t_{ij}=t_{i+1}$.
\end{Proof}

\possiblyAcceptedTypesAreNecessary*

\begin{Proof}
	We prove the statement for $\Vdash$ in place of $:$, in both instances. By
	Lemma~\ref{lem:guard-exclusion}, $v\Vdash t_i$. Since $p_ig_i$ accepts $v$,
	$\sigma=v/p_i\neq\fail$, so $v\Vdash\acc{p_i}{\Gamma_i}$ by
	Lemma~\ref{lem:guard-clauses}, and
	$g_i\sigma\seqreducesG\smalltrue$. If the pair $p_ig_i$ were analyzed by
	rule \RuleRef{guard:accepted:fail}{[fail]}, the result of the analysis of
	$g_i$ would be $\omega$ or would have no success environment, and
	Lemma~\ref{lem:safe-success-environment} would give that $g_i\sigma$ fails,
	or does not reduce to $\smalltrue$. So it is analyzed by
	\RuleRef{guard:accepted:accept}{[accept]}, and
	Lemma~\ref{lem:safe-success-environment}(G3) gives $j$ such that
	$v\Vdash\acc{p_i}{\Delta_{ij}}$. Since $v\Vdash t_i$, (V1) gives
	$v\Vdash\dynSup{t_i}$, and (V2) yields
	$v\Vdash\dynSup{t_i}\land\acc{p_i}{\Delta_{ij}}=t_{ij}$.
	With ``$:$'' read as $\vds$, the
	statement follows: if $\JudgmentTypeS{\varnothing}{v}{t}$, then $v\Vdash t$
	in the static instance, and $v\Vdash t_{ij}$ gives
	$\JudgmentTypeS{\varnothing}{v}{t_{ij}}$ by
	Lemma~\ref{lem:generators}(6).
\end{Proof}

\begin{lem}[Exhaustiveness]\label{lem:guard-exhaustive}
	If $v\Vdash t$ and $t\leq\bigvee_{\{(i,j)\mid\BoolB_{ij}\}}t_{ij}$, then some
	pair $p_ig_i$ accepts $v$.
\end{lem}

\begin{Proof}
	Otherwise, by Lemma~\ref{lem:guard-exclusion},
	$v\Vdash\neg\dynSup{t_{ij}}$ for all $i,j$ with $\BoolB_{ij}=\smalltrue$, so
	$v\Vdash\neg\bigvee\dynSup{t_{ij}}$ by (V2); and
	$v\Vdash\dynSup{t}\leq\bigvee\dynSup{t_{ij}}$ by (V1). By (V2),
	$v\Vdash\bott$, which contradicts (V4).
\end{Proof}

\subsection{Typing with Guards: Safety}\label{app:guard-safety}

\begin{figure}[t]
	\begin{align*}
		\begin{array}{l}
			\typingruleExtra{\RuleDef{typing:case:case}{\text{(case)}}}
			{\!\!\Gamma \vd e : t
				\quad
				(\forall i{\leq}n) \,\, (\forall j{\leq}m_i)
				\,\,\;(t_{i j} \nleq \bott\,\,\,\Rightarrow\,\,\,\Gamma, \TypeEnv{t_{i j}}{p_i} \vd e_i : s)\!\!\!\!}
			{\Gamma \vd \caseExprI{} : s}
			{t \leq \biglor_{i\leq n} \sAccType{p_ig_i}{}}
		\end{array} \\
		\begin{array}{l}
			\typingruleExtra{\RuleDef{typing:case:case-omega}{\text{(case$_\omega$)}}}
			{\!\!\Gamma \vd e : t
				\quad
				(\forall i{\leq}n) \,\, (\forall j{\leq}m_i)
				\,\,\;(t_{i j} \nleq \bott\,\,\,\Rightarrow\,\,\,\Gamma, \TypeEnv{t_{i j}}{p_i} \vd e_i : s)\!\!\!\!}
			{\Gamma \vd \caseExprI{} : s}
			{t \leq \biglor_{i\leq n} \pAccType{p_ig_i}{}}
		\end{array} \\
		\begin{array}{l}
			\typingruleExtra{\RuleDef{typing:case:case-star}{\text{(case$_\star$)}}}
			{\!\!\Gamma \vd e : t
				\qquad
				(\forall i{\leq}n) \,\, (\forall j{\leq}m_i)
				\,\,\;(t_{i j} \nleq \bott\,\,\,\Rightarrow\,\,\,\Gamma, \TypeEnv{t_{i j}}{p_i} \vd e_i : s)\!\!\!\!}
			{\Gamma \vd \caseExprI{} : \dyn \land s}
			{t \consist \biglor_{i\leq n} \pAccType{p_ig_i}{}}
		\end{array} \\
		\begin{array}{l}
			\TRuleCasePgWeak\hspace*{15mm}
		\end{array} \\[3mm]
		\begin{array}{l}
			\textit{where }
			\Gamma\,;t \vdash ({p_i}{g_i})_{i\leq n}
			\leadsto (t_{ij},\BoolB_{ij})_{i\leq n, j\leq m_i}
			\!\!\text{ and }\!
			\pAccType{p_i g_i} \!=\! \bigvee_{j\leq m_i} t_{i j}
			\!\!\text{ and }
			\sAccType{p_i g_i} = \bigvee_{\{j\leq m_i\,\mid\, \BoolB_{i j}\}} t_{i j}
		\end{array}
	\end{align*}
	\caption{Case Typing Rules}\label{app:fig:case-typing-rule}
\end{figure}

\noindent
We extend the safety results of Appendix~\ref{app:soundness} to the
expressions of Section~\ref{sec:guard-analysis}: the expressions of
Sections~\ref{sec:safe-erasure} and~\ref{sec:arity}, together with the guarded
case expressions $\caseExprI{}$, where $n\geq1$, and the expressions
$\GuardSize{e}$. In $\caseExprI{}$, the pattern $p_i$ binds its capture
variables in $g_i$ and $e_i$; up to $\alpha$-renaming, they are distinct from
the variables of a given environment and from the free variables of a given
expression. The new expressions reduce by rules
\RuleRef{reduction:guard-case}{\textsc{[Case]}},
\RuleRef{reduction:guard-case-omega}{\textsc{[Case$_\omega$]}},
\RuleRef{reduction:size}{\textsc{[Size]}} and
\RuleRef{reduction:size-omega}{\textsc{[Size$_\omega$]}} of
Figure~\ref{app:fig:standard-and-failure-reductions}, in the evaluation
contexts $\GuardSize{\Context}$ and $\CasePat{\Context}$. Every guard
$g\sigma$, where $\sigma$ maps the variables of $g$ to values, reduces to
exactly one of $\smalltrue$, $\smallfalse$ and $\fail$
(Appendix~\ref{app:core-elixir}). So every expression
$\texttt{case}\,\,v\,\p{p_ig_i\rightarrow e_i}_{i\leq n}$ reduces: by rule
\RuleRef{reduction:guard-case}{\textsc{[Case]}}, with the first pair $p_jg_j$
that accepts $v$, if there is one, and by rule
\RuleRef{reduction:guard-case-omega}{\textsc{[Case$_\omega$]}} to $\omegaCase$
otherwise; and $\GuardSize{v}$ reduces by rule
\RuleRef{reduction:size}{\textsc{[Size]}} if $v$ is a tuple, and to
$\omegaSize$ otherwise. Marked expressions (Definition~\ref{def:marked}) may
contain guarded case expressions and size expressions. Rule
\RuleRef{reduction:guard-case}{\textsc{[Case]}} substitutes the values captured by
the pattern as they are, without marking them, and equality and order on
marked values are those of their erasures. So no reduction depends on the
marks, and the correspondence between the reductions of a marked expression
and those of its erasure, stated after Definition~\ref{def:marked}, still
holds.

Guarded case expressions are typed by the rules of
Figure~\ref{app:fig:case-typing-rule}: rules
\RuleRef{typing:case:case}{(case)} and
\RuleRef{typing:case:case-omega}{(case$_\omega$)}, which is $\omega$-marked,
belong to the static and to the gradual system, rule
\RuleRef{typing:case:case-star}{(case$_\star$)} to the gradual system, and
rule \RuleRef{typing:case:case-circ}{(case$^\circ$)} to the weak system.
Expressions $\GuardSize{e}$ are typed by rule
\RuleRef{typing:static:size}{(size)} in the static and the gradual systems, by
rule \RuleRef{typing:gradual:size-star}{(size$_\star$)} in the gradual system,
and by rule \RuleRef{typing:weak:size-circ}{(size$^\circ$)} in the weak system
(Section~\ref{sec:guard-analysis}). The enriched systems $\vdsn$ and $\vdwn$
(Definitions~\ref{def:enriched-static} and~\ref{def:enriched-weak}) contain
the rules of the static, respectively weak, system, with their own judgment on
expressions. In every system, these rules use the same guarded pattern
sequence analysis, whose typing premises are judgments
$\JudgmentTypeS{\Gamma}{a}{s}$ on atoms derived without $\omega$-marked rules
(Appendix~\ref{app:guardanalysis}). The $\omega$-rules of the guard analysis
(Figure~\ref{app:fig:guard-analysis-false-failure-rules}) are not typing
rules: the hypothesis that a typing derivation uses no $\omega$-marked rule
does not concern them. Since the free variables of a guard are capture
variables of its pattern (Figure~\ref{app:fig:guard-syntax}), the analysis
does not depend on the environment.

\begin{lem}[Analysis and environments]\label{lem:guard-environment}
	Let $t$ be a type and $(p_ig_i)_{i\leq n}$ a sequence of guarded patterns.
	\begin{enumerate}
		\item For all environments $\Gamma$ and $\Gamma'$,
		      $\Gamma\,;t\vdash(p_ig_i)_{i\leq n}\leadsto(t_{ij},\BoolB_{ij})_{i\leq n,j\leq m_i}$
		      if and only if
		      $\Gamma'\,;t\vdash(p_ig_i)_{i\leq n}\leadsto(t_{ij},\BoolB_{ij})_{i\leq n,j\leq m_i}$.
		\item For every environment $\Gamma$, there are types $t_{ij}$ and flags
		      $\BoolB_{ij}$ such that
		      $\Gamma\,;t\vdash(p_ig_i)_{i\leq n}\leadsto(t_{ij},\BoolB_{ij})_{i\leq n,j\leq m_i}$.
	\end{enumerate}
\end{lem}

\begin{Proof}
	\begin{enumerate}
		\item For a pattern $p$, an environment $\Gamma$ and a map $\theta$ from
		      $\vars{p}$ to types, $\Gamma,\theta$ is the environment that maps the
		      variables of $p$ as $\theta$ does and the other variables as
		      $\Gamma$ does. Let the free variables of $g$ be in $\vars{p}$. We show,
		      by induction on the derivation, that if
		      $\JudgmentGuard{\Gamma,\theta}{g}{\GuardResult}$, then every
		      environment of $\GuardResult$ has the form $\Gamma,\theta'$, and
		      $\JudgmentGuard{\Gamma',\theta}{g}{\GuardResult'}$, where
		      $\GuardResult'$ is obtained from $\GuardResult$ by replacing every
		      environment $\Gamma,\theta'$ with $\Gamma',\theta'$. The rules
		      analyze their premises under the environment of their conclusion,
		      under environments of the results of their other premises, and, in
		      rule \RuleRef{guard:analysis:or}{(or)}, under
		      $(\Gamma,\theta),\TypeEnv{t'}{p}$, which is $\Gamma,\TypeEnv{t'}{p}$;
		      the environments of their results are these and, in rule
		      \RuleRef{guard:analysis:var}{(var)}, the environment
		      $(\Gamma,\theta)[x\refine\tau]_p$, which differs from
		      $\Gamma,\theta$ only on $\vars{p}$, since $x\in\vars{p}$
		      (Figure~\ref{app:def:environment-updates}). Their side conditions
		      depend on the environments only through the types of the variables
		      of $p$: the typing premises concern atoms of $g$, whose variables are
		      in $\vars{p}$ (Lemma~\ref{app:lemma:term_context_agreement}, whose
		      proof re-applies the rules of the given derivation, so that the
		      premises stay derived without $\omega$-marked rules); the
		      accepted types $\acc{p}{\cdot}$ and the maps $\TypeEnv{\cdot}{p}$
		      involve only these types; and environments are compared pointwise,
		      so $\Gamma,\theta_1\simeq\Gamma,\theta_2$ if and only if
		      $\theta_1\simeq\theta_2$, if and only if
		      $\Gamma',\theta_1\simeq\Gamma',\theta_2$. Rule
		      \RuleRef{guard:accepted:accept}{[accept]} analyzes the guard of $pg$
		      under $\Gamma,\TypeEnv{t}{p}$ and produces
		      $\dynSup{t}\land\acc{p}{\Delta}$, unchanged under $\Gamma'$ since
		      $t$ is fixed and $\acc{p}{\Delta}$ depends only on the types of the
		      variables of $p$. Rule \RuleRef{guard:accepted:fail}{[fail]} also
		      has the same result, and rule \RuleRef{guard:accepted:seq}{[seq]}
		      therefore passes the same residual types to its premises.
		      The converse is symmetric.
		\item By rule \RuleRef{guard:analysis:approx}{(approx)}, every guard has,
		      under every environment, a result with a success environment. So
		      rule \RuleRef{guard:accepted:accept}{[accept]} applies to every pair
		      $p_ig_i$ under every type, since $\TypeEnv{t}{p}$ is defined for
		      every type $t$ (Figure~\ref{app:fig:typing-environments}), and rule
		      \RuleRef{guard:accepted:seq}{[seq]}, by induction on $n$,
		      concludes.\qedhere
	\end{enumerate}
\end{Proof}

\paragraph{Static safety.}
Lemmas~\ref{lem:generators}, \ref{lem:dichotomy} and~\ref{lem:canonical-forms}
and Corollary~\ref{cor:selection} concern values, which do not change; their
proofs consider only the rules that type values, and they hold as stated.

\begin{lem}\label{lem:guard-static}
	Lemmas~\ref{app:lemma:term_context_agreement}, \ref{app:static_substitution},
	\ref{lem:progress} and~\ref{lem:preservation} hold for the marked
	expressions of Section~\ref{sec:guard-analysis}, where, in
	Lemma~\ref{lem:progress}, $e$ may also reduce to $\omegaCase$, which is
	impossible if the derivation does not use rule
	\RuleRef{typing:case:case-omega}{(case$_\omega$)}.
\end{lem}

\begin{Proof}
	We give the new cases of the proofs. When rule
	\RuleRef{typing:case:case}{(case)} or rule
	\RuleRef{typing:case:case-omega}{(case$_\omega$)} types a guarded case
	expression $\texttt{case}\,\,e_0\,\p{p_ig_i\rightarrow e_i}_{i\leq n}$ under
	an environment $\Gamma$, with the type $s$, the premises are
	\[
		\JudgmentTypeSN{\Gamma}{e_0}{u}
		\qquad
		\Gamma\,;u\vdash(p_ig_i)_{i\leq n}\leadsto(t_{ij},\BoolB_{ij})_{i\leq n,j\leq m_i}
	\]
	and $\JudgmentTypeSN{\Gamma,\TypeEnv{t_{ij}}{p_i}}{e_i}{s}$ for all $i,j$
	such that $t_{ij}\not\leq\bott$.
	\begin{description}
		\item[Lemma~\ref{app:lemma:term_context_agreement}] Rule
		      \RuleRef{typing:static:size}{(size)}: by the induction hypothesis.
		      Rules (case) and (case$_\omega$): the induction hypothesis applies to
		      $e_0$, whose free variables are free in the case expression; by
		      Lemma~\ref{lem:guard-environment}(1), the analysis holds under
		      $\Gamma'$; the environments $\Gamma,\TypeEnv{t_{ij}}{p_i}$ and
		      $\Gamma',\TypeEnv{t_{ij}}{p_i}$ agree on the free variables of $e_i$,
		      since those that are not in $\vars{p_i}$ are free in the case
		      expression, so the induction hypothesis applies to the branches. The
		      same rule concludes.
		\item[Lemma~\ref{app:static_substitution}] Rule (size): by the induction
		      hypothesis. Rules (case) and (case$_\omega$), under $\Gamma,x:t_1$:
		      up to $\alpha$-renaming, the variables of the patterns $p_i$ are not
		      $x$, are not in $\dom{\Gamma}$, and are not free in $e_1$. So the
		      substitution does not change the guards, whose free variables are
		      variables of the patterns, the environment of a branch is
		      $\Gamma,\TypeEnv{t_{ij}}{p_i},x:t_1$, and
		      $\JudgmentTypeSN{\Gamma,\TypeEnv{t_{ij}}{p_i}}{e_1}{t_1}$ by
		      Lemma~\ref{app:lemma:term_context_agreement}. The induction
		      hypothesis on $e_0$ and on the branches,
		      Lemma~\ref{lem:guard-environment}(1) for the analysis under $\Gamma$,
		      and the same rule conclude.
		\item[Lemma~\ref{lem:progress}] The redexes $r$ also include
		      $\GuardSize{v}$ and
		      $\texttt{case}\,\,v\,\p{p_ig_i\rightarrow e_i}_{i\leq n}$: the
		      argument of \texttt{size} and the matched expression are in
		      evaluation position, the branches are not, and both redexes reduce.
		      For $\GuardSize{v}$, the rule is (size), so $v$ has type
		      $\tupleTop$; it is a tuple by Lemma~\ref{lem:generators}(4), and $r$
		      reduces by rule \RuleRef{reduction:size}{\textsc{[Size]}}. For a
		      case expression, the rule is (case) or (case$_\omega$), and
		      $\JudgmentTypeSN{\varnothing}{v}{u}$, that is, $v\Vdash u$ in the
		      static instance of Appendix~\ref{app:guard-analysis-correctness}. If
		      the rule is (case), then
		      $u\leq\bigvee_{\{(i,j)\mid\BoolB_{ij}\}}t_{ij}$, so some pair accepts
		      $v$ (Lemma~\ref{lem:guard-exhaustive}), and $r$ reduces by rule
		      \RuleRef{reduction:guard-case}{\textsc{[Case]}}. If the rule is
		      (case$_\omega$), $r$ may reduce to $\omegaCase$.
		\item[Lemma~\ref{lem:preservation}] Rule
		      \RuleRef{reduction:context}{\textsc{[Context]}}: the general argument
		      applies, since the matched expression keeps its type and the
		      analysis depends only on this type and on the environment, which
		      does not change, so the same branches are typed. Rule \RuleRef{reduction:size}{\textsc{[Size]}}:
		      $e=\GuardSize{\tuple{v_1,\ldotsTwo,v_n}}$, the rule is (size),
		      $t=\intTop$, and $e'$ is the integer $n$, of type $\intTop$. Rule
		      \RuleRef{reduction:guard-case}{\textsc{[Case]}}:
		      $e=\texttt{case}\,\,v\,\p{p_ig_i\rightarrow e_i}_{i\leq n}$ and
		      $e'=e_k\sigma$, where $\sigma=v/(p_kg_k)$, so that $\sigma=v/p_k$, and
		      $v/(p_hg_h)=\fail$ for every $h<k$; here $\Gamma=\varnothing$ and
		      $s=t$. By Lemma~\ref{lem:guard-necessary}, in the static instance,
		      $v\Vdash t_{kj}$ for some $j\leq m_k$. So $t_{kj}\not\leq\bott$, by
		      (V1) and (V4), and $\JudgmentTypeSN{\TypeEnv{t_{kj}}{p_k}}{e_k}{t}$.
		      Since $t_{kj}$ is not the type $\bott$ of rule
		      \RuleRef{guard:accepted:fail}{[fail]}, it satisfies
		      $t_{kj}\leq\acc{p_k}{\Delta_{kj}}\leq\acc{p_k}{\varnothing}$
		      for a success environment $\Delta_{kj}$
		      (Lemma~\ref{lem:guard-patterns}(2)), and
		      Lemma~\ref{lem:guard-patterns}(6) gives
		      $\JudgmentTypeSN{\varnothing}{\sigma(x)}{(\TypeEnv{t_{kj}}{p_k})(x)}$
		      for every $x\in\vars{p_k}$. Up to $\alpha$-renaming, the variables of
		      $p_k$ are not free in $v$, hence not in the values $\sigma(x)$. The
		      Substitution Lemma~\ref{app:static_substitution}, applied to the
		      variables of $p_k$ one at a time, with
		      Lemma~\ref{app:lemma:term_context_agreement} to type each value
		      $\sigma(x)$ under the environment of the variables not yet
		      substituted, gives $\JudgmentTypeSN{\varnothing}{e_k\sigma}{t}$. The
		      derivations for the values $\sigma(x)$, which
		      Lemma~\ref{lem:generators}(1) builds from generators, use, besides
		      the rules that deduce generators, only premises of the derivation for
		      $v$, and Lemmas~\ref{app:static_substitution}
		      and~\ref{app:lemma:term_context_agreement} re-apply the rules of the
		      derivations they are given: no new $\omega$-marked rule is
		      used.\qedhere
	\end{description}
\end{Proof}

\guardStaticSoundness*

\begin{Proof}
	By Lemma~\ref{lem:guard-static}, the proof of
	Theorem~\ref{app:thm:static-soundness} applies to the expressions of
	Section~\ref{sec:guard-analysis}: a derivation that uses no $\omega$-marked
	rule uses neither rule \RuleRef{typing:case:case-omega}{(case$_\omega$)} nor
	the $\omega$-marked projection rules, so every expression of the reduction of
	$e$ is a value or reduces to an expression. We obtain the conclusion of
	Theorem~\ref{app:thm:static-soundness}: either $e$ diverges, or
	$e\seqreduces v$ and $\JudgmentTypeSN{\varnothing}{\hat{v}}{t}$ for a
	marking $\hat{v}$ of $v$, which is $\varnothing\vds v:t$ if $v$ is a
	constant. As for Theorem~\ref{thm:static-soundness}
	(Remark~\ref{app:rem:static-theorems}), this is the meaning of $v:t$ in the
	statement.
\end{Proof}

\guardOmegaSoundness*

\begin{Proof}
	As for Theorem~\ref{thm:guard-soundness}, along the proof of
	Theorem~\ref{app:thm:omega-static-soundness}. By
	Lemma~\ref{lem:guard-static}, the only failures that an expression typed in
	$\vdsn$ can reach are $\omegaOutOfRange$, through rules
	\RuleRef{typing:static:proj-omega}{(proj$_\omega$)} and
	\RuleRef{typing:static:proj-top-omega}{(proj$^\topp_\omega$)}, and
	$\omegaCase$, through rule \RuleRef{typing:case:case-omega}{(case$_\omega$)},
	which checks only the possibly accepted types, so that no pair may accept the
	matched value. The other failures are excluded as in the proof of
	Lemma~\ref{lem:progress}; in particular, the arguments of \texttt{size} are
	tuples.
\end{Proof}

\paragraph{Gradual safety.}
Lemmas~\ref{lem:weak-generators} and~\ref{lem:weak-dichotomy} concern values
and hold as stated.

\begin{lem}\label{lem:guard-weak}
	Lemmas~\ref{lem:gradual-progress}, \ref{lem:context-agreement-weak},
	\ref{lem:static-typing-imply-dynamic-typing}, \ref{lem:weak-totality},
	\ref{lem:weak-substitution}, \ref{lem:well-typedness-after-substitution},
	\ref{lem:subject-reduction} and~\ref{lem:gradual-to-strong}, and
	Corollary~\ref{cor:dynamic-typing}, hold for the expressions of
	Section~\ref{sec:guard-analysis}.
\end{lem}

\begin{Proof}
	Lemma~\ref{lem:well-typedness-after-substitution} and
	Corollary~\ref{cor:dynamic-typing} follow from the other lemmas, as before.
	We give the new cases of the other proofs. When a guarded case expression is
	typed in $\vdwn$ by rule \RuleRef{typing:case:case-circ}{(case$^\circ$)}, the
	premises are those of the proof of Lemma~\ref{lem:guard-static}, with
	$\vdwn$ in place of $\vdsn$, and the type is $s\land\dyn$.
	\begin{description}
		\item[Lemma~\ref{lem:gradual-progress}] The redexes also include
		      $\GuardSize{v}$ and
		      $\texttt{case}\,\,v\,\p{p_ig_i\rightarrow e_i}_{i\leq n}$, which
		      reduce.
		\item[Lemmas~\ref{lem:context-agreement-weak} and~\ref{lem:weak-substitution}]
		      Rule \RuleRef{typing:weak:size-circ}{(size$^\circ$)}: by the
		      induction hypothesis. Rule (case$^\circ$): as rules (case) and
		      (case$_\omega$) in the proof of Lemma~\ref{lem:guard-static}, with
		      Lemma~\ref{lem:context-agreement-weak} in place of
		      Lemma~\ref{app:lemma:term_context_agreement}.
		\item[Lemma~\ref{lem:static-typing-imply-dynamic-typing}] Rule
		      (size$^\circ$) deduces $\intTop\land\dyn\leq\dyn$, and subsumption
		      concludes. Rule (case$^\circ$): by the induction hypothesis, every
		      branch has type $\dyn$, and the rule with $s=\dyn$ deduces
		      $\dyn\land\dyn\simeq\dyn$.
		\item[Lemma~\ref{lem:weak-totality}] If $e=\GuardSize{e_0}$: the
		      induction hypothesis, rule (size$^\circ$) and subsumption. If
		      $e=\texttt{case}\,\,e_0\,\p{p_ig_i\rightarrow e_i}_{i\leq n}$: by the
		      induction hypothesis, $\Gamma\vdw e_0\tc\topp$; next by
		      Lemma~\ref{lem:guard-environment}(2),
		      $\Gamma\,;\topp\vdash(p_ig_i)_{i\leq n}\leadsto(t_{ij},\BoolB_{ij})_{i\leq n,j\leq m_i}$
		      for some $t_{ij}$ and $\BoolB_{ij}$; the free variables of $e_i$ are
		      in $\dom{\Gamma}\cup\vars{p_i}$, the domain of
		      $\Gamma,\TypeEnv{t_{ij}}{p_i}$, so the induction hypothesis gives
		      $\Gamma,\TypeEnv{t_{ij}}{p_i}\vdw e_i\tc\topp$ for all $i,j$; rule
		      (case$^\circ$), with $s=\topp$, and subsumption conclude.
		\item[Lemma~\ref{lem:subject-reduction}] Rule
		      \RuleRef{reduction:context}{\textsc{[Context]}}: the general argument
		      applies, as in \autoref{lem:guard-static}'s proof. Rule
		      \RuleRef{reduction:size}{\textsc{[Size]}}: the rule is
		      (size$^\circ$), so $t=\intTop\land\dyn$, and $e'$ is an integer $c$,
		      of type $c\land\dyn\leq\intTop\land\dyn$. Rule
		      \RuleRef{reduction:guard-case}{\textsc{[Case]}}: with the notation of the
		      proof of Lemma~\ref{lem:guard-static}, $t=s\land\dyn$. As there, with
		      the weak instance of Lemmas~\ref{lem:guard-necessary}
		      and~\ref{lem:guard-patterns}(6), and with
		      Lemmas~\ref{lem:weak-substitution}
		      and~\ref{lem:context-agreement-weak},
		      $\varnothing\vdwn e_k\sigma\tc s$ (the values $\sigma(x)$ are closed,
		      since $e$ is); Corollary~\ref{cor:dynamic-typing} gives
		      $\varnothing\vdwn e_k\sigma\tc s\land\dyn$.
		\item[Lemma~\ref{lem:gradual-to-strong}] Rules (case), (case$_\omega$)
		      and (case$_\star$) correspond to rule (case$^\circ$), which has the
		      same analysis and, by the induction hypothesis, the premises on the
		      matched expression and on the branches; it concludes $s\land\dyn$,
		      which is a subtype of $s$ and equivalent to the conclusion
		      $\dyn\land s$ of (case$_\star$). Rules (size) and (size$_\star$)
		      correspond to rule (size$^\circ$), whose premise follows from theirs
		      by subsumption, and whose conclusion $\intTop\land\dyn$ is a subtype
		      of theirs.\qedhere
	\end{description}
\end{Proof}

\guardGradualSoundness*

\begin{Proof}
	By Lemma~\ref{lem:guard-weak}, Theorem~\ref{app:thm:gradual-soundness} and
	Corollary~\ref{app:cor:gradual-soundness} hold for the expressions of
	Section~\ref{sec:guard-analysis}, with the same proofs. The statement is the
	corollary for $\JudgmentTypeG{\varnothing}{e}{t}$: if $e$ neither diverges nor
	reduces to a failure, then $e\seqreduces v$ with $\varnothing\vdwn v\tc t$,
	and $s\land t\not\leq\bott$ for every type $s$ such that
	$\varnothing\vdw v\tc s$.
\end{Proof}

\begin{remark}[Inference]\label{app:rem:inference}\em
	Section~\ref{sec:inference} replaces a non-annotated $\lambda$-abstraction
	$\lamPat[]{\,\ov{p g \rarr e}\,}$ by
	$\lambda^{\mathbb{I}}x.\texttt{case\,}x\,(\ov{p g \rarr e})$, where
	$\mathbb{I}=\{t_{ij}\rarr t'_{ij}\}_{ij}$ is the interface inferred by
	\RuleRef{typing:inference:infer}{rule (\textsc{infer})}. This turns an
	instance of rule (\textsc{infer}) into an instance of
	\RuleRef{typing:static:lambda}{rule ($\lambda$)} (of
	\RuleRef{typing:gradual:lambda-dyn}{rule ($\lambda_\dyn$)} for a domain
	containing `$\dyn$') with the same premises, in $\vdg$, and in $\vds$ for
	rule (\textsc{infer}) with $\vds$ in place of $\vdg$. Three points complete
	the argument.
	\begin{description}
		\item[No accepted type] If the analysis accepts no type, the family
		      $(t_{ij})$ is empty, and we take $\mathbb{I}=\{\bott\rarr\bott\}$.
		      Its premise $\Gamma,x:\bott\vd\texttt{case\,}x\,(\ov{p g \rarr e}):\bott$
		      holds by rule \RuleRef{typing:case:case}{(case)}, with a derivation of
		      the analysis from $\bott$ that accepts only empty types, so that no
		      branch is typed: for a pattern with capture variables,
		      $\TypeEnv{\bott}{p}$ assigns them $\bott$, and so does every
		      environment of every analysis of its guard
		      (Lemma~\ref{lem:guard-hulls}(2)); for a pattern without capture
		      variables, we take the derivation from $\topp$ used by rule
		      (\textsc{infer}), which accepts no type and does not depend on the
		      initial type. Subsumption gives the conclusion of rule
		      (\textsc{infer}), an empty intersection, which is $\topp$ (or
		      $\function_1$, if read as an intersection of arrows), and that of
		      rule (\textsc{infer}$_\star$), since moreover
		      $\bott\rarr\bott\leq\bott\rarr\dyn$.
		\item[Rule (\textsc{infer}$_\star$)] For
		      \RuleRef{typing:inference:infer-star}{rule (\textsc{infer}$_\star$)},
		      $\mathbb{I}$ also contains $t_{ij}\rarr\dyn$ for all $i,j$. The
		      derivation of the premise
		      $\Gamma,x:t_{ij}\vdg\texttt{case\,}x\,(\ov{p g \rarr e}):t'_{ij}$ ends
		      with rule (case), (case$_\omega$) or (case$_\star$), followed by
		      subsumption; \RuleRef{typing:case:case-star}{rule (case$_\star$)}
		      applies to the same premises, since $t\leq t'$ implies
		      $t\consist t'$, and concludes $\dyn\land s\leq\dyn$. Since
		      $\bigwedge_{ij}(t_{ij}\rarr\dyn)\leq(\bigvee_{ij}t_{ij})\rarr\dyn$
		      (compare the minimal and the maximal materializations), subsumption
		      gives the conclusion of rule (\textsc{infer}$_\star$).
		\item[One interface per $\lambda$-abstraction] Interfaces play no role
		      at run time (test types contain no arrows), so the annotated program
		      reduces as the original one. Hence, if a derivation gives each
		      non-annotated $\lambda$-abstraction a single interface wherever it
		      types it, annotating them yields a program typed by the rules of
		      Sections~\ref{sec:safe-erasure} and~\ref{sec:guard-analysis}, to
		      which their soundness theorems apply. We do not prove soundness
		      without this condition, which may fail: rule (\textsc{infer}) types
		      $\PatLam{\When{y}{(\GuardOr{y \isof \intTop}{y \isof \boolTop})} \rarr \PatLam{z \rarr y}}$
		      with
		      $(\intTop\rarr(\topp\rarr\intTop))\land(\boolTop\rarr(\topp\rarr\boolTop))$,
		      whereas an interface of the inner $\lambda$-abstraction must be
		      checked both under $y:\intTop$ and under $y:\boolTop$, so all its
		      result types contain $\intTop\lor\boolTop$.
	\end{description}
\end{remark}

\section{Semantic subtyping: multi-arity functions}
\label{app:functions}

In the original formulation of semantic subtyping, function types are of single arity $(t \to s)$, and multi-arity functions are represented by taking $t$ as a tuple of argument types. For instance, to declare that the binary function \elix{subtract} of Section \ref{sub:prop} takes two integers and returns an integer (cf. line~\ref{lnsubtracttype}) we would use the type $\{\integer,\integer\} \to \integer$. This has two problems:
\begin{itemize}
	\item It does not allow for representing the types of all functions of a given arity, which is a necessity in Elixir which has guards that can check the arity of functions. Consider for instance the type of functions of arity 3. In the interpretation of~\citet{FCB08}, every function has type $\bott \to \topp$, so we may consider the type $\{\bott,\bott,\bott\} \to \topp$. But
	the type that appears in the domain is in fact equivalent to $\bott$ : since no value of type $\bott$ exists, we cannot build a triple of them and the type is empty.  Note that type $\{\topp,\topp,\topp\}$ is not valid either since $\{\topp,\topp,\topp\}\to\topp$  is the type of total functions on triples, and not the top type of ternary functions.
	\item When mixing with the dynamic type, it derails the representation of types that contain dynamic as arguments. Consider the function
	\[
		\fun{\pc{\dyn,\integer}}{\integer}
	\]
	which in this representation is the type of functions that, given any first argument and an integer, return an integer.
	The theorem for representing gradual types tells us that this type is equivalent to
	\[
		\fun{\pc{\topp,\integer}}{\integer} \lor (\dynamic \land \p{\fun{\pc{\bott,\integer}}{\integer}})
	\]
	where we obtain the left term by replacing all covariant (resp. contravariant) occurrences of $\dyn$ by $\bott$ (resp. $\topp$), and the right term by doing the opposite of that. We notice that this states that
	such a function can be either a total function on pairs of $\topp$ and $\integer$, or a subtype of all functions that take some value, and an integer, and return an integer. But the latter type, in the original formulation of semantic subtyping, is not representable: it is equivalent
	to $\bott\to\integer$ since $\pc{\bott,\integer}$ is equivalent to $\bott$.
\end{itemize}
\noindent
Our solution is to introduce a new multi-arity function type $\fun{\p{t_1,\ldotsTwo,t_n}}{s}$ which is the type of functions that take $n$ arguments of types $t_1,\ldotsTwo,t_n$ and return a value of type $s$, and to extend the set-theoretic semantics to this type. The first problem is then solved in both interpretations below, since the type of all functions of arity $n$ is the base type $\function_n$; the second one only by the refinement of~\citet{lanvin2021semantic}, since in the interpretation of Section~\ref{subsec:multi-arity} the type $\fun{\p{\bott,\integer}}{\integer}$ is equivalent to $\function_2$. We do it for two interpretations at once: the interpretation of Section~\ref{subsec:multi-arity}, which extends the one of~\citet{FCB08}, and its refinement by~\citet{lanvin2021semantic}, which induces the subtyping relation of this work (Remark~\ref{rem:two-models}). The two differ only in the inputs of relations, and every definition and result of this appendix and of Appendix~\ref{sec:strong-subtyping} is stated and proved for both.

\subsection{Set-theoretic interpretation}


First, we formally extend the syntax of types to include multi-arity functions (including zero-arity functions for which $n=0$):
\begin{align*}
	t & ::= \cdots \mid \fun{\p{t_1,\ldotsTwo,t_n}}{t}
\end{align*}
and we add to base types an explicit top type for functions, and one for the functions of each arity $n$:
\begin{align*}
	b & ::= \cdots \mid \function \mid \function_n
\end{align*}
The interpretation of types with single arity arrows is given in Definition 3 of \cite{castagna2022programming}. The definition first inductively  defines a relation $d:t$ between the elements $d$ of a domain $\Dom$ and the types $t$ of the language, and then it interprets each type $t$ into a subset of $\Dom$ as follows: $\sem t = \{d \in\Dom \bnfor (d:t)\}$.  In particular the predicate-based interpretation of function types in~\cite[Definition 3]{castagna2022programming} suppose that $\fparts{\product{\Dom}{\DOmega}}\subseteq\Dom$, and is defined as follows:\\
 given $t,s$ types,
for all $R \in \fparts{\product{\Dom}{\DOmega}}$,
\begin{align*}
	(R : t \to s)
	\quad\Leftrightarrow\quad
	\forall (d, \delta) \in R. \,\, (d : t) \implies (\delta : s)
\end{align*}
In this interpretation, an arrow type whose domain is empty contains every relation, whatever its codomain. \citet{lanvin2021semantic} refines it by adding to the inputs of relations a new element $\invOmega$, which does not belong to $\DOmega$ and which is accepted by every domain: a relation that contains the pair $(\invOmega,\delta)$ belongs to $t\to s$ only if $\delta$ belongs to (the interpretation of) $s$, even when $t$ is empty.

We extend both interpretations to multi-arity functions at once. Let $\OO$ be either the empty set or the singleton $\{\invOmega\}$; we write $\DO$ for $\Dom\cup\OO$, and $\wO{X}$ for $X\cup\OO$ when $X$ is a subset of $\Dom$. The choice $\OO=\varnothing$ gives the interpretation of Section~\ref{subsec:multi-arity}; the choice $\OO=\{\invOmega\}$ gives the interpretation of~\citet{lanvin2021semantic}, extended to multi-arity functions. We suppose that $\Dom$ contains the disjoint union $\biguplus_{n\in\mathbb{N}}\fparts{\product{\DO^n}{\DOmega}}$, whose $n$-th summand, which we still write $\fparts{\product{\DO^n}{\DOmega}}$, contains the relations of arity $n$. A relation is thus tagged by its arity: relations of different arities, for instance the empty ones, are distinct elements of $\Dom$. Given $n\in\mathbb{N}$, for all $R \in \fparts{\product{\DO^n}{\DOmega}}$,
\begin{align*}
	(R : \fun{\p{t_1,\ldotsTwo,t_n}}{s})
	\quad\Leftrightarrow\quad
	& \forall ((d_1,\ldotsTwo,d_n), \delta) \in R. \\
	& \quad (\forall i\in\{1,...,n\}.\,\, d_i=\invOmega \text{ or } d_i : t_i) \implies (\delta : s)
\end{align*}
and both $(R : \function_n)$ and $(R : \function)$ hold. So the interpretation of $\function_n$ is the set $\fparts{\product{\DO^n}{\DOmega}}$ of all relations of arity $n$, and the interpretation of the top type of all functions $\function$ is the union of these sets for $n\in\mathbb{N}$. On sets, multi-arity function spaces are interpreted as follows.
\begin{Definition}\label{app:def:multi-arity}
	Let $X_1,\ldotsTwo,X_n, Y$ be subsets of $\Dom$. We define
	\begin{align*}
		\fun{\p{X_1,\ldotsTwo,X_n}}{Y} = \{
		& R \in \fparts{\product{\DO^n}{\DOmega}} \mid \\
		& \forall ((d_1,\ldotsTwo,d_n), \delta) \in R.\,\,
		(\forall i=1..n.\,\, d_i \in \wO{X_i})
		\Rightarrow \delta \in Y \}
	\end{align*}
\end{Definition}
For $\OO=\varnothing$, this is Definition~\ref{def:multi-arity}. The two interpretations differ on the arrow types that have an empty domain component.
\begin{itemize}
	\item For $\OO=\varnothing$, as in the interpretation of~\citet{FCB08}, an arrow type with an empty domain component contains every relation of its arity: for $n\geq1$, the types $\fun{\p{\bott,\ldotsTwo,\bott}}{\topp}$ and $\fun{\p{\bott,\integer}}{\integer}$ (for $n=2$) are both equivalent to $\function_n$.
	\item For $\OO=\{\invOmega\}$, the type $\fun{\p{\bott,\integer}}{\integer}$ contains the relations whose pairs $((d_1,d_2),\delta)$ with $d_1=\invOmega$ and $d_2\in\wO{\sem{\integer}}$ have $\delta\in\sem{\integer}$, such as $\{((\invOmega,1),2)\}$, but not $\{((\invOmega,1),\tttrue)\}$; so it is strictly contained in $\fun{\p{\bott,\bott}}{\topp}$. The latter is the largest arrow type of arity $n=2$, and it is strictly contained in $\function_2$, since the relation $\{((\invOmega,\invOmega),\Omega)\}$ belongs to $\function_2$ and to no arrow type of arity 2. More generally, for $n\geq1$, $\fun{\p{\bott,\ldotsTwo,\bott}}{\topp}$ is the largest arrow type of arity $n$, and $\function_n$ is strictly larger than every non-empty intersection of arrow types of arity $n$.
\end{itemize}
For $n=0$, in both interpretations, $\fun{\p{}}{\topp}$ is strictly contained in $\function_0$, since the relation $\{((),\Omega)\}$ belongs to $\function_0$ and not to $\fun{\p{}}{\topp}$. This is why $\function_n$ is a base type.

The first step is to show that this definition provides a form for functions, that is similar to the one we had for single-arity functions, as the finite parts of an analytic expression of the domain and codomain sets.

\begin{lem}\label{lemma:function-interpretation}
	For all $X_1,\ldotsTwo,X_n, Y$ subsets of $\Dom$, we have
	\[  \fun{\p{X_1,\ldotsTwo,X_n}}{Y} =
		\fparts{
			\ov{\wO{X_1}\stimes\cdots\stimes \wO{X_n}
				\stimes\ov{Y}^{\,\DOmega}
			}^{\,\DO^n\stimes\DOmega}}
	\]
	where the finite subsets of $\DO^n\stimes\DOmega$ are taken as relations of arity $n$.
\end{lem}

\begin{Proof}
	By reciprocal inclusion:
	\begin{itemize}
		\item $(\subset)$. Let $R \in \fun{\p{X_1,\ldotsTwo,X_n}}{Y}$.
		Consider any element $((d_1,\ldotsTwo,d_n), \delta) \in R$. Two cases:

		Case 1: There exists at least one $i$ such that $d_i \not\in \wO{X_i}$. Then $((d_1,\ldotsTwo,d_n), \delta)$ is not in $\wO{X_1}\stimes\cdots\stimes \wO{X_n}\stimes\ov{Y}^{\,\DOmega}$, so it belongs to its complement.

		Case 2: For all $i$, $d_i \in \wO{X_i}$. By definition of $R$, this implies $\delta \in Y$, which means $\delta \not\in \ov{Y}^{\,\DOmega}$. Therefore, $((d_1,\ldotsTwo,d_n), \delta)$ is not in $\wO{X_1}\stimes\cdots\stimes \wO{X_n}\stimes\ov{Y}^{\,\DOmega}$, and thus belongs to the complement.

		In both cases, $((d_1,\ldotsTwo,d_n), \delta) \in \ov{\wO{X_1}\stimes\cdots\stimes \wO{X_n}\stimes\ov{Y}^{\,\DOmega}}^{\,\DO^n\stimes\DOmega}$. Since $R$ is finite, we have $R \in \fparts{\ov{\wO{X_1}\stimes\cdots\stimes \wO{X_n}\stimes\ov{Y}^{\,\DOmega}}^{\,\DO^n\stimes\DOmega}}$.

		\item $(\supset)$. Let $R \in \fparts{\ov{\wO{X_1}\stimes\cdots\stimes \wO{X_n}\stimes\ov{Y}^{\,\DOmega}}^{\,\DO^n\stimes\DOmega}}$.
		Let $((d_1,\ldotsTwo,d_n), \delta) \in R$.

		Assume that $\forall i=1..n, d_i \in \wO{X_i}$. We need to show that $\delta \in Y$.

		By definition of $R$, we have $((d_1,\ldotsTwo,d_n), \delta) \in \ov{\wO{X_1}\stimes\cdots\stimes \wO{X_n}\stimes\ov{Y}^{\,\DOmega}}^{\,\DO^n\stimes\DOmega}$.

		This means $((d_1,\ldotsTwo,d_n), \delta) \not\in \wO{X_1}\stimes\cdots\stimes \wO{X_n}\stimes\ov{Y}^{\,\DOmega}$.

		Given our assumption that $\forall i=1..n, d_i \in \wO{X_i}$, the only way for $((d_1,\ldotsTwo,d_n), \delta)$ to not be in $\wO{X_1}\stimes\cdots\stimes \wO{X_n}\stimes\ov{Y}^{\,\DOmega}$ is if $\delta \not\in \ov{Y}^{\,\DOmega}$.

		Therefore, $\delta \in Y$, which proves that $R \in \fun{\p{X_1,\ldotsTwo,X_n}}{Y}$.\qedhere
	\end{itemize}
\end{Proof}

\newcommand{\Set}[1]{\{#1\}}

\noindent
Then, we want to give a formal proof of a result that is a generalization
of Lemma 4.6 of~\citet{frisch2004theorie}:

\begin{lem}\label{lem:multi-arity-diff}
	Let $X^{(1)},\ldotsTwo,X^{(n)}$ and, for $i\in P$, $Y_i^{(1)},\ldotsTwo,Y_i^{(n)}$ be sets (for instance, subsets of $\Dom$ or of $\DOmega$). Then, with the convention that an intersection over an empty $P$ is $X^{(1)}\stimes\ldotsTwo\stimes X^{(n)}$,
	\[\bigcap_{i\in P} (X^{(1)}\stimes\ldotsTwo\stimes X^{(n)}) \diff (Y_i^{(1)}\stimes\ldotsTwo\stimes Y_i^{(n)}) = \!\!\!\!
	\bigcup_{\iota:P\rightarrow[1,n]}
	\prod_{k=1}^n\,\, (X^{(k)} \diff\!\!\!\!\!\!\!\! \bigcup_{\Set{i\in P\mid \iota(i)=k}}\!\!\!\!\! Y_i^{(k)})
	\]
\end{lem}

\begin{Proof}
By reciprocal inclusion:
\begin{description}
	\item[$(\supseteq)$]
	Fix $\iota:P\to[1,n]$ and let $x=(x_1,\ldotsTwo,x_n)$ be in
	$\prod_{k=1}^n\big( X^{(k)} \diff \bigcup_{\{i\mid\iota(i)=k\}} Y_i^{(k)}\big)$.
	Then $x\in X^{(1)}\stimes\cdots\stimes X^{(n)}$.
	For every $i_0\in P$, set $k_0=\iota(i_0)$. Since
	$x_{k_0}\in X^{(k_0)}\,\diff\,\bigcup_{\{i\mid\iota(i)=k_0\}}Y_i^{(k_0)}$, we have
	$x_{k_0}\notin Y_{i_0}^{(k_0)}$, hence $x\notin Y_{i_0}^{(1)}\stimes\cdots\stimes Y_{i_0}^{(n)}$.
	Because $i_0$ was chosen arbitrary, and this property holds for all $i_0\in P$,we have
	$x\in\bigcap_{i\in P}\big((X^{(1)}\stimes\cdots\stimes X^{(n)})\diff (Y_i^{(1)}\stimes\cdots\stimes Y_i^{(n)})\big)$.
\item[$(\subseteq)$]
Let $x=(x_1,\ldotsTwo,x_n)$ belong to the left-hand side. Then for every $i\in P$,
$x\in (X^{(1)}\stimes\cdots\stimes X^{(n)})\diff (Y_i^{(1)}\stimes\cdots\stimes Y_i^{(n)})$,
so for every $i\in P$ there exists $k_i\in[1,n]$ with $x_{k_i}\notin Y_i^{(k_i)}$. Define $\iota_0$ to be the function that maps each $i\in P$ to the corresponding $k_i$, i.e., $\iota_0(i)=k_i$.
For every $k\in [1,n]$ we have $x_k\in X^{(k)}$ and for every $i$ with $\iota_0(i)=k$, we have $x_k\notin Y_i^{(k)}$; hence
$x_k\in X^{(k)}\diff\bigcup_{\{i\mid\iota_0(i)=k\}}Y_i^{(k)}$ (the union is $\varnothing$ if no such $i$).
Therefore, $x\in\prod_{k=1}^n\big(X^{(k)}\diff\bigcup_{\{i\mid\iota_0(i)=k\}}Y_i^{(k)}\big)$,
so $x$ belongs to the union of the right-hand side.\qedhere
\end{description}
\end{Proof}

\noindent
This allows us to prove the following theorem:

\begin{Theorem}\label{app:thm:set-containment-multi-arity}
  Let $n\in\mathbb{N}$, let $P$ and $N$ be finite sets, and let
  $(X_i^{(1)})_{i\in P}$, \ldots, ${(X_i^{(n)})}_{i\in P}$,
  $(X_i)_{i\in P}$, $(Y_i^{(1)})_{i\in N}$, \ldots, $(Y_i^{(n)})_{i\in N}$,
  $(Y_i)_{i\in N}$ be families of subsets of the domain $\Domain$.
  Then, for the function spaces of Definition~\ref{app:def:multi-arity},
\[\begin{array}{l}
 \displaystyle\Intersection{P}{\Function{\Arguments{X_i}}{X_i}}
        \subseteq
        \Union[j]{N}{\Function{\Arguments{Y_j}}{Y_j}}
    \quad \Leftrightarrow\quad  \exists j_0 \in N. \\
    \hspace*{8mm}\forall \iota : P \!\rightarrow\! [ 1,n+1 ].
    \left(\exists k{\in}[1,n].\,\, \wO{(Y_{j_0}^{(k)})} \!\subseteq
        \!\!\!\!\!\!\!\!\!\displaystyle
        \bigcup_{\{i\in P \,\mid\, \iota(i) = k\}}\!\!\!\!\!\! \wO{(X_i^{(k)})}\right)\\
    \hspace*{16mm}\textbf{or }\!\displaystyle
        \left(\bigcap_{\{i\in P \,\mid\, \iota(i) = n+1 \}} \!\!\!\!\!\!\!\!\!X_i \subseteq Y_{j_0}\right)
\end{array}\]
where a union over an empty set is $\varnothing$, an intersection of codomains
over an empty set is $\DOmega$, and the intersection of an empty family of
function spaces of arity $n$ is the set $\fparts{\product{\DO^n}{\DOmega}}$ of
all relations of arity $n$.
\end{Theorem}
For $\OO=\varnothing$, this is Theorem~\ref{thm:set-containment-multi-arity}. For $\OO=\{\invOmega\}$, since $\invOmega$ belongs to $\wO{(Y_{j_0}^{(k)})}$ and to every $\wO{(X_i^{(k)})}$, the condition $\wO{(Y_{j_0}^{(k)})} \subseteq \bigcup_{\{i\in P \,\mid\, \iota(i) = k\}} \wO{(X_i^{(k)})}$ holds if and only if some $i\in P$ has $\iota(i)=k$ and $Y_{j_0}^{(k)} \subseteq \bigcup_{\{i\in P \,\mid\, \iota(i) = k\}} X_i^{(k)}$: the first disjunct then requires $k$ to be in the image of $\iota$.


\begingroup
\setlength{\parindent}{0pt}
\noindent\textsc{Application: two binary arrows.} Before giving the proof of this theorem, we illustrate its main idea with a simple example: the containment between two binary arrow types.

Consider the statement of the theorem with these parameters:
\begin{itemize}
	\item $P = \{1\}$: We have a single function type on the left side of the containment
	\item $N = \{1\}$: We have a single function type on the right side
	\item $n = 2$: The functions have arity 2 (taking two arguments)
\end{itemize}

The theorem's statement provides a necessary and sufficient condition such that:
$$(X_{1}^{(1)}, X_{1}^{(2)}) \rightarrow X_1 \subseteq (Y_{1}^{(1)}, Y_{1}^{(2)}) \rightarrow Y_1$$

\noindent According to the theorem, this containment holds if and only if:
$$\exists j_0 \in N \text{ such that } \forall \iota: P \rightarrow [1,n+1]$$
and one of these conditions is satisfied:
\begin{itemize}

	\item $\exists k \in [1,n] \text{ such that } \wO{(Y_{j_0}^{(k)})} \subseteq \bigcup_{\{i \in P \mid \iota(i)=k\}} \wO{(X_{i}^{(k)})}$
	\item OR $\bigcap_{\{i \in P \mid \iota(i)=n+1\}} X_i \subseteq Y_{j_0}$
\end{itemize}

Since $N = 1$ (one function on the right), we have only one choice for $j_0$, which is 1. When no $i$ satisfies $\iota(i)=k$, the union is empty and the first condition for $k$ is $\wO{(Y_{1}^{(k)})}=\varnothing$, which holds if and only if $\OO=\varnothing$ and $Y_{1}^{(k)}=\varnothing$.

Since $P = 1$ (one function on the left), the functions to check are $\iota: \{1\} \rightarrow [1,3]$. There are three possibilities:

\begin{itemize}
	\setlength{\parindent}{0pt}

	\item Case 1: $\iota(1) = 1$

			\noindent First condition: $\exists k \in [1,2]$ such that $\wO{(Y_{1}^{(k)})} \subseteq \bigcup_{\{i \in P \mid \iota(i)=k\}} \wO{(X_{i}^{(k)})}$. For $k=1$ this is $Y_{1}^{(1)} \subseteq X_{1}^{(1)}$. For $k=2$, no $i$ satisfies $\iota(i)=2$, so this is $\OO=\varnothing$ and $Y_{1}^{(2)} = \varnothing$.

			\noindent
			Second condition: $\bigcap_{\{i \in P \mid \iota(i)=3\}} X_i \subseteq Y_1$.
			Since no $i$ satisfies $\iota(i) = 3$, this intersection is over an empty set, so this condition is false because the intersection is
			thus $\DOmega$ by convention, and $Y_1$ does not contain $\Omega$.

			For this case, the whole condition simplifies to: $(Y_{1}^{(1)} \subseteq X_{1}^{(1)}) \text{ OR } (\OO=\varnothing \text{ AND } Y_{1}^{(2)} = \varnothing)$

\item Case 2: $\iota(1) = 2$

First condition: symmetrically, it is $(\OO=\varnothing \text{ AND } Y_{1}^{(1)} = \varnothing) \text{ OR } (Y_{1}^{(2)} \subseteq X_{1}^{(2)})$.

			Second condition: Similar to Case 1, evaluates to false

			For this case, the whole condition simplifies to: $(\OO=\varnothing \text{ AND } Y_{1}^{(1)} = \varnothing) \text{ OR } (Y_{1}^{(2)} \subseteq X_{1}^{(2)})$

\item Case 3: $\iota(1) = 3$

First condition: no $i$ satisfies $\iota(i)\leq 2$, so both unions are empty and this condition is $\OO=\varnothing \text{ AND } (Y_{1}^{(1)} = \varnothing \text{ OR } Y_{1}^{(2)} = \varnothing)$.

Second condition: $\bigcap_{\{i \in P \mid \iota(i)=3\}} X_i \subseteq Y_1$. Since $\iota(1) = 3$, this becomes $X_1 \subseteq Y_1$

For this case, the condition simplifies to:
\[(\OO=\varnothing \text{ AND } (Y_{1}^{(1)} = \varnothing \text{ OR } Y_{1}^{(2)} = \varnothing)) \text{ OR } (X_1 \subseteq Y_1)\]
\end{itemize}

\textsc{Conclusion}

For the containment to hold, all three cases must be satisfied. If $\OO=\varnothing$ and $Y_{1}^{(1)}$ or $Y_{1}^{(2)}$ is empty, they all are; otherwise they reduce to three inclusions. Therefore:

$$(X_{1}^{(1)}, X_{1}^{(2)}) \rightarrow X_1 \subseteq (Y_{1}^{(1)}, Y_{1}^{(2)}) \rightarrow Y_1$$

if and only if:
$$\begin{array}{l}\bigl((Y_{1}^{(1)} \subseteq X_{1}^{(1)}) \text{ AND } (Y_{1}^{(2)} \subseteq X_{1}^{(2)}) \text{ AND } (X_1 \subseteq Y_1)\bigr)\\ \text{OR } \bigl(\OO=\varnothing \text{ AND } (Y_{1}^{(1)} = \varnothing \text{ OR } Y_{1}^{(2)} = \varnothing)\bigr)\end{array}$$

For $\OO=\{\invOmega\}$, and for $\OO=\varnothing$ when the domains $Y_{1}^{(1)}$ and $Y_{1}^{(2)}$ of the arrow on the right-hand side are not empty, this result confirms the standard contravariant/covariant subtyping rule for function types:
\begin{itemize}
	\item The parameter types must be contravariant (the $Y$ types must be subtypes of the $X$ types)
	\item The return type must be covariant ($X_1$ must be a subtype of $Y_1$)
\end{itemize}
For $\OO=\varnothing$, when one of the domains on the right-hand side is empty, the arrow on the right-hand side contains every relation of arity 2, and the containment holds whatever the codomains. For $\OO=\{\invOmega\}$, the input $\invOmega$ belongs to every domain, so the codomains are compared even when a domain is empty.
\endgroup

\begin{proof} Since all the arrows in the statement have arity $n$, their interpretations are subsets of the summand $\fparts{\product{\DO^n}{\DOmega}}$ of the relations of arity $n$, and we reason within this summand (in particular, the intersection over an empty $P$ is the whole summand). We use Theorems (4.7) and (4.8) from~\cite{frisch2004theorie}, the notation $\iota^{-1}(k) = \{i \in P \mid \iota(i)=k\}$, and Lemma~\ref{lemma:function-interpretation}.

	\begin{align*}
		& \bigcap_{i\in P} \left(X_{i}^{(1)},\ldots,X_{i}^{(n)}\right) \rightarrow X_i \subseteq \bigcup_{j\in N} \left( Y_{j}^{(1)},\ldots,Y_{j}^{(n)}\right) \rightarrow Y_j \\
		\overset{(\ref{lemma:function-interpretation})}{\Leftrightarrow} \, & \bigcap_{i\in P} \fparts{\ov{\wO{(X_{i}^{(1)})}\!\stimes\ldots\stimes\!\wO{(X_{i}^{(n)})}\!\stimes\!\ov{X_i}^{\DOmega}}^{\DO^n\stimes\DOmega}} \\
		& \qquad \subseteq \bigcup_{j\in N} \fparts{\ov{\wO{(Y_{j}^{(1)})}\!\stimes\ldots\stimes\!\wO{(Y_{j}^{(n)})}\!\stimes\!\ov{Y_j}^{\DOmega}}^{\DO^n\stimes\DOmega}} \\[3mm]
		\overset{(4.8)}{\Leftrightarrow}\, & \exists j_0 \in N.\,\, \bigcap_{i\in P} \ov{\wO{(X_{i}^{(1)})}\!\stimes\ldots\stimes\!\wO{(X_{i}^{(n)})}\!\stimes\!\ov{X_i}^{\DOmega}}^{\DO^n\stimes\DOmega} \\
		& \qquad \subseteq \ov{\wO{(Y_{j_0}^{(1)})}\!\stimes\ldots\stimes\!\wO{(Y_{j_0}^{(n)})}\!\stimes\!\ov{Y_{j_0}}^{\DOmega}}^{\DO^n\stimes\DOmega} \\
		\overset{(\ref{lem:multi-arity-diff})}{\Leftrightarrow}\, & \exists j_0\!\!\in\! N.\!\!\!\!\!\!\!\!\!\!\! \bigcup_{\iota\,: \,P \rightarrow [1,n+1]} \left( \prod_{k=1}^{n} \left( \ov{\bigcup_{i\in\iota^{-1}(k)}\!\!\!\!\! \wO{(X_{i}^{(k)})}}^{\DO}\right)\stimes\!\!\! \ov{\bigcup_{i\in \iota^{-1}(n+1)}\!\!\!\!\!\!\ov{X_{i}}^{\DOmega}}^{\DOmega}\!\! \,\right) \\
		& \qquad \subseteq \ov{\wO{(Y_{j_0}^{(1)})}\!\stimes\ldots\stimes\!\wO{(Y_{j_0}^{(n)})}\!\stimes\!\ov{Y_{j_0}}^{\DOmega}}^{\DO^n\stimes\DOmega} \\[2mm]
		\Leftrightarrow\, & \exists j_0 \in N.\!\!\!\!\!\!\!\! \bigcup_{\iota\,: \,P \rightarrow [1,n+1]}\left(\prod_{k=1}^{n} \Bigl(\wO{(Y_{j_0}^{(k)})}\cap\!\!\!\!\bigcap_{i\in\iota^{-1}(k)}\!\!\!\!\ov{\wO{(X_{i}^{(k)})}}^{\DO}\Bigr) \right.\\
		& \qquad \left. \stimes \Bigl(\ov{Y_{j_0}}^{\DOmega} \cap\!\!\!\!\!\!\bigcap_{i\in \iota^{-1}(n+1)}\!\!\!\!\!\! \!\!X_{i}\Bigr) \right) = \emptyset
	\end{align*}
	The equivalence above labeled \ref{lem:multi-arity-diff} is the application of Lemma~\ref{lem:multi-arity-diff}, with $n+1$ components, to the set
	$$\bigcap_{i\in P} (\overbrace{\DO\times\ldots\times\DO}^{n \text{ times}}\times\DOmega)\setminus(\wO{(X_{i}^{(1)})}\times\ldots\times \wO{(X_{i}^{(n)})}\times \ov{X_i}^{\DOmega})$$
	while the last equivalence is obtained by bringing the right-hand side set to the left-hand side, applying De Morgan's law on the negated unions, and applying the intersections of products component-wise. A union of products is empty if and only if each product is, and a product is empty if and only if one of its components is.

	Let a map $\iota : P \to [1,n+1]$, and $k \in [1,n]$.
	\begin{itemize}
		\setlength{\parindent}{0pt}
		\item \underline{How can the set $\wO{(Y_{j_0}^{(k)})} \cap
	\bigcap_{\{i\in P\,;\,\iota(i)=k\}}\ov{\wO{(X_{i}^{(k)})}}^{\DO}$
	be empty?}

	By De Morgan's law, this set is empty iff $\wO{(Y_{j_0}^{(k)})} \subseteq
	\bigcup_{\{i\in P\mid\iota(i)=k\}}\wO{(X_{i}^{(k)})}$. If $\{i \in P\mid\iota(i)=k\}$ is empty, the intersection is on an empty set and is $\DO$, the union is $\varnothing$, and the condition reads $\wO{(Y_{j_0}^{(k)})} = \varnothing$, that is, $\OO=\varnothing$ and $Y_{j_0}^{(k)} = \varnothing$. Since $k$ ranges over $[1,n]$, we obtain the first \textbf{or} clause of the theorem.

	\item \underline{How can the set $\ov{Y_{j_0}}^{\DOmega} \cap
	\bigcap_{\{i\in P\,;\,\iota(i)=n+1\}} X_{i}$
	be empty?}

	First, note that if there is no $i\in P$ such that $\iota(i)=n+1$, then the intersection is on an empty set and is $\DOmega$. Hence
	its intersection with $\ov{Y_{j_0}}^{\DOmega}$ is never empty, as it contains $\Omega$.

	If that is not the case, then this intersection is empty iff $\bigcap_{\{i\in P\mid\iota(i)=n+1\}} X_{i} \subseteq Y_{j_0}$.

	Finally, note that the convention that an intersection over the empty set over subsets of $\DOmega$ (which is the case for codomain sets) is $\DOmega$ takes into account the special case mentioned above.\qedhere
\end{itemize}

\end{proof}

\subsection{Subtyping algorithm}

In this subsection, $\leq$ is the subtyping relation induced by the interpretation of Definition~\ref{app:def:multi-arity}, for a fixed $\OO$, and we write $\beta$ for the Boolean $\tttrue$ if $\OO=\varnothing$ and $\ttfalse$ if $\OO=\{\invOmega\}$.
From the proof of Theorem~\ref{app:thm:set-containment-multi-arity} we see
that, if $P$ indexes arrows of arity $n$, the subtyping problem
\begin{equation}\label{app:eq:multi-arity}
	\bigwedge_{i\in P} \left(s_{1}^{(i)},\ldots,s_{n}^{(i)}\right)\rarr s^{(i)}
	\leq \bigvee_{j\in N} \left(t_{1}^{(j)},\ldots,t_{n}^{(j)}\right)\rarr t^{(j)}
\end{equation}
is decided by finding a single arrow (let us say it is $j=1$) on the right hand side such that
\begin{equation}\label{app:algo:multi-arity}
	\bigwedge_{i\in P} \left(s_{1}^{(i)},\ldots,s_{n}^{(i)}\right)\rarr s^{(i)} \,\,\leq\,\, \left(t_{1}^{(j)},\ldots,t_{n}^{(j)}\right)\rarr t^{(j)}
\end{equation}
Let us then define a function that for all $n\in\mathbb{N}$ decides $\eqref{app:algo:multi-arity}$.
This is expressed by function $\Phi_n$ of $n+2$ arguments: the first $n+1$ arguments are triples of a Boolean and two types, and the last one is a finite set of arrow types. The function is defined as follows:

\[
\scalebox{0.95}{$
	\begin{array}{lll}
         \Phi_n((b_1, t_1, s_1), \ldots , (b_n, t_n, s_n), (b, t, s), \emptyset) =
		 	(\exists i \in [1, n].\,\, b_i \textbf{ and } t_i \leq s_i) \textbf{ or } (b \textbf{ and } s \leq t) \\[1mm]
		 \Phi_n((b_1, t_1, s_1), \ldots , (b_n, t_n, s_n), (b, t, s), \{ (t_1',\ldotsTwo,t_n') \rightarrow t' \} \cup P) = \\[1mm]
		 \hspace{3.5cm}\Phi_n((b_1, t_1, s_1), \ldots , (b_n, t_n, s_n), (\tttrue, t, s \land t'), P) \textbf{ and} \\[1mm]
		 \hspace{3.5cm}\hfill{}\forall j =1..n.\, \Phi_n((b_1, t_1, s_1), \ldotsTwo ,(\tttrue, t_j, s_j \lor t_j' ), \ldotsTwo , (b_n, t_n, s_n), (b, t, s), P)
	\end{array}
$}
\]
\noindent
To motivate the definition of $\Phi_n$, the $t_j$ are meant to represent the $Y_{j_0}^{(j)}$ and, on the final call of the recursion, the $s_j$ hold each
$\bigcup_{\{i\in P\mid\iota(i)=j\}}X_{i}^{(j)}$, where $\iota$ represents the choice that was made at each recursive call. When no arrow was chosen for $j$, this union is empty, and $s_j$ is its initial value $\bott$. For $\OO=\varnothing$ the condition $t_j\leq s_j$ then holds exactly when $t_j$ is empty, as the first clause of Theorem~\ref{app:thm:set-containment-multi-arity} requires. For $\OO=\{\invOmega\}$ that clause cannot hold for $j$, since $\wO{(Y_{j_0}^{(j)})}$ contains $\invOmega$ and the empty union does not: the Boolean $b_j$, initially $\beta$, records whether some arrow was chosen for $j$, and the condition on $j$ is checked only when $b_j$ is true.

For $t$ and $s$ we consider instead the convention that the intersection over an empty set is $\DOmega$, and we add a Boolean variable $b$ to keep track of whether $n+1$ was chosen in a recursive call. If $b = \ttfalse$, that means that $n+1$ was never picked, and thus the condition $\bigcap_{\{i\in P\mid\iota(i)=n+1\}} X_{i} \subseteq Y_{j_0}$ is equal to $\DOmega \subseteq Y_{j_0}$, which is always false, since $\Omega$ is not in $Y_{j_0}$.

\begin{Theorem}\label{thm:arity-decide-acc} For all $n\in \mathbb{N}$, for $P$ a finite set of arrows of arity $n$ (when $P$ is empty, the intersection below is $\function_n$),
	\[
		\bigwedge_{f \in P} f \leq \left(t_1,\ldotsTwo,t_n\right) \rightarrow t
		\, \Leftrightarrow \,
		\Phi_n((\beta, t_1, \bott), \ldotsTwo, (\beta, t_n, \bott), (\ttfalse, t, \topp), P)
	\]
\end{Theorem}

\textsc{Application}. For $P = \{ (s_1, \ldotsTwo, s_n) \rightarrow s \}$, the call that consumes the only arrow in $P$ produces $n+1$ final calls: one for each
of the $n$ arguments, and one for the result type. The result call leads to the condition $(\exists i\in[1,n].\,\, \beta \textbf{ and } t_i \leq \bott) \textbf{ or } (s \leq t)$, and the call for the $j$-th argument to the condition $(\exists i\neq j.\,\, \beta \textbf{ and } t_i \leq \bott) \textbf{ or } (t_j \leq s_j)$.
Thus, we have:
\[\bigl((s \leq t) \textbf{ and } \forall i\in[1,n].\,\, (t_i \leq s_i)\bigr) \textbf{ or } \bigl(\beta \textbf{ and } \exists i\in[1,n].\,\, t_i \leq \bott\bigr)\]
which, for $n=2$, is the result of the application of Theorem~\ref{app:thm:set-containment-multi-arity} to two binary arrows.

\begin{Proof}
	\setlength{\parindent}{0pt}
	Write each arrow $f\in P$ as $(s_f^{(1)},\ldotsTwo,s_f^{(n)})\rightarrow s_f$, and, for a map $\iota : P \to [1,n+1]$, write $\iota^{-1}(k) = \{f\in P\mid \iota(f)=k\}$; an empty union of types is $\bott$ and an empty intersection of types is $\topp$. We prove, by induction on the size of $P$, that for all types $t_1,s_1,\ldotsTwo,t_n,s_n,t,s$ and all Booleans $b_1,\ldotsTwo,b_n,b$, the call $\Phi_n((b_1, t_1, s_1), \ldots , (b_n, t_n, s_n), (b, t, s), P)$ holds if and only if, for every map $\iota : P \to [1,n+1]$,
	\begin{equation}\tag{I}\label{app:eq:arity-invariant}
	\begin{array}{l}
	\Bigl(\exists k\in[1,n].\,\, \bigl(b_k \textbf{ or } \iota^{-1}(k)\neq\varnothing\bigr) \textbf{ and } t_k \leq s_k \lor \!\!\bigvee_{f\in\iota^{-1}(k)}\!\! s_f^{(k)}\Bigr)\\
	\textbf{or }
	\Bigl(\bigl(b \textbf{ or } \iota^{-1}(n+1)\neq\varnothing\bigr) \textbf{ and } s \land\!\!\!\! \bigwedge_{f\in\iota^{-1}(n+1)}\!\!\!\! s_f \leq t\Bigr)
	\end{array}
	\end{equation}

	\emph{Base case, $P=\varnothing$.} The only map is the empty one, all of whose preimages are empty: \eqref{app:eq:arity-invariant} is then the definition of $\Phi_n$ on $\varnothing$.

	\emph{Inductive case, $P = \{f'\}\cup P'$ with $f'\notin P'$ and $f' = (t'_1, \ldots, t'_n) \rightarrow t'$.} A map $\iota : P\to[1,n+1]$ is determined by the position $\iota(f')$ and by its restriction $\iota'$ to $P'$. If $\iota(f') = n+1$, then $\iota^{-1}(n+1)$ is not empty, $\bigwedge_{f\in\iota^{-1}(n+1)} s_f = t' \land \bigwedge_{f\in\iota'^{-1}(n+1)} s_f$, and $\iota^{-1}(k) = \iota'^{-1}(k)$ for $k\leq n$: so \eqref{app:eq:arity-invariant} for $\iota$ is \eqref{app:eq:arity-invariant} for $\iota'$ with the triple $(\tttrue, t, s\land t')$ in place of $(b,t,s)$. If $\iota(f') = j\in[1,n]$, then $\iota^{-1}(j)$ is not empty, $\bigvee_{f\in\iota^{-1}(j)} s_f^{(j)} = t'_j \lor \bigvee_{f\in\iota'^{-1}(j)} s_f^{(j)}$, and the other preimages are those of $\iota'$: so \eqref{app:eq:arity-invariant} for $\iota$ is \eqref{app:eq:arity-invariant} for $\iota'$ with the triple $(\tttrue, t_j, s_j\lor t'_j)$ in place of $(b_j,t_j,s_j)$. By the induction hypothesis, the first recursive call in the definition of $\Phi_n$ holds if and only if \eqref{app:eq:arity-invariant} holds for all the maps $\iota$ with $\iota(f')=n+1$, and the recursive call for $j\in[1,n]$ holds if and only if \eqref{app:eq:arity-invariant} holds for all the maps $\iota$ with $\iota(f')=j$. Since every map gives $f'$ exactly one position, the conjunction of these calls holds if and only if \eqref{app:eq:arity-invariant} holds for every map $\iota : P\to[1,n+1]$.

	\emph{Conclusion.} With $b_k=\beta$ and $s_k = \bott$ for every $k$, $b=\ttfalse$ and $s=\topp$, \eqref{app:eq:arity-invariant} reads: $\bigl(\exists k\in[1,n].\,\, (\beta$ or $\iota^{-1}(k)\neq\varnothing)$ and $t_k \leq \bigvee_{f\in\iota^{-1}(k)} s_f^{(k)}\bigr)$ or $\bigl(\iota^{-1}(n+1)\neq\varnothing$ and $\bigwedge_{f\in\iota^{-1}(n+1)} s_f \leq t\bigr)$. By Theorem~\ref{app:thm:set-containment-multi-arity} with $N$ a singleton, $\bigwedge_{f \in P} f \leq (t_1,\ldots,t_n) \rightarrow t$ holds if and only if this condition holds for every map $\iota$. Its first disjunct is the first clause of the theorem: for $\OO=\varnothing$ (and $\beta=\tttrue$) the clause is $t_k \leq \bigvee_{f\in\iota^{-1}(k)} s_f^{(k)}$, and for $\OO=\{\invOmega\}$ (and $\beta=\ttfalse$) it requires in addition $\iota^{-1}(k)\neq\varnothing$, as noted after the theorem. Its second disjunct is the second clause, which fails when $\iota^{-1}(n+1)=\varnothing$, since the intersection of an empty set of codomains is $\DOmega$.
\end{Proof}

Following~\citet{frisch2004theorie}, observe that arguments $s_i$ and $s$
are accumulators for the arguments and codomain of the arrows in $P$.
We can then define a function $\Phi'_n$ with fewer arguments (more precisely, where the first $n+1$ arguments are pairs of a Boolean and a type), by integrating those
into the parameters $t_i$ and $t$ using set operations.

Let us define
$\Phi'_n$:
\[
	\begin{array}{lll}
		\Phi'_n((b_1,t_1),\ldotsTwo,(b_n,t_n), (b, t), \varnothing)                                        =
		\left( \exists j\in\llbracket 1;n\rrbracket.\,\, b_j \textbf{ and } t_j \leq \bott \right) \textbf{ or }
 		\left( b \textbf{ and }  t \leq \bott \right)                                                                \\[1mm]
		\Phi'_n((b_1,t_1),\ldotsTwo,(b_n,t_n),(b,t),\{ (s_1,\ldotsTwo,s_n) \rightarrow s \} \cup P)  = \\
		\qquad\qquad ( \Phi'_n((b_1,t_1),\ldotsTwo,(b_n,t_n), (\tttrue,t \wedge s), P)\, \textbf{ and }                               \\
		\hspace{3.5cm}\hfill{}\forall j\in[1,n].\,\, \Phi'_n((b_1,t_1),\ldotsTwo,(\tttrue, t_j\smallsetminus s_j), \ldotsTwo, (b_n,t_n), (b, t), P)
		)
	\end{array}
\]
The link between $\Phi_n$ and $\Phi'_n$ is given by:
\[
	\Phi_n((b_1, t_1, s_1), \ldotsTwo, (b_n, t_n, s_n), (b, t, s), P) = \Phi'_n((b_1, t_1 \smallsetminus s_1), \ldotsTwo, (b_n, t_n \smallsetminus s_n), (b, s \smallsetminus t), P)
\]
which holds by induction on the size of $P$: for $P=\varnothing$, since $t_i\leq s_i$ if and only if $t_i\smallsetminus s_i\leq\bott$, and $s\leq t$ if and only if $s\smallsetminus t\leq\bott$; in the recursive case, since $(s\land t')\smallsetminus t = (s\smallsetminus t)\land t'$ and $t_j\smallsetminus(s_j\lor t'_j) = (t_j\smallsetminus s_j)\smallsetminus t'_j$, the Booleans being passed unchanged.

Thus we can use for the subtyping problem:

\begin{Theorem}\label{thm:arity-decide}
	For all $n\in \mathbb{N}$, for $P$ a finite set of arrows of arity $n$ (when $P$ is empty, the intersection below is $\function_n$),
	\[
		\bigwedge_{f \in P} f \leq (t_1,\ldotsTwo,t_n) \rightarrow t
		\,\, \Longleftrightarrow \,\,
		\Phi^\prime_n((\beta,t_1), \ldotsTwo, (\beta,t_n), (\ttfalse,\neg t), P)
	\]
\end{Theorem}

\begin{Proof}
	By Theorem~\ref{thm:arity-decide-acc} and the link between $\Phi_n$ and $\Phi'_n$, since $t_i\smallsetminus\bott = t_i$ and $\topp\smallsetminus t = \neg t$.
\end{Proof}

For $\OO=\varnothing$, $\beta=\tttrue$ and the Booleans $b_1,\ldotsTwo,b_n$ remain true in every recursive call; dropping them, $\Phi'_n$ is the function written $\Phi_n$ in Section~\ref{subsec:multi-arity}, and Theorem~\ref{thm:arity-decide} is the result cited there.

\section{Semantic subtyping: strong arrows}
\label{sec:strong-subtyping}

In this appendix we provide the necessary definitions and lemmas to extend the semantic subtyping framework of~\citet{frisch2004theorie} to strong arrows, according to the lines outlined in Section~\ref{sub:semantic-strong-arrows}.

As mentioned in the main text, we focus on the case for unary functions, as the extension to multi-arity functions is straightforward. As in Appendix~\ref{app:functions}, we treat at once the interpretation of Section~\ref{sub:semantic-strong-arrows} and its refinement by~\citet{lanvin2021semantic}: $\OO$ is either $\varnothing$ or $\{\invOmega\}$, $\DO=\Dom\cup\OO$, and $\wO{X}=X\cup\OO$. To this end we recall that, by Definition~\ref{app:def:multi-arity} with $n=1$, for $X,Y\subseteq \Domain$ we have
\[X\to Y = \{R\in \fparts{\DO\times\DOmega} \mid \forall (d,\delta)\in R.\,\, d\in \wO{X} \Rightarrow \delta\in Y\}\]
(for $\OO=\varnothing$, this is \cite[Definition 4.2]{frisch2004theorie}), and that by Lemma~\ref{lemma:function-interpretation} we have
\[X\rarr Y =\fparts{\ov{\wO{X}\times\ov{Y}^{\DOmega}}^{\DO\times\DOmega}}.\]

\subsection{Set Semantics}
\label{app:ss-strong}
We interpret strong arrows as in Section~\ref{sub:semantic-strong-arrows}.
\begin{Definition}\label{app:def:strong-arrow}
	Let $X$ and $Y$ be subsets of $\Domain$. The strong arrow $\StrongType{(X\rarr Y)}$ is the set of relations
	\[
		\StrongType{(X\rarr Y)} \eqdef (X\rarr Y)\cap\fparts{\DO\times(Y\cup\{\Omega\})}.
	\]
	For a set $\mathcal{X}$ of finite relations, we also write $\textsf{dom}(\mathcal{X}) = \{ d \in \DO \mid \forall R \in \mathcal{X}.\,\, (d,\Omega) \notin R \}$.
\end{Definition}
For $\OO=\varnothing$, this is the interpretation of Section~\ref{sub:semantic-strong-arrows}. For $\OO=\{\invOmega\}$, the second set of the intersection contains the relations with input $\invOmega$: so, in both interpretations, $\StrongType{(X\rarr Y)}\subseteq X\rarr Y$ and $\Dom\rarr Y\subseteq\StrongType{(\Dom\rarr Y)}$, since a relation of $\Dom\rarr Y$ has all its outputs in $Y$. In both interpretations, $\StrongType{(X\rarr Y)}$ is antitone in $X$ and monotone in $Y$, since $X\rarr Y$ is and the second set of the intersection is monotone in $Y$.

\newcommand{\CoDomain}{\DOmega}

\paragraph{Main subtyping decision problem (set formulation).}
Fix finite families of base sets $(X_i,Y_i)_{i\in I}$ and $(U_j,V_j)_{j\in J}$,
with $X_i,U_j\subseteq \Domain$ (domains) and $Y_i,V_j \subseteq \Domain$ (codomains).
Write $(X\!\to\!Y)$ for ordinary arrows and $\StrongType{(X\!\to\!Y)}$ for strong arrows.
By definition above, $\StrongType{(X\!\to\!Y)}$ is the set of relations
$R\subseteq \DO\times(Y\cup\{\Omega\})$ that are ``strong'' in the sense that on $d\in \wO{X}$ one must return in $Y$, on $d\notin \wO{X}$ one may return in $Y$, or be $\Omega$.
We build \emph{arrow literals} by allowing negation $\neg$ in front of either kind:
\[
\ell \;::=\; (X_i \rarr Y_i)
\;\mid\; \neg(X_i \rarr Y_i)
\;\mid\; \StrongType{(U_j \rarr V_j)}
\;\mid\; \neg\,\StrongType{(U_j \rarr V_j)}.
\]
A generic subtyping query $A \subseteq B$ is equivalent to testing the \emph{emptiness}
of a finite union of finite intersections of such literals (a DNF over arrow literals):
\[
\underbrace{\bigcup_{p=1}^{m}\;\bigcap_{q=1}^{n_p}\; \ell_{p,q}}_{\text{union of intersections of (strong/ordinary) arrows and their negations}}
\hspace{-2cm} \;=\; \emptyset,
\quad\text{each }\ell_{p,q}\text{ of the four forms above,}
\]
where $\neg$ denotes complement with respect to a fixed ambient universe of arrows.
All the lemmas that follow are devoted to rewriting such expressions—collapsing
intersections of strong arrows, translating mixed intersections/unions into base-set
conditions, and eliminating the remaining finite unions—until the emptiness test
reduces to inclusions between the underlying sets $X_i,Y_i,U_j,V_j$.

First, notice that the finite union of finite intersections above is empty if
and only if each of the intersections is empty. Thus, we reduce to solving the generic problem $\bigcap_{q=1}^{n}\; \ell_{q} = \emptyset$
which is thus, if we write the literals as:
\begin{equation}\tag{$\ast_0$}\label{eq:star0}
	\p{\bigcap_{i\in P} X_i \rarr Y_i}
	\cap
	\p{\bigcap_{i\in N} \neg (X_i \rarr Y_i)}
	\cap
	\p{\bigcap_{j\in Q} \StrongType{(U_j \rarr V_j)}}
	\cap
	\p{\bigcap_{j\in R} \neg \StrongType{(U_j \rarr V_j)}}
	= \emptyset
\end{equation}
Through De Morgan's laws, we know that each intersection of negated arrows is equivalent to the complement of the union of the arrows, thus:
\begin{align*}
	\bigcap_{i\in N} \neg\bigl(X_i \rarr Y_i\bigr)
	&= \fparts{\DO\times\CoDomain}
	   \smallsetminus
	   \Bigl(\,\bigcup_{i \in N} X_i \rarr Y_i\Bigr)\\[1ex]
	\intertext{and}
	\bigcap_{j\in R} \neg \StrongType{(U_j \rarr V_j)}
	&= \fparts{\DO\times\CoDomain}
	   \smallsetminus
	   \Bigl(\,\bigcup_{j \in R} \StrongType{(U_j \rarr V_j)}\Bigr)
	\end{align*}
Using this, we can rewrite equation \eqref{eq:star0} into the following set-inclusion problem:
\begin{equation}\tag{$\ast_1$}\label{eq:star1}
\p{\bigcap_{i\in P} X_i \rarr Y_i} \cap
\p{\bigcap_{j \in Q} \StrongType{(U_j \rarr V_j)}} \subseteq
\p{\bigcup_{i \in N}  X_i \rarr Y_i} \cup
\p{\bigcup_{j \in R} \StrongType{(U_j \rarr V_j)}}
\end{equation}

Our first step then is to introduce Lemma~\ref{lem:strong-intersection} as a \emph{normalisation step}: for finite non-empty $I$ it collapses
\(
\bigcap_{i\in I}\StrongType{(X_i\!\to\!Y_i)}
\)
into a single strong arrow
\(
\StrongType{(\bigcup_{i\in I}X_i \to \bigcap_{i\in I}Y_i)}
\).
This removes an outer intersection on strong arrows and leaves one arrow with a
union in the domain and an intersection in the codomain.

\begin{lem}\label{lem:strong-intersection}
	Let $I$ be a finite non-empty set, then we have:
	\begin{equation}
		\bigcap_{i \in I} \StrongType{\p{X_i \rarr Y_i}}
		\quad = \quad
		\StrongType{\p{\Function{\Union{I}{X_i}}{\Intersection{I}{Y_i}}}}
	\end{equation}
\end{lem}

\noindent
Proving both inclusions.
	\begin{description}
		\item[($\supseteq$)] Suppose $R \in \StrongSmall{\bigcup_{i\in I}X_i\rarr\bigcap_{i\in I} Y_i}$. By Definition~\ref{app:def:strong-arrow}, $R\subseteq\DO\times(\WithOmega{(\bigcap_{i\in I} Y_i)})$.
		      Let $(d,\delta)\in R$. Let $i_0\in I$.
		      \begin{itemize}
			      \item if $d\in \wO{X_{i_0}}$, then $d\in \wO{(\bigcup_{i\in I}X_i)}$ so
			            $\delta\in \bigcap_{i\in I}{Y_i} \subseteq Y_{i_0}$ by definition of $R$.
			      \item if $d\not\in \wO{X_{i_0}}$, then the strongness property of the arrow interpretation $R$ belongs to ensures
			            that $\delta \in \WithOmega{(\bigcap_{i\in I}Y_i)}\subseteq \WithOmega{Y_{i_0}}$.
		      \end{itemize}
		\item[($\subseteq$)] Now, suppose $R\in \bigcap_{i\in I}{\SmallStrong{\Function{X_i}{Y_i}}}$.
		      Let $(d,\delta)\in R$.
		      \begin{itemize}
			      \item if $d\in \wO{(\UnionI{X_i})}$, which, since $I$ is not empty, is $\bigcup_{i\in I}\wO{X_i}$. For all $i\in I$, either
			            $d\in \wO{X_i}$, thus $\delta\in Y_i$, or $d\not\in \wO{X_i}$, thus
			            $\delta\in \WithOmega{Y_i}$. Since there exists at least
			            one $j_0$ such that $d\in \wO{X_{j_0}}$, then we know that $\delta\in Y_{j_0}$. Since $Y_{j_0}$ does not contain
			            $\Omega$, then $\delta\not=\Omega$ and thus we have proven that $\delta \in \IntersectionI{Y_i}$.
			      \item if $d\not\in \wO{(\UnionI{X_i})}$, we can    apply the same reasoning as in the previous case,
			            except that we cannot deduce that $\delta\not=\Omega$. We thus
			            have $\delta\in \IntersectionI{\WithOmega{Y_i}}$.\qed%
		      \end{itemize}
\end{description}
For an empty family, the intersection is the set $\fparts{\DO\times\DOmega}$ of all relations of arity 1, while $\StrongType{(\varnothing\rarr\Dom)}$ is $\varnothing\rarr\Dom$, since $\Dom\cup\{\Omega\}=\DOmega$. For $\OO=\varnothing$ the two coincide, so the lemma also holds for $I=\varnothing$. For $\OO=\{\invOmega\}$ they do not: the relation $\{(\invOmega,\Omega)\}$ belongs to the first and not to the second. So, for $\OO=\{\invOmega\}$, the empty intersection of strong arrows, $\function_1$, is not a strong arrow, and we treat the case without strong arrows on the left separately (Lemma~\ref{lem:weak-product} and Theorem~\ref{thm:strong-no-strong-left} below).

When $Q$ is not empty, the application of this lemma to equation \eqref{eq:star1} means that instead
of having a finite intersection of strong arrows, the generic problem is reduced to a single strong arrow.
\begin{equation}\tag{$\ast_2$}\label{eq:star2}
		\p{\bigcap_{i\in P} X_i \rarr Y_i} \cap
		\StrongType{(U\rarr V)} \subseteq
		\p{\bigcup_{i \in N}  X_i \rarr Y_i} \cup
		\p{\bigcup_{j \in R} \StrongType{(U_j \rarr V_j)}}
\end{equation}
where we abbreviate
\[
U \;\coloneqq\; \bigcup_{j\in Q} U_j
\qquad\text{and}\qquad
V \;\coloneqq\; \bigcap_{j\in Q} V_j.
\]
When $Q$ is empty, the steps below apply in the same way to~\eqref{eq:star1}, with $\StrongType{(U\rarr V)}$ replaced by the set $\fparts{\DO\times\DOmega}$ of all relations of arity 1 (that is, with $E_0=E_1=\DO\times\DOmega$ below).
This latter form of~\eqref{eq:star2} can be seen as a set inclusion between (finite) intersections and unions of \emph{parts} (we recall that $\StrongType{(X\to Y)}$ is, by definition, the intersection of $X\to Y = \fparts{\ov{\wO{X}\times\ov{Y}^{\DOmega}}^{\DO\times\DOmega}}$ with $\fparts{\DO\times(\WithOmega{Y})}$). Indeed, by introducing sets $E_i, F_i, G_i, H_i$ via the identifications:
\begin{itemize}[nosep]
	\item for all $i \in P$, $X_i \rarr Y_i = \fparts{E_i}$;
	\item $\StrongType{(U \rarr V)} = \fparts{E_0} \cap \fparts{E_1}$;
	\item for all $i \in N$, $X_i \rarr Y_i = \fparts{F_i}$;
	\item for all $j \in R$, $\StrongType{(U_j \rarr V_j)} = \fparts{G_j} \cap \fparts{H_j}$.
\end{itemize}
we can rewrite~\eqref{eq:star2} as:
\[
\bigcap_{i\in P \cup \{0, 1\}}\fparts{E_i}
\;\subseteq\;
\bigcup_{i\in N}\fparts{F_i}
\;\cup\;
\bigcup_{j\in R}\fparts{G_j}\cap\fparts{H_j}
\]
There, Lemma~\ref{lem:exists-parts} will convert the global inclusion
in~\eqref{eq:star2} into an \emph{existential witness} on base sets:
\[
\exists i_0\in N:\ \bigcap_{i\in P\cup \{0, 1\}}E_i\subseteq F_{i_0}
\quad\text{or}\quad
\exists j_0\in R:\ \bigcap_{i\in P\cup \{0, 1\}}E_i\subseteq G_{j_0}\cap H_{j_0},
\]
or equivalently (by Lemma~\ref{lem:powerset-intersection-criterion}):
\[
\exists i_0\in N:\ \bigcap_{i\in P\cup \{0, 1\}}\fparts{E_i}\subseteq \fparts{F_{i_0}}
\quad\text{or}\quad
\exists j_0\in R:\ \bigcap_{i\in P\cup \{0, 1\}}\fparts{E_i}\subseteq \fparts{G_{j_0}}\cap \fparts{H_{j_0}}
\]
so, after these transformations, going back to arrow forms, we obtain
\begin{equation}\tag{$\ast_3$}\label{eq:star3}
	\begin{aligned}
	\exists i_0\in N:\ \p{\bigcap_{i\in P} X_i \rarr Y_i}\cap \StrongType{(U\rarr V)}
	&\subseteq X_{i_0} \rarr Y_{i_0}\\
	\text{or}\quad \exists j_0\in R:\ \p{\bigcap_{i\in P} X_i \rarr Y_i}\cap \StrongType{(U\rarr V)}
	&\subseteq \StrongType{(U_{j_0}\rarr V_{j_0})}
	\end{aligned}
\end{equation}
which means that we are left to decide that the intersection of arrows on the
left-hand side is contained in either
\begin{enumerate}
	\item[i)] a single weak arrow (Lemma~\ref{lem:strong-arrow}), or
	\item[ii)] a single strong arrow $\StrongType{(U_{j_0}\rarr V_{j_0})}$, which is in fact the
intersection of the weak arrow $(U_{j_0}\rarr V_{j_0})$ with the
powerset $\fparts{\DO\times(\WithOmega{V_{j_0}})}$.
Since a set is included in the intersection of two sets if and only if it
is included in both of them, the problem reduces to deciding both
inclusion in a weak arrow (already done) and a given powerset. The latter
is treated in Lemma~\ref{lem:strong-product}.
\end{enumerate}
When $Q$ is empty, Theorem~\ref{app:thm:set-containment-multi-arity} with $n=1$ and Lemma~\ref{lem:weak-product} play the roles of Lemmas~\ref{lem:strong-arrow} and~\ref{lem:strong-product}.

\begin{lem}[Powerset–intersection criterion]\label{lem:powerset-intersection-criterion}
	Let ${(A_i)}_{i\in I}$ and $B$ be abstract sets.
	\[
	\bigcap_{i\in I}\fparts{A_i}\ \subseteq\ \fparts{B}
	\quad\Longleftrightarrow\quad
	\bigcap_{i\in I} A_i\ \subseteq\ B.
	\]
	\end{lem}

	\begin{proof}
	Two basic facts:
	\[
	\text{(i)}\quad \bigcap_{i\in I}\fparts{A_i}=\fparts{\bigcap_{i\in I}A_i}
	\qquad
	\text{and}
	\qquad
	\text{(ii)}\quad \fparts{S}\subseteq \fparts{T}\ \Longleftrightarrow\ S\subseteq T.
	\]
	For (i): $X\in\bigcap_i\fparts{A_i}$ iff $X\subseteq A_i$ for all $i$, i.e. $X\subseteq\bigcap_i A_i$, i.e. $X\in\fparts{\bigcap_i A_i}$.

	\noindent
	For (ii): “$\Rightarrow$” take any $x\in S$, then $\{x\}\in\fparts{S}\subseteq\fparts{T}$ so $x\in T$; the converse is monotonicity of $\fparts{-}$.

	\noindent
	Now combine (i) and (ii):
	\[
	\bigcap_{i}\fparts{A_i}\subseteq \fparts{B}
	\ \Longleftrightarrow\
	\fparts{\bigcap_{i}A_i}\subseteq \fparts{B}
	\ \Longleftrightarrow\
	\bigcap_{i}A_i\subseteq B.\qedhere
	\]
	\end{proof}

\begin{lem}\label{lem:exists-parts}
	\begin{align*}
		 & \bc_{i\in I} \fparts{X_i} \sq \bu_{i\in P} \fparts{Y_i}
		\cup \bu_{i\in Q} \fparts{Z_i} \cap \fparts{W_i}               \\
		 & \iff
		\p{\exists i_0\in P.\, \bc_{i\in I} X_i \sq Y_{i_0}} \lor
		\p{\exists i_0\in Q.\, \bc_{i\in I} X_i \sq Z_{i_0} \cap W_{i_0}} \\
	\end{align*}
\end{lem}

\begin{proof}
	We spell out the proof using only the (finite) powerset reading of the
operators:
\[
  U \in \fparts{Y}\iff \bigl(U\subseteq Y\ \wedge\ U\text{ is finite}\bigr).
\]
Membership in (co)products of families is also expanded as usual:
\[
  U\in \bc_{i\in I}\mathcal{A}_i \iff \forall i\in I,\ U\in\mathcal{A}_i,
  \qquad
  U\in \bu_{k\in K}\mathcal{B}_k \iff \exists k\in K,\ U\in\mathcal{B}_k.
\]
We further use the facts (immediate from the two displays above):
\begin{itemize}
  \item[(F1)] $U\in \bc_{i\in I}\fparts{X_i}$ iff $U$ is finite and $U\subseteq \bigcap_{i\in I}X_i$.
  \item[(F2)] $U\in \bu_{i\in P}\fparts{Y_i}$ iff $\exists i\in P,\ U\subseteq Y_i$ and $U$ is finite.
  \item[(F3)] $U\in \bu_{i\in Q}\bigl(\fparts{Z_i}\cap\fparts{W_i}\bigr)$ iff
  $\exists i\in Q,\ \bigl(U\subseteq Z_i\ \wedge\ U\subseteq W_i\bigr)$ (and $U$ is finite).
\end{itemize}

\smallskip
\noindent\emph{($\Rightarrow$)} Assume
\begin{equation}\label{eq:dagger}
  \bc_{i\in I} \fparts{X_i} \ \sq\
  \Bigl(\,\bu_{i\in P} \fparts{Y_i}\Bigr)\ \cup\
  \Bigl(\,\bu_{i\in Q} \fparts{Z_i} \cap \fparts{W_i}\Bigr).
  \tag{$\dagger$}
\end{equation}
We prove the right-hand disjunction of the statement. We reason by contradiction.
Suppose
\begin{equation}\label{eq:negP}
  \neg\exists i_0\in P.\ \bigcap_{i\in I}X_i \subseteq Y_{i_0}
  \quad\text{and}\quad
  \neg\exists i_0\in Q.\ \bigcap_{i\in I}X_i \subseteq Z_{i_0}\cap W_{i_0}.
\end{equation}
From the first negation, for each $p\in P$ choose a witness
$x_p\in \bigl(\bigcap_{i\in I}X_i\bigr)\setminus Y_p$.
From the second negation, for each $q\in Q$ choose a witness
$t_q\in \bigl(\bigcap_{i\in I}X_i\bigr)\setminus (Z_q\cap W_q)$; i.e.,
for each $q$ we have either $t_q\notin Z_q$ or $t_q\notin W_q$.

Now set the \emph{finite} set
\[
  U \;\;:=\;\; \{x_p \mid p\in P\}\ \cup\ \{t_q \mid q\in Q\}.
\]
(Here we use that $P$ and $Q$ are finite index sets of the finite unions on the right-hand side.)
By construction $U\subseteq \bigcap_{i\in I}X_i$ and $U$ is finite, hence by (F1)
\[
  U\in \bc_{i\in I}\fparts{X_i}.
\]
We now show $U$ is \emph{not} in the right-hand side of~\eqref{eq:dagger}, contradicting~\eqref{eq:dagger}.

\begin{itemize}
  \item For any $p\in P$, we have $x_p\in U$ with $x_p\notin Y_p$, hence $U\not\subseteq Y_p$.
        Therefore, by (F2), $U\notin \fparts{Y_p}$ for all $p\in P$, and thus
        $U\notin \bu_{i\in P}\fparts{Y_i}$.
  \item For any $q\in Q$, by choice of $t_q$ we have $t_q\notin Z_q$ or $t_q\notin W_q$, hence
        $U\not\subseteq Z_q$ or $U\not\subseteq W_q$; therefore $U\notin \fparts{Z_q}$ or
        $U\notin \fparts{W_q}$, so $U\notin \fparts{Z_q}\cap\fparts{W_q}$.
        Consequently, by (F3), $U\notin \bu_{i\in Q}\bigl(\fparts{Z_i}\cap\fparts{W_i}\bigr)$.
\end{itemize}
Thus $U$ is in the left-hand side of~\eqref{eq:dagger} but in neither component of the right-hand side,
contradicting~\eqref{eq:dagger}. The contradiction shows that~\eqref{eq:negP} is false; hence
\[
  \bigl(\exists i_0\in P.\ \bc_{i\in I}X_i \subseteq Y_{i_0}\bigr)\ \ \lor\ \
  \bigl(\exists i_0\in Q.\ \bc_{i\in I}X_i \subseteq Z_{i_0}\cap W_{i_0}\bigr).
\]

\smallskip
\noindent\emph{($\Leftarrow$)} We prove each disjunct implies the inclusion.
\begin{itemize}
  \item If $\exists i_0\in P$ with $\bc_{i\in I}X_i \subseteq Y_{i_0}$, then by monotonicity of $\fparts{-}$ and (F1)--(F2),
        \[
          \bc_{i\in I}\fparts{X_i} \;=\; \fparts{\bc_{i\in I}X_i}
          \;\subseteq\; \fparts{Y_{i_0}} \;\subseteq\; \bu_{i\in P}\fparts{Y_i},
        \]
        hence the desired inclusion holds.
  \item If $\exists i_0\in Q$ with $\bc_{i\in I}X_i \subseteq Z_{i_0}\cap W_{i_0}$, then by
        monotonicity of $\fparts{-}$ and (F1) we have
        \[
          \bc_{i\in I}\fparts{X_i} \;\subseteq\; \fparts{\bc_{i\in I}X_i}
          \;\subseteq\; \fparts{Z_{i_0}\!\cap W_{i_0}}
          \;=\; \fparts{Z_{i_0}}\cap \fparts{W_{i_0}}
          \;\subseteq\; \bu_{i\in Q}\fparts{Z_i}\cap\fparts{W_i}
        \]
        yielding the desired inclusion.
\end{itemize}

\smallskip
Combining the two directions establishes the equivalence.
\end{proof}

\paragraph{Remark.}
The proof of the ``$\Rightarrow$'' direction only uses that the unions on the right are \emph{finite}
(so that one can collect finitely many counterexamples into a single finite set $U$) together with
the finite-character readings of $\parts{(-)}$ and $\fparts{(-)}$; this is precisely the
``strongness'' exploited in Lemmas~\ref{lem:strong-arrow} and~\ref{lem:strong-product}.

\begin{lem}\label{lem:strong-arrow}
	\begin{equation*}
		\Intersection{I}{\p{X_i\rarr Y_i}} \cap \StrongType{\p{U\rarr V}} \!
		\sq \p{W\rarr Z}
		\Leftrightarrow
		\begin{cases}
			\wO{W} \sq \bu_{i\in I} \wO{X_i} \cup \wO{U}                                  \\[4mm]
			\forall J \subseteq I.
			( \wO{W} \sq \displaystyle\bu_{j\in J} \wO{X_j} )
			\lor (\displaystyle\bc_{j \in I\setminus J} Y_j \cap V \sq Z ) \\[-4mm]
		\end{cases}
	\end{equation*}
	where a union over an empty $J$ is $\varnothing$.
\end{lem}
\begin{proof}
Write, for $A,B\subseteq\Domain$,
\[
A\!\to\!B \;\;=\;\; \fparts{\overline{\wO{A}\times \overline{B}^{\DOmega}}^{\DO\times\DOmega}}
\;\;=\;\; \fparts{\,(\overline{\wO{A}}^{\DO}\!\times\!\DOmega)\ \cup\ (\DO\!\times\!B)\,}.
\]
Thus, for every $d\in\DO$ and $\delta\in\CoDomain$,
\begin{equation}\label{eq:arrow-fiber}
(d,\delta)\in A\!\to\!B \;\text{iff}\; \bigl(d\notin \wO{A}\bigr)\ \text{or}\ \bigl(\delta\in B\bigr).
\end{equation}
By Definition~\ref{app:def:strong-arrow},
\[
\StrongType{(U\!\to\!V)}\;=\; (U\!\to\!V)\ \cap\ \fparts{\DO\times (V\cup\{\Omega\})}.
\]
Set
\[
\mathcal{L}\;\coloneqq\; \bigcap_{i\in I}(X_i\!\to\!Y_i)\ \cap\ \StrongType{(U\!\to\!V)}.
\]

\paragraph*{A useful computation (domain).}
For any arrow $A\!\to\!B$, using~\eqref{eq:arrow-fiber} we have
$(d,\Omega)\in A\!\to\!B$ iff $d\notin \wO{A}$; hence
\[
\dom{A\!\to\!B}=\{d\in\DO\mid (d,\Omega)\notin A\!\to\!B\}=\wO{A}.
\]
Since $\fparts{-}$ is closed under intersections,
\[
\dom{\mathcal{L}}
= \{d\in\DO\mid (d,\Omega)\notin \textstyle\bigcap_{i\in I}(X_i\!\to\!Y_i)\cap (U\!\to\!V)\}
= \Bigl(\,\bu_{i\in I} \wO{X_i}\Bigr) \cup \wO{U}. \tag{$\ast$}\label{eq:dom-L}
\]

\medskip
\noindent\textbf{($\Rightarrow$) Necessity.}
Assume $\mathcal{L}\sq W\!\to\!Z$.

\emph{(i) Domain side.}
Since $W\!\to\!Z$ forbids $(d,\Omega)$ for all $d\in \wO{W}$, inclusion $\mathcal{L}\sq W\!\to\!Z$
forces $\wO{W}\sq \dom{\mathcal{L}}$. By~\eqref{eq:dom-L}, this is
$\wO{W} \sq \bigl(\bu_{i\in I} \wO{X_i}\bigr)\cup \wO{U}$.

\emph{(ii) Codomain side.}
Fix $J\subseteq I$ and suppose, for contradiction, that
$\wO{W} \not\sq \bu_{j\in J} \wO{X_j}$ and $\bigl(\bc_{j\in I\setminus J}Y_j \cap V\bigr)\not\sq Z$.
Then pick $w\in \wO{W}\setminus \bigl(\bu_{j\in J} \wO{X_j}\bigr)$ (for $\OO=\{\invOmega\}$ and $J=\varnothing$, $w$ may be $\invOmega$) and
$z\in \bigl(\bc_{j\in I\setminus J}Y_j \cap V\bigr)\setminus Z$.
Consider the partial function $R=\{(w,z)\}$.
By construction and \eqref{eq:arrow-fiber}:
\begin{itemize}
\item For every $j\in J$, since $w\notin \wO{X_j}$, we have $\{(w,z)\}\in X_j\!\to\!Y_j$.
\item For every $i\in I\setminus J$, since $z\in Y_i$, we have $\{(w,z)\}\in X_i\!\to\!Y_i$.
\item As $z\in V$, $\{(w,z)\}\in U\!\to\!V$, and also $(w,z)\in \DO\times(V\cup\{\Omega\})$,
so $\{(w,z)\}\in\StrongType{(U\!\to\!V)}$.
\end{itemize}
Hence $R\in\mathcal{L}$. But $w\in \wO{W}$ and $z\notin Z$, so $R\notin W\!\to\!Z$, contradicting
$\mathcal{L}\sq W\!\to\!Z$. Therefore, for each $J\subseteq I$,
\[
\bigl(\wO{W} \sq \textstyle\bu_{j\in J} \wO{X_j}\bigr)\ \ \text{or}\ \
\bigl(\bc_{j\in I\setminus J} Y_j \cap V \sq Z\bigr).
\]

\medskip
\noindent\textbf{($\Leftarrow$) Sufficiency.}
Assume the two conditions in the statement. Let $R\in\mathcal{L}$ and fix $(w,o)\in R$
with $w\in \wO{W}$. We prove $o\neq\Omega$ and $o\in Z$, which by~\eqref{eq:arrow-fiber} gives $\{(w,o)\}\in W\!\to\!Z$.

First, $o\neq\Omega$: if $o=\Omega$, then $(w,\Omega)\in R\in\mathcal{L}\subseteq \bigcap_{i\in I}(X_i\!\to\!Y_i)\cap(U\!\to\!V)$
implies $w\notin \wO{X_i}$ for all $i\in I$ (by~\eqref{eq:arrow-fiber}) and $w\notin \wO{U}$,
hence $w\notin \bigl(\bu_{i\in I}\wO{X_i}\bigr)\cup \wO{U}$; this contradicts
$w\in \wO{W}\sq \bigl(\bu_{i\in I}\wO{X_i}\bigr)\cup \wO{U}$.

Now $o\in\Domain$. Define $J\subseteq I$ by $J\coloneqq\{\,i\in I \mid o\notin Y_i\,\}$.
Because by construction $(w,o)\in X_i\!\to\!Y_i$ for every $i\in I$,~\eqref{eq:arrow-fiber} yields
$w\notin \wO{X_j}$ for all $j\in J$, hence $w\notin \bu_{j\in J}\wO{X_j}$.
Applying the second condition (with this $J$), the left disjunct is impossible (since $w\in \wO{W}$),
so the right disjunct must hold:
\[
\bc_{i\in I\setminus J} Y_i \cap V \sq Z.
\]
But $(w,o)\in\StrongType{(U\!\to\!V)}$ forces $o\in V\cup\{\Omega\}$ and we already ruled out $o=\Omega$,
hence $o\in V$. Moreover, by the choice of $J$, $o\in Y_i$ for all $i\in I\setminus J$.
Therefore $o\in \bigl(\bc_{i\in I\setminus J} Y_i \cap V\bigr)\sq Z$.

Since this holds for every $(w,o)\in R$ with $w\in \wO{W}$, we have $R\in W\!\to\!Z$.
As $R\in\mathcal{L}$ was arbitrary, $\mathcal{L}\sq W\!\to\!Z$.

\medskip
Combining both directions yields the equivalence.
\end{proof}

\begin{remark}
This lemma can also be proved using a technique similar to Lemma 4.9 from p.73 of Thesis~\cite{frisch2004theorie}, that is, by directly decomposing the intersection of finite parts of set product complements as a union.
\end{remark}

\begin{lem}\label{lem:strong-product}
	\begin{align*}
		 & \bigcap_{i\in I} (X_i \rarr Y_i) \cap \StrongType{\p{U \rarr V}}
		\sq
		\fparts{\product{\DO}{(Z{\cup}\{\Omega\})}}                               \\[-2mm]
		 & \iff
		\scalebox{0.95}{$\displaystyle\p{
					\Intersection{I}{Y_i} \cap V \sq Z
				} \land
				\p{
					\forall \stackrel{J \neq \emptyset}{J \sq I}. \,
					\left( \Domain \sq \bu\limits_{j\in J} X_j \right)
					\lor \left( \bc\limits_{j\in I\setminus J} Y_j \cap V \sq Z \right)
				}
			$}
	\end{align*}
	where an intersection over an empty $I$ is $\Domain$.
\end{lem}
\begin{proof}
Recall the characterization~\eqref{eq:arrow-fiber} of arrows (valid for $d\in\DO$ and $o\in \Domain\cup\{\Omega\}$):
\begin{equation}\label{eq:fiber}
(d,o)\in (A\rarr B)\quad\Longleftrightarrow\quad (d\notin \wO{A})\ \lor\ (o\in B).
\end{equation}
By Definition~\ref{app:def:strong-arrow},
\[
\StrongType{(U\rarr V)} \;=\; (U\rarr V)\ \cap\ \fparts{\DO\times (V\cup\{\Omega\})}.
\]
Set
\[
\mathcal{L}\;\coloneqq\; \bigcap_{i\in I}(X_i\!\to\!Y_i)\ \cap\ \StrongType{(U\!\to\!V)}.
\]
Note that $R\in\mathcal{L}$ iff $R\subseteq \DO\times (V\cup\{\Omega\})$ and, for every $(d,o)\in R$,
\[
\forall i\in I.\ \bigl(d\notin \wO{X_i}\ \lor\ o\in Y_i\bigr)
\quad\text{and}\quad (d\notin \wO{U}\ \lor\ o\in V),
\]

\smallskip
\noindent\textbf{($\Rightarrow$)} Assume $\mathcal{L}\subseteq \fparts{\DO\times (Z\cup\{\Omega\})}$.

(1) Let $o\in \bigl(\bigcap_{i\in I}Y_i\bigr)\cap V$. For any $d\in \Domain$, by \eqref{eq:fiber} we have for all $i$ that
$\{(d,o)\}\in X_i\rarr Y_i$ since $o\in Y_i$, and $\{(d,o)\}\in \StrongType{(U\rarr V)}$ since $o\in V$; hence
$\{(d,o)\}\in \mathcal{L}$. The inclusion forces $o\in Z\cup\{\Omega\}$; but $o\in \Domain$, thus $o\in Z$.
Therefore $\bigl(\bigcap_{i\in I}Y_i\bigr)\cap V\subseteq Z$.

(2) Fix a nonempty $J\subseteq I$. Suppose, towards a contradiction, that
$\Domain \not\subseteq \bigcup_{j\in J} X_j$ and $(\bigcap_{j\in I\setminus J}Y_j)\cap V \not\subseteq Z$.
Choose $d\in \Domain\setminus \bigcup_{j\in J}X_j$ and
$o\in \bigl(\bigcap_{j\in I\setminus J}Y_j\bigr)\cap V\setminus Z$. Since $d\in\Domain$, $d\notin\wO{X_j}$ for every $j\in J$.
Then, using \eqref{eq:fiber}:
\[
\forall i\in I\setminus J,\ (d,o)\in X_i\rarr Y_i \text{ since } o\in Y_i;\qquad
\forall j\in J,\ (d,o)\in X_j\rarr Y_j \text{ since } d\notin \wO{X_j};
\]
and $(d,o)\in \StrongType{(U\rarr V)}$ because $o\in V$. Hence $\{(d,o)\}\in\mathcal{L}$ but $o\notin Z\cup\{\Omega\}$,
contradiction. Therefore for every nonempty $J\subseteq I$,
\[
\left( \Domain \sq \bu_{j\in J} X_j \right)\ \lor\ \left( \bc_{j\in I\setminus J} Y_j \cap V \sq Z \right).
\]

\smallskip
\noindent\textbf{($\Leftarrow$)} Assume the two conditions on the right-hand side hold. Let $R\in\mathcal{L}$ and
$(d,o)\in R$. We prove $o\in Z\cup\{\Omega\}$.

If $o=\Omega$ there is nothing to show. Otherwise $o\in \Domain$ and, since $R\in\StrongType{(U\rarr V)}$,
we have $o\in V$. If $o\in \bigcap_{i\in I}Y_i$, the first condition yields $o\in Z$. Otherwise set
\[
J\;\coloneqq\; \{\, i\in I \mid o\notin Y_i \,\}.
\]
Then $J\neq\emptyset$, and because $(d,o)\in X_i\rarr Y_i$ for all $i$,
\eqref{eq:fiber} gives $d\notin \wO{X_j}$ for every $j\in J$. In particular $d\neq\invOmega$ (for $\OO=\{\invOmega\}$, $\invOmega$ belongs to every $\wO{X_j}$, and $J$ is not empty), so $d\in\Domain\setminus \bigcup_{j\in J}X_j$.
By the second condition (applied to this $J$), we must have
$\bigl(\bigcap_{i\in I\setminus J}Y_i\bigr)\cap V\subseteq Z$ (the other disjunct fails at $d$).
But by definition of $J$, $o\in \bigcap_{i\in I\setminus J}Y_i$, and we already know $o\in V$; hence $o\in Z$.

Thus $o\in Z\cup\{\Omega\}$ for every $(d,o)\in R$, that is,
$R\in \fparts{\product{\DO}{(Z\cup\{\Omega\})}}$. Since $R\in\mathcal{L}$ was arbitrary,
$\mathcal{L}\subseteq \fparts{\product{\DO}{(Z\cup\{\Omega\})}}$.

\smallskip
Combining both directions establishes the equivalence.
\end{proof}

\noindent
When the left-hand side has no strong arrow, the same argument gives the following result.

\begin{lem}\label{lem:weak-product}
	For every finite set $I$,
	\[
		\bigcap_{i\in I} (X_i \rarr Y_i)
		\sq
		\fparts{\product{\DO}{(Z{\cup}\{\Omega\})}}
		\iff
		\forall J \sq I.\,
		\bigl( J\neq\varnothing \text{ and } \Domain \sq \bu\limits_{j\in J} X_j \bigr)
		\lor \bigl( \bc\limits_{j\in I\setminus J} Y_j \sq Z \bigr)
	\]
	where an intersection of subsets of $\Domain$ over an empty set is $\Domain$, and the intersection of an empty family of arrows is the set $\fparts{\DO\times\DOmega}$ of all relations of arity 1.
\end{lem}
\begin{proof}
Write $\mathcal{L}$ for the left-hand side: by~\eqref{eq:fiber}, $R\in\mathcal{L}$ iff for every $(d,o)\in R$ and every $i\in I$, $d\notin\wO{X_i}$ or $o\in Y_i$.

\noindent\textbf{($\Rightarrow$)} Fix $J\subseteq I$ and suppose that both disjuncts fail. Choose $o\in\bigl(\bigcap_{j\in I\setminus J}Y_j\bigr)\setminus Z$, which is an element of $\Domain$, and choose $d\in\DO$ with $d\notin\wO{X_j}$ for every $j\in J$: if $J$ is empty any $d$ will do, and otherwise any $d\in\Domain\setminus\bigcup_{j\in J}X_j$. Then $\{(d,o)\}\in\mathcal{L}$, since $d\notin\wO{X_j}$ for $j\in J$ and $o\in Y_i$ for $i\in I\setminus J$, but $o\notin Z\cup\{\Omega\}$: a contradiction.

\noindent\textbf{($\Leftarrow$)} Let $R\in\mathcal{L}$ and $(d,o)\in R$ with $o\neq\Omega$, and set $J\coloneqq\{i\in I\mid o\notin Y_i\}$. Then $d\notin\wO{X_j}$ for every $j\in J$. If $J$ is empty, the first disjunct fails; otherwise $d\neq\invOmega$, as in the proof of Lemma~\ref{lem:strong-product}, so $d\in\Domain\setminus\bigcup_{j\in J}X_j$ and the first disjunct fails again. So $\bigcap_{j\in I\setminus J}Y_j\subseteq Z$, and $o$, which belongs to this intersection by definition of $J$, belongs to $Z$.
\end{proof}

\noindent

Now that we have established the set-theoretic characterizations of weak and strong arrows, we derive the corresponding subtyping solutions.
We first establish those solutions for the case where the right-hand side
is either a single weak arrow, or a single strong arrow. As shown
previously for interpretation sets, this suffices to solve the subtyping
problem entirely.

\begin{theorem}\label{thm:strong-arrow-types}
	\begin{equation*}
		\begin{array}{l}
		\bint_{i\in I} (t_i \rarr s_i) \inter \StrongType{\p{c \rarr d}}
		\leq
		\p{a \rarr b}
		\,\,\iff \\[2mm]
		\qquad\begin{cases}
			a \leq \buni\limits_{i\in I} t_i \union c \\[2mm]
			\forall J \subseteq I. \quad
			\left( (J\neq\varnothing \text{ or } \OO=\varnothing) \text{ and } a \leq \buni\limits_{j \in J} t_j  \right) \lor
			\left( \bint\limits_{j \in I\setminus J} s_j \inter d \leq b \right)\\[-3mm]
		\end{cases}
		\end{array}
	\end{equation*}
\end{theorem}

\begin{proof} By application of Lemma~\ref{lem:strong-arrow}: since $\OO$ is contained in $\wO{\sem{c}}$, the inclusion $\wO{\sem{a}}\subseteq\bigcup_{i\in I}\wO{\sem{t_i}}\cup\wO{\sem{c}}$ holds iff $a \leq \buni_{i\in I} t_i \union c$; and $\wO{\sem{a}}\subseteq\bigcup_{j\in J}\wO{\sem{t_j}}$ holds iff $J\neq\varnothing$ or $\OO=\varnothing$, and $a \leq \buni_{j\in J} t_j$. \end{proof}

\noindent
For $\OO=\varnothing$, the condition ``$J\neq\varnothing$ or $\OO=\varnothing$'' always holds, and the theorem is the first equivalence of Section~\ref{sub:semantic-strong-arrows}. For $\OO=\{\invOmega\}$, it excludes $J=\varnothing$, so that $\bint_{i\in I} s_i\inter d\leq b$ is always required.

\begin{theorem}\label{thm:strong-strong}
	\begin{equation*}
		\bint_{i\in I} (t_i \rarr s_i) \inter \StrongType{\p{c \rarr d}}
		\leq
		\StrongType{(a \rarr b)}
		\,\, \iff \,\,
		\begin{cases}
			a \leq \buni\limits_{i\in I} t_i \union c \\[2mm]
			\forall J \subseteq I. \,\,
			\left( \buni\limits_{j \in J} t_j =  \mathds{1}  \right) \lor
			\left( \bint\limits_{j \in I\setminus J} s_j \inter d \leq b \right)\\[-3mm]
		\end{cases}\bigskip
	\end{equation*}
\end{theorem}

\begin{proof}
	Since $\StrongType{(a\rarr b)}$ is the intersection of $a\rarr b$ with $\fparts{\DO\times(\sem{b}\cup\{\Omega\})}$, the left-hand side is contained in it iff it is contained in both, which Lemmas~\ref{lem:strong-arrow} and~\ref{lem:strong-product} characterize: we take the conjunction of their right-hand side conditions. The first condition of Lemma~\ref{lem:strong-arrow} is the first condition of the statement, as in the proof of Theorem~\ref{thm:strong-arrow-types}. For $J=\varnothing$, the first condition of Lemma~\ref{lem:strong-product} is $\bint_{i\in I} s_i\inter d\leq b$, which is the condition of the statement for $J=\varnothing$ (an empty union is $\mathds{O}$) and implies the condition of Lemma~\ref{lem:strong-arrow} for $J=\varnothing$. For $J\neq\varnothing$, the condition $\Dom\subseteq\bigcup_{j\in J}\sem{t_j}$ of Lemma~\ref{lem:strong-product} is $\buni_{j \in J} t_j =  \mathds{1}$, and it implies $\wO{\sem{a}}\subseteq\bigcup_{j\in J}\wO{\sem{t_j}}$, the condition of Lemma~\ref{lem:strong-arrow}.
\end{proof}

\noindent
The statement is the same for both interpretations; for $\OO=\varnothing$ it is the second equivalence of Section~\ref{sub:semantic-strong-arrows}. When the left-hand side has no strong arrow, a strong target is handled as follows.

\begin{theorem}\label{thm:strong-no-strong-left}
	\begin{equation*}
		\bint_{i\in I} (t_i \rarr s_i)
		\leq
		\StrongType{(a \rarr b)}
		\,\, \iff \,\,
		\begin{cases}
			(I\neq\varnothing \text{ or } \OO=\varnothing) \text{ and } a \leq \buni\limits_{i\in I} t_i \\[2mm]
			\forall J \subseteq I. \,\,
			\left( \buni\limits_{j \in J} t_j =  \mathds{1}  \right) \lor
			\left( \bint\limits_{j \in I\setminus J} s_j \leq b \right)\\[-3mm]
		\end{cases}\bigskip
	\end{equation*}
	where the intersection of an empty family of arrows is $\function_1$, and an empty intersection of types is $\mathds{1}$.
\end{theorem}

\begin{proof}
	As in the proof of Theorem~\ref{thm:strong-strong}, the left-hand side is contained in $\StrongType{(a\rarr b)}$ iff it is contained in $a\rarr b$ and in $\fparts{\DO\times(\sem{b}\cup\{\Omega\})}$. By Theorem~\ref{app:thm:set-containment-multi-arity} with $n=1$, $P=I$, $N$ a singleton, and $J=\{i\in I\mid\iota(i)=1\}$, the first inclusion holds iff, for every $J\subseteq I$, $\wO{\sem{a}}\subseteq\bigcup_{j\in J}\wO{\sem{t_j}}$, or $J\neq I$ and $\bint_{j\in I\setminus J}s_j\leq b$. By Lemma~\ref{lem:weak-product}, the second inclusion holds iff, for every $J\subseteq I$, $J\neq\varnothing$ and $\buni_{j\in J}t_j=\mathds{1}$, or $\bint_{j\in I\setminus J}s_j\leq b$; since an empty union is $\mathds{O}$, this is the second condition of the statement. For $J=I$, the condition of the first inclusion is $\wO{\sem{a}}\subseteq\bigcup_{i\in I}\wO{\sem{t_i}}$, that is, the first condition of the statement. For $J\neq I$, it is implied by the condition of the second inclusion, since $J\neq\varnothing$ and $\buni_{j\in J}t_j=\mathds{1}$ imply $\wO{\sem{a}}\subseteq\bigcup_{j\in J}\wO{\sem{t_j}}$.
\end{proof}

\noindent
For $\OO=\varnothing$, this is the instance $c=\mathds{O}$, $d=\mathds{1}$ of Theorem~\ref{thm:strong-strong}, since $\StrongType{(\mathds{O}\rarr\mathds{1})}$ is then $\function_1$ (Lemma~\ref{lem:strong-intersection} and the remark after it). For $\OO=\{\invOmega\}$, it requires $I\neq\varnothing$: $\function_1$ is contained in no strong arrow, and, by Theorem~\ref{app:thm:set-containment-multi-arity} with $P=\varnothing$, in no arrow.

\subsection{Subtyping Algorithm}\label{app:algo:strong}

We can now derive the solutions to the subtyping problem for mixed weak and strong arrows on types.

\begin{theorem}[Subtyping with weak/strong arrows]\label{thm:algo-strong}
	Let $I,P,Q,R$ be finite, with $P$ non-empty, and let $\StrongType{(u_i \rarr w_i)}$, for $i\in P$, be the strong arrows of the left-hand side.
	Write $t \;=\; \buni_{i\in P} u_i$ and $s \;=\; \bint_{i\in P} w_i$, so that $\StrongType{(t \rarr s)}$ is the intersection of these strong arrows (Lemma~\ref{lem:strong-intersection}).
	Then the mixed entailment
	\[
	\Bigl(\;\bigwedge_{i \in I} (t_i \rarr s_i)\;\Bigr) \;\land\; (t \rarr s)^{\star}
	\;\;\leq\;\;
	\Bigl(\;\bigvee_{i \in R} (t_i \rarr s_i)\;\Bigr)
	\;\;\lor\;\;
	\Bigl(\;\bigvee_{i \in Q} (a_i \rarr b_i)^{\star}\;\Bigr)
	\]
	holds iff at least one of the following \emph{finite witnesses} exists:
	\begin{description}
	\item[\normalfont(R-branch)] $\exists\, i_0\in R$ such that
	\[
	\Bigl(\;\bigwedge_{i \in I} (t_i \rarr s_i)\;\Bigr) \land (t \rarr s)^{\star}
	\;\;\leq\;\; (t_{i_0} \rarr s_{i_0}),
	\]
	equivalently (by Theorem~\ref{thm:strong-arrow-types}, with $a\!=\!t_{i_0}$, $b\!=\!s_{i_0}$, $c\!=\!t$, $d\!=\!s$):
	\[
	\begin{cases}
	t_{i_0} \;\leq\; \buni\limits_{i\in I} t_i \;\union\; t,\\[1mm]
	\forall J\subseteq I.\;\; \Bigl((J\neq\varnothing \text{ or } \OO=\varnothing) \text{ and } t_{i_0}\leq\buni\limits_{j\in J} t_j\Bigr)\;\lor\;
	\Bigl(\bint\limits_{j\in I\setminus J} s_j \;\inter\; s \;\leq\; s_{i_0}\Bigr).
	\end{cases}
	\]

	\item[\normalfont(Q-branch)] $\exists\, j_0\in Q$ such that
	\[
	\Bigl(\;\bigwedge_{i \in I} (t_i \rarr s_i)\;\Bigr) \land (t \rarr s)^{\star}
	\;\;\leq\;\; (a_{j_0} \rarr b_{j_0})^{\star},
	\]
	equivalently (by Theorem~\ref{thm:strong-strong}, with $a\!=\!a_{j_0}$, $b\!=\!b_{j_0}$, $c\!=\!t$, $d\!=\!s$):
	\[
	\begin{cases}
	a_{j_0} \;\leq\; \buni\limits_{i\in I} t_i \;\union\; t,\\[1mm]
	\forall J\subseteq I.\;\; \Bigl(\buni\limits_{j\in J} t_j = \mathds{1}\Bigr)\;\lor\;
	\Bigl(\bint\limits_{j\in I\setminus J} s_j \;\inter\; s \;\leq\; b_{j_0}\Bigr).
	\end{cases}
	\]
	\end{description}
\emph{Proof.}
Apply the splitting Lemma~\ref{lem:exists-parts} to reduce the problem to an existential witness on one target ($R$ or $Q$); then discharge the resulting inclusion by
Theorems~\ref{thm:strong-arrow-types} (weak target) and~\ref{thm:strong-strong} (strong target).
\qed
	\end{theorem}

\noindent
When $P$ is empty, the same splitting reduces the problem to one target, and the resulting inclusion is decided by Theorem~\ref{app:thm:set-containment-multi-arity} with $n=1$ (weak target) and by Theorem~\ref{thm:strong-no-strong-left} (strong target). For $\OO=\varnothing$, one may equivalently apply the theorem above with $t=\mathds{O}$ and $s=\mathds{1}$, since $\StrongType{(\mathds{O}\rarr\mathds{1})}$ is then the set of all relations of arity 1.

	\begin{cor}[One-strong target]\label{cor:one-strong}
	\[
	(t \rarr s)\;\land\; \StrongType{(c \rarr d)}
	\;\;\leq\;\; \StrongType{(a \rarr b)}
	\quad\Longleftrightarrow\quad
	\begin{cases}
	a \;\leq\; t \;\union\; c,\\[1mm]
	s \;\inter\; d \;\leq\; b,\\[1mm]
	(t=\mathds{1})\;\lor\;(d\leq b).
	\end{cases}
	\]
	\emph{Proof.} Instance of the Q-branch with $I$ and $P$ singletons. \qed
	\end{cor}

\section{Dead-Code Commit-Scan Protocol}\label{app:dead-code-protocol}

The dead-code scan in Section~\ref{sec:eval} is a commit-message and
commit-body search followed by diff-based filtering. For each of the 16
repositories in the dead-code scan corpus, we scanned non-merge commits
reachable from the default branch after May~1, 2024, using keyword families
for the removed-code signal and for the type/compiler evidence.

\begin{table}[ht]
\centering
\small
\begin{tabular}{|l|p{0.68\textwidth}|}
\hline\rowcolor{violet2}
\textbf{Family} & \textbf{Terms} \\[-1pt]
\hline
Removed-code terms &
\texttt{dead code}, \texttt{unused}, \texttt{unused clause},
\texttt{unused clauses}, \texttt{unused branch}, \texttt{unused branches},
\texttt{unreachable}, \texttt{non-reachable}, \texttt{redundant},
\texttt{useless}, \texttt{unneeded}, \texttt{unnecessary} \\[-1.5pt]
\hline
Type/compiler terms &
\texttt{type system}, \texttt{typesystem}, \texttt{type-system},
\texttt{type checker}, \texttt{typechecker}, \texttt{type-checker},
\texttt{type warning}, \texttt{type warnings}, \texttt{compiler},
\texttt{warning}, \texttt{warnings}, \texttt{Elixir 1.17},
\texttt{Elixir 1.18}, \texttt{Elixir 1.19}, \texttt{Elixir 1.20},
\texttt{latest Elixir}, \texttt{is never used}, \texttt{will never match},
\texttt{can never succeed}, \texttt{will always evaluate} \\[-1.5pt]
\hline
\end{tabular}
\caption{Keyword families used for the dead-code commit scan}
\label{tab:dead-code-keywords}
\end{table}

A candidate was accepted only when the patch deleted the code identified by the
warning. We counted deleted lines in Elixir source files (\texttt{.ex},
\texttt{.exs}, \texttt{.heex}, and \texttt{.eex}) when the deletion removed
that code.
For mixed commits, such as the Ash regular-expression-support patch, we counted
only the hunks attributable to the warning. We used two evidence categories:
\emph{direct} for commits whose metadata explicitly names the type system, type
checker, compiler, or a relevant Elixir release; and \emph{warning-linked} for
commits whose metadata names an Elixir warning or dead-code cleanup, but not
the type-system pass itself.

\begin{table}[ht]
\centering
\small
\begin{tabular}{|l|l|l|r|l|}
\hline\rowcolor{violet2}
\textbf{Project} & \textbf{Commit} & \textbf{Bucket} & \textbf{Lines} & \textbf{Trigger} \\[-1pt]
\hline
Postgrex & 3308f27 & Direct & 15 & \texttt{typesystem} \\[-1.5pt]
\rowcolor{violet1}
Flame & 0c0c287 & Direct & 9 & \texttt{Elixir v1.18} \\[-1.5pt]
Phoenix LiveView & 6c6e2aa & Direct & 8 & \texttt{type system} \\[-1.5pt]
\rowcolor{violet1}
Phoenix LiveView & 4aa5c83 & Direct & 4 & \texttt{type system} \\[-1.5pt]
Ecto & 8d02c19 & Direct & 4 & \texttt{type system} \\[-1.5pt]
\rowcolor{violet1}
ExDoc & fc57bdc & Direct & 11 & \texttt{type system} \\[-1.5pt]
Livebook & 13e10ae & Direct & 4 & \texttt{type system} \\[-1.5pt]
\rowcolor{violet1}
Livebook & 80951e5 & Direct & 31 & \texttt{compiler} \\[-1.5pt]
Ash & ccd53d8 & Direct & 5 & \texttt{1.19 type checker} \\[-1.5pt]
\rowcolor{violet1}
Phoenix & b1228c4 & Warning-linked & 14 & \texttt{unused clause}, \texttt{Elixir 1.18} \\[-1.5pt]
Spitfire & 434521e & Warning-linked & 14 & \texttt{warnings}, \texttt{Elixir 1.18} \\[-1.5pt]
\rowcolor{violet1}
Postgrex & e4f7942 & Warning-linked & 45 & \texttt{unused clauses} \\[-1.5pt]
Phoenix LiveView & 004e5fa & Warning-linked & 4 & \texttt{dead code} \\[-1.5pt]
\rowcolor{violet1}
Phoenix LiveView & b311b7e & Warning-linked & 11 & \texttt{warnings}, \texttt{latest Elixir} \\[-1.5pt]
\hline
\textbf{Total} & & & \textbf{179} & \\[-1.5pt]
\hline
\end{tabular}
\caption{Accepted dead-code commits after diff filtering}
\label{tab:dead-code-accepted-commits}
\end{table}

\section{\emph{If-T} Benchmark Reproducibility}\label{app:ift-benchmark-repro}

The public experiment artifact~\cite{CoreElixirExperiments} records the Elixir port
of the thirteen core benchmark items of Guo and Greenman~\cite{GuoGreenman25}
and the exact result table used in Section~\ref{if-t-benchmark}. Its
\emph{If-T} directory is organized so that the benchmark code, the
result-generation harness, and the compiler revision used for the reported run
can be inspected independently.

The Elixir port uses the research compiler branch that supports asserted
function type forms. Since function annotations are not part of Elixir as of v1.20,
these assertions provide the benchmark's function boundaries while leaving the
program bodies as ordinary Elixir code. A benchmark item is counted as accepted
only when its positive program is accepted and its negative program is rejected.
For Elixir, a type warning emitted by the compiler is treated as a rejection,
because such warnings are the implementation's user-facing signal for a failed
type check.

The port follows the structure of the reference examples as closely as the
target language permits. Type tests are expressed with Elixir guards and
predicates, structured-data examples use Elixir tuples and maps, and operations
on strings and binaries use the corresponding typed standard-library operations.
The adaptations are therefore intended to test the same narrowing facts as the
reference benchmark, rather than to exploit Elixir-specific encodings.

The only failed Elixir core item is \texttt{alias}. As discussed in
Section~\ref{if-t-benchmark}, this is a positive-side precision limitation: the
checker does not connect a saved Boolean test result back to the value whose type
was tested. The negative alias witnesses are rejected, so the result records a
missing alias-tracking refinement rather than an accepted ill-typed program.

\section{Guard Exactness Corpus}\label{app:guard-exactness-corpus}

Table~\ref{tab:guard_exactness_full} reports the guard-exactness measurement for each project in the selected open-source Elixir repository set. Across these repositories, 141,755 of 164,701 analyzed pattern/guard pairs are exact, for a weighted exactness ratio of 86.07\%.

\begin{table}[ht]
\centering
\small
\begin{tabular}{|l|r|r|r|}
\hline\rowcolor{violet2}
\textbf{Project} & \textbf{Analyzed} & \textbf{Exact} & \textbf{Exactness} \\[-1pt]
\hline
HexPm & 12,048 & 10,595 & 87.94\% \\[-1.5pt]
\rowcolor{violet1}
Phoenix & 3,542 & 3,318 & 93.68\% \\[-1.5pt]
PhoenixLiveView & 5,290 & 4,480 & 84.69\% \\[-1.5pt]
\rowcolor{violet1}
Livebook & 15,459 & 14,105 & 91.24\% \\[-1.5pt]
Credo & 11,448 & 8,225 & 71.85\% \\[-1.5pt]
\rowcolor{violet1}
ExDoc & 1,800 & 1,533 & 85.17\% \\[-1.5pt]
Nerves & 1,112 & 1,029 & 92.54\% \\[-1.5pt]
\rowcolor{violet1}
Ecto & 5,207 & 4,415 & 84.79\% \\[-1.5pt]
Postgrex & 7,563 & 2,634 & 34.83\% \\[-1.5pt]
\rowcolor{violet1}
Flame & 686 & 648 & 94.46\% \\[-1.5pt]
Ash & 30,358 & 29,004 & 95.54\% \\[-1.5pt]
\rowcolor{violet1}
Spitfire & 1,226 & 1,069 & 87.19\% \\[-1.5pt]
SQL & 6,098 & 4,283 & 70.24\% \\[-1.5pt]
\rowcolor{violet1}
OpenApiSpex & 1,693 & 1,544 & 91.20\% \\[-1.5pt]
MixSBOM & 7,820 & 5,049 & 64.57\% \\[-1.5pt]
\rowcolor{violet1}
AbsintheFederation & 413 & 407 & 98.55\% \\[-1.5pt]
Blockscout & 52,938 & 49,417 & 93.35\% \\[-1.5pt]
\hline
\end{tabular}
\caption{Exact pattern/guard pairs for selected open-source Elixir projects}
\label{tab:guard_exactness_full}
\end{table}

\section{Arrow-Return Informativeness Corpus}\label{app:arrow-return-corpus}

Table~\ref{tab:arrow_return_full} reports the inferred-arrow return
informativeness measurement for each project in the same selected open-source
Elixir repository set. The experiment counts one inferred arrow as informative when its return
type is not exactly \texttt{dynamic()}. Gradual return types such as
\texttt{dynamic(integer())} therefore count as informative, since they preserve
the dynamic component required by safe erasure while still recording the static
shape recovered by inference. Across the corpus, 43,158 of 69,976 inferred
arrows have informative returns, for a weighted ratio of 61.68\%.

\begin{table}[ht]
\centering
\small
\begin{tabular}{|l|r|r|r|r|r|}
\hline\rowcolor{violet2}
\textbf{Project} & \textbf{LoC} & \textbf{Functions} & \textbf{Arrows} & \textbf{Informative} & \textbf{Ratio} \\[-1pt]
\hline
Blockscout & 183,142 & 20,182 & 26,384 & 15,493 & 58.72\% \\[-1.5pt]
\rowcolor{violet1}
Ash & 108,532 & 11,265 & 13,267 & 8,479 & 63.91\% \\[-1.5pt]
Livebook & 55,119 & 6,693 & 7,424 & 5,566 & 74.97\% \\[-1.5pt]
\rowcolor{violet1}
HexPm & 28,315 & 4,408 & 6,072 & 3,850 & 63.41\% \\[-1.5pt]
Ecto & 25,559 & 1,173 & 1,606 & 1,091 & 67.93\% \\[-1.5pt]
\rowcolor{violet1}
Credo & 23,712 & 3,998 & 4,441 & 2,274 & 51.20\% \\[-1.5pt]
PhoenixLiveView & 23,507 & 1,149 & 1,400 & 803 & 57.36\% \\[-1.5pt]
\rowcolor{violet1}
Phoenix & 17,863 & 797 & 956 & 535 & 55.96\% \\[-1.5pt]
MixSBOM & 13,328 & 5,324 & 5,696 & 3,397 & 59.64\% \\[-1.5pt]
\rowcolor{violet1}
Postgrex & 9,153 & 600 & 688 & 433 & 62.94\% \\[-1.5pt]
OpenApiSpex & 7,123 & 464 & 630 & 379 & 60.16\% \\[-1.5pt]
\rowcolor{violet1}
ExDoc & 5,882 & 357 & 422 & 275 & 65.17\% \\[-1.5pt]
Nerves & 4,831 & 254 & 274 & 126 & 45.99\% \\[-1.5pt]
\rowcolor{violet1}
Spitfire & 3,853 & 110 & 119 & 93 & 78.15\% \\[-1.5pt]
SQL & 3,538 & 142 & 174 & 87 & 50.00\% \\[-1.5pt]
\rowcolor{violet1}
Flame & 2,806 & 199 & 228 & 140 & 61.40\% \\[-1.5pt]
AbsintheFederation & 1,736 & 143 & 195 & 137 & 70.26\% \\[-1.5pt]
\hline
\textbf{Total} & \textbf{517,999} & \textbf{57,258} & \textbf{69,976} & \textbf{43,158} & \textbf{61.68\%} \\[-1.5pt]
\hline
\end{tabular}
\caption{Inferred-arrow return informativeness for selected open-source Elixir projects}
\label{tab:arrow_return_full}
\end{table}
\newpage
\end{document}